\documentclass[10pt]{article} 
\usepackage[preprint]{tmlr}

\usepackage{amsmath}
\usepackage{amssymb}
\usepackage{array}
\usepackage{booktabs}
\usepackage{caption}
\usepackage{colortbl}
\usepackage{enumitem}
\usepackage{etoolbox}
\usepackage[T1]{fontenc}
\usepackage{float}
\usepackage{graphicx}
\usepackage{hyperref}
\usepackage[noabbrev,capitalize,nameinlink]{cleveref}
\usepackage[varqu,varl]{inconsolata}
\usepackage{listings}
\usepackage{makecell}
\usepackage{microtype}
\usepackage{multirow}
\usepackage{newfloat}
\usepackage{pifont}
\usepackage{scalerel}
\usepackage{subcaption}
\usepackage{tikz}
\usepackage[most]{tcolorbox}
\usepackage{wrapfig}
\usepackage{xcolor}

\definecolor{forestgreen}{RGB}{34,139,34}
\definecolor{lightforestgreen}{RGB}{232,245,233}
\definecolor{translateBlue}{HTML}{4C8BF5}
\definecolor{present}{HTML}{1a1a1a}
\definecolor{absent}{HTML}{c8c8c8}

\definecolor{sqlkeyword}{RGB}{0,0,180}
\definecolor{sqlstring}{RGB}{163,21,21}
\definecolor{sqlcomment}{RGB}{0,128,0}
\definecolor{sqlnumber}{RGB}{128,0,128}
\definecolor{sqlfunction}{RGB}{56,118,29}

\lstdefinestyle{sqlstyle}{%
  language=SQL,
  basicstyle=\fontfamily{cmtt}\selectfont\small\color{white!20!black},
  keywordstyle=\bfseries\color{sqlkeyword},
  stringstyle=\color{sqlstring},
  commentstyle=\itshape\color{sqlcomment},
  numberstyle=\color{sqlnumber},
  morekeywords={OVER,PARTITION,RANK,DENSE_RANK,ROW_NUMBER,COALESCE,
                ISNULL,CTE,WITH,RECURSIVE,LATERAL,FILTER,WITHIN,
                FETCH,NEXT,ROWS,ONLY,NULLS,FIRST,LAST,BOOLEAN,
                BIGINT,SERIAL,TEXT,LIMIT,OFFSET,ILIKE,RETURNING},
  breaklines=true,
  columns=fullflexible,
  keepspaces=true,
  showstringspaces=false,
  aboveskip=0pt,
  belowskip=0pt,
  literate={--}{{{\color{sqlcomment}--}}}{2},
}

\tcbuselibrary{breakable, skins}

\lstdefinestyle{termstyle}{%
  basicstyle=\fontfamily{cmtt}\selectfont\small\scshape\color{white!20!black},
  breaklines=true,
  columns=fullflexible,
  keepspaces=true,
  showstringspaces=false,
  aboveskip=0pt,
  belowskip=0pt,
  xleftmargin=0pt,
  breakindent=0pt,
}

\newtcolorbox{codeframe}[1][]{%
  colback=teal!4!white,
  colframe=teal!55!black,
  coltitle=white,
  colbacktitle=teal!55!black,
  fonttitle=\ttfamily\bfseries,
  boxrule=0.8pt,
  arc=4pt,
  left=6pt,
  right=6pt,
  top=6pt,
  bottom=6pt,
  enhanced,
  #1,
}

\newtcolorbox{codeframetop}[1][]{%
  colback=teal!4!white,
  colframe=teal!55!black,
  coltitle=white,
  colbacktitle=teal!55!black,
  fonttitle=\ttfamily\bfseries,
  boxrule=0.8pt,
  arc=4pt,
  left=6pt,
  right=6pt,
  top=6pt,
  bottom=6pt,
  enhanced,
  sharp corners=south,         
  bottomrule=0pt,              
  after skip=0pt,              
  #1,
}

\newtcolorbox{codeframebottom}[1][]{%
  colback=teal!4!white,
  colframe=teal!55!black,
  coltitle=white,
  colbacktitle=teal!55!black,
  fonttitle=\ttfamily\bfseries,
  boxrule=0.8pt,
  arc=4pt,
  left=6pt,
  right=6pt,
  top=6pt,
  bottom=6pt,
  enhanced,
  sharp corners=north,         
  toprule=0pt,                 
  before skip=0pt,             
  #1,
}

\newfloat{codelisting}{htbp}{lol}
\floatname{codelisting}{Listing}

\crefname{codelisting}{listing}{listings}
\Crefname{codelisting}{Listing}{Listings}

\newlength{\iconht}
\newcommand{\inlineicon}[1]{\raisebox{-1.5pt}{\includegraphics[height=\iconht]{#1}}}

\newcommand{\iconFunctional}{\inlineicon{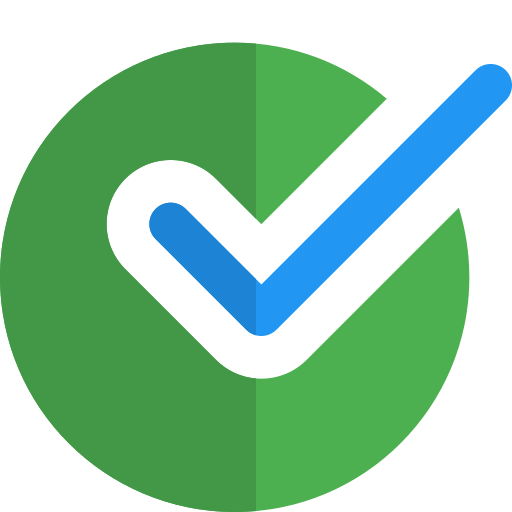}}
\newcommand{\iconRuntime}{\inlineicon{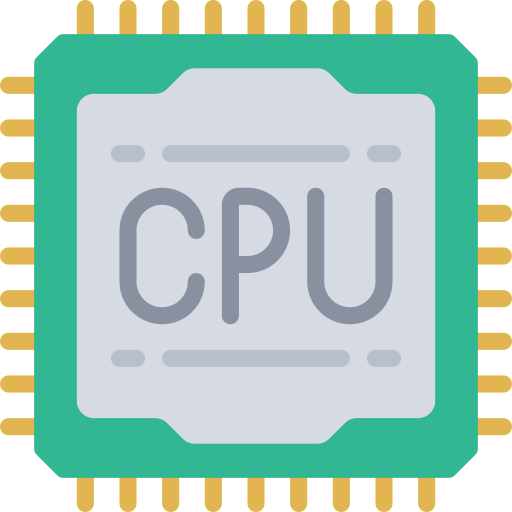}}
\newcommand{\iconMemory}{\inlineicon{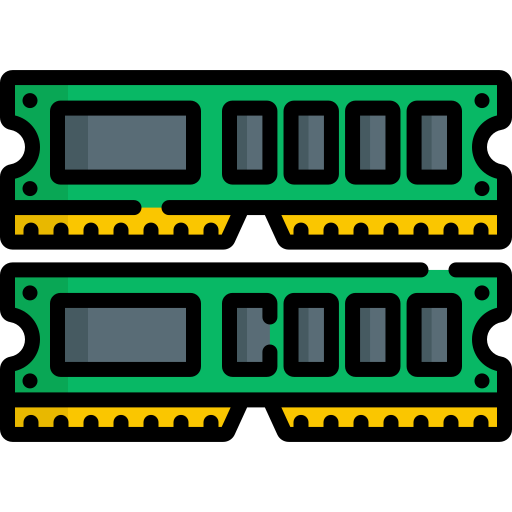}}
\newcommand{\iconRead}{\inlineicon{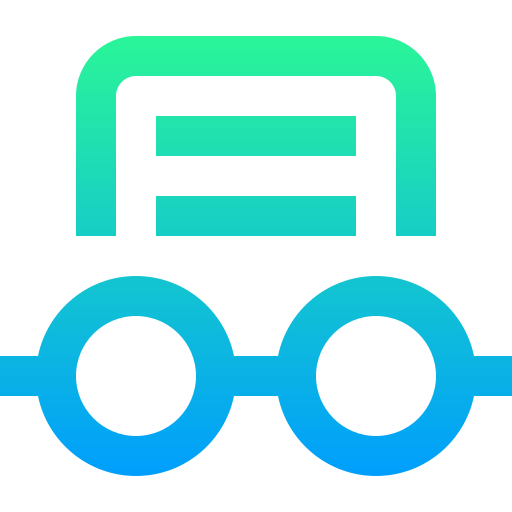}}
\newcommand{\iconSecurity}{\inlineicon{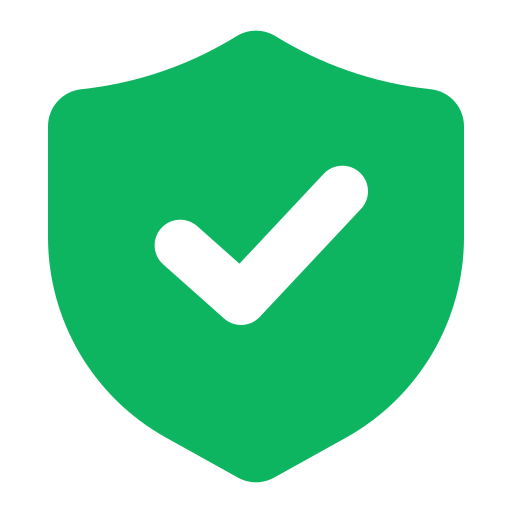}}
\newcommand{\iconHelp}{\inlineicon{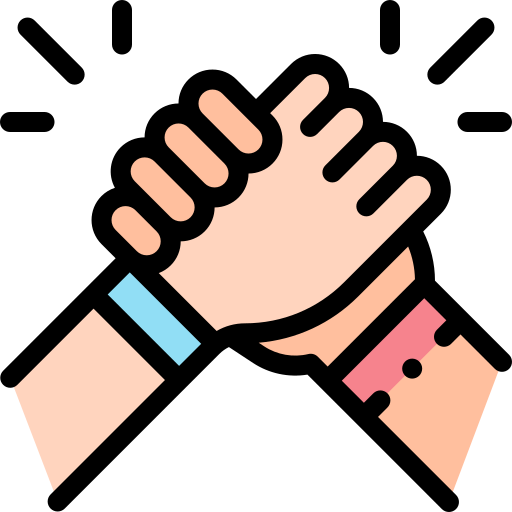}}
\newcommand{\iconHarm}{\inlineicon{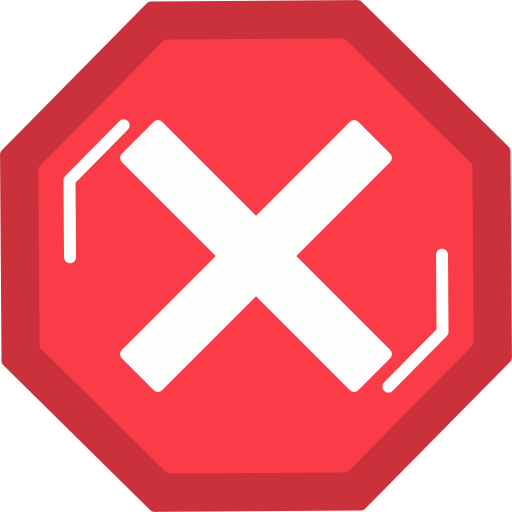}}

\newcommand{\iconC}{\inlineicon{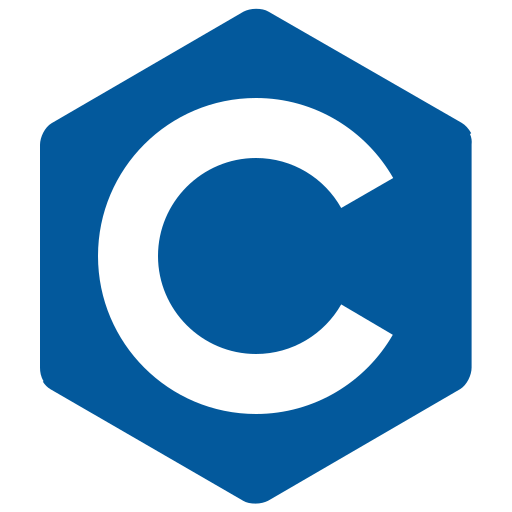}}
\newcommand{\iconCpp}{\inlineicon{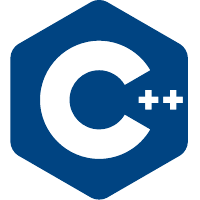}}
\newcommand{\iconCs}{\inlineicon{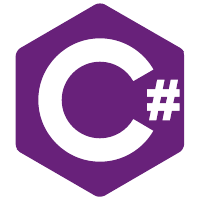}}
\newcommand{\iconGo}{\inlineicon{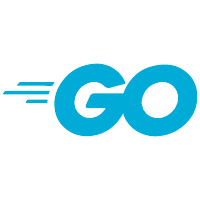}}
\newcommand{\iconJava}{\inlineicon{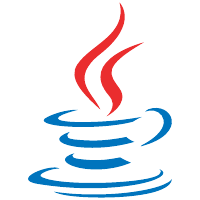}}
\newcommand{\iconJs}{\inlineicon{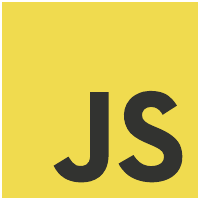}}
\newcommand{\iconPython}{\inlineicon{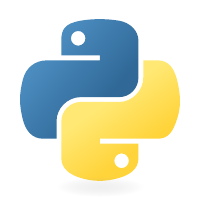}}
\newcommand{\iconRuby}{\inlineicon{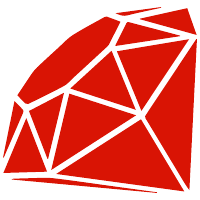}}
\newcommand{\iconNL}{\inlineicon{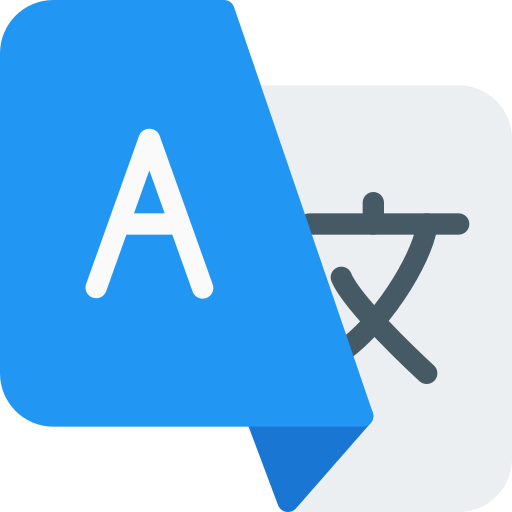}}

\newcommand{\iconAuxiliary}{\inlineicon{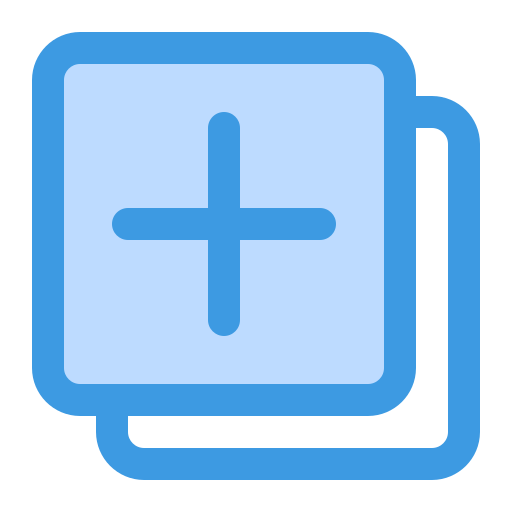}}
\newcommand{\iconCode}{\inlineicon{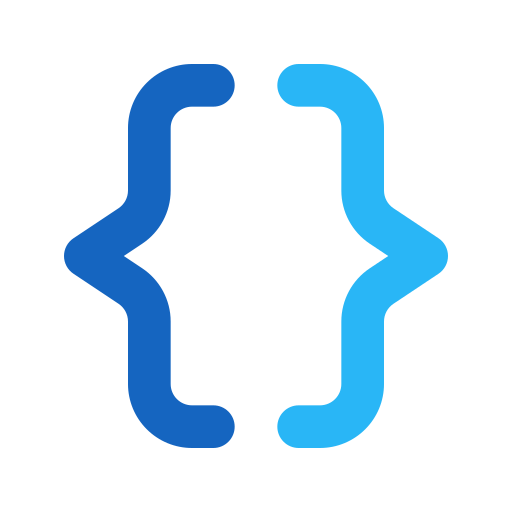}}
\newcommand{\iconGenerative}{\inlineicon{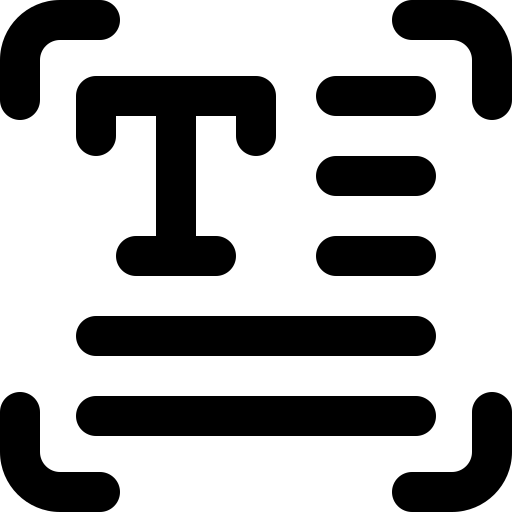}}
\newcommand{\iconMath}{\inlineicon{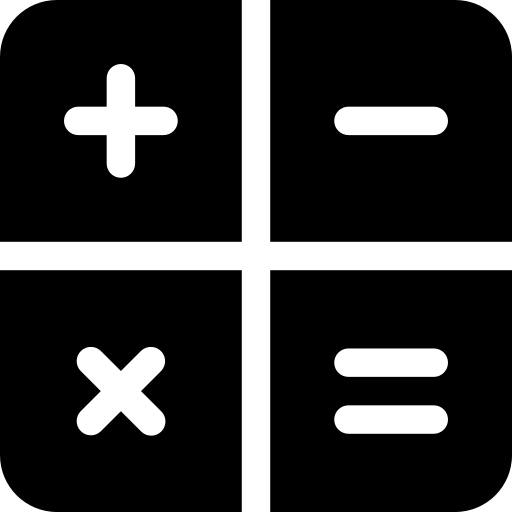}}
\newcommand{\iconPrinciple}{\inlineicon{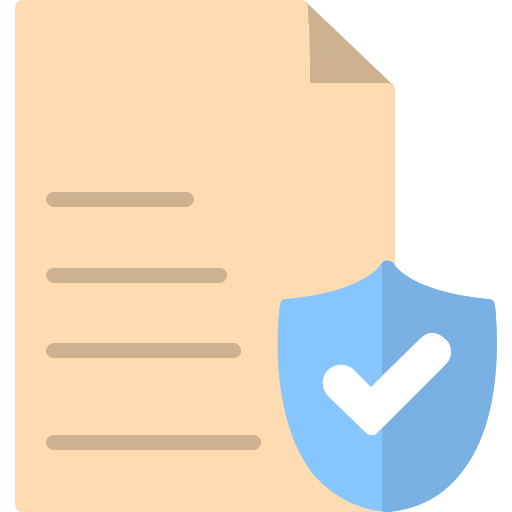}}
\newcommand{\iconReasoning}{\inlineicon{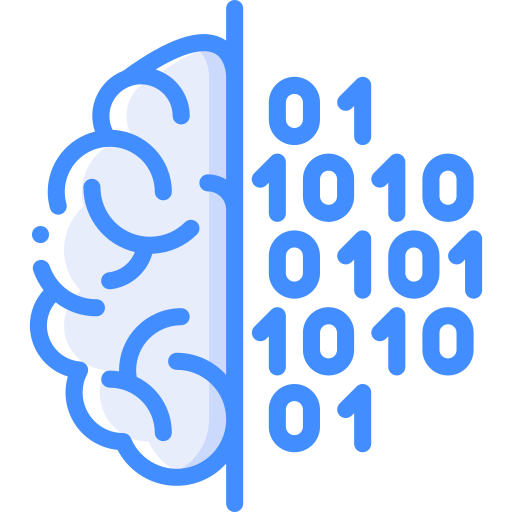}}

\newcommand{\iconDatabase}{\inlineicon{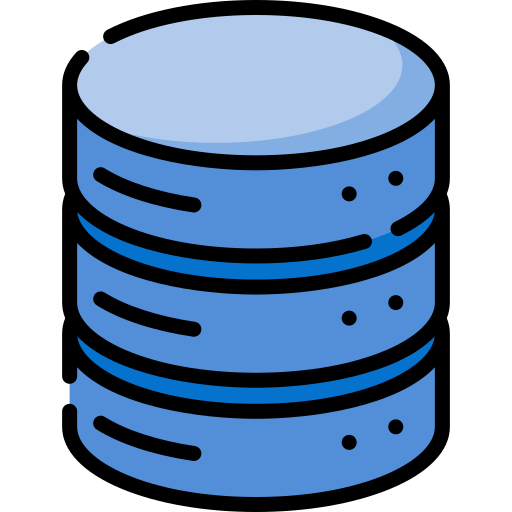}}

\newcommand{\hficon}{\inlineicon{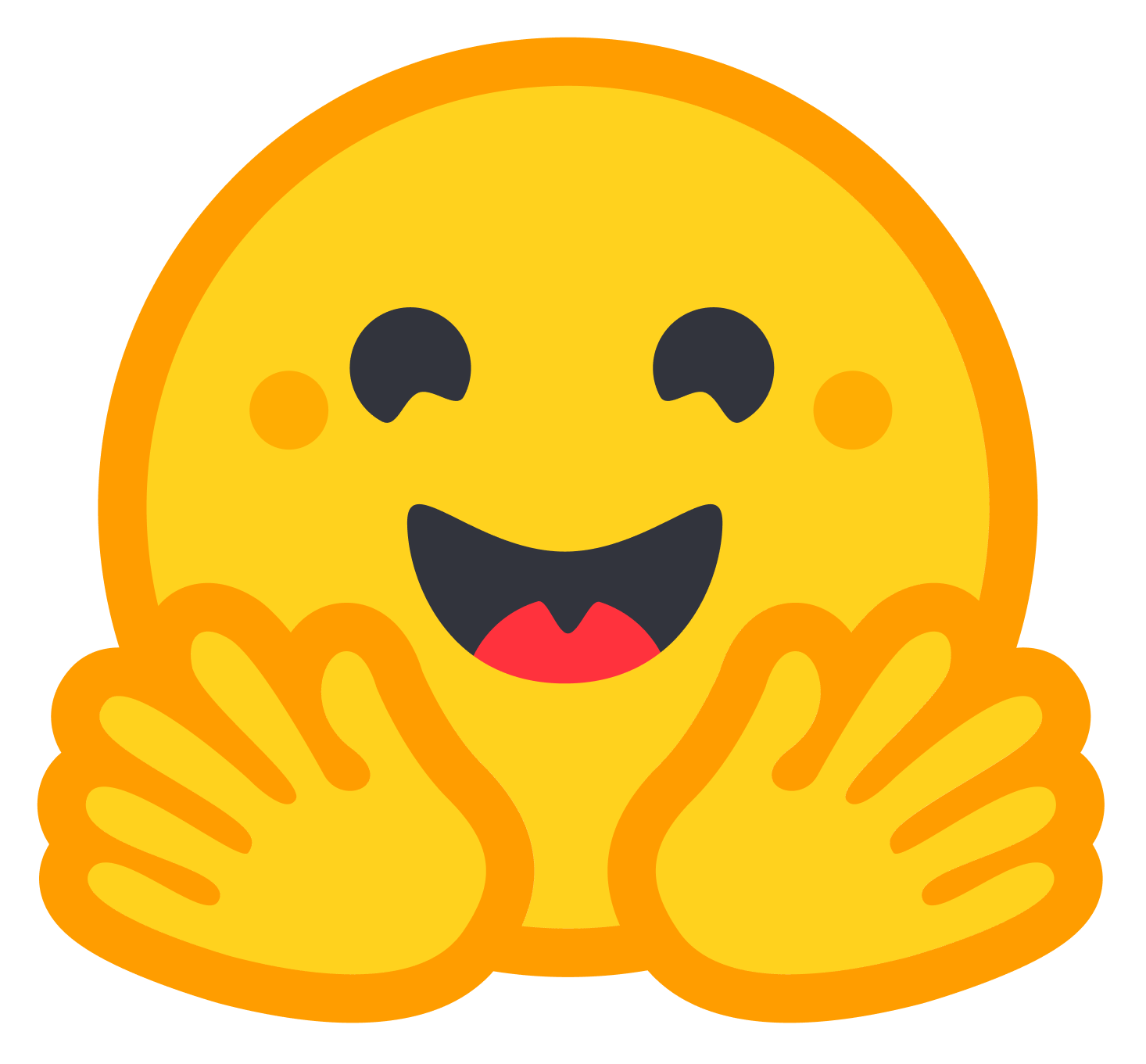}}
\newcommand{\giticon}{\inlineicon{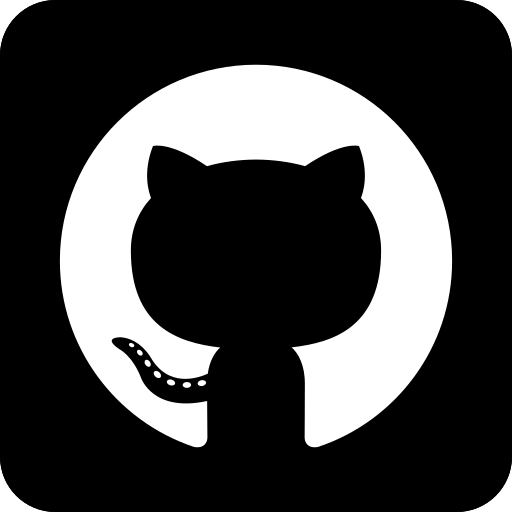}}
\newcommand{\bqicon}{\inlineicon{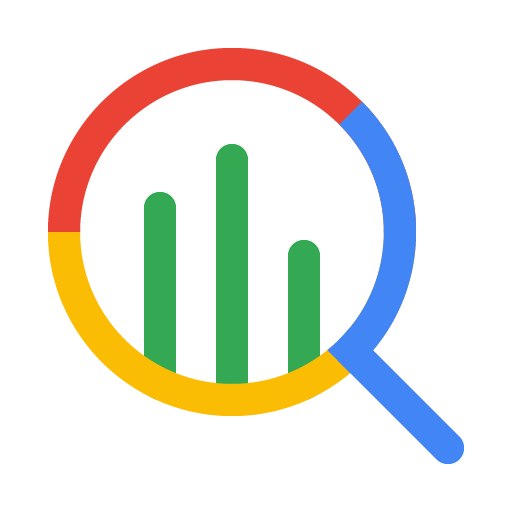}}
\newcommand{\dockericon}{\inlineicon{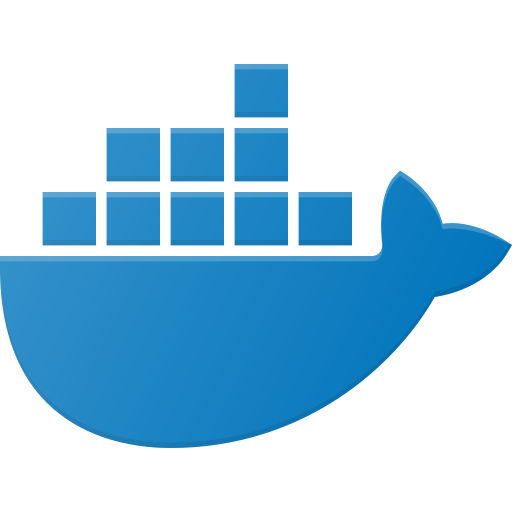}}

\newcommand{\criterion}[3]{%
  \ifnum#1=1
    {#2\;\textnormal{#3}}%
  \else
    {\tikz[baseline=(X.base)]{\node[opacity=0.15,inner sep=0pt](X){#2\;\textnormal{#3}};}}%
  \fi
}

\newcommand{\criterialine}[7]{%
  \criterion{#1}{\scriptsize\iconFunctional}{\scriptsize\texttt{Correctness}}\hspace{1em}
  \criterion{#2}{\scriptsize\iconRuntime}{\scriptsize\texttt{Runtime}}\hspace{1em}
  \criterion{#3}{\scriptsize\iconMemory}{\scriptsize\texttt{Memory}}\hspace{1em}
  \criterion{#4}{\scriptsize\iconRead}{\scriptsize\texttt{Readability}}\hspace{1em}
  \criterion{#5}{\scriptsize\iconSecurity}{\scriptsize\texttt{Security}}\hspace{1em}
  \criterion{#6}{\scriptsize\iconHelp}{\scriptsize\texttt{Helpfulness}}\hspace{1em}
  \criterion{#7}{\scriptsize\iconHarm}{\scriptsize\texttt{Harmlessness}}
}

\newcommand{\langline}[9]{%
  \criterion{#1}{\scriptsize\iconC}{\scriptsize\texttt{C}}\hspace{1em}
  \criterion{#2}{\scriptsize\iconCs}{\scriptsize\texttt{C\#}}\hspace{1em}
  \criterion{#3}{\scriptsize\iconCpp}{\scriptsize\texttt{C++}}\hspace{1em}
  \criterion{#4}{\scriptsize\iconGo}{\scriptsize\texttt{Go}}\hspace{1em}
  \criterion{#5}{\scriptsize\iconJava}{\scriptsize\texttt{Java}}\hspace{1em}
  \criterion{#6}{\scriptsize\iconJs}{\scriptsize\texttt{JS}}\hspace{1em}
  \criterion{#7}{\scriptsize\iconPython}{\scriptsize\texttt{Python}}\hspace{1em}
  \criterion{#8}{\scriptsize\iconRuby}{\scriptsize\texttt{Ruby}}\hspace{1em}
  \criterion{#9}{\scriptsize\iconNL}{\scriptsize\texttt{Natural Language}}
}

\newcommand{\hsep}{\;\;{\tiny$\blacksquare$}\;\;}

\newcommand{\hfmodel}[1]{\hficon\;\href{https://huggingface.co/#1}{\texttt{#1}}}
\newcommand{\gitrepo}[1]{\giticon\;\href{https://github.com/#1}{\texttt{#1}}}
\newcommand{\bqrepo}[1]{\bqicon\;\href{https://console.cloud.google.com/marketplace/product/#1}{\texttt{#1}}}
\newcommand{\dockerrepo}[1]{\dockericon\;\href{https://hub.docker.com/repository/docker/#1}{\texttt{#1}}}

\newlength{\samplenumwd}
\newcommand{\dataset}[7]{%
  \settowidth{\samplenumwd}{\small\texttt{000000}}
  \item {\parfillskip=0pt\relax
    \textbf{\texttt{#1}}%
    \ifnotempty{#2}{\hsep{\small\hficon\;\href{#2}{\texttt{#3}}}}%
    \hfill{\small\iconDatabase\texttt{\makebox[\samplenumwd][r]{#4} samples}}\par}%

  #5%

  {\scriptsize\textbf{\texttt{Pref. Criteria:}}\hspace{0.7em}}{\small #6}%
  \vspace{-1pt}%
  {\scriptsize\textbf{\texttt{Response Lang:}}\hspace{1.45em}}{\small #7}%
}

\newcommand{\gradientcell}[6]{%
    \ifdimcomp{#1pt}{>}{#3pt}{\cellcolor{#5!100.0!#4!#6}\texttt{#1}}{%
        \ifdimcomp{#1pt}{<}{#2pt}{\cellcolor{#5!0.0!#4!#6}\texttt{#1}}{%
            \pgfmathparse{int(round(100*(#1-#2)/(#3-#2)))}%
            \xdef\tempa{\pgfmathresult}%
            \cellcolor{#5!\tempa!#4!#6}\texttt{#1}%
        }}%
}

\newcommand{\mrbhl}[1]{\gradientcell{#1}{60}{93}{white}{translateBlue}{70}}
\newcommand{\mrbhlf}[1]{\gradientcell{#1}{0.2}{0.5}{white}{translateBlue}{70}}

\title{\begin{wrapfigure}[2]{l}{0.09\textwidth}
\vspace{-0.8\baselineskip}
  \includegraphics[width=0.09\textwidth]{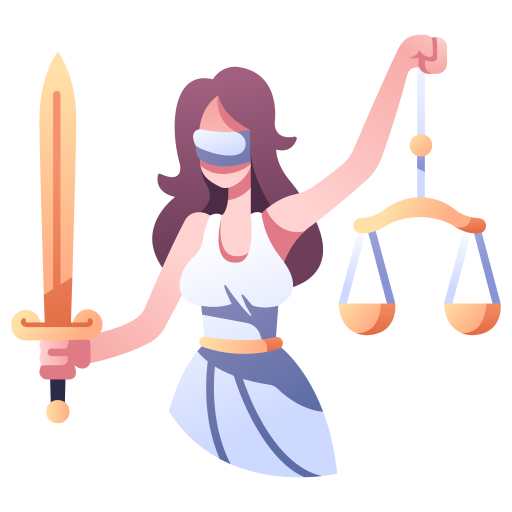}
\end{wrapfigure} 
\texttt{Themis}: Training Robust Multilingual Code Reward Models for Flexible Multi-Criteria Scoring 
}

\author{\name Indraneil Paul and Iryna Gurevych \email \texttt{\{name.lastname\}@tu-darmstadt.de} \\
\addr UKP Lab, TU Darmstadt and National Research Center for Applied Cybersecurity ATHENE
\AND
\name Goran Glava\v{s} \email \texttt{\{name.lastname\}@uni-wuerzburg.de} \\
\addr Center for Artificial Intelligence and Data Science, University of Würzburg}

\def\month{03}  
\def\year{2026} 
\def\openreview{\url{https://openreview.net/forum?id=XXXX}} 

\makeatletter
\AddToHook{cmd/appendix/before}{\def\cref@section@alias{appendix}}
\newcommand{\ifnotempty}[2]{%
  \edef\@tempa{#1}\ifx\@tempa\empty\else#2\fi
}
\newcommand{\deanon}[1]{\if@accepted #1\fi}
\if@accepted
  \hypersetup{%
    pdfauthor={Indraneil Paul},
    pdftitle={Themis: Training Robust Multilingual Code Reward Models for Flexible Multi-Criteria Scoring}}
\fi
\makeatother
\begin{document}

\maketitle

\begin{abstract}

Reward models (RMs) have become an indispensable fixture of the language model (LM) post-training playbook, enabling policy alignment and test-time scaling. Research on the application of RMs in code generation, however, has been comparatively sparse, with existing work largely focusing on execution feedback. This choice constrains post-training to optimizing functional correctness over self-contained executable code. In this work, we examine the training and evaluation of multilingual, multi-criteria code RMs. To this end, we first compile \textbf{\texttt{Themis-CodeRewardBench}}, a benchmark to evaluate code RMs across five preference dimensions (i.e., criteria) and eight programming languages, on which we profile \texttt{50+} code, math, and general-purpose RMs. Observing the limited proficiency of current RMs beyond scoring for functional correctness, we develop \textbf{\texttt{Themis-CodePreference}}, the largest open-source collection of code preferences to date (more than \texttt{350k} preference pairs), and use it to train \textbf{\texttt{Themis-RM}}, a suite of multilingual code reward models for flexible multi-criteria scoring, ranging in size from \texttt{0.6B} to \texttt{32B} parameters. Our experiments and ablations demonstrate positive scaling trends, strong cross-lingual transfer when training on diverse preferences, and the importance of multi-criteria training for reliable code reward modeling. 

\end{abstract}

\section{Introduction}
\label{sec:Introduction}

Language model (LM) post-training recipes commonly rely on extrinsic reward signals to score LM responses. Access to such a reward enables model developers to align model outputs to end-user preferences~\citep{DBLP:journals/corr/abs-1909-08593, DBLP:journals/corr/abs-2112-00861, DBLP:conf/nips/Ouyang0JAWMZASR22}, improve the reliability of human data labeling~\citep{DBLP:conf/emnlp/0005C0H0L24}, and efficiently allocate compute via test-time scaling~\citep{DBLP:journals/corr/abs-2408-03314, DBLP:journals/corr/abs-2404-00725}. Further, leveraging extrinsic rewards for on-policy training pushes the performance frontier~\citep{DBLP:conf/acl/AhmadianCGFKPUH24, DBLP:journals/corr/abs-2402-04792, DBLP:conf/nips/0001SSB024, DBLP:conf/icml/TajwarSSR0XEFK24} while unlocking superior generalization~\citep{DBLP:conf/iclr/HayesSPP25, DBLP:conf/iclr/KirkMNLHGR24, DBLP:conf/icml/ChuZYTXSLL025} and enabling learning from negative outcomes~\citep{DBLP:journals/corr/abs-2510-08696, DBLP:journals/corr/abs-2506-01347, DBLP:journals/tmlr/SinghCAAPGLH0XP24, DBLP:journals/tmlr/Dong0GZCPDZS023}. 

In general-domain settings, extrinsic rewards are usually obtained from a specialized reward model (RM)~\citep{DBLP:journals/corr/abs-2503-23829, DBLP:journals/corr/abs-2507-17746}, owing to their flexibility across disparate criteria~\citep{DBLP:conf/iclr/KimS0JLLYSKTS24} and their implicit regularizing effect in noisy and underspecified settings~\citep{DBLP:conf/icml/0015DYW0J0024, DBLP:journals/corr/abs-2503-01067}. Despite measures of code quality spanning multiple attributes such as correctness, wall time, memory efficiency, and security, post-training for code generation~\citep{DBLP:journals/corr/abs-2505-22312, DBLP:journals/corr/abs-2411-15124, DBLP:journals/corr/abs-2505-16400} avoids RMs and resorts to runtime rewards grounded in execution~\citep{DBLP:journals/tmlr/LiuZXF00Y23, DBLP:conf/icml/GehringZCMCS25, DBLP:conf/nips/Le0GSH22, DBLP:journals/tmlr/ShojaeeJTR23}. In this work, we posit that existing code reward sources are inadequate and that robust multi-criteria code RMs are crucial for improved code generation post-training that accounts for non-functional code quality criteria and removes the bottleneck of procuring readily executable code.

\paragraph{The limitations of execution feedback.}

Input-output matching via test-case execution is the dominant paradigm for sourcing rewards in code generation post-training~\citep{DBLP:journals/corr/abs-2505-21297, DBLP:conf/emnlp/ShiFGZW22, DBLP:journals/corr/abs-2509-07604}. However, it is limited to scenarios where model developers can procure their own test-cases. A common strategy to sidestep this bottleneck involves synthetically generating test-cases using LMs~\citep{DBLP:journals/corr/abs-2508-21107, DBLP:journals/corr/abs-2510-12803}. However, despite positive scaling characteristics~\citep{DBLP:conf/acl/MaZZYLT25}, test-case generation approaches often rely on circular consensus mechanisms~\citep{DBLP:conf/acl/ToNB24, DBLP:conf/iclr/ChenZNZLLC23, DBLP:journals/corr/abs-2411-13611, DBLP:conf/acl/XuLYZP25, DBLP:conf/acl/Zhang0DXZSLJ25, DBLP:conf/acl/ZengJ0NCC25, DBLP:conf/emnlp/WangLSDL25} that result in brittle or erroneous test-cases~\citep{DBLP:conf/naacl/WangYWHCSZCM25, DBLP:journals/corr/abs-2506-06821}. What is more, test-case feedback is only possible with self-contained executable code: this is highly restrictive as it limits models' post-training exposure to code devoid of external dependencies, which is very rare~\citep{DBLP:conf/icse/YuSRZZMLLWX24}, i.e., it is unrepresentative of the natural code distribution ``in the wild''. It also prevents the use of (most) code in programming languages with module-level compilation units, such as \texttt{Rust}, \texttt{Go}, and \texttt{Swift}. Attempts at mitigation via synthetic repository-level test-case generation are limited owing to the considerable difficulty of the task~\citep{DBLP:conf/nips/MundlerMHV24, DBLP:conf/iclr/JainSR25}. Finally, test-case verification conventionally relies on well-formed inputs derived from user-specified input preconditions and can thus miss subtle interface contract violations~\citep{Lim2025ContractEvalAB}. 

Alternatively, prior work has sought to expand runtime reward acquisition via auxiliary information such as execution duration and data flow~\citep{Duan2023PerfRLAS, DBLP:journals/corr/abs-2404-18864, DBLP:journals/tmlr/ShojaeeJTR23}. However, such rewards can be noisy unless the code execution occurs in a tightly controlled emulated environment~\citep{DBLP:conf/iclr/ShypulaMZ0GYHNR24}. Additionally, such rewards are often noisy proxies for an underlying objective and prone to reward hacking~\citep{DBLP:conf/iclr/PanBS22, DBLP:journals/alife/LehmanCMAABBBBB20, DBLP:conf/icml/GaoSH23}. For instance, logic invoked in a compile-time macro can throw off the use of a compilation reward as a proxy for syntactic correctness. In contrast, 
RMs derived from massively pre-trained LLMs, owing to their extensive knowledge, can complement code execution signals by providing extrinsic rewards across a wide range of programming languages in underspecified scenarios~\citep{DBLP:conf/ijcai/EverittKOL17}, and are not limited to functional correctness evaluation. 

\paragraph{The drawbacks of LMs as surrogate code executors.}

Recent work has adopted the LM-as-a-judge approach~\citep{DBLP:conf/emnlp/LiuIXWXZ23, DBLP:conf/iclr/LiSYF0024} to evaluate code via specialized prompting techniques~\citep{DBLP:conf/IEEEcloud/AggarwalCDMPBM24, DBLP:conf/emnlp/TongZ24, DBLP:conf/eacl/Zhuo24}, but general-purpose LMs tend to be poorly calibrated code evaluators in reference-free settings~\citep{DBLP:journals/corr/abs-2501-16655, DBLP:conf/emnlp/LyuHLSZ25}. In parallel, attempts have been made to train (i) specialized LM scorers as code execution surrogates~\citep{DBLP:journals/corr/abs-2601-13097}, (ii) quality estimators~\citep{DBLP:conf/icml/Ni0RSYWL23, DBLP:conf/emnlp/ShiFGZW22}, and (iii) re-rankers~\citep{DBLP:conf/icml/ZhangYHLYF023, DBLP:conf/nips/InalaWYCELMG22}. However, these efforts primarily target self-contained \texttt{Python} code, a setting in which the execution-based feedback excels. Furthermore, prior work shows that when applied to low-resource programming languages, surrogate code execution models suffer from non-trivial performance drops~\citep{DBLP:conf/emnlp/LyuHLSZ25}. We thus argue that the impact of post-training in code generation can be expanded via robust RMs capable of scoring code outputs across diverse scenarios, programming languages, and over disparate user-specified criteria.

\paragraph{The need for multilingual and multi-criteria code RM benchmarks.}

RM training losses are poor predictors of test-time regret~\citep{DBLP:conf/icml/FluriLAF0S25}, necessitating the creation of challenging and realistic RM preference evaluation benchmarks~\citep{DBLP:conf/naacl/LambertPMMLCDKZCSH25, DBLP:journals/corr/abs-2506-01937, DBLP:conf/iclr/Liu0M00L25, DBLP:conf/iclr/TanZMTC0PS25, DBLP:conf/iclr/FrickLCCAJZGS25}. Exhaustive evaluation enables the flagging of weak RMs before deployment, thereby preventing downstream reward hacking~\citep{DBLP:conf/iclr/0002G0L0K25}. Yet the diversity and difficulty of code preferences in RM evaluation datasets are extremely limited: these datasets consist entirely of valid-buggy \texttt{Python} solution pairs sourced from a select few elementary and dated competitive coding datasets~\citep{DBLP:journals/corr/abs-2108-07732, DBLP:journals/corr/abs-2107-03374, DBLP:conf/iclr/MuennighoffLZZH24}. Existing attempts at dedicated evaluations for general-domain code RMs are similarly limited in programming language and scenario coverage~\citep{DBLP:journals/corr/abs-2502-16614, DBLP:conf/coling/ZhaoLT0YL025, DBLP:journals/corr/abs-2507-10535, Ficek2025ScoringVE}. There is thus an acute need for code RM evaluation benchmarks that assess scoring across a varied mix of scenarios, in multiple programming languages, and on criteria beyond (just) functional correctness.

\paragraph{Contributions.} We address the abovementioned research gaps and make the following contributions:

\begin{itemize}[leftmargin=*]

\item We compile \texttt{Themis-CodeRewardBench}, an extensive code RM evaluation benchmark comprising \texttt{$\approx$8.9k} diverse code preferences across eight programming languages and five scoring dimensions. Subsequently, we study the failure modes of \texttt{45} existing code RMs (ref. \cref{sec:Code_Reward_Bench}).

\item We develop \texttt{Themis-GeneralPreference} and \texttt{Themis-CodePreference}, two extensive preference datasets containing \texttt{110k+} general-domain and \texttt{350k+} code-domain preferences, respectively; the latter covers five different dimensions/criteria of code quality: functional correctness, execution efficiency, memory efficiency, readability and maintainability, and security. 

\item We use the above preference datasets to train \texttt{Themis-RM}, a suite of multilingual open-source code RMs ranging from \texttt{0.6B} to \texttt{32B} parameters in size, capable of flexibly scoring code across arbitrary (i.e., user-defined) subsets of the five quality criteria (ref. \cref{sec:Themis_RM}).

\item We conduct an extensive investigation into the performance trends of the \texttt{Themis-RM} suite across model sizes, including on unseen criteria and programming languages (ref. \cref{sec:Experiments}). Our work constitutes the first test of the viability of using code RMs to score open-domain code under adversarial settings, i.e., on unseen criteria and out-of-distribution programming languages.

\end{itemize}

We structure our investigation into training state-of-the-art RMs for code post-training in \Cref{sec:Experiments} around the following set of research questions:

\noindent\textbf{\texttt{RQ1}:} How accurately can RMs score code across a set of diverse real-world code quality preferences?

\noindent\textbf{\texttt{RQ2}:} What is the best approach to minimize cross-criteria interference in multi-criteria code RMs?

\noindent\textbf{\texttt{RQ3}:} How well does preference learning for code RMs transfer to unseen programming languages?

\noindent\textbf{\texttt{RQ4}:} Do RMs trained on multi-criteria preferences exhibit the requisite adversarial robustness and listwise re-ranking competence necessary for downstream applications such as post-training and search?

\section{Related Work}
\label{sec:related_work}

We briefly summarize four lines of relevant existing research: \textbf{1.)} surrogate code execution modeling for alignment and quality estimation, \textbf{2.)} extrinsic rewards for code beyond functional correctness, \textbf{3.)} multi-criteria reward models, and \textbf{4.)} reward model evaluation benchmarks.

\paragraph{Surrogate code execution modeling for alignment and quality estimation.}

LM developers seeking to ensure quality control of model-generated code have traditionally relied on test-case feedback. However, in addition to restrictive executability requirements, existing approaches in this mold require users to either bring their own test-cases~\citep{DBLP:journals/tmlr/LiuZXF00Y23, DBLP:conf/icml/GehringZCMCS25, DBLP:journals/tmlr/TaoC0MR0M25}, generate synthetic test-cases~\citep{DBLP:conf/acl/ToNB24, DBLP:conf/iclr/ChenZNZLLC23, DBLP:conf/acl/Zhang0DXZSLJ25, DBLP:journals/corr/abs-2411-13611, DBLP:conf/nips/0003C0DJMVWG024} or source reference solutions~\citep{DBLP:journals/corr/abs-2412-13464}. Consequently, recent work has looked to repurpose neural surrogates (a.k.a. verifiers) for code execution simulation~\citep{DBLP:journals/jcphy/HesthavenU18, DBLP:journals/corr/abs-1901-05125, DBLP:conf/nips/YanSKRH20, DBLP:conf/nips/RodionovP23, DBLP:conf/emnlp/LyuHLSZ25} as a means of scoring code that is not readily executable. Early work in this vein trained quality estimators to simulate test-case execution~\citep{DBLP:conf/icml/Ni0RSYWL23, DBLP:conf/emnlp/Zhou0AN23, DBLP:conf/nips/InalaWYCELMG22}, but the resultant models can be poorly calibrated with respect to program semantics in reference-free settings~\citep{DBLP:conf/emnlp/ShiFGZW22}.

Subsequent work has leveraged pairwise preference learning~\citep{DBLP:conf/nips/Ouyang0JAWMZASR22, DBLP:journals/corr/abs-1909-08593} to improve robustness to noisy signals. Approaches in this paradigm have bootstrapped scalar RMs from execution feedback~\citep{DBLP:conf/acl/ZengJ0NCC25, Zhu2026CodeScalerSC} as well as synthetic feedback~\citep{DBLP:journals/corr/abs-2403-09032}. Pairwise RMs~\citep{DBLP:conf/acl/Jiang0L23, DBLP:journals/corr/abs-2505-10320, DBLP:journals/corr/abs-2504-00050, DBLP:journals/corr/abs-2506-03637, DBLP:journals/corr/abs-2505-11475, DBLP:journals/corr/abs-2504-02495, DBLP:journals/corr/abs-2410-03837} evolve this approach to its logical end by operating on pairs of code responses at both training and inference time. However, pairwise RMs can produce inconsistent and intransitive preferences~\citep{DBLP:journals/corr/abs-2403-16950}, which demand specialized prompting techniques in listwise re-ranking tasks~\citep{zhu2024starlingb, DBLP:conf/naacl/QinJHZWYSLLMWB24}. In parallel, attempts have been made to leverage the commonsense knowledge of frontier LMs~\citep{DBLP:conf/nips/ZhaoLH23} to provide free-text feedback that reasons about code semantics~\citep{DBLP:journals/tosem/DongDJLLJ25, DBLP:conf/eacl/Zhuo24, DBLP:journals/corr/abs-2311-07215}. The range of such textual feedback spans iterative self-reflection~\citep{DBLP:conf/nips/ShinnCGNY23, DBLP:conf/nips/MadaanTGHGW0DPY23}, refinement proposals~\citep{DBLP:journals/corr/abs-2502-09183, DBLP:conf/acl/ZhangLLLJ23, DBLP:journals/corr/abs-2303-16749}, deliberating compiler outputs~\citep{DBLP:conf/nips/GuptaCCS20, DBLP:conf/iclr/GouSGSYDC24, DBLP:conf/acl/WangWWMLZLWJL22, DBLP:conf/iclr/ChenLSZ24} and studying execution traces~\citep{DBLP:conf/icml/NiACDSSY24}. However, as we illustrate in \Cref{subsec:Experiments_RQ1}, generation-based surrogate code executors can exhibit low scoring fidelity, impeding their ability to resolve fine-grained preferences. Our experiments with \texttt{Themis-RM} show that scalar RMs trained on diverse code preferences can accurately score code across a variety of domains (ref. \cref{subsec:Experiments_RQ1}), while maintaining high scoring fidelity, thus enabling listwise re-ranking in challenging settings (ref. \cref{subsec:Experiments_RQ4}). 

\paragraph{Extrinsic rewards for code beyond functional correctness.}

Owing to the difficulty of multi-objective optimization of LMs~\citep{DBLP:conf/acl/NguyenPSB25, DBLP:conf/nips/YangLXHZA24, DBLP:conf/icml/0010PLQ00C24, DBLP:conf/acl/ZhouLS00O024, DBLP:journals/tcyb/LiZW21, DBLP:conf/iclr/DaiPSJXL0024} and its occasionally unclear tradeoffs for code generation~\citep{DBLP:conf/kbse/CoignionQR25}, LM post-training has largely avoided improving code quality along non-functional axes. However, as LM-generated code proliferates~\citep{DBLP:journals/corr/abs-2506-08945}, the potentially dire security implications of synthetically generated code smells~\citep{DBLP:journals/corr/abs-2510-03029, DBLP:journals/cacm/PearceATDK25, DBLP:journals/corr/abs-2502-01853} have led to a renewed interest. Prior work on enhancing the security of LM code outputs has incorporated static analyzer feedback~\citep{DBLP:journals/corr/abs-2508-14419, Siddiq2022SecurityEvalDM, DBLP:conf/ccs/HeV23, DBLP:journals/corr/abs-2601-01184}. But static analyzers broadly rely on predefined surface-level patterns in code and cannot catch vulnerabilities that require dependency context and runtime information~\citep{DBLP:journals/corr/abs-2408-13855}. Accordingly, recent work incorporated LM-based feedback to mine security preferences~\citep{DBLP:journals/corr/abs-2504-04699, DBLP:conf/icml/XuSG0W025}, constrain inference-time decoding~\citep{DBLP:conf/trustcom/QuLKLWYH25, DBLP:conf/emnlp/WangLHWH25, DBLP:journals/corr/abs-2405-00218, DBLP:conf/issta/Li0ZLL0C024}, or directly score fine-~\citep{Quan2026LearningTG} and coarse-grained~\citep{Wu2026SecureCG, DBLP:conf/icse/DingFISCAWR025, DBLP:journals/corr/abs-2507-16887, DBLP:journals/corr/abs-2401-07031} security hardness. In tandem, the reasoning capabilities of LMs have been leveraged towards iterative code hardening~\citep{DBLP:journals/corr/abs-2507-19060, DBLP:conf/kbse/Zhang0FFL24, DBLP:conf/nips/LeSZXS24}, security-focused synthetic environment generation~\citep{DBLP:journals/corr/abs-2508-00910}, and in-context learning of secure coding principles~\citep{DBLP:conf/emnlp/ZhangDT0CCWY24, DBLP:conf/icml/HeVKV24}.

Similarly, the scarcity of work optimizing wall-clock metrics has led to LM-generated code lagging behind developer-written code in efficiency~\citep{DBLP:journals/corr/abs-2507-12415, DBLP:journals/corr/abs-2509-09853, DBLP:journals/corr/abs-2511-06090}. Attempts to address this gap via post-training have resorted to retrieval-augmented generation~\citep{DBLP:journals/corr/abs-2512-22827}, compiler feedback~\citep{DBLP:conf/ics/LamouriAKB25}, or test-case execution feedback~\citep{DBLP:journals/corr/abs-2508-20124, DBLP:conf/emnlp/WaghjaleVWF24}. These efforts are, however, impeded by the need for executable code, pre-existing test-cases, and/or reference solutions. More recent approaches derive extrinsic rewards by generating synthetic test-cases~\citep{DBLP:journals/corr/abs-2502-18489}, using scalar reward models~\citep{DBLP:journals/corr/abs-2404-18864}, or directly from single- or multi-step improvements via textual reasoning~\citep{DBLP:conf/icml/0005ZDLWQCG025, DBLP:journals/corr/abs-2512-14018, DBLP:conf/forge/PengGLXSS25, DBLP:journals/corr/abs-2505-23387, DBLP:journals/corr/abs-2410-10209}. The use of extrinsic signals to improve code memory efficiency~\citep{DBLP:journals/corr/abs-2601-01215}, maintainability~\citep{DBLP:journals/corr/abs-2505-19442, DBLP:conf/saner/NunesFSNFS25}, and specification following~\citep{DBLP:journals/corr/abs-2409-12866} is even more sporadic. Our analyses with \texttt{Themis-RM} show that non-functional requirements do not necessarily interfere with functional correctness in reward modeling, and that specific design choices during training can enable scalar RMs to model preferences across multiple axes (ref. \cref{subsec:Experiments_RQ2}).

\paragraph{Multi-criteria reward models.}

Preference learning for RMs has typically been framed as the Nash equilibrium of an implicit two-player game~\citep{DBLP:conf/iclr/JacobSFA24, DBLP:conf/nips/0001L023, DBLP:conf/icml/MunosVCARGTGMFM24}. This formulation implies that training scalar RMs via naive multi-task learning~\citep{DBLP:conf/emnlp/VuKATFS24, DBLP:conf/iclr/YangKCPT24, DBLP:journals/corr/abs-2506-09183} across multilingual, multi-task, and personalized preferences is bound to incur interference~\citep{DBLP:conf/icml/ChakrabortyQYKM24} and distributional biases~\citep{DBLP:conf/fat/ChristianKTSD25}. Prior attempts to mitigate such interference involve model merging~\citep{DBLP:journals/corr/abs-2502-06876, DBLP:journals/corr/abs-2310-11564, DBLP:conf/emnlp/KimSLLSWNL0S24}, factorized representations~\citep{DBLP:journals/corr/abs-2503-06358, DBLP:conf/emnlp/CalderonER25}, steering vectors~\citep{DBLP:conf/icml/LinJXCC25}, Bayesian optimization~\citep{DBLP:conf/iclr/WinataASKW25} and specialized neural layers~\citep{DBLP:journals/corr/abs-2601-15968, DBLP:conf/emnlp/00030X0024}. Subsequent work has demonstrated that detailing diverse preference scenarios as textual contexts allows scalar~\citep{DBLP:conf/nips/LeePKS24, DBLP:conf/iclr/SunSZZCCYG24} and implicit RMs~\citep{DBLP:conf/acl/WangLXYDQZZ24, DBLP:conf/emnlp/DongWSWK23, DBLP:conf/emnlp/WangXZXJ25} to effectively disentangle them~\citep{DBLP:journals/corr/abs-2507-07375}.

Additionally, the body of work on generative~\citep{DBLP:conf/iclr/KwonXBS23, DBLP:journals/corr/abs-2212-08073, DBLP:journals/corr/abs-2509-06822, DBLP:journals/corr/abs-2509-21500, DBLP:conf/emnlp/Zhong0YMJLZJH22, DBLP:conf/naacl/LiuSX0CKGH24, DBLP:journals/tmlr/JiangLZHLC24} and reasoning RMs~\citep{DBLP:journals/corr/abs-2504-00050, DBLP:journals/corr/abs-2505-14674, DBLP:journals/corr/abs-2504-02495} also demonstrates the benefits of incorporating textual evaluation criteria via rubrics, towards both preference learning and downstream optimization~\cite{DBLP:journals/corr/abs-2505-13388, DBLP:journals/corr/abs-2506-03637, DBLP:journals/corr/abs-2511-10507, DBLP:journals/corr/abs-2510-07743, DBLP:journals/corr/abs-2508-12790}. More recent work outlines how the general knowledge of LMs can be used to self-propose evaluation criteria in underspecified scoring settings~\citep{DBLP:journals/corr/abs-2505-02387, DBLP:conf/icml/Saha0GWW25, DBLP:conf/emnlp/SaadFalconVBNFVSKM25, Zhou2026AutoChecklistCP, DBLP:journals/corr/abs-2408-11791}. Our evaluations with \texttt{Themis-RM} show that modeling evaluation criteria via constituting principles enables scalar RMs to effectively model multi-dimensional preferences with minimal interference (ref. \cref{subsec:Experiments_RQ2}).

\paragraph{Reward model evaluation benchmarks.}

Identifying coverage shortcomings~\citep{DBLP:journals/corr/abs-2401-06080} and distributional biases~\citep{DBLP:conf/emnlp/LiFZZLXWJH25, DBLP:conf/nips/KirkWRBMGCBW0VH24} of RMs before their downstream deployment is crucial for minimizing reward hacking at deployment time~\citep{DBLP:conf/iclr/0002G0L0K25, DBLP:conf/iclr/PanBS22}. Early work in this direction outlined the strong correlation between RM accuracy in modeling preferences and downstream post-training performance~\citep{DBLP:conf/iclr/FrickLCCAJZGS25}. Accordingly, a significant body of work evaluates RMs on general-domain preferences~\citep{DBLP:conf/iclr/TanZMTC0PS25, DBLP:conf/iclr/ZhouZWXDBSXFMZG25, DBLP:conf/naacl/LambertPMMLCDKZCSH25} as well as for pairwise rankings on a host of specialized domains such as instruction following~\citep{Wen2026IFRewardBenchBJ}, computer use~\citep{DBLP:journals/corr/abs-2510-18596}, multi-modal~\citep{DBLP:conf/cvpr/LiW0YSWALLLKL25}, and multilingual~\citep{DBLP:conf/acl/GurejaMIMSW0RHF25} settings. More recent efforts benchmark the pairwise re-ranking abilities of code RMs by repurposing existing competition code generation benchmarks~\citep{Ficek2025ScoringVE, DBLP:conf/coling/ZhaoLT0YL025, DBLP:journals/corr/abs-2507-10535, DBLP:journals/tacl/NiYZRFSYLYXJZRCC24, DBLP:journals/corr/abs-2502-16614}. Such benchmarks, however, risk testing RMs on a narrow distribution of prompts in a select few high-resource programming languages, and may be contaminated~\citep{DBLP:conf/emnlp/MattonSATAHMVGG24}. We contribute \texttt{Themis-CodeRewardBench} (ref. \cref{sec:Code_Reward_Bench}), a diverse, multi-criteria, and multilingual code RM evaluation benchmark that enables us to uncover critical gaps in the coverage of current code RMs (ref. \cref{subsec:Experiments_RQ1}). Further work has outlined the importance of holistic profiling of reward models on adversarial~\citep{DBLP:journals/corr/abs-2601-12186, DBLP:conf/iclr/Liu0M00L25, DBLP:journals/corr/abs-2505-16222} and listwise ranking settings~\citep{DBLP:journals/corr/abs-2506-01937, DBLP:conf/iclr/WenL0LXLHHZ025, DBLP:conf/acl/KimKKC0Y25} to detect over-optimization failure modes that preference-based accuracy evaluation can occasionally miss. Our robust evaluations on \texttt{Themis-RM} show how the advantages of open-domain code preference learning extend to such scenarios (ref. \cref{subsec:Experiments_RQ4}).

\section{\texttt{Themis-CodeRewardBench}: Multilingual And Multi-Criteria Code RM Evaluation}
\label{sec:Code_Reward_Bench}

\newlength{\datasetcol}
\setlength{\datasetcol}{4.3cm}
 
\newlength{\desccol}
\setlength{\desccol}{5cm}
 
\newlength{\langcol}
\setlength{\langcol}{0.8cm}
 
\newlength{\provcol}
\setlength{\provcol}{1.6cm}
 
\newlength{\langcolsep}
\setlength{\langcolsep}{\dimexpr\langcol+2\tabcolsep\relax}
 
\newlength{\provcolsep}
\setlength{\provcolsep}{\dimexpr\provcol+2\tabcolsep\relax}
 
\newlength{\ddcol}
\setlength{\ddcol}{\dimexpr\datasetcol+2\tabcolsep+\desccol+\tabcolsep\relax}
 
\newcommand{\zx}{\raisebox{0pt}[0pt][0pt]{{\color[HTML]{800000}\ding{55}}}}
 
\newcommand{\n}[1]{\texttt{#1}}
 
\newcommand{\thinrule}[1]{\rule{#1}{\cmidrulewidth}}
 
\newcommand{\group}[1]{%
  \renewcommand{\arraystretch}{1.3}%
  \begin{tabular}[c]{@{}
      p{\datasetcol}
      p{\desccol}
      *{8}{>{\centering\arraybackslash}p{\langcol}}
      *{2}{>{\centering\arraybackslash}p{\provcol}}
    @{}}
    #1
  \end{tabular}%
}

\begin{table}[t!]
  \centering
  \caption{Criteria-level dataset composition of the \texttt{Themis-CodeRewardBench} reward model evaluation benchmark introduced in \cref{sec:Code_Reward_Bench}, broken down across the eight constituent programming languages. For detailed evaluation results on this benchmark, refer to \cref{subsec:Experiments_RQ1} and \cref{subsec:Experiments_RQ3}.} 
  \label{tab:coderewardbench_breakdown}
  \renewcommand{\cellalign}{cc}
  \scalebox{0.6}{%
\def\arraystretch{1.12}
  \begin{tabular}{>{\centering\arraybackslash}m{2.4cm} c}
    \toprule
    \multirow{3}{*}{\texttt{\textbf{Criteria}}}
    & \makebox[\ddcol][l]{%
        \makebox[\dimexpr\datasetcol+2\tabcolsep\relax][l]{%
          \multirow{3}{*}{\texttt{\textbf{Dataset}}}}%
        \multirow{3}{*}{\texttt{\textbf{Short Description}}}}%
      \makebox[\dimexpr8\langcolsep\relax][c]{%
        \texttt{\textbf{Programming Language}}}%
      \makebox[\dimexpr2\provcolsep\relax][c]{%
        \texttt{\textbf{Provenance}}} \\[-5pt]
    & \makebox[\ddcol][l]{}%
      \makebox[\dimexpr8\langcolsep\relax][c]{%
        \thinrule{\dimexpr8\langcolsep-2\tabcolsep\relax}}%
      \makebox[\dimexpr2\provcolsep\relax][c]{%
        \thinrule{\dimexpr2\provcolsep-2\tabcolsep\relax}} \\[2pt]
    & \makebox[\ddcol][l]{}%
      \makebox[\langcolsep][c]{\texttt{\textbf{C}}}%
      \makebox[\langcolsep][c]{\texttt{\textbf{C++}}}%
      \makebox[\langcolsep][c]{\texttt{\textbf{C\#}}}%
      \makebox[\langcolsep][c]{\texttt{\textbf{Go}}}%
      \makebox[\langcolsep][c]{\texttt{\textbf{Java}}}%
      \makebox[\langcolsep][c]{\texttt{\textbf{JS}}}%
      \makebox[\langcolsep][c]{\texttt{\textbf{Python}}}%
      \makebox[\langcolsep][c]{\texttt{\textbf{Ruby}}}%
      \makebox[\provcolsep][c]{\texttt{\textbf{Prompts}}}%
      \makebox[\provcolsep][c]{\texttt{\textbf{Responses}}} \\
    \midrule
    \makecell{\textbf{\texttt{Functional}}\\\textbf{\texttt{Correctness}}\\\textbf{\texttt{(FC)}}}
    &
    \group{
      {\ttfamily Commit Preference\newline Correctness}
        & Multi-LLM verified functionality bug-fixes 
expressed in GitHub commits.
        & \n{11} & \n{25} & \n{42} & \n{59}
        & \n{134}  & \n{241}       & \n{216}       & \n{97}
        & \ttfamily LLM\newline Generated  & \ttfamily Human\newline Written \\
      {\ttfamily HumanEvalPack\newline \cite{DBLP:conf/iclr/MuennighoffLZZH24}}
        & Manually injected bugs in 
translated HumanEval solutions.
        & \zx & \n{147} & \zx       & \n{117}
        & \n{132}       & \n{113} & \n{119}       & \zx
        & \ttfamily Human\newline Written & \ttfamily Human\newline Written  \\
      {\ttfamily MBPPPlusFix-Hard\newline \cite{DBLP:journals/corr/abs-2502-01619}}
        & Discerning LLM generations that satisfy a high
proportion, but not all the test-cases.
        & \zx & \zx & \zx & \zx
        & \zx       & \zx & \n{37} & \zx
        & \ttfamily Human\newline Written & \n{Mixed}  \\
      {\ttfamily MDEval\newline \cite{DBLP:journals/corr/abs-2411-02310}}
        & Manually annotated buggy-fixed pairs
inspired by GitHub code.
        & \n{12} & \n{11} & \zx & \n{24}
        & \zx  & \n{30} & \n{22}       & \n{35}
        & \ttfamily Human\newline Written & \ttfamily Human\newline Written \\
      {\ttfamily DebugEval\newline \cite{DBLP:conf/naacl/YangWLLYWGYLY25}}
        & DebugBench and LiveCodeBench 
buggy-fixed pairs (LeetCode).
        & \zx & \n{211} & \zx & \zx
        & \n{194}       & \zx & \n{319}     & \zx
        & \ttfamily Human\newline Written & \ttfamily Human\newline Written  \\
      {\ttfamily RunBugRun-V1\newline \cite{DBLP:journals/corr/abs-2304-01102}}
        & User grouped fast-slow pairs from 
CodeNet (AtCoder and Aizu).
        & \n{279} & \n{501} & \zx & \n{352}
        & \n{453} &	\n{164}	& \n{305} &	\n{376}
        & \ttfamily Human\newline Written & \ttfamily Human\newline Written
    } \\
    \midrule
    \makecell{\textbf{\texttt{Execution}}\\\textbf{\texttt{Efficiency}}\\\textbf{\texttt{(EE)}}}
    &
    \group{
      {\ttfamily Commit Preference\newline Runtime}
        & Multi-LLM verified runtime improving code changes 
expressed in GitHub commits.
       & \n{4} & \n{6} & \n{21} & \n{15}
        & \n{55}  & \n{39}       & \n{56}       & \n{42}
        & \ttfamily LLM\newline Generated & \ttfamily Human\newline Written \\
      {\ttfamily Pie4Perf\newline \cite{DBLP:conf/iclr/ShypulaMZ0GYHNR24}}
        & Emulator verified user-grouped curated pairs of slow-fast 
solutions from CodeNet.
        & \zx & \n{460} & \zx & \zx
        & \zx       & \zx & \zx & \zx
        & \ttfamily Human\newline Written & \ttfamily Human\newline Written  \\
      {\ttfamily ECCO\newline \cite{DBLP:conf/emnlp/WaghjaleVWF24}}
        & Execution verified curated pairs of slow-fast 
solutions from CodeNet.
        & \zx & \zx & \zx & \zx
        & \zx & \zx & \n{399} & \zx
        & \ttfamily Human\newline Written & \ttfamily Human\newline Written  \\
      {\ttfamily EvalPerf\newline \cite{DBLP:journals/corr/abs-2408-06450}}
        & LLM completions of varying 
efficiency to EvalPlus.
        & \zx & \zx & \zx & \zx
        & \zx       & \zx & \n{212} & \zx
        & \ttfamily Human\newline Written & \ttfamily LLM\newline Generated
    } \\
    \midrule
    \makecell{\textbf{\texttt{Memory}}\\\textbf{\texttt{Efficiency}}\\\textbf{\texttt{(ME)}}}
    &
    \group{
      {\ttfamily Commit Preference\newline Memory}
        & Multi-LLM verified memory usage improving code changes 
expressed in GitHub commits.
        & \n{73} & \n{43} & \n{8} & \n{12}
        & \n{52}  & \n{28} & \n{26} & \n{10}
        & \ttfamily LLM\newline Generated & \ttfamily Human\newline Written  \\
      {\ttfamily NoFunEval Memory\newline \cite{DBLP:journals/corr/abs-2401-15963}}
        & Human validated mined memory 
efficiency-enhancing commits.
        & \n{2} & \zx & \zx & \zx
        & \n{35} & \zx & \zx & \zx
        & \ttfamily LLM\newline Generated & \ttfamily Human\newline Written
    } \\
    \midrule
    \makecell{\textbf{\texttt{Readability And}}\\\textbf{\texttt{Maintainability}}\\\textbf{\texttt{(R\&M)}}}
    &
    \group{
      {\ttfamily Commit Preference\newline CodeStyle}
        & Multi-LLM verified code stylistic improvements 
expressed in GitHub commits.
        & \n{13} & \n{24} & \n{80} & \n{101}
        & \n{257}  & \n{325} & \n{365} & \n{206}
        & \ttfamily LLM\newline Generated & \ttfamily Human\newline Written  \\
      {\ttfamily NoFunEval Maintain\newline \cite{DBLP:journals/corr/abs-2401-15963}}
        & CodeQL-verified maintainability fixes linked to GitHub commits.
        & \zx & \zx & \zx & \zx
        & \zx & \zx & \n{128} & \zx
        & \ttfamily LLM\newline Generated & \ttfamily Human\newline Written
    } \\
    \midrule
        \makecell{\textbf{\texttt{Security}}\\\textbf{\texttt{Hardness}}\\\textbf{\texttt{(SH)}}}
    &
    \group{
      {\ttfamily Commit Preference\newline Security}
        & Multi-LLM verified vulnerability fixes 
expressed in GitHub commits.
        & \n{31} & \n{42} & \n{37} & \n{46}
        & \n{143}  & \n{144} & \n{172} & \n{154}
        & \ttfamily LLM\newline Generated & \ttfamily Human\newline Written  \\
      {\ttfamily CodePrefBench Security\newline \cite{DBLP:journals/corr/abs-2410-03837}}
        & Synthetically fixed CyberSecEval 
samples using LLMs.
        & \zx & \zx & \zx & \zx
        & \zx       & \zx & \n{173}       & \zx
        & \ttfamily LLM\newline Generated & \n{Mixed}  \\
      {\ttfamily Vul4J\newline \cite{DBLP:conf/msr/BuiSF22}}
        & Human and test validated commits 
mined from SAP's Project KB.
        & \zx & \zx & \zx & \zx
        & \n{8}       & \zx & \zx      & \zx
        & \ttfamily LLM\newline Generated  & \ttfamily Human\newline Written \\
      {\ttfamily SecBench\newline \cite{DBLP:journals/corr/abs-2412-20787}}
        & Fusion of multiple security-fix 
commit datasets mined from CWEs.
        & \n{2} & \n{1} & \zx & \zx
        & \n{4}       & \zx & \n{2}       & \n{5}
        & \ttfamily LLM\newline Generated & \ttfamily Human\newline Written \\
      {\ttfamily NoFunEval Security\newline \cite{DBLP:journals/corr/abs-2401-15963}}
        & Asleep at the Keyboard completions 
by CoPilot that trigger CodeQL rules.
        & \n{12} & \zx & \zx & \zx
        & \zx       & \zx & \n{15}       & \zx
        & \ttfamily LLM\newline Generated & \ttfamily LLM\newline Generated
    } \\
    \bottomrule
  \end{tabular}%
  }
\end{table}

To evaluate RMs across multilingual and multi-dimensional code preferences, we first curate the \texttt{Themis-CodeRewardBench} benchmark, a collection comprising \texttt{13} distinct pre-existing and newly constructed code preference datasets. 
Our multi-criteria benchmark evaluates RMs \textbf{(1)} on five code quality dimensions: \texttt{Functional Correctness (FC)}, \texttt{Execution Efficiency (EE)}, \texttt{Memory Efficiency (ME)}, \texttt{Readability And Maintainability (R\&M)}, and \texttt{Security Hardness (SH)}, and (2) for eight high- and medium-resource programming languages: \texttt{C}, \texttt{C\#}, \texttt{C++}, \texttt{Go}, \texttt{Java}, \texttt{JavaScript}, \texttt{Python}, and \texttt{Ruby}. 
Deferring to prior findings~\citep{DBLP:conf/iclr/FrickLCCAJZGS25} and the strong precedent in existing RM evaluation, we use preference accuracy as the evaluation metric on \texttt{Themis-CodeRewardBench}.

\begin{figure}[t!]
    \centering
    \includegraphics[width=\linewidth]{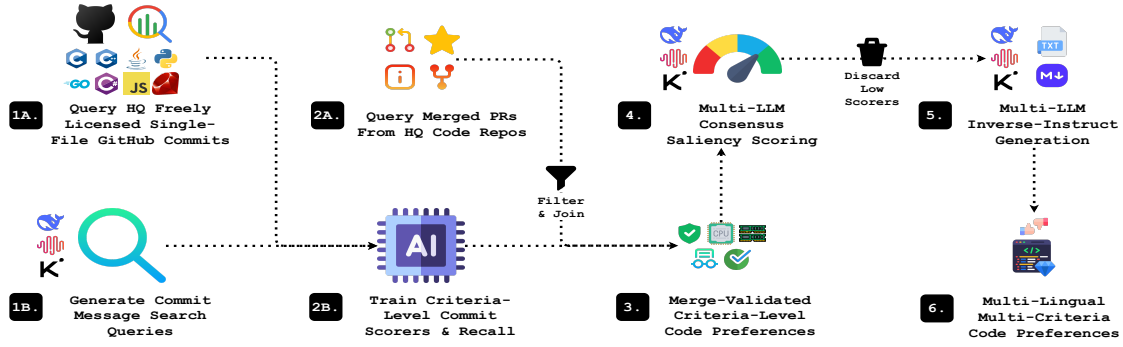}
    \caption{Overview of our pipeline for mining multi-programming-language multi-criteria code preferences from single-file merged GitHub commits. Used both in \texttt{Themis-CodeRewardBench} and for \texttt{Themis-CodePreference}.}
    \label{fig:Commit_Pref}
\end{figure}

We start by procuring the test splits of diverse existing datasets that contain explicit or implicit expressions of code preference. Such datasets range from human-annotated code changes to execution-validated code metrics computed for human- or model-generated code. This curation yields pairs of code with disparate functional correctness~\citep{DBLP:conf/iclr/MuennighoffLZZH24, DBLP:journals/corr/abs-2502-01619, DBLP:journals/corr/abs-2411-02310, DBLP:conf/naacl/YangWLLYWGYLY25, DBLP:journals/corr/abs-2304-01102}, execution efficiency~\citep{DBLP:conf/iclr/ShypulaMZ0GYHNR24, DBLP:conf/emnlp/WaghjaleVWF24, DBLP:journals/corr/abs-2408-06450}, security~\citep{DBLP:journals/corr/abs-2410-03837, DBLP:journals/corr/abs-2412-20787, DBLP:conf/msr/BuiSF22}, and code-style compliance or maintainability~\citep{DBLP:journals/corr/abs-2401-15963}. 
For datasets lacking explicit queries or instructions, we synthetically generate inverse instructions (see \cref{subsec:Commit_Inverse} for the prompt). 
The resulting collection built from existing datasets, albeit varied in provenance, is rather thinly spread across most combinations of programming languages and evaluation criteria (especially for non-functional criteria). To address this limitation, we develop a multi-step data-collection pipeline to extract implicit code preferences from GitHub commits, as described next.

\begin{figure}[t!]
    \centering
    \hfill
    \begin{subfigure}[b]{0.225\textwidth}
        \centering
        \includegraphics[width=\textwidth,height=7cm]{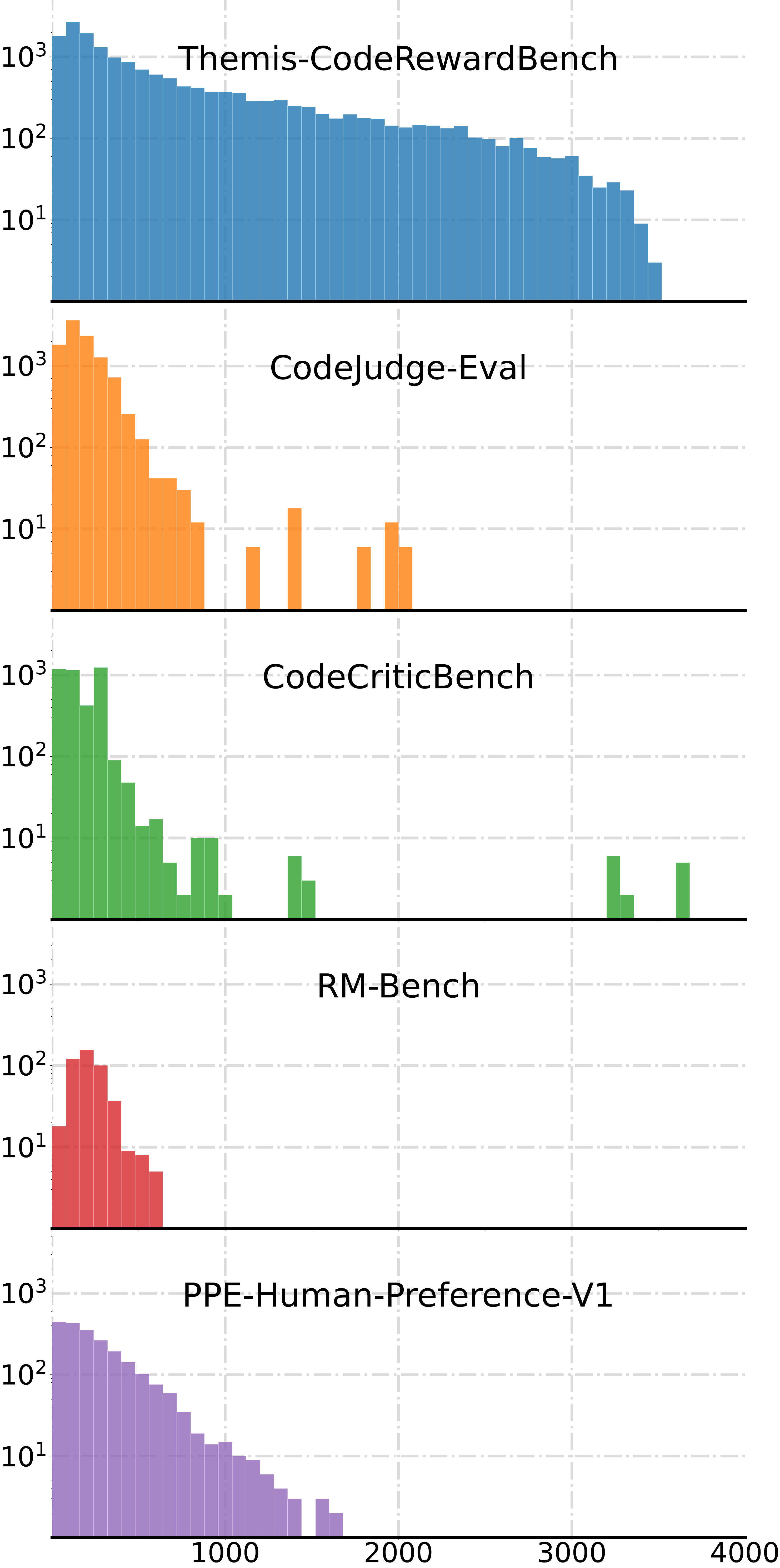}
        \caption{Response code seq. length distribution.}
        \label{subfig:CRB_a}
    \end{subfigure}%
    \hfill
    \begin{subfigure}[b]{0.225\textwidth}
        \centering
        \includegraphics[width=\textwidth,height=7cm]{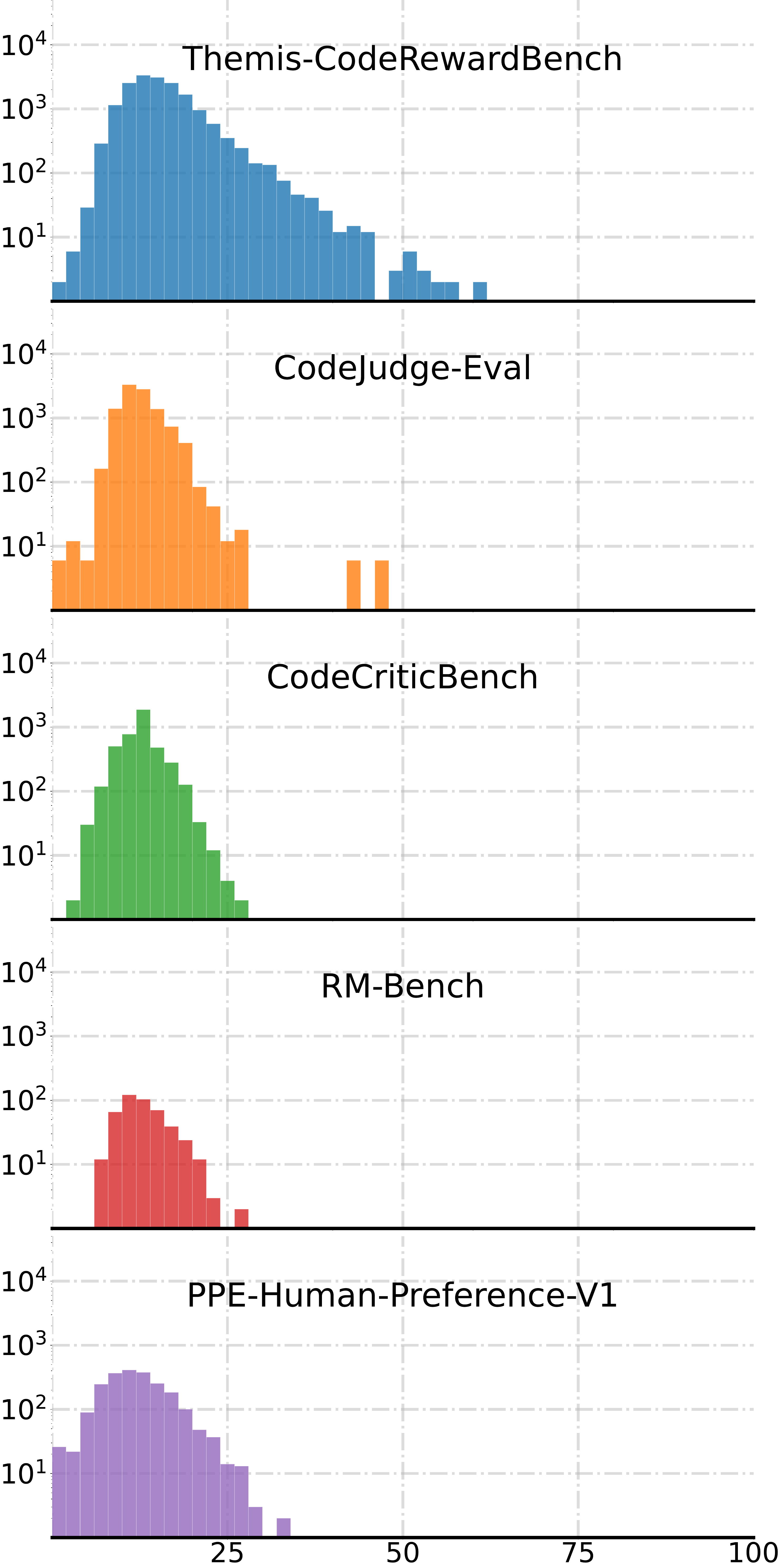}
        \caption{Response code AST depth distribution.}
        \label{subfig:CRB_b}
    \end{subfigure}%
    \hfill
    \begin{subfigure}[b]{0.5\textwidth}
        \centering
        \includegraphics[width=\textwidth,height=7cm]{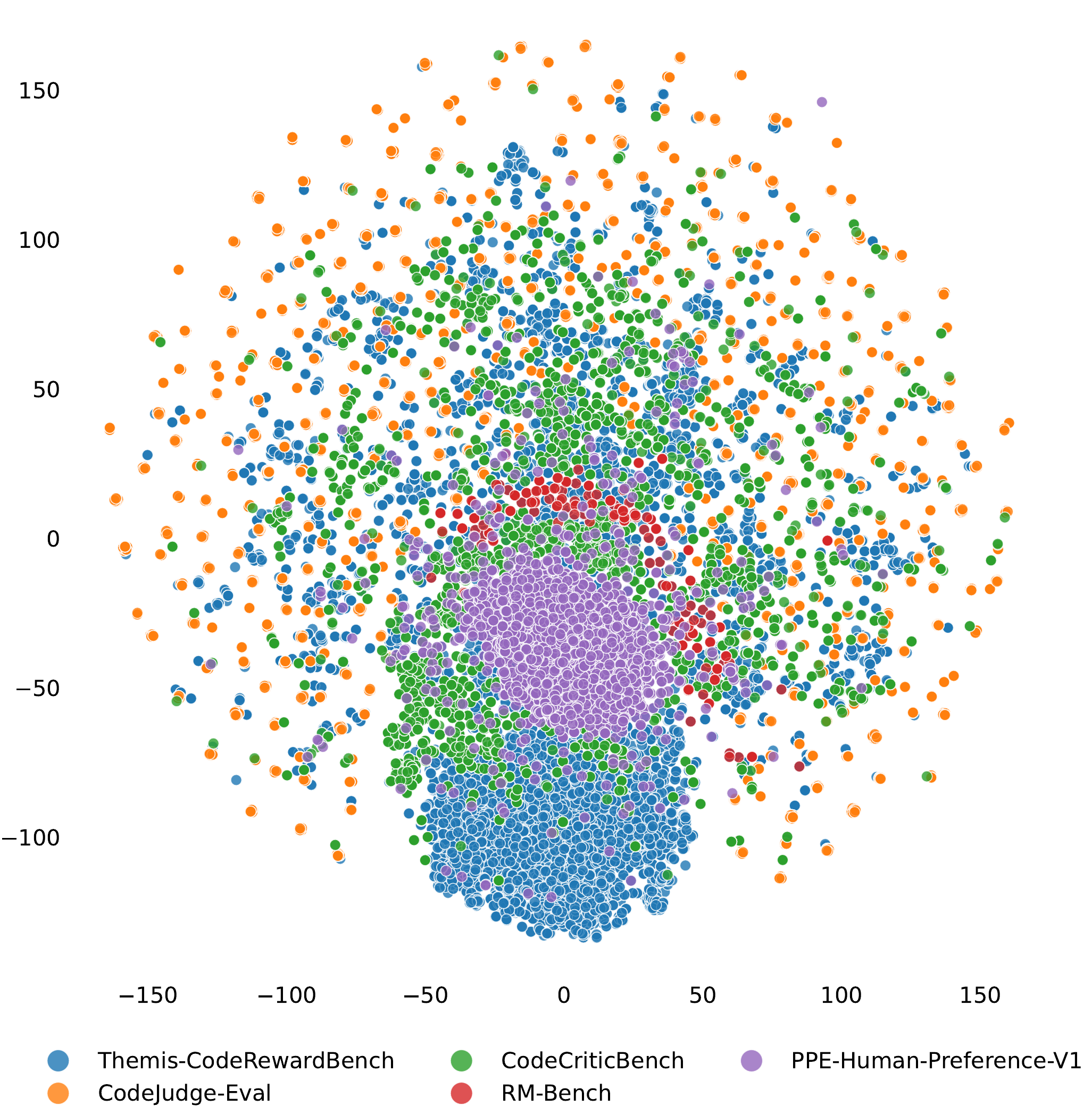}
        \caption{2D t-SNE plot of prompt embeddings, as generated by \\
        \href{https://huggingface.co/Alibaba-NLP/gte-Qwen2-1.5B-instruct}{\texttt{\scalerel*{\includegraphics{Figures/hf.png}}{\texttt{I}} Alibaba-NLP/gte-Qwen2-1.5B-instruct}}~\cite{DBLP:journals/corr/abs-2308-03281}.}
        \label{subfig:CRB_c}
    \end{subfigure}
    \hfill
    \caption{Comparison of \texttt{Themis-CodeRewardBench} against the code subsets of popular existing RM evaluation benchmarks. \texttt{Themis-CodeRewardBench} judges RMs on (a) longer and (b) more complex code responses, over a (c) largely novel distribution of prompts.}
    \label{fig:CodeRewardBench_Dist}
\end{figure}

We begin our commit mining by querying the public GitHub archives\footnote{\bqrepo{github/github-repos}} using a modified version of the pipeline detailed in~\citet{DBLP:conf/iclr/MuennighoffLZZH24} (see \cref{subsec:BigQuery_SQL} for the consolidated query), searching for single-file code changes in openly licensed repositories. We then generate search terms for each of our five preference criteria (detailed in \cref{subsec:Commit_Mining}) using frontier open-LMs~\citep{DBLP:journals/corr/abs-2602-02276, DBLP:journals/corr/abs-2512-02556, minimaxMiniMaxM25}. We leverage samples retrieved with these search terms to train criteria-specialized ModernBERT~\citep{DBLP:conf/acl/WarnerCCWHTGBLA25} commit classifiers, which we use to recall high-confidence code changes corresponding to the criteria. 
Subsequently, we incorporate pull request data from reputable repositories\footnote{We deem repositories with \texttt{15+} GitHub stars, \texttt{5+} contributors, and \texttt{10+} GitHub issues as a reasonable proxy for reputability and commitment to good software engineering practices.} using \texttt{GHTorrent}\footnote{\gitrepo{ghtorrent/ghtorrent.org}} and only retain non-reverted commits authored between \texttt{June 2019} and \texttt{January 2021} that are part of merged pull requests, ensuring implicit human validation of intents. Next, we solicit a consensus of multiple frontier open-LMs to weed out multi-purpose code changes or changes without a clear intent (see \cref{subsec:Commit_Saliency} for the annotation prompt). Finally, we synthetically generate realistic inverse instructions across a multitude of styles for the remaining code change pairs (we provide the prompt in \cref{subsec:Commit_Inverse}).
\cref{fig:Commit_Pref} illustrates this commit-mining workflow. We provide the complete dataset-, criteria-, and language-level make-up of \texttt{Themis-CodeRewardBench} in \cref{tab:coderewardbench_breakdown}. \cref{fig:CodeRewardBench_Dist} compares \texttt{Themis-CodeRewardBench} against code subsets of existing RM benchmarks, showing that \texttt{Themis-CodeRewardBench} introduces a largely novel distribution of code preferences (\cref{subfig:CRB_c}), for code of increased complexity (\cref{subfig:CRB_a} and \cref{subfig:CRB_b}).

\section{\texttt{Themis-RM}: Training Multilingual Criteria-Following Code RMs}
\label{sec:Themis_RM}

To establish the benefits of diverse code-specific reward modeling, we train the \texttt{Themis-RM} suite of RMs. We train \texttt{Themis-RM} models in two stages: (1) a preference model pre-training (PT) stage, followed by (2) a preference modeling (PM) stage. In this section, we detail the training data and exact setup for both stages.

\subsection{Training Data Mixture}
\label{subsec:Train_Datamix}

\paragraph{Preference Model Pre-Training (PT):} 

Scoring code on non-functional and stylistic axes demands a general understanding of human preferences and conventions. Hence, we first pre-train the \texttt{Themis-RM} suite to instill common human-inspired notions of preference evaluation such as relevance, helpfulness, and harmlessness. To this end, we curate \texttt{Themis-GeneralPreference}, a \texttt{110k+} sample mix of natural language and code preferences curated from popular existing preference and retrieval datasets (ref. \cref{subsec:Data_Comp} for the detailed composition). This approach mirrors prior work, which demonstrated the benefits of training LMs for approximating open-domain human preferences~\citep{DBLP:journals/corr/abs-2112-00861, DBLP:journals/corr/abs-2505-10527, DBLP:conf/icml/KorbakSCBBPBP23}.

\paragraph{Preference Modeling (PM):}

Our second training stage centers on \texttt{Themis-CodePreference}, a \texttt{350k+} preference dataset comprising preferences across the same five dimensions (i.e., criteria) of code evaluation and the eight programming languages that our \texttt{Themis-CodeRewardBench} benchmark evaluates. In the interest of training on diverse and complex scenarios, we source GitHub commit preferences and synthetic prompts using the pipeline described in \cref{sec:Code_Reward_Bench}. We ensure that all commits in our training data are pushed before \texttt{March 2019} and come from a disjoint set of repositories, vis-à-vis the commit data in \texttt{Themis-CodeRewardBench}. Having single-intent single-file-changing commits enables us to source preference pairs for each of the criteria of interest, which curtails training noise~\citep{DBLP:conf/iclr/YangKCPT24, DBLP:journals/corr/abs-2407-16008, DBLP:journals/tacl/DOosterlinckXDDSPKM25} and impedes reward hacking~\citep{DBLP:conf/acl/KimKKC0Y25}. Additionally, acquiring criteria-specific commit preferences allows us to train \texttt{Themis-RM} models on non-functional preferences that are otherwise rare in traditional instruction-tuning datasets~\citep{DBLP:conf/icml/XuSG0W025, tdcommonsUsingCode}.

Aiming to deliver complete and well-rounded RMs for code, we also, naturally, train \texttt{Themis-RM} on diverse multi-dimensional preferences---functional correctness, wall time, and memory usage preferences---sourced from a range of existing code contest datasets~\citep{DBLP:conf/nips/Puri0JZDZD0CDTB21, DBLP:conf/emnlp/WaghjaleVWF24, DBLP:journals/corr/abs-2505-23387, DBLP:journals/corr/abs-2304-01102}. 
Finally, we source a mix of functional and non-functional code preferences from LM-generated responses to user prompts. Training on synthetic data improves the utility of the \texttt{Themis-RM} suite in downstream post-training settings~\citep{DBLP:conf/icml/TajwarSSR0XEFK24}. We source model-generated preferences by synthetically introducing algorithmic and syntactic errors (i.e., bugs) into responses in existing instruction-tuning datasets~\citep{DBLP:conf/icml/XuSG0W025, Zhu2026CodeScalerSC}. \cref{subsec:Data_Comp} details the criteria- and language-level composition of \texttt{Themis-CodePreference} and the data filtering steps.

\subsection{Model Training}
\label{subsec:ModelTraining}

\texttt{Themis-RM} represents a suite of code-specialized RMs ranging from \texttt{0.6B} to \texttt{32B} parameters in size that we derive from Qwen3~\citep{DBLP:journals/corr/abs-2505-09388} dense models using the aforementioned two-phase training. Our training incorporates optional system prompts $p$ that allow the specification of custom evaluation criteria, with \texttt{15\%} of the training samples unaccompanied by any criteria (i.e., without $p$), a further \texttt{20\%} accompanied by a generic formulation that outlines all the criteria, and the rest of the samples coupled with the evaluation instruction that captures exactly one criterion (ref. \cref{subsec:Themis_Principles} for the evaluation instructions). Our experimental results in \cref{subsec:Experiments_RQ2} show that this prompt strategy effectively disentangles multi-criteria preferences, eliminating the need for criteria-specific modules~\citep{DBLP:conf/emnlp/00030X0024}, ensembling~\citep{DBLP:conf/iclr/CosteAK024}, model-merging~\citep{DBLP:journals/corr/abs-2310-11564, DBLP:conf/nips/RameCDGSSC23}, or data augmentation~\citep{DBLP:conf/iclr/000200CWJGS0YSM25, DBLP:journals/corr/abs-2309-16155, DBLP:conf/emnlp/WuYCKCG25}.

\texttt{Themis-RM} models are trained using the Bradley-Terry~\citep{DBLP:journals/corr/abs-2507-07375} reward modeling objective on preference tuples $(p, x, y_c, y_r)\in\mathcal{D}$ sourced from the aforementioned datasets. Owing to the largely offline nature of our training data acquisition and the heightened risk of over-optimization~\citep{DBLP:journals/corr/abs-2505-20556}, we adopt additional regularization for the RM hidden states. Prior work has regularized RM representations via a range of auxiliary information, including uncertainty~\citep{DBLP:journals/corr/abs-2410-00847, DBLP:journals/tmlr/ShekharS025, DBLP:journals/corr/abs-2403-17297}, distributional data~\citep{DBLP:journals/corr/abs-2409-10164}, and adversarial generators~\citep{DBLP:journals/corr/abs-2504-06141}. We opt for a minimally invasive conditional language modeling loss~\citep{DBLP:conf/nips/YangDLZZ24, DBLP:journals/corr/abs-2310-13639} over the preferred response $y_c$ as our regularizer. Finally, owing to the poor calibration of scalar RMs beyond their normal scoring range~\citep{DBLP:journals/ploscb/BaysD17}, we adopt a reward magnitude regularizer~\citep{DBLP:journals/corr/abs-2312-09244}. Our overall training objective is specified as follows:
\vspace{-0.2em}
\begin{equation*}
\begin{gathered}
\mathcal{L} = -\mathbb{E}_{([p],\, x,\, y_c,\, y_r) \sim \mathcal{D}} \Big[ 
  \log \sigma \!\left( r_\theta(p, x, y_c) - r_\theta(p, x, y_r) \right)
  + \lambda \cdot \log p_\theta(y_c \mid p, x)
  + \mu \cdot \left( r_\theta(p, x, y_c) + r_\theta(p, x, y_r) \right)^2
\Big] \\
{\footnotesize \text{where } \sigma(\cdot) \text{: sigmoid function; } r_\theta \text{: scalar RM; } p \text{: criteria prompt; } x \text{: task prompt; } y_c \text{: chosen response; } y_r \text{: rejected response.}}
\end{gathered}
\end{equation*}
\vspace{-0.7em}

For simplicity, we utilize the above objective in both training phases, which we verify is extremely effective~\citep{DBLP:journals/corr/abs-2410-02229} (ref. \cref{subsec:Experiments_RQ2}) and removes the need for bespoke objectives for the PT phase~\citep{DBLP:journals/corr/abs-2507-05197, DBLP:journals/corr/abs-2112-00861}. We train on \texttt{Themis-GeneralPreference} for two epochs during the PT phase and on \texttt{Themis-CodePreference} for one epoch during the PM phase (we train using a modified fork of the \texttt{trl}\footnote{\gitrepo{huggingface/trl}} framework). We leverage the AdamW optimizer~\citep{DBLP:conf/iclr/LoshchilovH19} and a cosine scheduler with a \texttt{5\%} warmup in each of the phases. We outline the complete training configuration and architectural heritage of the \texttt{Themis-RM} suite in \cref{appdx:1_Arch_Train}.

\section{Experimental Research Questions And Results}
\label{sec:Experiments}

\subsection{\texttt{RQ1}: Judging Reward Models On Multilingual Multi-Criteria Code Preferences}
\label{subsec:Experiments_RQ1}

We begin our inquiry with a rigorous evaluation of the competence of existing RMs in scoring code. To this end, we assess a diverse mix of the most competitive existing general-purpose, code-specialized, scalar, generative, and reasoning RMs and compare them against \texttt{Themis-RM} on \texttt{Themis-CodeRewardBench} (ref. \cref{tab:Aspect_Results}). 
Our assessments render current RMs largely unusable for scoring code along non-functional axes such as efficiency and security, often degenerating to purely random scoring. \texttt{Themis-RM}, in contrast, displays strong performance on all axes across all model sizes. Our smallest model, \texttt{Themis-RM-0.6B}, outscores multiple \texttt{>100x} larger general-purpose RMs. In contrast, our largest model, \texttt{Themis-RM-32B}, establishes itself as the clear state-of-the-art (including on functional correctness). 
Within each evaluated model family (including \texttt{Themis-RM}), we observe a strong positive scaling trend in scoring for functional correctness. This trend is in line with prior findings on the correlation between RMs' verification and generation abilities~\citep{DBLP:journals/corr/abs-2509-17995}, both conforming to the well-known scaling laws~\citep{DBLP:journals/corr/abs-2001-08361, DBLP:journals/corr/abs-2203-15556}. Profiling the results for the functional correctness scoring (see \cref{tab:FC_Results}), which is the usual operating mode for existing code RMs, across individual datasets uncovers glaring limitations of existing RMs. Concretely, most RMs fare poorly when required to grade code outside the common distribution of code contest data, trailing substantially on code sourced from commit preferences. 
Similarly, a look at the scores on the \texttt{MBPP+Fix (Hard)} dataset~\citep{DBLP:journals/corr/abs-2502-01619} shows that RMs largely cannot discern between varying degrees of partial correctness; only larger \texttt{Themis-RM} models and a handful of existing RMs can capture these subtle differences. Existing RM benchmarks~\citep{DBLP:conf/naacl/LambertPMMLCDKZCSH25, DBLP:conf/iclr/Liu0M00L25} fail to uncover these weaknesses because they primarily rely on a small set of simple and dated datasets such as \texttt{HumanEvalPack}~\citep{DBLP:conf/iclr/MuennighoffLZZH24}, which, as our results show, are too saturated to serve as a useful indicator of RMs' scoring ability.

\def\arraystretch{1.12}
\begin{table}[!thbp]
\centering
\caption{Detailed criteria-level preference accuracy of extant RMs and the \texttt{Themis-RM} suite on \texttt{Themis-CodeRewardBench} (\textbf{\texttt{RQ1}}: ref. to \cref{subsec:Experiments_RQ1}) followed by \texttt{Themis-RM-8B} training ablations (\textbf{\texttt{RQ2}}: ref. to \cref{subsec:Experiments_RQ2}). For detailed dataset-level scores per RM, consult \cref{tab:FC_Results,tab:EM_Results,tab:RS_Results} in \cref{appdx:4_Detailed_Results}.}
\label{tab:Aspect_Results}
\scalebox{0.6}{%
\begin{tabular}{@{}c l l r c ccccc@{}}
\toprule
& \multirow{4}{*}{\texttt{\textbf{Model}}} & & \multirow{4}{*}{\texttt{\textbf{Size}}}  & \multirow{4}{*}{\texttt{\textbf{Average}}} & \multicolumn{5}{c}{\textbf{\texttt{Criteria-Level Accuracy}}} \\
\cmidrule(lr){6-10}
&  & &  &  & \textbf{\texttt{Functional}} & \textbf{\texttt{Execution}} & \textbf{\texttt{Memory}} & \textbf{\texttt{Readability And}} & \textbf{\texttt{Security}} \\
&  & &  &  & \textbf{\texttt{Correctness}} & \textbf{\texttt{Efficiency}} & \textbf{\texttt{Efficiency}} & \textbf{\texttt{Maintainability}} & \textbf{\texttt{Hardness}} \\
&  & &  &  & \textbf{\texttt{(FC)}} & \textbf{\texttt{(EE)}} & \textbf{\texttt{(ME)}} & \textbf{\texttt{(R\&M)}} & \textbf{\texttt{(SH)}} \\
\midrule
\multirow{13}{*}{\rotatebox[origin=c]{90}{\textbf{\texttt{XL}}}}
& \hfmodel{Qwen/Qwen2.5-Math-RM-72B}         & \iconMath & \texttt{72B} & \mrbhl{69.06} & \mrbhl{79.89} & \mrbhl{55.92} & \mrbhl{57.09} & \mrbhl{57.44} & \mrbhl{55.30} \\
& \hfmodel{nvidia/AceMath-72B-RM}     & \iconMath & \texttt{72B}  & \mrbhl{72.83} & \mrbhl{84.76} & \mrbhl{61.73} & \mrbhl{53.29} & \mrbhl{59.24} & \mrbhl{56.21} \\
& \hfmodel{Qwen/WorldPM-72B-RLHFLow}     &              & \texttt{72B}  & \mrbhl{76.96} & \mrbhl{86.02} & \mrbhl{60.73} & \mrbhl{63.67} & \mrbhl{70.38} & \mrbhl{68.52} \\
& \hfmodel{ContextualAI/LMUnit-qwen2.5-72b}         & \iconAuxiliary\;\iconGenerative\;\iconPrinciple\;\iconReasoning & \texttt{72B} & \mrbhl{27.89} & \mrbhl{39.10} & \mrbhl{20.17} & \mrbhl{6.92} & \mrbhl{11.81} & \mrbhl{14.53} \\
& \hfmodel{nvidia/Llama-3.3-Nemotron-70B-Reward}          &  & \texttt{70B}  & \mrbhl{78.39} & \mrbhl{88.76} & \mrbhl{61.30} & \mrbhl{66.44} & \mrbhl{65.78} & \mrbhl{73.26} \\
& \hfmodel{infly/INF-ORM-Llama3.1-70B}         & \iconGenerative   & \texttt{70B}  &  \mrbhl{74.84} & \mrbhl{82.88} & \mrbhl{62.03} & \mrbhl{62.63} & \mrbhl{68.18} & \mrbhl{66.60} \\
& \hfmodel{allenai/Llama-3.1-70B-Instruct-RM-RB2}     &             & \texttt{70B}  & \mrbhl{78.23} & \mrbhl{86.23} & \mrbhl{64.02} & \mrbhl{64.71} & \mrbhl{71.25} & \mrbhl{72.96} \\
& \hfmodel{allenai/Llama-3.1-Tulu-3-70B-SFT-RM-RB2}     &  & \texttt{70B}  & \mrbhl{78.96} & \mrbhl{86.23} & \mrbhl{65.16} & \mrbhl{67.13} & \mrbhl{73.58} & \mrbhl{73.76} \\
& \hfmodel{Nexusflow/Athene-RM-70B}     &              & \texttt{70B}  & \mrbhl{81.19} & \mrbhl{87.69} & \mrbhl{67.07} & \mrbhl{77.16} & \mrbhl{75.45} & \mrbhl{78.30} \\
& \hfmodel{Nexusflow/Starling-RM-34B}         &  & \texttt{34B} & \mrbhl{70.63} & \mrbhl{77.75} & \mrbhl{54.55} & \mrbhl{63.67} & \mrbhl{64.58} & \mrbhl{68.72} \\
& \textbf{\texttt{Themis-RM-32B}}          & \iconAuxiliary\;\iconCode\;\iconPrinciple & \texttt{32B}  & \mrbhl{91.82} & \mrbhl{94.27} & \mrbhl{84.95} & \mrbhl{95.16} & \mrbhl{87.59} & \mrbhl{94.55} \\
& \hfmodel{TIGER-Lab/AceCodeRM-32B}         & \iconCode   & \texttt{32B}  & \mrbhl{62.95} & \mrbhl{73.67} & \mrbhl{57.91} & \mrbhl{40.48} & \mrbhl{46.36} & \mrbhl{49.55} \\
& \hfmodel{nvidia/Qwen3-Nemotron-32B-GenRM-Principle}     &   \iconGenerative\;\iconPrinciple           & \texttt{32B}  & \mrbhl{71.67} & \mrbhl{79.34} & \mrbhl{66.77} & \mrbhl{68.17} & \mrbhl{57.17} & \mrbhl{64.08} \\
\midrule
\multirow{7}{*}{\rotatebox[origin=c]{90}{\textbf{\texttt{L}}}}
& \hfmodel{nicolinho/QRM-Gemma-2-27B}         & \iconAuxiliary & \texttt{27B} & \mrbhl{52.48} & \mrbhl{53.89} & \mrbhl{52.71} & \mrbhl{47.06} & \mrbhl{50.57} & \mrbhl{48.85} \\
& \hfmodel{ShikaiChen/LDL-Reward-Gemma-2-27B-v0.1}     & \iconAuxiliary & \texttt{27B}  & \mrbhl{75.57} & \mrbhl{84.28} & \mrbhl{65.47} & \mrbhl{62.63} & \mrbhl{67.31} & \mrbhl{63.17} \\
& \hfmodel{Skywork/Skywork-Reward-Gemma-2-27B-v0.2}     &              & \texttt{27B}  & \mrbhl{54.16} & \mrbhl{56.74} & \mrbhl{56.74} & \mrbhl{49.13} & \mrbhl{50.77} & \mrbhl{51.36} \\
& \hfmodel{internlm/internlm2-20b-reward}         & \iconAuxiliary\;\iconPrinciple & \texttt{20B} & \mrbhl{73.13} & \mrbhl{81.38} & \mrbhl{60.20} & \mrbhl{65.74} & \mrbhl{64.11} & \mrbhl{66.26} \\
& \textbf{\texttt{Themis-RM-14B}}          & \iconAuxiliary\;\iconCode\;\iconPrinciple & \texttt{14B}  & \mrbhl{91.19} & \mrbhl{92.74} & \mrbhl{86.10} & \mrbhl{94.46} & \mrbhl{87.32} & \mrbhl{95.36} \\
& \hfmodel{rubricreward/R3-Qwen3-14B-14k}         & \iconGenerative\;\iconPrinciple\;\iconReasoning   & \texttt{14B}  & \mrbhl{45.41} & \mrbhl{56.8} & \mrbhl{41.79} & \mrbhl{31.83} & \mrbhl{24.95} & \mrbhl{30.17} \\
& \hfmodel{openbmb/UltraRM-13b}     &             & \texttt{13B}  & \mrbhl{70.43} & \mrbhl{74.17} & \mrbhl{55.23} & \mrbhl{66.44} & \mrbhl{71.18} & \mrbhl{72.45} \\
\midrule
\multirow{19}{*}{\rotatebox[origin=c]{90}{\textbf{\texttt{M}}}}
& \textbf{\texttt{Themis-RM-8B}}          & \iconAuxiliary\;\iconCode\;\iconPrinciple & \texttt{8B} & \mrbhl{89.78} & \mrbhl{91.75} & \mrbhl{83.65} & \mrbhl{92.04} & \mrbhl{86.06} & \mrbhl{93.34} \\
& \hfmodel{LARK-Lab/CodeScaler-8B}     & \iconCode & \texttt{8B}  & \mrbhl{79.12} & \mrbhl{87.42} & \mrbhl{61.27} & \mrbhl{72.32} & \mrbhl{73.45} & \mrbhl{73.26} \\
& \hfmodel{rubricreward/R3-Qwen3-8B-14k}         & \iconGenerative\;\iconPrinciple\;\iconReasoning & \texttt{8B} & \mrbhl{43.23} & \mrbhl{52.91} & \mrbhl{39.88} & \mrbhl{29.07} & \mrbhl{27.22} & \mrbhl{29.36} \\
& \hfmodel{Skywork/Skywork-Reward-V2-Qwen3-8B}     &              & \texttt{8B}  & \mrbhl{79.97} & \mrbhl{87.25} & \mrbhl{64.63} & \mrbhl{71.63} & \mrbhl{75.05} & \mrbhl{74.97} \\
& \hfmodel{RLHFlow/ArmoRM-Llama3-8B-v0.1}         & \iconAuxiliary & \texttt{8B} & \mrbhl{71.76} & \mrbhl{79.59} & \mrbhl{61.96} & \mrbhl{58.13} & \mrbhl{64.71} & \mrbhl{61.55} \\
& \hfmodel{nicolinho/QRM-Llama3.1-8B-v2}     & \iconAuxiliary & \texttt{8B}  & \mrbhl{71.81} & \mrbhl{80.16} & \mrbhl{59.74} & \mrbhl{66.78} & \mrbhl{62.98} & \mrbhl{62.36} \\
& \hfmodel{allenai/Llama-3.1-8B-Base-RM-RB2}     &             & \texttt{8B}  & \mrbhl{75.77} & \mrbhl{82.57} & \mrbhl{61.42} & \mrbhl{69.90} & \mrbhl{70.78} & \mrbhl{71.24} \\
& \hfmodel{Ray2333/GRM-llama3-8B-sftreg}          & \iconAuxiliary & \texttt{8B} & \mrbhl{71.81} & \mrbhl{79.05} & \mrbhl{58.75} & \mrbhl{57.44} & \mrbhl{66.11} & \mrbhl{67.00} \\
& \hfmodel{LxzGordon/URM-LLaMa-3.1-8B}     & \iconAuxiliary & \texttt{8B}  & \mrbhl{71.62} & \mrbhl{79.91} & \mrbhl{60.73} & \mrbhl{65.05} & \mrbhl{62.31} & \mrbhl{62.16} \\
& \hfmodel{NCSOFT/Llama-3-OffsetBias-RM-8B}     &              & \texttt{8B}  & \mrbhl{71.76} & \mrbhl{79.80} & \mrbhl{59.05} & \mrbhl{59.17} & \mrbhl{63.91} & \mrbhl{65.29} \\
& \hfmodel{sfairXC/FsfairX-LLaMA3-RM-v0.1}         &  & \texttt{8B} & \mrbhl{73.52} & \mrbhl{81.02} & \mrbhl{59.74} & \mrbhl{67.82} & \mrbhl{67.31} & \mrbhl{66.60} \\
& \hfmodel{Nexusflow/Athene-RM-8B}     &  & \texttt{8B}  & \mrbhl{76.58} & \mrbhl{82.08} & \mrbhl{63.18} & \mrbhl{70.24} & \mrbhl{73.18} & \mrbhl{74.77} \\
& \hfmodel{TIGER-Lab/AceCodeRM-7B}     & \iconCode  & \texttt{7B}  & \mrbhl{71.11} & \mrbhl{82.48} & \mrbhl{55.54} & \mrbhl{60.55} & \mrbhl{59.44} & \mrbhl{57.62} \\
& \hfmodel{eth-dl-rewards/internlm2-7b-reward-code-100k}          & \iconCode & \texttt{7B} & \mrbhl{70.48} & \mrbhl{79.82} & \mrbhl{56.07} & \mrbhl{59.17} & \mrbhl{64.91} & \mrbhl{60.54} \\
& \hfmodel{eth-dl-rewards/internlm2-7b-reward-math-100k}     & \iconMath & \texttt{7B}  & \mrbhl{70.30} & \mrbhl{77.75} & \mrbhl{56.15} & \mrbhl{57.79} & \mrbhl{66.18} & \mrbhl{62.97} \\
& \hfmodel{reciprocate/mistral-7b-gsm8k-code-rm}        & \iconMath & \texttt{7B} & \mrbhl{63.60} & \mrbhl{71.68} & \mrbhl{48.51} & \mrbhl{53.29} & \mrbhl{58.91} & \mrbhl{54.69} \\
& \hfmodel{nvidia/AceMath-7B-RM}     & \iconMath             & \texttt{7B}  & \mrbhl{67.41} & \mrbhl{77.40} & \mrbhl{59.28} & \mrbhl{55.63} & \mrbhl{53.24} & \mrbhl{55.50} \\
& \hfmodel{openbmb/Eurus-RM-7b}     &  & \texttt{7B}  & \mrbhl{67.37} & \mrbhl{74.09} & \mrbhl{54.93} & \mrbhl{58.13} & \mrbhl{61.44} & \mrbhl{63.07} \\
& \hfmodel{internlm/internlm2-7b-reward}     &    \iconAuxiliary\;\iconPrinciple         & \texttt{7B}  & \mrbhl{71.73} & \mrbhl{79.30} & \mrbhl{57.37} & \mrbhl{62.28} & \mrbhl{67.38} & \mrbhl{63.57} \\
\midrule
\multirow{6}{*}{\rotatebox[origin=c]{90}{\textbf{\texttt{S}}}}
& \textbf{\texttt{Themis-RM-4B}}          & \iconAuxiliary\;\iconCode\;\iconPrinciple & \texttt{4B} & \mrbhl{88.39} & \mrbhl{89.72} & \mrbhl{83.65} & \mrbhl{92.39} & \mrbhl{84.52} & \mrbhl{92.94} \\
& \hfmodel{LARK-Lab/CodeScaler-4B}     & \iconCode & \texttt{4B}  & \mrbhl{77.97} & \mrbhl{85.58} & \mrbhl{63.94} & \mrbhl{71.28} & \mrbhl{72.05} & \mrbhl{70.74} \\
& \hfmodel{PKU-ONELab/CE-RM-4B}     & \iconAuxiliary\;\iconGenerative\;\iconPrinciple\;\iconReasoning             & \texttt{4B}  & \mrbhl{57.16} & \mrbhl{66.64} & \mrbhl{52.79} & \mrbhl{39.10} & \mrbhl{44.16} & \mrbhl{42.18} \\
& \hfmodel{rubricreward/R3-Qwen3-4B-14k}         & \iconGenerative\;\iconPrinciple\;\iconReasoning & \texttt{4B} & \mrbhl{41.10} & \mrbhl{50.52} & \mrbhl{41.10} & \mrbhl{21.80} & \mrbhl{23.95} & \mrbhl{27.25} \\
& \hfmodel{Skywork/Skywork-Reward-V2-Qwen3-4B}     &  & \texttt{4B}  & \mrbhl{79.27} & \mrbhl{86.40} & \mrbhl{64.78} & \mrbhl{72.66} & \mrbhl{74.38} & \mrbhl{73.36} \\
& \hfmodel{Ray2333/GRM-llama3.2-3B-sftreg}     &  \iconAuxiliary           & \texttt{3B}  & \mrbhl{71.28} & \mrbhl{78.30} & \mrbhl{57.07} & \mrbhl{62.98} & \mrbhl{65.98} & \mrbhl{66.70} \\
\midrule
\multirow{4}{*}{\rotatebox[origin=c]{90}{\textbf{\texttt{XS}}}}
& \hfmodel{internlm/internlm2-1\_8b-reward}  & \iconAuxiliary\;\iconPrinciple & \texttt{1.8B}  & \mrbhl{63.84} & \mrbhl{67.06} & \mrbhl{52.56} & \mrbhl{62.28} & \mrbhl{64.71} & \mrbhl{62.36} \\
& \textbf{\texttt{Themis-RM-1.7B}}  & \iconAuxiliary\;\iconCode\;\iconPrinciple & \texttt{1.7B}   & \mrbhl{83.04} & \mrbhl{82.82} & \mrbhl{81.89} & \mrbhl{86.85} & \mrbhl{80.05} & \mrbhl{89.00} \\
& \hfmodel{LARK-Lab/CodeScaler-1.7B}  & \iconCode & \texttt{1.7B}  & \mrbhl{73.75} & \mrbhl{80.85} & \mrbhl{61.65} & \mrbhl{69.90} & \mrbhl{67.38} & \mrbhl{66.30} \\
& \hfmodel{Skywork/Skywork-Reward-V2-Qwen3-1.7B}  &  & \texttt{1.7B}   & \mrbhl{75.60} & \mrbhl{82.36} & \mrbhl{64.17} & \mrbhl{69.20} & \mrbhl{68.78} & \mrbhl{70.33} \\
\midrule
\multirow{2}{*}{\rotatebox[origin=c]{90}{\textbf{\texttt{XXS}}}}
& \textbf{\texttt{Themis-RM-0.6B}}   & \iconAuxiliary\;\iconCode\;\iconPrinciple            & \texttt{0.6B}  & \mrbhl{79.26} & \mrbhl{77.00} & \mrbhl{79.98} & \mrbhl{86.85} & \mrbhl{78.79} & \mrbhl{87.69} \\
&  \hfmodel{Skywork/Skywork-Reward-V2-Qwen3-0.6B}   &  & \texttt{0.6B}  & \mrbhl{72.77} & \mrbhl{78.51} & \mrbhl{62.80} & \mrbhl{62.98} & \mrbhl{69.45} & \mrbhl{66.20} \\
\midrule
\midrule
\multirow{10}{*}{\rotatebox[origin=c]{90}{\textbf{\texttt{Ablations}}}}
& \textbf{\texttt{Themis-RM-8B w/o AuxLoss}}          & \iconCode\;\iconPrinciple & \texttt{8B} & \mrbhl{87.23} & \mrbhl{89.43} & \mrbhl{81.98} & \mrbhl{88.96} & \mrbhl{82.55} & \mrbhl{90.21} \\
& \textbf{\texttt{Themis-RM-8B w/o AuxLoss And PT}}          & \iconCode\;\iconPrinciple & \texttt{8B} & \mrbhl{85.14} & \mrbhl{86.43} & \mrbhl{81.94} & \mrbhl{88.79} & \mrbhl{80.58} & \mrbhl{88.98} \\
& \textbf{\texttt{Themis-RM-8B w/ All-Criteria Prompt (Inference-Time Only)}}          & \iconAuxiliary\;\iconCode\;\iconPrinciple & \texttt{8B} & \mrbhl{88.16} & \mrbhl{90.94} & \mrbhl{81.13} & \mrbhl{92.04} & \mrbhl{82.52} & \mrbhl{91.42} \\
& \textbf{\texttt{Themis-RM-8B w/o Criteria System Prompt}}          & \iconAuxiliary\;\iconCode & \texttt{8B} & \mrbhl{87.88} & \mrbhl{90.60} & \mrbhl{80.83} & \mrbhl{91.00} & \mrbhl{82.05} & \mrbhl{91.93} \\
& \textbf{\texttt{Themis-RM-8B w/ Criteria-Level Model-Merge}}          & \iconAuxiliary\;\iconCode\;\iconPrinciple & \texttt{8B} & \mrbhl{77.07} & \mrbhl{83.58} & \mrbhl{70.03} & \mrbhl{73.18} & \mrbhl{62.36} & \mrbhl{78.41} \\
& \textbf{\texttt{Themis-RM-8B w/ FC Prefs. Only}}          & \iconAuxiliary\;\iconCode\;\iconPrinciple & \texttt{8B} & \mrbhl{78.61} & \mrbhl{90.31} & \mrbhl{58.76} & \mrbhl{70.11} & \mrbhl{62.34} & \mrbhl{75.58} \\
& \textbf{\texttt{Themis-RM-8B w/ EE Prefs. Only}}          & \iconAuxiliary\;\iconCode\;\iconPrinciple & \texttt{8B} & \mrbhl{72.67} & \mrbhl{74.91} & \mrbhl{82.67} & \mrbhl{66.48} & \mrbhl{60.23} & \mrbhl{69.32} \\
& \textbf{\texttt{Themis-RM-8B w/ ME Prefs. Only}}          & \iconAuxiliary\;\iconCode\;\iconPrinciple & \texttt{8B} & \mrbhl{67.65} & \mrbhl{69.97} & \mrbhl{67.79} & \mrbhl{83.74} & \mrbhl{56.31} & \mrbhl{68.72} \\
& \textbf{\texttt{Themis-RM-8B w/ R\&M Prefs. Only}}          & \iconAuxiliary\;\iconCode\;\iconPrinciple & \texttt{8B} & \mrbhl{72.04} & \mrbhl{76.56} & \mrbhl{60.19} & \mrbhl{70.61} & \mrbhl{71.11} & \mrbhl{67.71} \\
& \textbf{\texttt{Themis-RM-8B w/ SH Prefs. Only}}          & \iconAuxiliary\;\iconCode\;\iconPrinciple & \texttt{8B} & \mrbhl{73.35} & \mrbhl{78.67} & \mrbhl{60.09} & \mrbhl{73.12} & \mrbhl{58.77} & \mrbhl{87.32} \\
\bottomrule
\end{tabular}%
}
\scalebox{0.75}{\footnotesize
\iconAuxiliary\;\textbf{\texttt{Auxiliary Training Objectives}} \quad \iconCode\;\textbf{\texttt{Code RM}} \quad \iconPrinciple\;\textbf{\texttt{Criteria-Following RM}} \quad \iconMath\;\textbf{\texttt{Math RM}} \quad \iconGenerative\;\textbf{\texttt{Generative RM}} \quad  \iconReasoning\;\textbf{\texttt{Reasoning RM}}}
\end{table}

We further observe that generative~\citep{DBLP:journals/corr/abs-2410-12832, DBLP:conf/iclr/ZhangHBKKA25} and reasoning-enabled~\citep{DBLP:journals/corr/abs-2505-13388} RMs are unable to leverage inference-time compute effectively for pointwise reference-free code evaluation and often suffer due to the low resolution of text-based scoring. This effect is even more pronounced for generative RMs that self-propose evaluation rubrics~\citep{DBLP:journals/corr/abs-2601-20327, DBLP:conf/emnlp/SaadFalconVBNFVSKM25}. Conversely, we find that scalar reward modeling is well-suited to reference-free code evaluation. Besides the clear dominance of the \texttt{Themis-RM} models, the most competitive existing RMs in our assessment are \textit{all} scalar RMs~\citep{DBLP:journals/corr/abs-2507-01352, Zhu2026CodeScalerSC}. This, we believe, is in line with the findings that sequence-level representations are effective for both functional and non-functional code understanding~\citep{DBLP:conf/aaai/LinWYZX26, Ribeiro2025OnLR, DBLP:journals/corr/abs-2509-26476}. 
Overall, these results show that training on diverse, multi-criteria code preferences is crucial for a strong code RM. Our results indicate an above-average performance from math RMs, which confirms prior findings of positive transfer~\citep{DBLP:journals/corr/abs-2506-14965, DBLP:journals/corr/abs-2410-02229}: this transfer, however, is limited only to judging functional correctness in algorithmic code.

\subsection{\texttt{RQ2}: Training Choices To Minimize Multi-Criteria Interference In Code RMs}
\label{subsec:Experiments_RQ2}

Multi-criteria RMs require minimization of cross-criteria interference. We thus perform a series of ablations on \texttt{Themis-RM-8B} to assess the effects of components of the \texttt{Themis-RM} training recipe and to investigate how the various code evaluation criteria interact. Our results, shown at the bottom of \cref{tab:Aspect_Results}, indicate that both priming the RM with PT and including auxiliary training losses improve performance across the board. The PT phase enables RMs to learn general notions of human preference~\citep{DBLP:journals/corr/abs-2410-02229} and mitigate value biases learned in LM pre-training~\citep{DBLP:journals/corr/abs-2601-20838}. Training with auxiliary losses helps mitigate generator-validator inconsistencies~\citep{DBLP:conf/iclr/LiSLHL24}, thereby improving scoring accuracy.

Given the consistently weak performance of existing RMs on non-functional preferences (\cref{subsec:Experiments_RQ1}), we next investigate the role of criteria prompting in disentangling multi-dimensional preferences. To this end, we train a multi-task baseline model \textit{without} any criteria-defining prompts. We observe that, while learning multi-dimensional preferences via multi-task learning is viable, including explicit criteria prompts has a markedly positive effect: this is in line with results from rubric-enabled evaluators~\citep{DBLP:journals/corr/abs-2507-17746, DBLP:journals/corr/abs-2509-21500, DBLP:journals/corr/abs-2510-07743}. 
The main benefit of RM training with criterion prompts (see \S\ref{subsec:ModelTraining}), however, emerges with criterion-specific prompting at inference time, as inference with a prompt that lists all criteria only marginally improves over the multi-task baseline (88.16 vs. 87.88\%). 
We next test cross-criteria transfer, where we train using single-criteria preferences (\{\texttt{FC}, \texttt{EE}, \texttt{ME}, \texttt{R\&M}, \texttt{SH}\} \texttt{Prefs.\,Only} in \cref{tab:Aspect_Results}) and evaluate for all other criteria. The results show strong generalization of \texttt{Themis-RM}: even in zero-shot cross-criteria transfer, it outperforms most existing RMs (cf.\, the ``M'' section of \cref{tab:Aspect_Results}).        
We find that training on FC preferences transfers better to non-functional requirements than vice versa, owing to non-functional code improvements frequently coinciding with functional code changes. 
None of the criterion-specific models match the full \texttt{Themis-RM}, not even for their respective criterion: this suggests positive transfer between the quality criteria. 
Finally, we find that merging~\citep{DBLP:conf/nips/YadavTCRB23} of criterion-specific models is surprisingly ineffective, with the merged RM underperforming the multi-task baseline by over 10 points. This finding is likely due to the under-specified nature of the Bradley-Terry training objective, which can lead to individual RMs with very different reward scales (and our weak regularization is unable to mitigate this effect fully).\footnote{We thus experimented with higher values of $\mu \in \{0.1, 0.5\}$, but that overly constrained reward variance and led to lower accuracy, particularly harming the RMs' performance in listwise re-ranking~\citep{DBLP:conf/emnlp/ChenZSCZS24, DBLP:journals/corr/abs-2503-15477}.}

\def\arraystretch{1.12}
\begin{table}[!thbp]
\centering
\caption{Detailed programming-language-level preference accuracy of extant RMs and the \texttt{Themis-RM} suite on \texttt{Themis-CodeRewardBench} (\textbf{\texttt{RQ3}}: ref. to \cref{subsec:Experiments_RQ3}) followed by \texttt{Themis-RM-8B} cross-lingual transfer ablations (\textbf{\texttt{RQ3}}: ref. to \cref{subsec:Experiments_RQ3}). For detailed dataset-level scores per RM, consult \cref{tab:FC_Results,tab:EM_Results,tab:RS_Results} in \cref{appdx:4_Detailed_Results}.}
\label{tab:Language_Results}
\scalebox{0.6}{%
\begin{tabular}{@{}c l l r c cccccccc@{}}
\toprule
& \multirow{2}{*}{\texttt{\textbf{Model}}} & & \multirow{2}{*}{\texttt{\textbf{Size}}}  & \multirow{2}{*}{\texttt{\textbf{Average}}} & \multicolumn{8}{c}{\textbf{\texttt{Programming-Language-Level Accuracy}}} \\
\cmidrule(lr){6-13}
&  & &  &  & \textbf{\texttt{C}} & \textbf{\texttt{C\#}} & \textbf{\texttt{C++}} & \textbf{\texttt{Go}} & \textbf{\texttt{Java}} & \textbf{\texttt{JS}} & \textbf{\texttt{Python}} & \textbf{\texttt{Ruby}} \\
\midrule
\multirow{13}{*}{\rotatebox[origin=c]{90}{\textbf{\texttt{XL}}}}
& \hfmodel{Qwen/Qwen2.5-Math-RM-72B}         & \iconMath & \texttt{72B} & \mrbhl{69.06} & \mrbhl{74.94} & \mrbhl{60.11} & \mrbhl{70.7} & \mrbhl{75.62} & \mrbhl{71.23} & \mrbhl{66.51} & \mrbhl{65.94} & \mrbhl{68.54} \\
& \hfmodel{nvidia/AceMath-72B-RM}     & \iconMath & \texttt{72B}  & \mrbhl{72.83} & \mrbhl{77.45} & \mrbhl{61.17} & \mrbhl{77.7} & \mrbhl{78.93} & \mrbhl{76.21} & \mrbhl{68.17} & \mrbhl{69.33} & \mrbhl{70.27} \\
& \hfmodel{Qwen/WorldPM-72B-RLHFLow}     &              & \texttt{72B}  & \mrbhl{76.96} & \mrbhl{79.73} & \mrbhl{73.40} & \mrbhl{78.72} & \mrbhl{85.12} & \mrbhl{81.12} & \mrbhl{75.00} & \mrbhl{72.37} & \mrbhl{75.57} \\
& \hfmodel{ContextualAI/LMUnit-qwen2.5-72b}         & \iconAuxiliary\;\iconGenerative\;\iconPrinciple\;\iconReasoning & \texttt{72B} & \mrbhl{27.89} & \mrbhl{15.95} & \mrbhl{7.98} & \mrbhl{31.41} & \mrbhl{21.76} & \mrbhl{34.70} & \mrbhl{23.89} & \mrbhl{32.46} & \mrbhl{18.05} \\
& \hfmodel{nvidia/Llama-3.3-Nemotron-70B-Reward}          &  & \texttt{70B}  & \mrbhl{78.39} & \mrbhl{82.69} & \mrbhl{70.21} & \mrbhl{79.61} & \mrbhl{84.99} & \mrbhl{83.57} & \mrbhl{77.12} & \mrbhl{73.23} & \mrbhl{78.49} \\
& \hfmodel{infly/INF-ORM-Llama3.1-70B}         & \iconGenerative   & \texttt{70B}  & \mrbhl{74.84} & \mrbhl{76.77} & \mrbhl{67.55} & \mrbhl{74.85} & \mrbhl{79.34} & \mrbhl{78.73} & \mrbhl{73.52} & \mrbhl{72.76} & \mrbhl{72.97} \\
& \hfmodel{allenai/Llama-3.1-70B-Instruct-RM-RB2}     &             & \texttt{70B}  & \mrbhl{78.23} & \mrbhl{80.64} & \mrbhl{75.00} & \mrbhl{78.25} & \mrbhl{81.13} & \mrbhl{81.80} & \mrbhl{77.95} & \mrbhl{75.68} & \mrbhl{77.19} \\
& \hfmodel{allenai/Llama-3.1-Tulu-3-70B-SFT-RM-RB2}     &  & \texttt{70B}  & \mrbhl{78.96} & \mrbhl{79.04} & \mrbhl{75.00} & \mrbhl{77.91} & \mrbhl{83.47} & \mrbhl{83.30} & \mrbhl{79.61} & \mrbhl{75.92} & \mrbhl{78.70} \\
& \hfmodel{Nexusflow/Athene-RM-70B}     &              & \texttt{70B}  & \mrbhl{81.19} & \mrbhl{81.78} & \mrbhl{76.60} & \mrbhl{78.86} & \mrbhl{86.36} & \mrbhl{85.48} & \mrbhl{83.30} & \mrbhl{78.37} & \mrbhl{80.00} \\
& \hfmodel{Nexusflow/Starling-RM-34B}         &  & \texttt{34B} & \mrbhl{70.63} & \mrbhl{75.17} & \mrbhl{61.70} & \mrbhl{69.20} & \mrbhl{73.55} & \mrbhl{73.89} & \mrbhl{71.40} & \mrbhl{67.07} & \mrbhl{74.05} \\
& \textbf{\texttt{Themis-RM-32B}}          & \iconAuxiliary\;\iconCode\;\iconPrinciple & \texttt{32B}  & \mrbhl{91.82} & \mrbhl{92.94} & \mrbhl{94.15} & \mrbhl{89.67} & \mrbhl{93.94} & \mrbhl{95.16} & \mrbhl{91.42} & \mrbhl{90.34} & \mrbhl{91.89} \\
& \hfmodel{TIGER-Lab/AceCodeRM-32B}         & \iconCode   & \texttt{32B}  & \mrbhl{62.95} & \mrbhl{58.54} & \mrbhl{47.87} & \mrbhl{68.05} & \mrbhl{64.88} & \mrbhl{66.05} & \mrbhl{57.56} & \mrbhl{61.46} & \mrbhl{64.00} \\\
& \hfmodel{nvidia/Qwen3-Nemotron-32B-GenRM-Principle}     &   \iconGenerative\;\iconPrinciple           & \texttt{32B}  & \mrbhl{71.67} & \mrbhl{71.07} & \mrbhl{60.11} & \mrbhl{75.80} & \mrbhl{74.79} & \mrbhl{75.19} & \mrbhl{66.79} & \mrbhl{71.43} & \mrbhl{66.05} \\
\midrule
\multirow{7}{*}{\rotatebox[origin=c]{90}{\textbf{\texttt{L}}}}
& \hfmodel{nicolinho/QRM-Gemma-2-27B}         & \iconAuxiliary & \texttt{27B} & \mrbhl{52.48} & \mrbhl{52.16} & \mrbhl{46.81} & \mrbhl{52.48} & \mrbhl{51.10} & \mrbhl{54.87} & \mrbhl{51.38} & \mrbhl{53.59} & \mrbhl{49.30} \\
& \hfmodel{ShikaiChen/LDL-Reward-Gemma-2-27B-v0.1}     & \iconAuxiliary & \texttt{27B}  & \mrbhl{75.57} & \mrbhl{76.99} & \mrbhl{69.15} & \mrbhl{77.57} & \mrbhl{80.58} & \mrbhl{79.35} & \mrbhl{72.51} & \mrbhl{72.53} & \mrbhl{75.14} \\
& \hfmodel{Skywork/Skywork-Reward-Gemma-2-27B-v0.2}     &              & \texttt{27B}  & \mrbhl{54.16} & \mrbhl{57.63} & \mrbhl{50.53} & \mrbhl{54.49} & \mrbhl{56.34} & \mrbhl{53.10} & \mrbhl{52.03} & \mrbhl{54.72} & \mrbhl{53.51} \\
& \hfmodel{internlm/internlm2-20b-reward}         & \iconAuxiliary\;\iconPrinciple & \texttt{20B} & \mrbhl{73.13} & \mrbhl{77.45} & \mrbhl{70.05} & \mrbhl{74.29} & \mrbhl{77.10} & \mrbhl{78.94} & \mrbhl{72.11} & \mrbhl{69.29} & \mrbhl{69.41} \\
& \textbf{\texttt{Themis-RM-14B}}          & \iconAuxiliary\;\iconCode\;\iconPrinciple & \texttt{14B}  & \mrbhl{91.19} & \mrbhl{94.53} & \mrbhl{92.02} & \mrbhl{89.60} & \mrbhl{92.56} & \mrbhl{93.73} & \mrbhl{91.05} & \mrbhl{89.63} & \mrbhl{91.35} \\
& \hfmodel{rubricreward/R3-Qwen3-14B-14k}         & \iconGenerative\;\iconPrinciple\;\iconReasoning   & \texttt{14B}  & \mrbhl{45.41} & \mrbhl{47.84} & \mrbhl{29.26} & \mrbhl{55.61} & \mrbhl{49.45} & \mrbhl{49.69} & \mrbhl{38.93} & \mrbhl{43.76} & \mrbhl{33.51} \\
& \hfmodel{openbmb/UltraRM-13b}     &             & \texttt{13B}  & \mrbhl{70.43} & \mrbhl{68.79} & \mrbhl{74.47} & \mrbhl{68.12} & \mrbhl{72.04} & \mrbhl{74.78} & \mrbhl{74.45} & \mrbhl{66.95} & \mrbhl{70.81} \\
\midrule
\multirow{19}{*}{\rotatebox[origin=c]{90}{\textbf{\texttt{M}}}}
& \textbf{\texttt{Themis-RM-8B}}          & \iconAuxiliary\;\iconCode\;\iconPrinciple & \texttt{8B} & \mrbhl{89.78} & \mrbhl{90.21} & \mrbhl{90.43} & \mrbhl{88.24} & \mrbhl{91.60} & \mrbhl{92.43} & \mrbhl{89.94} & \mrbhl{88.82} & \mrbhl{88.76} \\
& \hfmodel{LARK-Lab/CodeScaler-8B}     & \iconCode & \texttt{8B}  & \mrbhl{79.12} & \mrbhl{84.74} & \mrbhl{78.19} & \mrbhl{76.55} & \mrbhl{84.85} & \mrbhl{83.98} & \mrbhl{78.14} & \mrbhl{76.46} & \mrbhl{77.08} \\
& \hfmodel{rubricreward/R3-Qwen3-8B-14k}         & \iconGenerative\;\iconPrinciple\;\iconReasoning & \texttt{8B} & \mrbhl{43.23} & \mrbhl{38.72} & \mrbhl{29.79} & \mrbhl{46.91} & \mrbhl{46.83} & \mrbhl{48.19} & \mrbhl{38.84} & \mrbhl{43.92} & \mrbhl{34.81} \\
& \hfmodel{Skywork/Skywork-Reward-V2-Qwen3-8B}     &              & \texttt{8B}  & \mrbhl{79.97} & \mrbhl{84.74} & \mrbhl{79.79} & \mrbhl{77.50} & \mrbhl{81.62} & \mrbhl{84.46} & \mrbhl{79.89} & \mrbhl{77.71} & \mrbhl{79.46} \\
& \hfmodel{RLHFlow/ArmoRM-Llama3-8B-v0.1}         & \iconAuxiliary & \texttt{8B} & \mrbhl{71.76} & \mrbhl{73.58} & \mrbhl{62.23} & \mrbhl{76.00} & \mrbhl{72.31} & \mrbhl{74.64} & \mrbhl{70.20} & \mrbhl{69.06} & \mrbhl{70.38} \\
& \hfmodel{nicolinho/QRM-Llama3.1-8B-v2}     & \iconAuxiliary & \texttt{8B}  & \mrbhl{71.81} & \mrbhl{75.40} & \mrbhl{68.09} & \mrbhl{71.31} & \mrbhl{72.87} & \mrbhl{76.07} & \mrbhl{71.49} & \mrbhl{70.30} & \mrbhl{68.65} \\
& \hfmodel{allenai/Llama-3.1-8B-Base-RM-RB2}     &             & \texttt{8B}  & \mrbhl{75.77} & \mrbhl{78.82} & \mrbhl{72.34} & \mrbhl{74.17} & \mrbhl{79.06} & \mrbhl{78.94} & \mrbhl{77.40} & \mrbhl{78.31} & \mrbhl{73.51} \\
& \hfmodel{Ray2333/GRM-llama3-8B-sftreg}          & \iconAuxiliary & \texttt{8B} & \mrbhl{71.81} & \mrbhl{71.30} & \mrbhl{68.09} & \mrbhl{73.96} & \mrbhl{72.59} & \mrbhl{75.94} & \mrbhl{71.86} & \mrbhl{69.33} & \mrbhl{69.08} \\
& \hfmodel{LxzGordon/URM-LLaMa-3.1-8B}     & \iconAuxiliary & \texttt{8B}  & \mrbhl{71.62} & \mrbhl{76.08} & \mrbhl{73.62} & \mrbhl{65.96} & \mrbhl{73.00} & \mrbhl{74.98} & \mrbhl{71.96} & \mrbhl{68.32} & \mrbhl{69.84} \\
& \hfmodel{NCSOFT/Llama-3-OffsetBias-RM-8B}     &              & \texttt{8B}  & \mrbhl{71.76} & \mrbhl{71.75} & \mrbhl{63.83} & \mrbhl{74.10} & \mrbhl{76.03} & \mrbhl{75.53} & \mrbhl{73.25} & \mrbhl{67.81} & \mrbhl{69.51} \\
& \hfmodel{sfairXC/FsfairX-LLaMA3-RM-v0.1}         &  & \texttt{8B} & \mrbhl{73.52} & \mrbhl{77.45} & \mrbhl{71.28} & \mrbhl{74.44} & \mrbhl{75.90} & \mrbhl{77.03} & \mrbhl{75.18} & \mrbhl{69.60} & \mrbhl{72.11} \\
& \hfmodel{Nexusflow/Athene-RM-8B}     &  & \texttt{8B}  & \mrbhl{76.58} & \mrbhl{77.90} & \mrbhl{77.66} & \mrbhl{74.03} & \mrbhl{79.06} & \mrbhl{80.44} & \mrbhl{80.26} & \mrbhl{74.12} & \mrbhl{74.27} \\
& \hfmodel{TIGER-Lab/AceCodeRM-7B}     & \iconCode  & \texttt{7B}  & \mrbhl{71.11} & \mrbhl{76.77} & \mrbhl{62.77} & \mrbhl{72.47} & \mrbhl{75.90} & \mrbhl{77.78} & \mrbhl{68.54} & \mrbhl{67.23} & \mrbhl{67.46} \\
& \hfmodel{eth-dl-rewards/internlm2-7b-reward-code-100k}          & \iconCode & \texttt{7B} & \mrbhl{70.48} & \mrbhl{70.84} & \mrbhl{67.02} & \mrbhl{72.60} & \mrbhl{76.45} & \mrbhl{75.12} & \mrbhl{72.42} & \mrbhl{66.45} & \mrbhl{64.54} \\
& \hfmodel{eth-dl-rewards/internlm2-7b-reward-math-100k}     & \iconMath & \texttt{7B}  & \mrbhl{70.30} & \mrbhl{70.62} & \mrbhl{69.68} & \mrbhl{70.84} & \mrbhl{77.00} & \mrbhl{74.71} & \mrbhl{73.71} & \mrbhl{65.28} & \mrbhl{67.14} \\
& \hfmodel{reciprocate/mistral-7b-gsm8k-code-rm}        & \iconMath & \texttt{7B} & \mrbhl{63.60} & \mrbhl{63.78} & \mrbhl{63.30} & \mrbhl{65.87} & \mrbhl{65.15} & \mrbhl{67.83} & \mrbhl{63.84} & \mrbhl{59.31} & \mrbhl{63.68} \\
& \hfmodel{nvidia/AceMath-7B-RM}     & \iconMath             & \texttt{7B}  & \mrbhl{67.41} & \mrbhl{68.56} & \mrbhl{53.72} & \mrbhl{74.58} & \mrbhl{71.49} & \mrbhl{70.07} & \mrbhl{62.64} & \mrbhl{65.98} & \mrbhl{60.43} \\
& \hfmodel{openbmb/Eurus-RM-7b}     &  & \texttt{7B}  & \mrbhl{67.37} & \mrbhl{67.88} & \mrbhl{68.09} & \mrbhl{68.52} & \mrbhl{69.56} & \mrbhl{72.67} & \mrbhl{65.68} & \mrbhl{63.37} & \mrbhl{68.11} \\
& \hfmodel{internlm/internlm2-7b-reward}     &     \iconAuxiliary\;\iconPrinciple        & \texttt{7B}  & \mrbhl{71.73} & \mrbhl{74.03} & \mrbhl{68.62} & \mrbhl{72.13} & \mrbhl{78.37} & \mrbhl{75.39} & \mrbhl{74.72} & \mrbhl{66.91} & \mrbhl{69.51} \\
\midrule
\multirow{6}{*}{\rotatebox[origin=c]{90}{\textbf{\texttt{S}}}}
& \textbf{\texttt{Themis-RM-4B}}          & \iconAuxiliary\;\iconCode\;\iconPrinciple & \texttt{4B} &  \mrbhl{88.39} & \mrbhl{86.79} & \mrbhl{88.83} & \mrbhl{88.17} & \mrbhl{90.36} & \mrbhl{90.66} & \mrbhl{90.50} & \mrbhl{86.94} & \mrbhl{85.84} \\
& \hfmodel{LARK-Lab/CodeScaler-4B}     & \iconCode & \texttt{4B}  & \mrbhl{77.97} & \mrbhl{81.78} & \mrbhl{71.28} & \mrbhl{77.63} & \mrbhl{83.20} & \mrbhl{82.75} & \mrbhl{78.97} & \mrbhl{74.36} & \mrbhl{74.27} \\
& \hfmodel{PKU-ONELab/CE-RM-4B}     & \iconAuxiliary\;\iconGenerative\;\iconPrinciple\;\iconReasoning             & \texttt{4B}  & \mrbhl{57.16} & \mrbhl{57.63} & \mrbhl{43.09} & \mrbhl{64.58} & \mrbhl{58.40} & \mrbhl{61.76} & \mrbhl{53.84} & \mrbhl{55.53} & \mrbhl{49.08} \\
& \hfmodel{rubricreward/R3-Qwen3-4B-14k}         & \iconGenerative\;\iconPrinciple\;\iconReasoning & \texttt{4B} & \mrbhl{41.10} & \mrbhl{36.67} & \mrbhl{22.87} & \mrbhl{48.33} & \mrbhl{42.56} & \mrbhl{46.08} & \mrbhl{32.93} & \mrbhl{40.84} & \mrbhl{36.65} \\
& \hfmodel{Skywork/Skywork-Reward-V2-Qwen3-4B}     &  & \texttt{4B}  & \mrbhl{79.27} & \mrbhl{81.09} & \mrbhl{81.91} & \mrbhl{78.72} & \mrbhl{84.85} & \mrbhl{83.71} & \mrbhl{79.70} & \mrbhl{76.23} & \mrbhl{75.24} \\
& \hfmodel{Ray2333/GRM-llama3.2-3B-sftreg}     &  \iconAuxiliary           & \texttt{3B}  & \mrbhl{71.28} & \mrbhl{74.72} & \mrbhl{60.11} & \mrbhl{71.72} & \mrbhl{73.55} & \mrbhl{75.26} & \mrbhl{71.68} & \mrbhl{69.21} & \mrbhl{68.43} \\
\midrule
\multirow{4}{*}{\rotatebox[origin=c]{90}{\textbf{\texttt{XS}}}}
& \hfmodel{internlm/internlm2-1\_8b-reward}  & \iconAuxiliary\;\iconPrinciple & \texttt{1.8B}  & \mrbhl{63.84} & \mrbhl{64.24} & \mrbhl{59.57} & \mrbhl{63.77} & \mrbhl{69.70} & \mrbhl{64.89} & \mrbhl{66.70} & \mrbhl{61.57} & \mrbhl{61.30} \\
& \textbf{\texttt{Themis-RM-1.7B}}  & \iconAuxiliary\;\iconCode\;\iconPrinciple & \texttt{1.7B}   & \mrbhl{83.04} & \mrbhl{80.64} & \mrbhl{82.45} & \mrbhl{83.82} & \mrbhl{84.99} & \mrbhl{84.66} & \mrbhl{83.49} & \mrbhl{81.80} & \mrbhl{81.84} \\
& \hfmodel{LARK-Lab/CodeScaler-1.7B}  & \iconCode & \texttt{1.7B}  & \mrbhl{73.75} & \mrbhl{76.31} & \mrbhl{75.00} & \mrbhl{73.01} & \mrbhl{81.27} & \mrbhl{77.30} & \mrbhl{73.34} & \mrbhl{71.47} & \mrbhl{68.76} \\
& \hfmodel{Skywork/Skywork-Reward-V2-Qwen3-1.7B}  &  & \texttt{1.7B}   & \mrbhl{75.60} & \mrbhl{76.54} & \mrbhl{77.13} & \mrbhl{74.92} & \mrbhl{80.30} & \mrbhl{78.19} & \mrbhl{75.83} & \mrbhl{73.85} & \mrbhl{72.76} \\
\midrule
\multirow{2}{*}{\rotatebox[origin=c]{90}{\textbf{\texttt{XXS}}}}
& \textbf{\texttt{Themis-RM-0.6B}}   & \iconAuxiliary\;\iconCode\;\iconPrinciple            & \texttt{0.6B}  & \mrbhl{79.26} & \mrbhl{79.73} & \mrbhl{77.63} & \mrbhl{78.31} & \mrbhl{75.90} & \mrbhl{82.07} & \mrbhl{80.07} & \mrbhl{79.35} & \mrbhl{77.84} \\
&  \hfmodel{Skywork/Skywork-Reward-V2-Qwen3-0.6B}   &  & \texttt{0.6B}  & \mrbhl{72.77} & \mrbhl{74.26} & \mrbhl{72.87} & \mrbhl{72.26} & \mrbhl{73.97} & \mrbhl{76.01} & \mrbhl{73.52} & \mrbhl{71.63} & \mrbhl{68.08} \\
\midrule
\midrule
\multirow{4}{*}{\rotatebox[origin=c]{90}{\textbf{\texttt{Ablations}}}}
& \textbf{\texttt{Themis-RM-8B w/ 50k All Language Prefs. Only}}          & \iconAuxiliary\;\iconCode\;\iconPrinciple & \texttt{8B} & \mrbhl{85.31} & \mrbhl{84.63} & \mrbhl{82.91} & \mrbhl{83.57} & \mrbhl{87.95} & \mrbhl{87.11} & \mrbhl{83.58} & \mrbhl{85.24} & \mrbhl{86.19} \\
& \textbf{\texttt{Themis-RM-8B w/ 50k High Resource Prefs. Only}}          & \iconAuxiliary\;\iconCode\;\iconPrinciple & \texttt{8B} & \mrbhl{84.87} & \mrbhl{82.46} & \mrbhl{81.04} & \mrbhl{82.25} & \mrbhl{83.84} & \mrbhl{88.42} & \mrbhl{85.79} & \mrbhl{85.29} & \mrbhl{83.94} \\
& \textbf{\texttt{Themis-RM-8B w/ 50k Python Prefs. Only}}          & \iconAuxiliary\;\iconCode\;\iconPrinciple & \texttt{8B} & \mrbhl{78.62} & \mrbhl{72.09} & \mrbhl{69.52} & \mrbhl{69.75} & \mrbhl{72.56} & \mrbhl{73.51} & \mrbhl{83.44} & \mrbhl{86.01} & \mrbhl{84.43} \\
& \textbf{\texttt{Themis-RM-8B w/ 50k Java Prefs. Only}}          & \iconAuxiliary\;\iconCode\;\iconPrinciple & \texttt{8B} & \mrbhl{81.42} & \mrbhl{80.32} & \mrbhl{78.51} & \mrbhl{79.55} & \mrbhl{75.36} & \mrbhl{87.91} & \mrbhl{82.11} & \mrbhl{81.40} & \mrbhl{79.31} \\
\bottomrule
\end{tabular}%
}
\scalebox{0.75}{\footnotesize
\iconAuxiliary\;\textbf{\texttt{Auxiliary Training Objectives}} \quad \iconCode\;\textbf{\texttt{Code RM}} \quad \iconPrinciple\;\textbf{\texttt{Criteria-Following RM}} \quad \iconMath\;\textbf{\texttt{Math RM}} \quad \iconGenerative\;\textbf{\texttt{Generative RM}} \quad \iconReasoning\;\textbf{\texttt{Reasoning RM}}}
\end{table}

\subsection{\texttt{RQ3}: Cross-Lingual Transfer In Code RMs}
\label{subsec:Experiments_RQ3}

We next examine the multilingual and cross-lingual characteristics of code RMs. Multilingual interference is a well-documented phenomenon~\citep{DBLP:conf/emnlp/WangLT20, DBLP:conf/acl/ConneauKGCWGGOZ20, DBLP:journals/corr/abs-1907-05019, DBLP:journals/corr/abs-2005-00633}, but prior work on RMs has largely focused on natural language, with mixed findings ranging from strong interference~\citep{DBLP:conf/emnlp/LaiNNNDRN23, DBLP:conf/acl/GurejaMIMSW0RHF25} to positive transfer~\citep{DBLP:conf/emnlp/WuBKEB24, DBLP:conf/naacl/HongLMRT25, DBLP:conf/emnlp/DangAMKUH24}. 
The question is arguably even more pertinent for code LMs, which often exhibit large performance disparities even across high-resource programming languages~\citep{DBLP:conf/iclr/AthiwaratkunGWL23, DBLP:conf/acl/PaulGG24, DBLP:journals/corr/abs-2504-08703}. 
We first compare per-programming-language accuracy of \texttt{Themis-RM} with that of existing RMs on \texttt{Themis-CodeRewardBench} in \cref{tab:Language_Results}. Barring the rare poor performance of some RMs in a few specific languages (attributable to their respective training distributions), we find that most RMs exhibit modest performance differences between high- and medium-resource programming languages. 
We next perform controlled cross-lingual transfer experiments with \texttt{Themis-RM-8B} (see bottom of \cref{tab:Language_Results}), mirroring the setup presented in~\citet{DBLP:conf/emnlp/DangAMKUH24}. We run size-controlled comparisons between all-language, high-resource-language, and Python- and Java-only training runs. We observe the following transfer patterns. Python RMs transfer better to other dynamically typed languages, whereas Java RMs transfer better to statically typed ones. Training on all languages yields the best performance, suggesting net-positive cross-lingual transfer effects. Interestingly, all four models show small performance differences across languages, suggesting that training code RMs on diverse, multi-criteria preferences yields stable multilingual reward modeling. We believe this is due to (1) the extensive oversampling of multilingual code data in the training of modern LMs~\citep{DBLP:journals/corr/abs-2602-12237, DBLP:conf/iclr/LiuZMZDPJL25} and (2) 
situational coverage in high-resource languages being more beneficial than data volume in low-resource ones~\citep{DBLP:journals/corr/abs-2409-11239}. 

\subsection{\texttt{RQ4}: Downstream Robustness Evaluation}
\label{subsec:Experiments_RQ4}

\def\arraystretch{1.12}
\begin{table}[!htbp]
\centering
\caption{Re-ranking and adversarial robustness evaluation of the \texttt{Themis-RM} suite (\textbf{\texttt{RQ4}}: ref. to \cref{subsec:Experiments_RQ4}). We benchmark against a selection of the highest scoring extant RMs on \texttt{Themis-CodeRewardBench}.}
\label{tab:Robustness_Results}
\scalebox{0.6}{%
\begin{tabular}{@{}c l l r ccccccccc@{}}
\toprule
& \multirow{3}{*}{\texttt{\textbf{Model}}} & & \multirow{3}{*}{\texttt{\textbf{Size}}}  & \multicolumn{6}{c}{\textbf{\texttt{Re-Ranking}}} & \multicolumn{3}{c}{\textbf{\texttt{Adversarial}}} \\
\cmidrule(lr){5-10} \cmidrule(lr){11-13}
&  &  &  & \multicolumn{3}{c}{\textbf{\texttt{Hits@10 (Mean)}}} & \multicolumn{3}{c}{\textbf{\texttt{Rank Corr.@40 (Median)}}} & \multicolumn{3}{c}{\textbf{\texttt{Accuracy}}} \\
\cmidrule(lr){5-7} \cmidrule(lr){8-10} \cmidrule(lr){11-13}
&  & &  & \textbf{\texttt{C++}} & \textbf{\texttt{Java}} & \textbf{\texttt{Python}} & \textbf{\texttt{C++}} & \textbf{\texttt{Java}} & \textbf{\texttt{Python}} & \textbf{\texttt{C++}} & \textbf{\texttt{Java}} & \textbf{\texttt{Python}} \\
\midrule
\multirow{5}{*}{\rotatebox[origin=c]{90}{\textbf{\texttt{XL}}}}
& \hfmodel{Qwen/WorldPM-72B-RLHFLow}     &              & \texttt{72B}  & \mrbhl{43.12} & \mrbhl{40.14} & \mrbhl{45.76} & \mrbhlf{0.0591} & \mrbhlf{0.0656} & \mrbhlf{0.1644} & \mrbhl{57.88} & \mrbhl{56.45} & \mrbhl{59.88} \\
& \hfmodel{nvidia/Llama-3.3-Nemotron-70B-Reward}          &  & \texttt{70B}  & \mrbhl{71.91} & \mrbhl{73.14} & \mrbhl{78.59} & \mrbhlf{0.2568} & \mrbhlf{0.2644} & \mrbhlf{0.3012} & \mrbhl{63.48} & \mrbhl{62.12} & \mrbhl{55.91} \\
& \hfmodel{Nexusflow/Athene-RM-70B}     &              & \texttt{70B}  & \mrbhl{24.88} & \mrbhl{27.78} & \mrbhl{29.34} & \mrbhlf{-0.0567} & \mrbhlf{-0.0245} & \mrbhlf{0.1281} & \mrbhl{53.19} & \mrbhl{52.44} & \mrbhl{54.47} \\
& \textbf{\texttt{Themis-RM-32B}}          & \iconAuxiliary\;\iconCode\;\iconPrinciple & \texttt{32B} & \mrbhl{97.65} & \mrbhl{98.59} & \mrbhl{97.44} & \mrbhlf{0.5067} & \mrbhlf{0.5352} & \mrbhlf{0.5018} & \mrbhl{81.43} & \mrbhl{82.09} & \mrbhl{83.02} \\
& \hfmodel{TIGER-Lab/AceCodeRM-32B}         & \iconCode   & \texttt{32B}  & \mrbhl{21.76} & \mrbhl{19.23} & \mrbhl{22.34} & \mrbhlf{0.0256} & \mrbhlf{0.067} & \mrbhlf{0.3415} & \mrbhl{62.41} & \mrbhl{65.33} & \mrbhl{64.76} \\
\midrule
\multirow{1}{*}{\rotatebox[origin=c]{90}{\textbf{\texttt{L}}}}
& \textbf{\texttt{Themis-RM-14B}}          & \iconAuxiliary\;\iconCode\;\iconPrinciple & \texttt{14B} & \mrbhl{95.88} & \mrbhl{95.91} & \mrbhl{96.84} & \mrbhlf{0.4649} & \mrbhlf{0.4711} & \mrbhlf{0.4619} & \mrbhl{76.48} & \mrbhl{77.56} & \mrbhl{79.29} \\
\midrule
\multirow{5}{*}{\rotatebox[origin=c]{90}{\textbf{\texttt{M}}}}
& \textbf{\texttt{Themis-RM-8B}}          & \iconAuxiliary\;\iconCode\;\iconPrinciple & \texttt{8B} & \mrbhl{95.29} & \mrbhl{94.75} & \mrbhl{96.26} & \mrbhlf{0.4448} & \mrbhlf{0.4404} & \mrbhlf{0.4387} & \mrbhl{72.04} & \mrbhl{73.48} & \mrbhl{74.54} \\
& \hfmodel{LARK-Lab/CodeScaler-8B}     & \iconCode & \texttt{8B}  & \mrbhl{94.65} & \mrbhl{94.87} & \mrbhl{96.43} & \mrbhlf{0.4431} & \mrbhlf{0.4486} & \mrbhlf{0.4403} & \mrbhl{73.28} & \mrbhl{71.7} & \mrbhl{75.43} \\
& \hfmodel{Skywork/Skywork-Reward-V2-Qwen3-8B}     &              & \texttt{8B}  & \mrbhl{91.45} & \mrbhl{90.45} & \mrbhl{91.41} & \mrbhlf{0.3765} & \mrbhlf{0.3734} & \mrbhlf{0.3922} & \mrbhl{56.81} & \mrbhl{56.78} & \mrbhl{57.65} \\
& \hfmodel{Nexusflow/Athene-RM-8B}     &  & \texttt{8B}  & \mrbhl{33.97} & \mrbhl{34.82} & \mrbhl{34.98} & \mrbhlf{0.3492} & \mrbhlf{0.3522} & \mrbhlf{0.3411} & \mrbhl{50.89} & \mrbhl{44.5} & \mrbhl{47.04} \\
& \hfmodel{TIGER-Lab/AceCodeRM-7B}     & \iconCode  & \texttt{7B}  & \mrbhl{34.21} & \mrbhl{32.09} & \mrbhl{35.46} & \mrbhlf{-0.0426} & \mrbhlf{0.0398} & \mrbhlf{-0.0131} & \mrbhl{62.21} & \mrbhl{63.14} & \mrbhl{63.02} \\
\midrule
\multirow{3}{*}{\rotatebox[origin=c]{90}{\textbf{\texttt{S}}}}
& \textbf{\texttt{Themis-RM-4B}}          & \iconAuxiliary\;\iconCode\;\iconPrinciple & \texttt{4B} & \mrbhl{93.17} & \mrbhl{93.45} & \mrbhl{95.1} & \mrbhlf{0.4148} & \mrbhlf{0.4192} & \mrbhlf{0.4197} & \mrbhl{68.73} & \mrbhl{71.18} & \mrbhl{69.45} \\
& \hfmodel{LARK-Lab/CodeScaler-4B}     & \iconCode & \texttt{4B}  & \mrbhl{93.33} & \mrbhl{93.11} & \mrbhl{93.78} & \mrbhlf{0.4255} & \mrbhlf{0.4214} & \mrbhlf{0.4216} & \mrbhl{71.33} & \mrbhl{71.06} & \mrbhl{74.19} \\
& \hfmodel{Skywork/Skywork-Reward-V2-Qwen3-4B}     &  & \texttt{4B}  & \mrbhl{89.45} & \mrbhl{88.32} & \mrbhl{90.19} & \mrbhlf{0.3677} & \mrbhlf{0.3594} & \mrbhlf{0.3687} & \mrbhl{55.96} & \mrbhl{54.76} & \mrbhl{57.16} \\
\midrule
\multirow{3}{*}{\rotatebox[origin=c]{90}{\textbf{\texttt{XS}}}}
& \textbf{\texttt{Themis-RM-1.7B}}  & \iconAuxiliary\;\iconCode\;\iconPrinciple & \texttt{1.7B} & \mrbhl{89.45} & \mrbhl{88.98} & \mrbhl{90.77} & \mrbhlf{0.3978} & \mrbhlf{0.4067} & \mrbhlf{0.4092} & \mrbhl{61.39} & \mrbhl{56.92} & \mrbhl{64.26} \\
& \hfmodel{LARK-Lab/CodeScaler-1.7B}  & \iconCode & \texttt{1.7B} & \mrbhl{90.86} & \mrbhl{89.85} & \mrbhl{91.72} & \mrbhlf{0.4046} & \mrbhlf{0.4095} & \mrbhlf{0.4113} & \mrbhl{63.06} & \mrbhl{60.54} & \mrbhl{64.97} \\
& \hfmodel{Skywork/Skywork-Reward-V2-Qwen3-1.7B}  &  & \texttt{1.7B}   & \mrbhl{84.52} & \mrbhl{82.89} & \mrbhl{84.54} & \mrbhlf{0.2219} & \mrbhlf{0.2106} & \mrbhlf{0.2384} & \mrbhl{51.12} & \mrbhl{49.69} & \mrbhl{52.69} \\
\midrule
\multirow{2}{*}{\rotatebox[origin=c]{90}{\textbf{\texttt{XXS}}}}
& \textbf{\texttt{Themis-RM-0.6B}}   & \iconAuxiliary\;\iconCode\;\iconPrinciple            & \texttt{0.6B} & \mrbhl{84.65} & \mrbhl{83.98} & \mrbhl{86.76} & \mrbhlf{0.1769} & \mrbhlf{0.1878} & \mrbhlf{0.2171} & \mrbhl{53.11} & \mrbhl{50.16} & \mrbhl{53.14} \\
&  \hfmodel{Skywork/Skywork-Reward-V2-Qwen3-0.6B}   &  & \texttt{0.6B} & \mrbhl{79.53} & \mrbhl{78.55} & \mrbhl{79.87} & \mrbhlf{0.0971} & \mrbhlf{0.1042} & \mrbhlf{0.1065} & \mrbhl{47.07} & \mrbhl{42.69} & \mrbhl{44.26} \\
\bottomrule
\end{tabular}%
}
\scalebox{0.75}{\footnotesize
\iconAuxiliary\;\textbf{\texttt{Auxiliary Training Objectives}} \quad \iconCode\;\textbf{\texttt{Code RM}} \quad \iconPrinciple\;\textbf{\texttt{Criteria-Following RM}} \quad \iconMath\;\textbf{\texttt{Math RM}} \quad \iconGenerative\;\textbf{\texttt{Generative RM}} \quad \iconReasoning\;\textbf{\texttt{Reasoning RM}}}
\end{table}

The most prevalent downstream applications of RMs, such as post-training~\citep{DBLP:journals/corr/abs-2501-12948, DBLP:journals/tmlr/Dong0GZCPDZS023, DBLP:journals/tmlr/SinghCAAPGLH0XP24, DBLP:journals/corr/abs-1910-00177} and score-guided inference-time search~\citep{DBLP:conf/acl/WangLSXDLCWS24, DBLP:journals/corr/abs-2504-16084} demand accurate listwise re-ranking of LM responses. Lists of model-generated responses can typically contain model-specific artifacts~\citep{DBLP:journals/corr/abs-2507-05197, DBLP:conf/emnlp/OrelPGN25} and attempts at verifier-hacking~\citep{DBLP:conf/eacl/MoonHLKKJ26}, to which the RMs with evaluation criteria prompts are more susceptible~\citep{DBLP:journals/corr/abs-2509-03419}. To verify the downstream effectiveness of \texttt{Themis-RM}, we directly evaluate its listwise re-ranking performance and adversarial preference accuracy on code. 

For listwise re-ranking, we source \texttt{40} solutions each in \texttt{C++}, \texttt{Java}, and \texttt{Python} to problems in \texttt{CodeContests+}~\citep{DBLP:conf/emnlp/WangLSDL25} from a mix of open-source LMs~\citep{DBLP:journals/corr/abs-2505-09388, DBLP:journals/corr/abs-2408-00118, DBLP:journals/corr/abs-2308-12950, DBLP:journals/corr/abs-2401-14196}. We then execute these solutions against the predefined test-cases using a code sandbox\footnote{\gitrepo{bytedance/sandboxfusion}}, and observe how well RMs can retrieve completely correct solutions (\texttt{Hits@10}) and how well their listwise scores correlate with the ground truth (\texttt{Rank Corr.@40}). We then evaluate robustness to adversarial perturbations using accuracy on the \texttt{Aletheia-Adv}~\citep{DBLP:journals/corr/abs-2601-12186} suite, which tests RM accuracy on correct-incorrect code pairs modified using a variety of judge-hacking behaviors~\citep{DBLP:journals/corr/abs-2504-09946, DBLP:journals/corr/abs-2504-14119}. 
Overall, the results (in \cref{tab:Robustness_Results}) demonstrate the competitiveness of \texttt{Themis-RM} models in realistic downstream scenarios. Our models (i) achieve state-of-the-art adversarial robustness and (ii) match the performance of the best RMs specialized for correctness of contest-type code in re-ranking, a task proven to predict downstream post-training utility~\citep{DBLP:journals/corr/abs-2512-16041, DBLP:conf/iclr/WenL0LXLHHZ025, DBLP:conf/acl/KimKKC0Y25}. 
Most encouragingly, and in contrast to prior work~\citep{DBLP:conf/icml/BartoldsonDPK24}, we find that \texttt{Themis-RM} exhibits better scaling trends in downstream robustness than in our pairwise accuracy tests.

\section{Conclusion}
\label{sec:Conclusion}

We present the \texttt{Themis} project, an investigation into the viability of model-based scoring of LM-generated code for multi-criteria alignment, quality estimation, and re-ranking. We first construct \texttt{Themis-CodeRewardBench}, an exhaustive preference evaluation suite that tests RMs' scoring competence across eight typologically diverse programming languages and their ability to score code preferences across different functional and non-functional criteria accurately. Subsequently, we outline the creation of \texttt{Themis-GeneralPreference} and \texttt{Themis-CodePreference}, two diverse multi-dimensional preference datasets that we use to train \texttt{Themis-RM}, a suite of state-of-the-art multilingual multi-criteria code RMs. Our opening evaluations demonstrate that current RMs struggle to model code preferences beyond a very narrow mode of operation, namely, judging the functional correctness of code contest solutions. In subsequent experiments, we illustrate that code RMs trained on a variety of code preferences and evaluation criteria exhibit clear-cut positive scaling trends with minimal cross-criteria interference, non-trivial cross-lingual transfer, and robust performance in adversarial settings, establishing them as a viable reward source for code post-training. We hope that our findings, coupled with open-source artifacts, will expand the use of RMs in post-training beyond executable code.

\deanon{%
\subsubsection*{Acknowledgments}

We gratefully acknowledge support from the hessian.AI Service Center (funded by the German Federal Ministry of Research, Technology, and Space, BMFTR, grant no. 16IS22091) and the hessian.AI Innovation Lab (funded by the Hessian Ministry for Digital Strategy and Innovation, grant no. S-DIW04/0013/003). Additionally, our work is supported by the German Federal Ministry of Education and Research and the Hessian Ministry of Higher Education, Research, Science and the Arts within their joint support of the National Research Center for Applied Cybersecurity ATHENE.%
}

\bibliography{tmlr}

@inproceedings{DBLP:conf/nips/Ouyang0JAWMZASR22,
  author       = {Long Ouyang and
                  Jeffrey Wu and
                  Xu Jiang and
                  Diogo Almeida and
                  Carroll L. Wainwright and
                  Pamela Mishkin and
                  Chong Zhang and
                  Sandhini Agarwal and
                  Katarina Slama and
                  Alex Ray and
                  John Schulman and
                  Jacob Hilton and
                  Fraser Kelton and
                  Luke Miller and
                  Maddie Simens and
                  Amanda Askell and
                  Peter Welinder and
                  Paul F. Christiano and
                  Jan Leike and
                  Ryan Lowe},
  editor       = {Sanmi Koyejo and
                  S. Mohamed and
                  A. Agarwal and
                  Danielle Belgrave and
                  K. Cho and
                  A. Oh},
  title        = {Training language models to follow instructions with human feedback},
  booktitle    = {Advances in Neural Information Processing Systems 35: Annual Conference
                  on Neural Information Processing Systems 2022, NeurIPS 2022, New Orleans,
                  LA, USA, November 28 - December 9, 2022},
  year         = {2022},
  url          = {http://papers.nips.cc/paper\_files/paper/2022/hash/b1efde53be364a73914f58805a001731-Abstract-Conference.html},
  bibsource    = {dblp computer science bibliography, https://dblp.org}
}

@article{DBLP:journals/corr/abs-2112-00861,
  author       = {Amanda Askell and
                  Yuntao Bai and
                  Anna Chen and
                  Dawn Drain and
                  Deep Ganguli and
                  Tom Henighan and
                  Andy Jones and
                  Nicholas Joseph and
                  Benjamin Mann and
                  Nova DasSarma and
                  Nelson Elhage and
                  Zac Hatfield{-}Dodds and
                  Danny Hernandez and
                  Jackson Kernion and
                  Kamal Ndousse and
                  Catherine Olsson and
                  Dario Amodei and
                  Tom B. Brown and
                  Jack Clark and
                  Sam McCandlish and
                  Chris Olah and
                  Jared Kaplan},
  title        = {A General Language Assistant as a Laboratory for Alignment},
  journal      = {CoRR},
  volume       = {abs/2112.00861},
  year         = {2021},
  url          = {https://arxiv.org/abs/2112.00861},
  eprinttype    = {arXiv},
  eprint       = {2112.00861},
  bibsource    = {dblp computer science bibliography, https://dblp.org}
}

@article{DBLP:journals/corr/abs-1909-08593,
  author       = {Daniel M. Ziegler and
                  Nisan Stiennon and
                  Jeffrey Wu and
                  Tom B. Brown and
                  Alec Radford and
                  Dario Amodei and
                  Paul F. Christiano and
                  Geoffrey Irving},
  title        = {Fine-Tuning Language Models from Human Preferences},
  journal      = {CoRR},
  volume       = {abs/1909.08593},
  year         = {2019},
  url          = {http://arxiv.org/abs/1909.08593},
  eprinttype    = {arXiv},
  eprint       = {1909.08593},
  bibsource    = {dblp computer science bibliography, https://dblp.org}
}

@inproceedings{DBLP:conf/acl/AhmadianCGFKPUH24,
  author       = {Arash Ahmadian and
                  Chris Cremer and
                  Matthias Gall{\'{e}} and
                  Marzieh Fadaee and
                  Julia Kreutzer and
                  Olivier Pietquin and
                  Ahmet {\"{U}}st{\"{u}}n and
                  Sara Hooker},
  editor       = {Lun{-}Wei Ku and
                  Andre Martins and
                  Vivek Srikumar},
  title        = {Back to Basics: Revisiting REINFORCE-Style Optimization for Learning
                  from Human Feedback in LLMs},
  booktitle    = {Proceedings of the 62nd Annual Meeting of the Association for Computational
                  Linguistics (Volume 1: Long Papers), {ACL} 2024, Bangkok, Thailand,
                  August 11-16, 2024},
  pages        = {12248--12267},
  publisher    = {Association for Computational Linguistics},
  year         = {2024},
  url          = {https://doi.org/10.18653/v1/2024.acl-long.662},
  doi          = {10.18653/V1/2024.ACL-LONG.662},
  bibsource    = {dblp computer science bibliography, https://dblp.org}
}

@article{DBLP:journals/corr/abs-2402-04792,
  author       = {Shangmin Guo and
                  Biao Zhang and
                  Tianlin Liu and
                  Tianqi Liu and
                  Misha Khalman and
                  Felipe Llinares and
                  Alexandre Ram{\'{e}} and
                  Thomas Mesnard and
                  Yao Zhao and
                  Bilal Piot and
                  Johan Ferret and
                  Mathieu Blondel},
  title        = {Direct Language Model Alignment from Online {AI} Feedback},
  journal      = {CoRR},
  volume       = {abs/2402.04792},
  year         = {2024},
  url          = {https://doi.org/10.48550/arXiv.2402.04792},
  doi          = {10.48550/ARXIV.2402.04792},
  eprinttype    = {arXiv},
  eprint       = {2402.04792},
  bibsource    = {dblp computer science bibliography, https://dblp.org}
}

@inproceedings{DBLP:conf/nips/0001SSB024,
  author       = {Yuda Song and
                  Gokul Swamy and
                  Aarti Singh and
                  J. Andrew Bagnell and
                  Wen Sun},
  editor       = {Amir Globersons and
                  Lester Mackey and
                  Danielle Belgrave and
                  Angela Fan and
                  Ulrich Paquet and
                  Jakub M. Tomczak and
                  Cheng Zhang},
  title        = {The Importance of Online Data: Understanding Preference Fine-tuning
                  via Coverage},
  booktitle    = {Advances in Neural Information Processing Systems 38: Annual Conference
                  on Neural Information Processing Systems 2024, NeurIPS 2024, Vancouver,
                  BC, Canada, December 10 - 15, 2024},
  year         = {2024},
  url          = {http://papers.nips.cc/paper\_files/paper/2024/hash/16c628ab12dc4caca8e7712affa6c767-Abstract-Conference.html},
  bibsource    = {dblp computer science bibliography, https://dblp.org}
}

@inproceedings{DBLP:conf/icml/TajwarSSR0XEFK24,
  author       = {Fahim Tajwar and
                  Anikait Singh and
                  Archit Sharma and
                  Rafael Rafailov and
                  Jeff Schneider and
                  Tengyang Xie and
                  Stefano Ermon and
                  Chelsea Finn and
                  Aviral Kumar},
  editor       = {Ruslan Salakhutdinov and
                  Zico Kolter and
                  Katherine A. Heller and
                  Adrian Weller and
                  Nuria Oliver and
                  Jonathan Scarlett and
                  Felix Berkenkamp},
  title        = {Preference Fine-Tuning of LLMs Should Leverage Suboptimal, On-Policy
                  Data},
  booktitle    = {Forty-first International Conference on Machine Learning, {ICML} 2024,
                  Vienna, Austria, July 21-27, 2024},
  series       = {Proceedings of Machine Learning Research},
  volume       = {235},
  pages        = {47441--47474},
  publisher    = {{PMLR} / OpenReview.net},
  year         = {2024},
  url          = {https://proceedings.mlr.press/v235/tajwar24a.html},
  bibsource    = {dblp computer science bibliography, https://dblp.org}
}

@article{DBLP:journals/corr/abs-2408-03314,
  author       = {Charlie Snell and
                  Jaehoon Lee and
                  Kelvin Xu and
                  Aviral Kumar},
  title        = {Scaling {LLM} Test-Time Compute Optimally can be More Effective than
                  Scaling Model Parameters},
  journal      = {CoRR},
  volume       = {abs/2408.03314},
  year         = {2024},
  url          = {https://doi.org/10.48550/arXiv.2408.03314},
  doi          = {10.48550/ARXIV.2408.03314},
  eprinttype    = {arXiv},
  eprint       = {2408.03314},
  bibsource    = {dblp computer science bibliography, https://dblp.org}
}

@article{DBLP:journals/corr/abs-2404-00725,
  author       = {Michael Hassid and
                  Tal Remez and
                  Jonas Gehring and
                  Roy Schwartz and
                  Yossi Adi},
  title        = {The Larger the Better? Improved {LLM} Code-Generation via Budget Reallocation},
  journal      = {CoRR},
  volume       = {abs/2404.00725},
  year         = {2024},
  url          = {https://doi.org/10.48550/arXiv.2404.00725},
  doi          = {10.48550/ARXIV.2404.00725},
  eprinttype    = {arXiv},
  eprint       = {2404.00725},
  bibsource    = {dblp computer science bibliography, https://dblp.org}
}

@inproceedings{DBLP:conf/emnlp/0005C0H0L24,
  author       = {Sen Yang and
                  Leyang Cui and
                  Deng Cai and
                  Xinting Huang and
                  Shuming Shi and
                  Wai Lam},
  editor       = {Yaser Al{-}Onaizan and
                  Mohit Bansal and
                  Yun{-}Nung Chen},
  title        = {Not All Preference Pairs Are Created Equal: {A} Recipe for Annotation-Efficient
                  Iterative Preference Learning},
  booktitle    = {Findings of the Association for Computational Linguistics: {EMNLP}
                  2024, Miami, Florida, USA, November 12-16, 2024},
  series       = {Findings of {ACL}},
  volume       = {{EMNLP} 2024},
  pages        = {6549--6561},
  publisher    = {Association for Computational Linguistics},
  year         = {2024},
  url          = {https://doi.org/10.18653/v1/2024.findings-emnlp.382},
  doi          = {10.18653/V1/2024.FINDINGS-EMNLP.382},
  bibsource    = {dblp computer science bibliography, https://dblp.org}
}

@inproceedings{DBLP:conf/iclr/HayesSPP25,
  author       = {Jamie Hayes and
                  Ilia Shumailov and
                  William P. Porter and
                  Aneesh Pappu},
  title        = {Measuring memorization in {RLHF} for code completion},
  booktitle    = {The Thirteenth International Conference on Learning Representations,
                  {ICLR} 2025, Singapore, April 24-28, 2025},
  publisher    = {OpenReview.net},
  year         = {2025},
  url          = {https://openreview.net/forum?id=Tg8RLxpMDu},
  bibsource    = {dblp computer science bibliography, https://dblp.org}
}

@inproceedings{DBLP:conf/iclr/KirkMNLHGR24,
  author       = {Robert Kirk and
                  Ishita Mediratta and
                  Christoforos Nalmpantis and
                  Jelena Luketina and
                  Eric Hambro and
                  Edward Grefenstette and
                  Roberta Raileanu},
  title        = {Understanding the Effects of {RLHF} on {LLM} Generalisation and Diversity},
  booktitle    = {The Twelfth International Conference on Learning Representations,
                  {ICLR} 2024, Vienna, Austria, May 7-11, 2024},
  publisher    = {OpenReview.net},
  year         = {2024},
  url          = {https://openreview.net/forum?id=PXD3FAVHJT},
  bibsource    = {dblp computer science bibliography, https://dblp.org}
}

@inproceedings{DBLP:conf/icml/ChuZYTXSLL025,
  author       = {Tianzhe Chu and
                  Yuexiang Zhai and
                  Jihan Yang and
                  Shengbang Tong and
                  Saining Xie and
                  Dale Schuurmans and
                  Quoc V. Le and
                  Sergey Levine and
                  Yi Ma},
  editor       = {Aarti Singh and
                  Maryam Fazel and
                  Daniel Hsu and
                  Simon Lacoste{-}Julien and
                  Felix Berkenkamp and
                  Tegan Maharaj and
                  Kiri Wagstaff and
                  Jerry Zhu},
  title        = {{SFT} Memorizes, {RL} Generalizes: {A} Comparative Study of Foundation
                  Model Post-training},
  booktitle    = {Forty-second International Conference on Machine Learning, {ICML}
                  2025, Vancouver, BC, Canada, July 13-19, 2025},
  series       = {Proceedings of Machine Learning Research},
  volume       = {267},
  publisher    = {{PMLR} / OpenReview.net},
  year         = {2025},
  url          = {https://proceedings.mlr.press/v267/chu25c.html},
  bibsource    = {dblp computer science bibliography, https://dblp.org}
}

@article{DBLP:journals/corr/abs-2510-08696,
  author       = {Yunzhen Feng and
                  Parag Jain and
                  Anthony Hartshorn and
                  Yaqi Duan and
                  Julia Kempe},
  title        = {Don't Waste Mistakes: Leveraging Negative RL-Groups via Confidence
                  Reweighting},
  journal      = {CoRR},
  volume       = {abs/2510.08696},
  year         = {2025},
  url          = {https://doi.org/10.48550/arXiv.2510.08696},
  doi          = {10.48550/ARXIV.2510.08696},
  eprinttype    = {arXiv},
  eprint       = {2510.08696},
  bibsource    = {dblp computer science bibliography, https://dblp.org}
}

@article{DBLP:journals/corr/abs-2506-01347,
  author       = {Xinyu Zhu and
                  Mengzhou Xia and
                  Zhepei Wei and
                  Wei{-}Lin Chen and
                  Danqi Chen and
                  Yu Meng},
  title        = {The Surprising Effectiveness of Negative Reinforcement in {LLM} Reasoning},
  journal      = {CoRR},
  volume       = {abs/2506.01347},
  year         = {2025},
  url          = {https://doi.org/10.48550/arXiv.2506.01347},
  doi          = {10.48550/ARXIV.2506.01347},
  eprinttype    = {arXiv},
  eprint       = {2506.01347},
  bibsource    = {dblp computer science bibliography, https://dblp.org}
}

@article{DBLP:journals/corr/abs-2503-23829,
  author       = {Yi Su and
                  Dian Yu and
                  Linfeng Song and
                  Juntao Li and
                  Haitao Mi and
                  Zhaopeng Tu and
                  Min Zhang and
                  Dong Yu},
  title        = {Crossing the Reward Bridge: Expanding {RL} with Verifiable Rewards
                  Across Diverse Domains},
  journal      = {CoRR},
  volume       = {abs/2503.23829},
  year         = {2025},
  url          = {https://doi.org/10.48550/arXiv.2503.23829},
  doi          = {10.48550/ARXIV.2503.23829},
  eprinttype    = {arXiv},
  eprint       = {2503.23829},
  bibsource    = {dblp computer science bibliography, https://dblp.org}
}

@article{DBLP:journals/corr/abs-2507-17746,
  author       = {Anisha Gunjal and
                  Anthony Wang and
                  Elaine Lau and
                  Vaskar Nath and
                  Bing Liu and
                  Sean Hendryx},
  title        = {Rubrics as Rewards: Reinforcement Learning Beyond Verifiable Domains},
  journal      = {CoRR},
  volume       = {abs/2507.17746},
  year         = {2025},
  url          = {https://doi.org/10.48550/arXiv.2507.17746},
  doi          = {10.48550/ARXIV.2507.17746},
  eprinttype    = {arXiv},
  eprint       = {2507.17746},
  bibsource    = {dblp computer science bibliography, https://dblp.org}
}

@inproceedings{DBLP:conf/iclr/KimS0JLLYSKTS24,
  author       = {Seungone Kim and
                  Jamin Shin and
                  Yejin Choi and
                  Joel Jang and
                  Shayne Longpre and
                  Hwaran Lee and
                  Sangdoo Yun and
                  Seongjin Shin and
                  Sungdong Kim and
                  James Thorne and
                  Minjoon Seo},
  title        = {Prometheus: Inducing Fine-Grained Evaluation Capability in Language
                  Models},
  booktitle    = {The Twelfth International Conference on Learning Representations,
                  {ICLR} 2024, Vienna, Austria, May 7-11, 2024},
  publisher    = {OpenReview.net},
  year         = {2024},
  url          = {https://openreview.net/forum?id=8euJaTveKw},
  bibsource    = {dblp computer science bibliography, https://dblp.org}
}

@article{DBLP:journals/tmlr/Dong0GZCPDZS023,
  author       = {Hanze Dong and
                  Wei Xiong and
                  Deepanshu Goyal and
                  Yihan Zhang and
                  Winnie Chow and
                  Rui Pan and
                  Shizhe Diao and
                  Jipeng Zhang and
                  Kashun Shum and
                  Tong Zhang},
  title        = {{RAFT:} Reward rAnked FineTuning for Generative Foundation Model Alignment},
  journal      = {Trans. Mach. Learn. Res.},
  volume       = {2023},
  year         = {2023},
  url          = {https://openreview.net/forum?id=m7p5O7zblY},
  bibsource    = {dblp computer science bibliography, https://dblp.org}
}

@inproceedings{DBLP:conf/icml/0015DYW0J0024,
  author       = {Wei Xiong and
                  Hanze Dong and
                  Chenlu Ye and
                  Ziqi Wang and
                  Han Zhong and
                  Heng Ji and
                  Nan Jiang and
                  Tong Zhang},
  editor       = {Ruslan Salakhutdinov and
                  Zico Kolter and
                  Katherine A. Heller and
                  Adrian Weller and
                  Nuria Oliver and
                  Jonathan Scarlett and
                  Felix Berkenkamp},
  title        = {Iterative Preference Learning from Human Feedback: Bridging Theory
                  and Practice for {RLHF} under KL-constraint},
  booktitle    = {Forty-first International Conference on Machine Learning, {ICML} 2024,
                  Vienna, Austria, July 21-27, 2024},
  series       = {Proceedings of Machine Learning Research},
  volume       = {235},
  pages        = {54715--54754},
  publisher    = {{PMLR} / OpenReview.net},
  year         = {2024},
  url          = {https://proceedings.mlr.press/v235/xiong24a.html},
  bibsource    = {dblp computer science bibliography, https://dblp.org}
}

@article{DBLP:journals/corr/abs-2503-01067,
  author       = {Gokul Swamy and
                  Sanjiban Choudhury and
                  Wen Sun and
                  Zhiwei Steven Wu and
                  J. Andrew Bagnell},
  title        = {All Roads Lead to Likelihood: The Value of Reinforcement Learning
                  in Fine-Tuning},
  journal      = {CoRR},
  volume       = {abs/2503.01067},
  year         = {2025},
  url          = {https://doi.org/10.48550/arXiv.2503.01067},
  doi          = {10.48550/ARXIV.2503.01067},
  eprinttype    = {arXiv},
  eprint       = {2503.01067},
  bibsource    = {dblp computer science bibliography, https://dblp.org}
}

@article{DBLP:journals/corr/abs-2505-22312,
  author       = {Jujie He and
                  Jiacai Liu and
                  Chris Yuhao Liu and
                  Rui Yan and
                  Chaojie Wang and
                  Peng Cheng and
                  Xiaoyu Zhang and
                  Fuxiang Zhang and
                  Jiacheng Xu and
                  Wei Shen and
                  Siyuan Li and
                  Liang Zeng and
                  Tianwen Wei and
                  Cheng Cheng and
                  Bo An and
                  Yang Liu and
                  Yahui Zhou},
  title        = {Skywork Open Reasoner 1 Technical Report},
  journal      = {CoRR},
  volume       = {abs/2505.22312},
  year         = {2025},
  url          = {https://doi.org/10.48550/arXiv.2505.22312},
  doi          = {10.48550/ARXIV.2505.22312},
  eprinttype    = {arXiv},
  eprint       = {2505.22312},
  bibsource    = {dblp computer science bibliography, https://dblp.org}
}

@article{DBLP:journals/corr/abs-2505-16400,
  author       = {Yang Chen and
                  Zhuolin Yang and
                  Zihan Liu and
                  Chankyu Lee and
                  Peng Xu and
                  Mohammad Shoeybi and
                  Bryan Catanzaro and
                  Wei Ping},
  title        = {AceReason-Nemotron: Advancing Math and Code Reasoning through Reinforcement
                  Learning},
  journal      = {CoRR},
  volume       = {abs/2505.16400},
  year         = {2025},
  url          = {https://doi.org/10.48550/arXiv.2505.16400},
  doi          = {10.48550/ARXIV.2505.16400},
  eprinttype    = {arXiv},
  eprint       = {2505.16400},
  bibsource    = {dblp computer science bibliography, https://dblp.org}
}

@article{DBLP:journals/tmlr/LiuZXF00Y23,
  author       = {Jiate Liu and
                  Yiqin Zhu and
                  Kaiwen Xiao and
                  Qiang Fu and
                  Xiao Han and
                  Wei Yang and
                  Deheng Ye},
  title        = {{RLTF:} Reinforcement Learning from Unit Test Feedback},
  journal      = {Trans. Mach. Learn. Res.},
  volume       = {2023},
  year         = {2023},
  url          = {https://openreview.net/forum?id=hjYmsV6nXZ},
  bibsource    = {dblp computer science bibliography, https://dblp.org}
}

@inproceedings{DBLP:conf/icml/GehringZCMCS25,
  author       = {Jonas Gehring and
                  Kunhao Zheng and
                  Jade Copet and
                  Vegard Mella and
                  Taco Cohen and
                  Gabriel Synnaeve},
  editor       = {Aarti Singh and
                  Maryam Fazel and
                  Daniel Hsu and
                  Simon Lacoste{-}Julien and
                  Felix Berkenkamp and
                  Tegan Maharaj and
                  Kiri Wagstaff and
                  Jerry Zhu},
  title        = {{RLEF:} Grounding Code LLMs in Execution Feedback with Reinforcement
                  Learning},
  booktitle    = {Forty-second International Conference on Machine Learning, {ICML}
                  2025, Vancouver, BC, Canada, July 13-19, 2025},
  series       = {Proceedings of Machine Learning Research},
  volume       = {267},
  publisher    = {{PMLR} / OpenReview.net},
  year         = {2025},
  url          = {https://proceedings.mlr.press/v267/gehring25a.html},
  bibsource    = {dblp computer science bibliography, https://dblp.org}
}

@inproceedings{DBLP:conf/nips/Le0GSH22,
  author       = {Hung Le and
                  Yue Wang and
                  Akhilesh Deepak Gotmare and
                  Silvio Savarese and
                  Steven Chu{-}Hong Hoi},
  editor       = {Sanmi Koyejo and
                  S. Mohamed and
                  A. Agarwal and
                  Danielle Belgrave and
                  K. Cho and
                  A. Oh},
  title        = {CodeRL: Mastering Code Generation through Pretrained Models and Deep
                  Reinforcement Learning},
  booktitle    = {Advances in Neural Information Processing Systems 35: Annual Conference
                  on Neural Information Processing Systems 2022, NeurIPS 2022, New Orleans,
                  LA, USA, November 28 - December 9, 2022},
  year         = {2022},
  url          = {http://papers.nips.cc/paper\_files/paper/2022/hash/8636419dea1aa9fbd25fc4248e702da4-Abstract-Conference.html},
  bibsource    = {dblp computer science bibliography, https://dblp.org}
}

@article{DBLP:journals/tmlr/ShojaeeJTR23,
  author       = {Parshin Shojaee and
                  Aneesh Jain and
                  Sindhu Tipirneni and
                  Chandan K. Reddy},
  title        = {Execution-based Code Generation using Deep Reinforcement Learning},
  journal      = {Trans. Mach. Learn. Res.},
  volume       = {2023},
  year         = {2023},
  url          = {https://openreview.net/forum?id=0XBuaxqEcG},
  bibsource    = {dblp computer science bibliography, https://dblp.org}
}

@article{DBLP:journals/corr/abs-2505-21297,
  author       = {Yifei Liu and
                  Li Lyna Zhang and
                  Yi Zhu and
                  Bingcheng Dong and
                  Xudong Zhou and
                  Ning Shang and
                  Fan Yang and
                  Mao Yang},
  title        = {rStar-Coder: Scaling Competitive Code Reasoning with a Large-Scale
                  Verified Dataset},
  journal      = {CoRR},
  volume       = {abs/2505.21297},
  year         = {2025},
  url          = {https://doi.org/10.48550/arXiv.2505.21297},
  doi          = {10.48550/ARXIV.2505.21297},
  eprinttype    = {arXiv},
  eprint       = {2505.21297},
  bibsource    = {dblp computer science bibliography, https://dblp.org}
}

@article{DBLP:journals/corr/abs-2509-07604,
  author       = {Zhoujun Cheng and
                  Richard Fan and
                  Shibo Hao and
                  Taylor W. Killian and
                  Haonan Li and
                  Suqi Sun and
                  Hector Ren and
                  Alexander Moreno and
                  Daqian Zhang and
                  Tianjun Zhong and
                  Yuxin Xiong and
                  Yuanzhe Hu and
                  Yutao Xie and
                  Xudong Han and
                  Yuqi Wang and
                  Varad Pimpalkhute and
                  Yonghao Zhuang and
                  Aaryamonvikram Singh and
                  Xuezhi Liang and
                  Anze Xie and
                  Jianshu She and
                  Desai Fan and
                  Chengqian Gao and
                  Liqun Ma and
                  Mikhail Yurochkin and
                  John Maggs and
                  Xuezhe Ma and
                  Guowei He and
                  Zhiting Hu and
                  Zhengzhong Liu and
                  Eric P. Xing},
  title        = {K2-Think: {A} Parameter-Efficient Reasoning System},
  journal      = {CoRR},
  volume       = {abs/2509.07604},
  year         = {2025},
  url          = {https://doi.org/10.48550/arXiv.2509.07604},
  doi          = {10.48550/ARXIV.2509.07604},
  eprinttype    = {arXiv},
  eprint       = {2509.07604},
  bibsource    = {dblp computer science bibliography, https://dblp.org}
}

@inproceedings{DBLP:conf/acl/ToNB24,
  author       = {Hung To and
                  Minh Nguyen and
                  Nghi Bui},
  editor       = {Lun{-}Wei Ku and
                  Andre Martins and
                  Vivek Srikumar},
  title        = {Functional Overlap Reranking for Neural Code Generation},
  booktitle    = {Findings of the Association for Computational Linguistics, {ACL} 2024,
                  Bangkok, Thailand and virtual meeting, August 11-16, 2024},
  series       = {Findings of {ACL}},
  volume       = {{ACL} 2024},
  pages        = {3686--3704},
  publisher    = {Association for Computational Linguistics},
  year         = {2024},
  url          = {https://doi.org/10.18653/v1/2024.findings-acl.220},
  doi          = {10.18653/V1/2024.FINDINGS-ACL.220},
  bibsource    = {dblp computer science bibliography, https://dblp.org}
}

@inproceedings{DBLP:conf/iclr/ChenZNZLLC23,
  author       = {Bei Chen and
                  Fengji Zhang and
                  Anh Nguyen and
                  Daoguang Zan and
                  Zeqi Lin and
                  Jian{-}Guang Lou and
                  Weizhu Chen},
  title        = {CodeT: Code Generation with Generated Tests},
  booktitle    = {The Eleventh International Conference on Learning Representations,
                  {ICLR} 2023, Kigali, Rwanda, May 1-5, 2023},
  publisher    = {OpenReview.net},
  year         = {2023},
  url          = {https://openreview.net/forum?id=ktrw68Cmu9c},
  bibsource    = {dblp computer science bibliography, https://dblp.org}
}

@article{DBLP:journals/corr/abs-2411-13611,
  author       = {Zhihan Liu and
                  Shenao Zhang and
                  Zhaoran Wang},
  title        = {{DSTC:} Direct Preference Learning with Only Self-Generated Tests
                  and Code to Improve Code LMs},
  journal      = {CoRR},
  volume       = {abs/2411.13611},
  year         = {2024},
  url          = {https://doi.org/10.48550/arXiv.2411.13611},
  doi          = {10.48550/ARXIV.2411.13611},
  eprinttype    = {arXiv},
  eprint       = {2411.13611},
  bibsource    = {dblp computer science bibliography, https://dblp.org}
}

@inproceedings{DBLP:conf/acl/XuLYZP25,
  author       = {Zhangchen Xu and
                  Yang Liu and
                  Yueqin Yin and
                  Mingyuan Zhou and
                  Radha Poovendran},
  editor       = {Wanxiang Che and
                  Joyce Nabende and
                  Ekaterina Shutova and
                  Mohammad Taher Pilehvar},
  title        = {KodCode: {A} Diverse, Challenging, and Verifiable Synthetic Dataset
                  for Coding},
  booktitle    = {Findings of the Association for Computational Linguistics, {ACL} 2025,
                  Vienna, Austria, July 27 - August 1, 2025},
  series       = {Findings of {ACL}},
  volume       = {{ACL} 2025},
  pages        = {6980--7008},
  publisher    = {Association for Computational Linguistics},
  year         = {2025},
  url          = {https://doi.org/10.18653/v1/2025.findings-acl.365},
  doi          = {10.18653/V1/2025.FINDINGS-ACL.365},
  bibsource    = {dblp computer science bibliography, https://dblp.org}
}

@article{DBLP:journals/corr/abs-2510-12803,
  author       = {Shang Zhou and
                  Zihan Zheng and
                  Kaiyuan Liu and
                  Zeyu Shen and
                  Zerui Cheng and
                  Zexing Chen and
                  Hansen He and
                  Jianzhu Yao and
                  Huanzhi Mao and
                  Qiuyang Mang and
                  Tianfu Fu and
                  Beichen Li and
                  Dongruixuan Li and
                  Wenhao Chai and
                  Zhuang Liu and
                  Aleksandra Korolova and
                  Peter Henderson and
                  Natasha Jaques and
                  Pramod Viswanath and
                  Saining Xie and
                  Jingbo Shang},
  title        = {AutoCode: LLMs as Problem Setters for Competitive Programming},
  journal      = {CoRR},
  volume       = {abs/2510.12803},
  year         = {2025},
  url          = {https://doi.org/10.48550/arXiv.2510.12803},
  doi          = {10.48550/ARXIV.2510.12803},
  eprinttype    = {arXiv},
  eprint       = {2510.12803},
  bibsource    = {dblp computer science bibliography, https://dblp.org}
}

@inproceedings{DBLP:conf/acl/Zhang0DXZSLJ25,
  author       = {Kechi Zhang and
                  Ge Li and
                  Yihong Dong and
                  Jingjing Xu and
                  Jun Zhang and
                  Jing Su and
                  Yongfei Liu and
                  Zhi Jin},
  editor       = {Wanxiang Che and
                  Joyce Nabende and
                  Ekaterina Shutova and
                  Mohammad Taher Pilehvar},
  title        = {CodeDPO: Aligning Code Models with Self Generated and Verified Source
                  Code},
  booktitle    = {Proceedings of the 63rd Annual Meeting of the Association for Computational
                  Linguistics (Volume 1: Long Papers), {ACL} 2025, Vienna, Austria,
                  July 27 - August 1, 2025},
  pages        = {15854--15871},
  publisher    = {Association for Computational Linguistics},
  year         = {2025},
  url          = {https://aclanthology.org/2025.acl-long.771/},
  bibsource    = {dblp computer science bibliography, https://dblp.org}
}

@inproceedings{DBLP:conf/acl/ZengJ0NCC25,
  author       = {Huaye Zeng and
                  Dongfu Jiang and
                  Haozhe Wang and
                  Ping Nie and
                  Xiaotong Chen and
                  Wenhu Chen},
  editor       = {Wanxiang Che and
                  Joyce Nabende and
                  Ekaterina Shutova and
                  Mohammad Taher Pilehvar},
  title        = {{ACECODER:} Acing Coder {RL} via Automated Test-Case Synthesis},
  booktitle    = {Proceedings of the 63rd Annual Meeting of the Association for Computational
                  Linguistics (Volume 1: Long Papers), {ACL} 2025, Vienna, Austria,
                  July 27 - August 1, 2025},
  pages        = {12023--12040},
  publisher    = {Association for Computational Linguistics},
  year         = {2025},
  url          = {https://aclanthology.org/2025.acl-long.587/},
  bibsource    = {dblp computer science bibliography, https://dblp.org}
}

@article{DBLP:journals/corr/abs-2508-21107,
  author       = {Dongjun Lee and
                  Changho Hwang and
                  Kimin Lee},
  title        = {Learning to Generate Unit Test via Adversarial Reinforcement Learning},
  journal      = {CoRR},
  volume       = {abs/2508.21107},
  year         = {2025},
  url          = {https://doi.org/10.48550/arXiv.2508.21107},
  doi          = {10.48550/ARXIV.2508.21107},
  eprinttype    = {arXiv},
  eprint       = {2508.21107},
  bibsource    = {dblp computer science bibliography, https://dblp.org}
}

@inproceedings{DBLP:conf/acl/MaZZYLT25,
  author       = {Zeyao Ma and
                  Xiaokang Zhang and
                  Jing Zhang and
                  Jifan Yu and
                  Sijia Luo and
                  Jie Tang},
  editor       = {Wanxiang Che and
                  Joyce Nabende and
                  Ekaterina Shutova and
                  Mohammad Taher Pilehvar},
  title        = {Dynamic Scaling of Unit Tests for Code Reward Modeling},
  booktitle    = {Proceedings of the 63rd Annual Meeting of the Association for Computational
                  Linguistics (Volume 1: Long Papers), {ACL} 2025, Vienna, Austria,
                  July 27 - August 1, 2025},
  pages        = {6917--6935},
  publisher    = {Association for Computational Linguistics},
  year         = {2025},
  url          = {https://aclanthology.org/2025.acl-long.343/},
  bibsource    = {dblp computer science bibliography, https://dblp.org}
}

@inproceedings{DBLP:conf/emnlp/WangLSDL25,
  author       = {Zihan Wang and
                  Siyao Liu and
                  Yang Sun and
                  Ming Ding and
                  Hongyan Li},
  editor       = {Christos Christodoulopoulos and
                  Tanmoy Chakraborty and
                  Carolyn Rose and
                  Violet Peng},
  title        = {CodeContests+: High-Quality Test Case Generation for Competitive Programming},
  booktitle    = {Findings of the Association for Computational Linguistics: {EMNLP}
                  2025, Suzhou, China, November 4-9, 2025},
  pages        = {5576--5600},
  publisher    = {Association for Computational Linguistics},
  year         = {2025},
  url          = {https://aclanthology.org/2025.findings-emnlp.299/},
  bibsource    = {dblp computer science bibliography, https://dblp.org}
}

@inproceedings{DBLP:conf/naacl/WangYWHCSZCM25,
  author       = {Wenhan Wang and
                  Chenyuan Yang and
                  Zhijie Wang and
                  Yuheng Huang and
                  Zhaoyang Chu and
                  Da Song and
                  Lingming Zhang and
                  An Ran Chen and
                  Lei Ma},
  editor       = {Luis Chiruzzo and
                  Alan Ritter and
                  Lu Wang},
  title        = {{TESTEVAL:} Benchmarking Large Language Models for Test Case Generation},
  booktitle    = {Findings of the Association for Computational Linguistics: {NAACL}
                  2025, Albuquerque, New Mexico, USA, April 29 - May 4, 2025},
  series       = {Findings of {ACL}},
  volume       = {{NAACL} 2025},
  pages        = {3547--3562},
  publisher    = {Association for Computational Linguistics},
  year         = {2025},
  url          = {https://doi.org/10.18653/v1/2025.findings-naacl.197},
  doi          = {10.18653/V1/2025.FINDINGS-NAACL.197},
  bibsource    = {dblp computer science bibliography, https://dblp.org}
}

@article{DBLP:journals/corr/abs-2506-06821,
  author       = {Yuhan Cao and
                  Zian Chen and
                  Kun Quan and
                  Ziliang Zhang and
                  Yu Wang and
                  Xiaoning Dong and
                  Yeqi Feng and
                  Guanzhong He and
                  Jingcheng Huang and
                  Jianhao Li and
                  Yixuan Tan and
                  Jiafu Tang and
                  Yilin Tang and
                  Junlei Wu and
                  Qianyu Xiao and
                  Can Zheng and
                  Shouchen Zhou and
                  Yuxiang Zhu and
                  Yiming Huang and
                  Tian Xie and
                  Tianxing He},
  title        = {Can LLMs Generate Reliable Test Case Generators? {A} Study on Competition-Level
                  Programming Problems},
  journal      = {CoRR},
  volume       = {abs/2506.06821},
  year         = {2025},
  url          = {https://doi.org/10.48550/arXiv.2506.06821},
  doi          = {10.48550/ARXIV.2506.06821},
  eprinttype    = {arXiv},
  eprint       = {2506.06821},
  bibsource    = {dblp computer science bibliography, https://dblp.org}
}

@inproceedings{DBLP:conf/icse/YuSRZZMLLWX24,
  author       = {Hao Yu and
                  Bo Shen and
                  Dezhi Ran and
                  Jiaxin Zhang and
                  Qi Zhang and
                  Yuchi Ma and
                  Guangtai Liang and
                  Ying Li and
                  Qianxiang Wang and
                  Tao Xie},
  title        = {CoderEval: {A} Benchmark of Pragmatic Code Generation with Generative
                  Pre-trained Models},
  booktitle    = {Proceedings of the 46th {IEEE/ACM} International Conference on Software
                  Engineering, {ICSE} 2024, Lisbon, Portugal, April 14-20, 2024},
  pages        = {37:1--37:12},
  publisher    = {{ACM}},
  year         = {2024},
  url          = {https://doi.org/10.1145/3597503.3623316},
  doi          = {10.1145/3597503.3623316},
  bibsource    = {dblp computer science bibliography, https://dblp.org}
}

@inproceedings{DBLP:conf/nips/MundlerMHV24,
  author       = {Niels M{\"{u}}ndler and
                  Mark Niklas M{\"{u}}ller and
                  Jingxuan He and
                  Martin T. Vechev},
  editor       = {Amir Globersons and
                  Lester Mackey and
                  Danielle Belgrave and
                  Angela Fan and
                  Ulrich Paquet and
                  Jakub M. Tomczak and
                  Cheng Zhang},
  title        = {SWT-Bench: Testing and Validating Real-World Bug-Fixes with Code Agents},
  booktitle    = {Advances in Neural Information Processing Systems 38: Annual Conference
                  on Neural Information Processing Systems 2024, NeurIPS 2024, Vancouver,
                  BC, Canada, December 10 - 15, 2024},
  year         = {2024},
  url          = {http://papers.nips.cc/paper\_files/paper/2024/hash/94f093b41fc2666376fb1f667fe282f3-Abstract-Conference.html},
  bibsource    = {dblp computer science bibliography, https://dblp.org}
}

@inproceedings{DBLP:conf/iclr/JainSR25,
  author       = {Kush Jain and
                  Gabriel Synnaeve and
                  Baptiste Rozi{\`{e}}re},
  title        = {TestGenEval: {A} Real World Unit Test Generation and Test Completion
                  Benchmark},
  booktitle    = {The Thirteenth International Conference on Learning Representations,
                  {ICLR} 2025, Singapore, April 24-28, 2025},
  publisher    = {OpenReview.net},
  year         = {2025},
  url          = {https://openreview.net/forum?id=7o6SG5gVev},
  bibsource    = {dblp computer science bibliography, https://dblp.org}
}

@inproceedings{Lim2025ContractEvalAB,
  title={ContractEval: A Benchmark for Evaluating Contract-Satisfying Assertions in Code Generation},
  author={Soohan Lim and Joonghyuk Hahn and Hyunwoo Park and Sang-Ki Ko and Yo-Sub Han},
  booktitle={Findings of the Association for Computational Linguistics, {ACL} 2026},
  pages={42549--42566},
  publisher={Association for Computational Linguistics},
  year={2026},
  url={https://doi.org/10.18653/v1/2026.findings-acl.2112}
}

@article{Duan2023PerfRLAS,
  title={PerfRL: A Small Language Model Framework for Efficient Code Optimization},
  author={Shukai Duan and Nikos Kanakaris and Xiongye Xiao and Heng Ping and Chenyu Zhou and Nesreen K. Ahmed and Guixiang Ma and Mihai Capotă and Theodore L. Willke and Shahin Nazarian and Paul Bogdan},
  journal={CoRR},
  volume={abs/2312.05657},
  year={2023},
  url={https://doi.org/10.48550/arXiv.2312.05657},
  doi={10.48550/arXiv.2312.05657},
  eprinttype={arXiv},
  eprint={2312.05657}
}

@article{DBLP:journals/corr/abs-2404-18864,
  author       = {Daniel Nichols and
                  Pranav Polasam and
                  Harshitha Menon and
                  Aniruddha Marathe and
                  Todd Gamblin and
                  Abhinav Bhatele},
  title        = {Performance-Aligned LLMs for Generating Fast Code},
  journal      = {CoRR},
  volume       = {abs/2404.18864},
  year         = {2024},
  url          = {https://doi.org/10.48550/arXiv.2404.18864},
  doi          = {10.48550/ARXIV.2404.18864},
  eprinttype    = {arXiv},
  eprint       = {2404.18864},
  bibsource    = {dblp computer science bibliography, https://dblp.org}
}

@inproceedings{DBLP:conf/iclr/ShypulaMZ0GYHNR24,
  author       = {Alexander Shypula and
                  Aman Madaan and
                  Yimeng Zeng and
                  Uri Alon and
                  Jacob R. Gardner and
                  Yiming Yang and
                  Milad Hashemi and
                  Graham Neubig and
                  Parthasarathy Ranganathan and
                  Osbert Bastani and
                  Amir Yazdanbakhsh},
  title        = {Learning Performance-Improving Code Edits},
  booktitle    = {The Twelfth International Conference on Learning Representations,
                  {ICLR} 2024, Vienna, Austria, May 7-11, 2024},
  publisher    = {OpenReview.net},
  year         = {2024},
  url          = {https://openreview.net/forum?id=ix7rLVHXyY},
  bibsource    = {dblp computer science bibliography, https://dblp.org}
}

@inproceedings{DBLP:conf/iclr/PanBS22,
  author       = {Alexander Pan and
                  Kush Bhatia and
                  Jacob Steinhardt},
  title        = {The Effects of Reward Misspecification: Mapping and Mitigating Misaligned
                  Models},
  booktitle    = {The Tenth International Conference on Learning Representations, {ICLR}
                  2022, Virtual Event, April 25-29, 2022},
  publisher    = {OpenReview.net},
  year         = {2022},
  url          = {https://openreview.net/forum?id=JYtwGwIL7ye},
  bibsource    = {dblp computer science bibliography, https://dblp.org}
}

@article{DBLP:journals/alife/LehmanCMAABBBBB20,
  author       = {Joel Lehman and
                  Jeff Clune and
                  Dusan Misevic and
                  Christoph Adami and
                  Lee Altenberg and
                  Julie Beaulieu and
                  Peter J. Bentley and
                  Samuel Bernard and
                  Guillaume Beslon and
                  David M. Bryson and
                  Nick Cheney and
                  Patryk Chrabaszcz and
                  Antoine Cully and
                  St{\'{e}}phane Doncieux and
                  Fred C. Dyer and
                  Kai Olav Ellefsen and
                  Robert Feldt and
                  Stephan Fischer and
                  Stephanie Forrest and
                  Antoine Fr{\'{e}}noy and
                  Christian Gagn{\'{e}} and
                  L{\'{e}}ni K. Le Goff and
                  Laura M. Grabowski and
                  Babak Hodjat and
                  Frank Hutter and
                  Laurent Keller and
                  Carole Knibbe and
                  Peter Krcah and
                  Richard E. Lenski and
                  Hod Lipson and
                  Robert MacCurdy and
                  Carlos Maestre and
                  Risto Miikkulainen and
                  Sara Mitri and
                  David E. Moriarty and
                  Jean{-}Baptiste Mouret and
                  Anh Nguyen and
                  Charles Ofria and
                  Marc Parizeau and
                  David P. Parsons and
                  Robert T. Pennock and
                  William F. Punch and
                  Thomas S. Ray and
                  Marc Schoenauer and
                  Eric Schulte and
                  Karl Sims and
                  Kenneth O. Stanley and
                  Fran{\c{c}}ois Taddei and
                  Danesh Tarapore and
                  Simon Thibault and
                  Richard A. Watson and
                  Westley Weimer and
                  Jason Yosinski},
  title        = {The Surprising Creativity of Digital Evolution: {A} Collection of
                  Anecdotes from the Evolutionary Computation and Artificial Life Research
                  Communities},
  journal      = {Artif. Life},
  volume       = {26},
  number       = {2},
  pages        = {274--306},
  year         = {2020},
  url          = {https://doi.org/10.1162/artl\_a\_00319},
  doi          = {10.1162/ARTL\_A\_00319},
  bibsource    = {dblp computer science bibliography, https://dblp.org}
}

@inproceedings{DBLP:conf/icml/GaoSH23,
  author       = {Leo Gao and
                  John Schulman and
                  Jacob Hilton},
  editor       = {Andreas Krause and
                  Emma Brunskill and
                  Kyunghyun Cho and
                  Barbara Engelhardt and
                  Sivan Sabato and
                  Jonathan Scarlett},
  title        = {Scaling Laws for Reward Model Overoptimization},
  booktitle    = {International Conference on Machine Learning, {ICML} 2023, 23-29 July
                  2023, Honolulu, Hawaii, {USA}},
  series       = {Proceedings of Machine Learning Research},
  volume       = {202},
  pages        = {10835--10866},
  publisher    = {{PMLR}},
  year         = {2023},
  url          = {https://proceedings.mlr.press/v202/gao23h.html},
  bibsource    = {dblp computer science bibliography, https://dblp.org}
}

@inproceedings{DBLP:conf/ijcai/EverittKOL17,
  author       = {Tom Everitt and
                  Victoria Krakovna and
                  Laurent Orseau and
                  Shane Legg},
  editor       = {Carles Sierra},
  title        = {Reinforcement Learning with a Corrupted Reward Channel},
  booktitle    = {Proceedings of the Twenty-Sixth International Joint Conference on
                  Artificial Intelligence, {IJCAI} 2017, Melbourne, Australia, August
                  19-25, 2017},
  pages        = {4705--4713},
  publisher    = {ijcai.org},
  year         = {2017},
  url          = {https://doi.org/10.24963/ijcai.2017/656},
  doi          = {10.24963/IJCAI.2017/656},
  bibsource    = {dblp computer science bibliography, https://dblp.org}
}

@inproceedings{DBLP:conf/emnlp/LiuIXWXZ23,
  author       = {Yang Liu and
                  Dan Iter and
                  Yichong Xu and
                  Shuohang Wang and
                  Ruochen Xu and
                  Chenguang Zhu},
  editor       = {Houda Bouamor and
                  Juan Pino and
                  Kalika Bali},
  title        = {G-Eval: {NLG} Evaluation using Gpt-4 with Better Human Alignment},
  booktitle    = {Proceedings of the 2023 Conference on Empirical Methods in Natural
                  Language Processing, {EMNLP} 2023, Singapore, December 6-10, 2023},
  pages        = {2511--2522},
  publisher    = {Association for Computational Linguistics},
  year         = {2023},
  url          = {https://doi.org/10.18653/v1/2023.emnlp-main.153},
  doi          = {10.18653/V1/2023.EMNLP-MAIN.153},
  bibsource    = {dblp computer science bibliography, https://dblp.org}
}

@inproceedings{DBLP:conf/iclr/LiSYF0024,
  author       = {Junlong Li and
                  Shichao Sun and
                  Weizhe Yuan and
                  Run{-}Ze Fan and
                  Hai Zhao and
                  Pengfei Liu},
  title        = {Generative Judge for Evaluating Alignment},
  booktitle    = {The Twelfth International Conference on Learning Representations,
                  {ICLR} 2024, Vienna, Austria, May 7-11, 2024},
  publisher    = {OpenReview.net},
  year         = {2024},
  url          = {https://openreview.net/forum?id=gtkFw6sZGS},
  bibsource    = {dblp computer science bibliography, https://dblp.org}
}

@inproceedings{DBLP:conf/IEEEcloud/AggarwalCDMPBM24,
  author       = {Pooja Aggarwal and
                  Oishik Chatterjee and
                  Ting Dai and
                  Prateeti Mohapatra and
                  Brent Paulovicks and
                  Brad Blancett and
                  Arthur De Magalhaes},
  editor       = {Rong N. Chang and
                  Carl K. Chang and
                  Jingwei Yang and
                  Nimanthi L. Atukorala and
                  Zhi Jin and
                  Michael Sheng and
                  Jing Fan and
                  Kenneth Fletcher and
                  Qiang He and
                  Tevfik Kosar and
                  Santonu Sarkar and
                  Sreekrishnan Venkateswaran and
                  Shangguang Wang and
                  Xuanzhe Liu and
                  Seetharami Seelam and
                  Chandra Narayanaswami and
                  Ziliang Zong},
  title        = {CodeSift: An LLM-Based Reference-Less Framework for Automatic Code
                  Validation},
  booktitle    = {17th {IEEE} International Conference on Cloud Computing, {CLOUD} 2024,
                  Shenzhen, China, July 7-13, 2024},
  pages        = {404--410},
  publisher    = {{IEEE}},
  year         = {2024},
  url          = {https://doi.org/10.1109/CLOUD62652.2024.00052},
  doi          = {10.1109/CLOUD62652.2024.00052},
  bibsource    = {dblp computer science bibliography, https://dblp.org}
}

@inproceedings{DBLP:conf/emnlp/TongZ24,
  author       = {Weixi Tong and
                  Tianyi Zhang},
  editor       = {Yaser Al{-}Onaizan and
                  Mohit Bansal and
                  Yun{-}Nung Chen},
  title        = {CodeJudge: Evaluating Code Generation with Large Language Models},
  booktitle    = {Proceedings of the 2024 Conference on Empirical Methods in Natural
                  Language Processing, {EMNLP} 2024, Miami, FL, USA, November 12-16,
                  2024},
  pages        = {20032--20051},
  publisher    = {Association for Computational Linguistics},
  year         = {2024},
  url          = {https://doi.org/10.18653/v1/2024.emnlp-main.1118},
  doi          = {10.18653/V1/2024.EMNLP-MAIN.1118},
  bibsource    = {dblp computer science bibliography, https://dblp.org}
}

@inproceedings{DBLP:conf/eacl/Zhuo24,
  author       = {Terry Yue Zhuo},
  editor       = {Yvette Graham and
                  Matthew Purver},
  title        = {ICE-Score: Instructing Large Language Models to Evaluate Code},
  booktitle    = {Findings of the Association for Computational Linguistics: {EACL}
                  2024, St. Julian's, Malta, March 17-22, 2024},
  series       = {Findings of {ACL}},
  volume       = {{EACL} 2024},
  pages        = {2232--2242},
  publisher    = {Association for Computational Linguistics},
  year         = {2024},
  url          = {https://aclanthology.org/2024.findings-eacl.148},
  bibsource    = {dblp computer science bibliography, https://dblp.org}
}

@article{DBLP:journals/corr/abs-2501-16655,
  author       = {Aashish Yadavally and
                  Hoan Nguyen and
                  Laurent Callot and
                  Gauthier Guinet},
  title        = {Large Language Model Critics for Execution-Free Evaluation of Code
                  Changes},
  journal      = {CoRR},
  volume       = {abs/2501.16655},
  year         = {2025},
  url          = {https://doi.org/10.48550/arXiv.2501.16655},
  doi          = {10.48550/ARXIV.2501.16655},
  eprinttype    = {arXiv},
  eprint       = {2501.16655},
  bibsource    = {dblp computer science bibliography, https://dblp.org}
}

@inproceedings{DBLP:conf/emnlp/LyuHLSZ25,
  author       = {Bohan Lyu and
                  Siqiao Huang and
                  Zichen Liang and
                  Qian Sun and
                  Jiaming Zhang},
  editor       = {Christos Christodoulopoulos and
                  Tanmoy Chakraborty and
                  Carolyn Rose and
                  Violet Peng},
  title        = {Surge: On the Potential of Large Language Models as General-Purpose
                  Surrogate Code Executors},
  booktitle    = {Proceedings of the 2025 Conference on Empirical Methods in Natural
                  Language Processing, {EMNLP} 2025, Suzhou, China, November 4-9, 2025},
  pages        = {3268--3308},
  publisher    = {Association for Computational Linguistics},
  year         = {2025},
  url          = {https://doi.org/10.18653/v1/2025.emnlp-main.162},
  doi          = {10.18653/V1/2025.EMNLP-MAIN.162},
  bibsource    = {dblp computer science bibliography, https://dblp.org}
}

@inproceedings{DBLP:conf/icml/Ni0RSYWL23,
  author       = {Ansong Ni and
                  Srini Iyer and
                  Dragomir Radev and
                  Veselin Stoyanov and
                  Wen{-}Tau Yih and
                  Sida I. Wang and
                  Xi Victoria Lin},
  editor       = {Andreas Krause and
                  Emma Brunskill and
                  Kyunghyun Cho and
                  Barbara Engelhardt and
                  Sivan Sabato and
                  Jonathan Scarlett},
  title        = {{LEVER:} Learning to Verify Language-to-Code Generation with Execution},
  booktitle    = {International Conference on Machine Learning, {ICML} 2023, 23-29 July
                  2023, Honolulu, Hawaii, {USA}},
  series       = {Proceedings of Machine Learning Research},
  volume       = {202},
  pages        = {26106--26128},
  publisher    = {{PMLR}},
  year         = {2023},
  url          = {https://proceedings.mlr.press/v202/ni23b.html},
  bibsource    = {dblp computer science bibliography, https://dblp.org}
}

@inproceedings{DBLP:conf/emnlp/ShiFGZW22,
  author       = {Freda Shi and
                  Daniel Fried and
                  Marjan Ghazvininejad and
                  Luke Zettlemoyer and
                  Sida I. Wang},
  editor       = {Yoav Goldberg and
                  Zornitsa Kozareva and
                  Yue Zhang},
  title        = {Natural Language to Code Translation with Execution},
  booktitle    = {Proceedings of the 2022 Conference on Empirical Methods in Natural
                  Language Processing, {EMNLP} 2022, Abu Dhabi, United Arab Emirates,
                  December 7-11, 2022},
  pages        = {3533--3546},
  publisher    = {Association for Computational Linguistics},
  year         = {2022},
  url          = {https://doi.org/10.18653/v1/2022.emnlp-main.231},
  doi          = {10.18653/V1/2022.EMNLP-MAIN.231},
  bibsource    = {dblp computer science bibliography, https://dblp.org}
}

@inproceedings{DBLP:conf/icml/ZhangYHLYF023,
  author       = {Tianyi Zhang and
                  Tao Yu and
                  Tatsunori Hashimoto and
                  Mike Lewis and
                  Wen{-}Tau Yih and
                  Daniel Fried and
                  Sida Wang},
  editor       = {Andreas Krause and
                  Emma Brunskill and
                  Kyunghyun Cho and
                  Barbara Engelhardt and
                  Sivan Sabato and
                  Jonathan Scarlett},
  title        = {Coder Reviewer Reranking for Code Generation},
  booktitle    = {International Conference on Machine Learning, {ICML} 2023, 23-29 July
                  2023, Honolulu, Hawaii, {USA}},
  series       = {Proceedings of Machine Learning Research},
  volume       = {202},
  pages        = {41832--41846},
  publisher    = {{PMLR}},
  year         = {2023},
  url          = {https://proceedings.mlr.press/v202/zhang23av.html},
  bibsource    = {dblp computer science bibliography, https://dblp.org}
}

@inproceedings{DBLP:conf/nips/InalaWYCELMG22,
  author       = {Jeevana Priya Inala and
                  Chenglong Wang and
                  Mei Yang and
                  Andr{\'{e}}s Codas and
                  Mark Encarnaci{\'{o}}n and
                  Shuvendu K. Lahiri and
                  Madanlal Musuvathi and
                  Jianfeng Gao},
  editor       = {Sanmi Koyejo and
                  S. Mohamed and
                  A. Agarwal and
                  Danielle Belgrave and
                  K. Cho and
                  A. Oh},
  title        = {Fault-Aware Neural Code Rankers},
  booktitle    = {Advances in Neural Information Processing Systems 35: Annual Conference
                  on Neural Information Processing Systems 2022, NeurIPS 2022, New Orleans,
                  LA, USA, November 28 - December 9, 2022},
  year         = {2022},
  url          = {http://papers.nips.cc/paper\_files/paper/2022/hash/5762c579d09811b7639be2389b3d07be-Abstract-Conference.html},
  bibsource    = {dblp computer science bibliography, https://dblp.org}
}

@article{DBLP:journals/corr/abs-2601-13097,
  author       = {Elena Bruches and
                  Daniil Grebenkin and
                  Mikhail Klementev and
                  Vadim Alperovich and
                  Roman Derunets and
                  Dari Baturova and
                  Georgy Mkrtchyan and
                  Oleg Sedukhin and
                  Ivan Bondarenko and
                  Nikolay Bushkov and
                  Stanislav Moiseev},
  title        = {{RM} -RF: Reward Model for Run-Free Unit Test Evaluation},
  journal      = {CoRR},
  volume       = {abs/2601.13097},
  year         = {2026},
  url          = {https://doi.org/10.48550/arXiv.2601.13097},
  doi          = {10.48550/ARXIV.2601.13097},
  eprinttype    = {arXiv},
  eprint       = {2601.13097},
  bibsource    = {dblp computer science bibliography, https://dblp.org}
}

@inproceedings{DBLP:conf/icml/FluriLAF0S25,
  author       = {Lukas Fluri and
                  Leon Lang and
                  Alessandro Abate and
                  Patrick Forr{\'{e}} and
                  David Krueger and
                  Joar Max Viktor Skalse},
  editor       = {Aarti Singh and
                  Maryam Fazel and
                  Daniel Hsu and
                  Simon Lacoste{-}Julien and
                  Felix Berkenkamp and
                  Tegan Maharaj and
                  Kiri Wagstaff and
                  Jerry Zhu},
  title        = {The Perils of Optimizing Learned Reward Functions: Low Training Error
                  Does Not Guarantee Low Regret},
  booktitle    = {Forty-second International Conference on Machine Learning, {ICML}
                  2025, Vancouver, BC, Canada, July 13-19, 2025},
  series       = {Proceedings of Machine Learning Research},
  volume       = {267},
  publisher    = {{PMLR} / OpenReview.net},
  year         = {2025},
  url          = {https://proceedings.mlr.press/v267/fluri25a.html},
  bibsource    = {dblp computer science bibliography, https://dblp.org}
}

@inproceedings{DBLP:conf/naacl/LambertPMMLCDKZCSH25,
  author       = {Nathan Lambert and
                  Valentina Pyatkin and
                  Jacob Morrison and
                  Lester James V. Miranda and
                  Bill Yuchen Lin and
                  Khyathi Raghavi Chandu and
                  Nouha Dziri and
                  Sachin Kumar and
                  Tom Zick and
                  Yejin Choi and
                  Noah A. Smith and
                  Hannaneh Hajishirzi},
  editor       = {Luis Chiruzzo and
                  Alan Ritter and
                  Lu Wang},
  title        = {RewardBench: Evaluating Reward Models for Language Modeling},
  booktitle    = {Findings of the Association for Computational Linguistics: {NAACL}
                  2025, Albuquerque, New Mexico, USA, April 29 - May 4, 2025},
  series       = {Findings of {ACL}},
  volume       = {{NAACL} 2025},
  pages        = {1755--1797},
  publisher    = {Association for Computational Linguistics},
  year         = {2025},
  url          = {https://doi.org/10.18653/v1/2025.findings-naacl.96},
  doi          = {10.18653/V1/2025.FINDINGS-NAACL.96},
  bibsource    = {dblp computer science bibliography, https://dblp.org}
}

@inproceedings{DBLP:conf/iclr/Liu0M00L25,
  author       = {Yantao Liu and
                  Zijun Yao and
                  Rui Min and
                  Yixin Cao and
                  Lei Hou and
                  Juanzi Li},
  title        = {RM-Bench: Benchmarking Reward Models of Language Models with Subtlety
                  and Style},
  booktitle    = {The Thirteenth International Conference on Learning Representations,
                  {ICLR} 2025, Singapore, April 24-28, 2025},
  publisher    = {OpenReview.net},
  year         = {2025},
  url          = {https://openreview.net/forum?id=QEHrmQPBdd},
  bibsource    = {dblp computer science bibliography, https://dblp.org}
}

@inproceedings{DBLP:conf/iclr/TanZMTC0PS25,
  author       = {Sijun Tan and
                  Siyuan Zhuang and
                  Kyle Montgomery and
                  William Yuan Tang and
                  Alejandro Cuadron and
                  Chenguang Wang and
                  Raluca A. Popa and
                  Ion Stoica},
  title        = {JudgeBench: {A} Benchmark for Evaluating LLM-Based Judges},
  booktitle    = {The Thirteenth International Conference on Learning Representations,
                  {ICLR} 2025, Singapore, April 24-28, 2025},
  publisher    = {OpenReview.net},
  year         = {2025},
  url          = {https://openreview.net/forum?id=G0dksFayVq},
  bibsource    = {dblp computer science bibliography, https://dblp.org}
}

@inproceedings{DBLP:conf/iclr/FrickLCCAJZGS25,
  author       = {Evan Frick and
                  Tianle Li and
                  Connor Chen and
                  Wei{-}Lin Chiang and
                  Anastasios Nikolas Angelopoulos and
                  Jiantao Jiao and
                  Banghua Zhu and
                  Joseph E. Gonzalez and
                  Ion Stoica},
  title        = {How to Evaluate Reward Models for {RLHF}},
  booktitle    = {The Thirteenth International Conference on Learning Representations,
                  {ICLR} 2025, Singapore, April 24-28, 2025},
  publisher    = {OpenReview.net},
  year         = {2025},
  url          = {https://openreview.net/forum?id=cbttLtO94Q},
  bibsource    = {dblp computer science bibliography, https://dblp.org}
}

@inproceedings{DBLP:conf/iclr/0002G0L0K25,
  author       = {Zhihui Xie and
                  Jiahui Gao and
                  Lei Li and
                  Zhenguo Li and
                  Qi Liu and
                  Lingpeng Kong},
  title        = {Jailbreaking as a Reward Misspecification Problem},
  booktitle    = {The Thirteenth International Conference on Learning Representations,
                  {ICLR} 2025, Singapore, April 24-28, 2025},
  publisher    = {OpenReview.net},
  year         = {2025},
  url          = {https://openreview.net/forum?id=uBnM3EFovQ},
  bibsource    = {dblp computer science bibliography, https://dblp.org}
}

@article{DBLP:journals/corr/abs-2108-07732,
  author       = {Jacob Austin and
                  Augustus Odena and
                  Maxwell I. Nye and
                  Maarten Bosma and
                  Henryk Michalewski and
                  David Dohan and
                  Ellen Jiang and
                  Carrie J. Cai and
                  Michael Terry and
                  Quoc V. Le and
                  Charles Sutton},
  title        = {Program Synthesis with Large Language Models},
  journal      = {CoRR},
  volume       = {abs/2108.07732},
  year         = {2021},
  url          = {https://arxiv.org/abs/2108.07732},
  eprinttype    = {arXiv},
  eprint       = {2108.07732},
  bibsource    = {dblp computer science bibliography, https://dblp.org}
}

@article{DBLP:journals/corr/abs-2107-03374,
  author       = {Mark Chen and
                  Jerry Tworek and
                  Heewoo Jun and
                  Qiming Yuan and
                  Henrique Pond{\'{e}} de Oliveira Pinto and
                  Jared Kaplan and
                  Harri Edwards and
                  Yuri Burda and
                  Nicholas Joseph and
                  Greg Brockman and
                  Alex Ray and
                  Raul Puri and
                  Gretchen Krueger and
                  Michael Petrov and
                  Heidy Khlaaf and
                  Girish Sastry and
                  Pamela Mishkin and
                  Brooke Chan and
                  Scott Gray and
                  Nick Ryder and
                  Mikhail Pavlov and
                  Alethea Power and
                  Lukasz Kaiser and
                  Mohammad Bavarian and
                  Clemens Winter and
                  Philippe Tillet and
                  Felipe Petroski Such and
                  Dave Cummings and
                  Matthias Plappert and
                  Fotios Chantzis and
                  Elizabeth Barnes and
                  Ariel Herbert{-}Voss and
                  William Hebgen Guss and
                  Alex Nichol and
                  Alex Paino and
                  Nikolas Tezak and
                  Jie Tang and
                  Igor Babuschkin and
                  Suchir Balaji and
                  Shantanu Jain and
                  William Saunders and
                  Christopher Hesse and
                  Andrew N. Carr and
                  Jan Leike and
                  Joshua Achiam and
                  Vedant Misra and
                  Evan Morikawa and
                  Alec Radford and
                  Matthew Knight and
                  Miles Brundage and
                  Mira Murati and
                  Katie Mayer and
                  Peter Welinder and
                  Bob McGrew and
                  Dario Amodei and
                  Sam McCandlish and
                  Ilya Sutskever and
                  Wojciech Zaremba},
  title        = {Evaluating Large Language Models Trained on Code},
  journal      = {CoRR},
  volume       = {abs/2107.03374},
  year         = {2021},
  url          = {https://arxiv.org/abs/2107.03374},
  eprinttype    = {arXiv},
  eprint       = {2107.03374},
  bibsource    = {dblp computer science bibliography, https://dblp.org}
}

@article{DBLP:journals/corr/abs-2502-16614,
  author       = {Alexander Zhang and
                  Marcus Dong and
                  Jiaheng Liu and
                  Wei Zhang and
                  Yejie Wang and
                  Jian Yang and
                  Ge Zhang and
                  Tianyu Liu and
                  Zhongyuan Peng and
                  Yingshui Tan and
                  Yuanxing Zhang and
                  Zhexu Wang and
                  Weixun Wang and
                  Yancheng He and
                  Ken Deng and
                  Wangchunshu Zhou and
                  Wenhao Huang and
                  Zhaoxiang Zhang},
  title        = {CodeCriticBench: {A} Holistic Code Critique Benchmark for Large Language
                  Models},
  journal      = {CoRR},
  volume       = {abs/2502.16614},
  year         = {2025},
  url          = {https://doi.org/10.48550/arXiv.2502.16614},
  doi          = {10.48550/ARXIV.2502.16614},
  eprinttype    = {arXiv},
  eprint       = {2502.16614},
  bibsource    = {dblp computer science bibliography, https://dblp.org}
}

@inproceedings{DBLP:conf/coling/ZhaoLT0YL025,
  author       = {Yuwei Zhao and
                  Ziyang Luo and
                  Yuchen Tian and
                  Hongzhan Lin and
                  Weixiang Yan and
                  Annan Li and
                  Jing Ma},
  editor       = {Owen Rambow and
                  Leo Wanner and
                  Marianna Apidianaki and
                  Hend Al{-}Khalifa and
                  Barbara Di Eugenio and
                  Steven Schockaert},
  title        = {CodeJudge-Eval: Can Large Language Models be Good Judges in Code Understanding?},
  booktitle    = {Proceedings of the 31st International Conference on Computational
                  Linguistics, {COLING} 2025, Abu Dhabi, UAE, January 19-24, 2025},
  pages        = {73--95},
  publisher    = {Association for Computational Linguistics},
  year         = {2025},
  url          = {https://aclanthology.org/2025.coling-main.7/},
  bibsource    = {dblp computer science bibliography, https://dblp.org}
}

@article{DBLP:journals/corr/abs-2507-10535,
  author       = {Hongchao Jiang and
                  Yiming Chen and
                  Yushi Cao and
                  Hung{-}yi Lee and
                  Robby T. Tan},
  title        = {CodeJudgeBench: Benchmarking LLM-as-a-Judge for Coding Tasks},
  journal      = {CoRR},
  volume       = {abs/2507.10535},
  year         = {2025},
  url          = {https://doi.org/10.48550/arXiv.2507.10535},
  doi          = {10.48550/ARXIV.2507.10535},
  eprinttype    = {arXiv},
  eprint       = {2507.10535},
  bibsource    = {dblp computer science bibliography, https://dblp.org}
}

@article{Ficek2025ScoringVE,
  title={Scoring Verifiers: Evaluating Synthetic Verification in Code and Reasoning},
  author={Aleksander Ficek and Somshubra Majumdar and Vahid Noroozi and Boris Ginsburg},
  journal={ArXiv},
  year={2025},
  volume={abs/2502.13820},
  url={https://api.semanticscholar.org/CorpusID:277502118}
}

@article{DBLP:journals/tmlr/TaoC0MR0M25,
  author       = {Leitian Tao and
                  Xiang Chen and
                  Tong Yu and
                  Tung Mai and
                  Ryan A. Rossi and
                  Yixuan Li and
                  Saayan Mitra},
  title        = {CodeLutra: Boosting {LLM} Code Generation via Preference-Guided Refinement},
  journal      = {Trans. Mach. Learn. Res.},
  volume       = {2025},
  year         = {2025},
  url          = {https://openreview.net/forum?id=IGsEgWM4to},
  bibsource    = {dblp computer science bibliography, https://dblp.org}
}

@article{DBLP:journals/corr/abs-2412-13464,
  author       = {Nan Wang and
                  Yafei Liu and
                  Chen Chen and
                  Haonan Lu},
  title        = {GenX: Mastering Code and Test Generation with Execution Feedback},
  journal      = {CoRR},
  volume       = {abs/2412.13464},
  year         = {2024},
  url          = {https://doi.org/10.48550/arXiv.2412.13464},
  doi          = {10.48550/ARXIV.2412.13464},
  eprinttype    = {arXiv},
  eprint       = {2412.13464},
  bibsource    = {dblp computer science bibliography, https://dblp.org}
}

@article{DBLP:journals/jcphy/HesthavenU18,
  author       = {Jan S. Hesthaven and
                  Stefano Ubbiali},
  title        = {Non-intrusive reduced order modeling of nonlinear problems using neural
                  networks},
  journal      = {J. Comput. Phys.},
  volume       = {363},
  pages        = {55--78},
  year         = {2018},
  url          = {https://doi.org/10.1016/j.jcp.2018.02.037},
  doi          = {10.1016/J.JCP.2018.02.037},
  bibsource    = {dblp computer science bibliography, https://dblp.org}
}

@article{DBLP:journals/corr/abs-1901-05125,
  author       = {Dan Lu and
                  Daniel M. Ricciuto},
  title        = {Efficient surrogate modeling methods for large-scale Earth system
                  models based on machine learning techniques},
  journal      = {CoRR},
  volume       = {abs/1901.05125},
  year         = {2019},
  url          = {http://arxiv.org/abs/1901.05125},
  eprinttype    = {arXiv},
  eprint       = {1901.05125},
  bibsource    = {dblp computer science bibliography, https://dblp.org}
}

@inproceedings{DBLP:conf/nips/YanSKRH20,
  author       = {Yujun Yan and
                  Kevin Swersky and
                  Danai Koutra and
                  Parthasarathy Ranganathan and
                  Milad Hashemi},
  editor       = {Hugo Larochelle and
                  Marc'Aurelio Ranzato and
                  Raia Hadsell and
                  Maria{-}Florina Balcan and
                  Hsuan{-}Tien Lin},
  title        = {Neural Execution Engines: Learning to Execute Subroutines},
  booktitle    = {Advances in Neural Information Processing Systems 33: Annual Conference
                  on Neural Information Processing Systems 2020, NeurIPS 2020, December
                  6-12, 2020, virtual},
  year         = {2020},
  url          = {https://proceedings.neurips.cc/paper/2020/hash/c8b9abffb45bf79a630fb613dcd23449-Abstract.html},
  bibsource    = {dblp computer science bibliography, https://dblp.org}
}

@inproceedings{DBLP:conf/nips/RodionovP23,
  author       = {Gleb Rodionov and
                  Liudmila Prokhorenkova},
  editor       = {Alice Oh and
                  Tristan Naumann and
                  Amir Globerson and
                  Kate Saenko and
                  Moritz Hardt and
                  Sergey Levine},
  title        = {Neural Algorithmic Reasoning Without Intermediate Supervision},
  booktitle    = {Advances in Neural Information Processing Systems 36: Annual Conference
                  on Neural Information Processing Systems 2023, NeurIPS 2023, New Orleans,
                  LA, USA, December 10 - 16, 2023},
  year         = {2023},
  url          = {http://papers.nips.cc/paper\_files/paper/2023/hash/a2370db7c99791ad5d9f3ef48ad6d464-Abstract-Conference.html},
  bibsource    = {dblp computer science bibliography, https://dblp.org}
}

@inproceedings{DBLP:conf/nips/0003C0DJMVWG024,
  author       = {Yuxiang Wei and
                  Federico Cassano and
                  Jiawei Liu and
                  Yifeng Ding and
                  Naman Jain and
                  Zachary Mueller and
                  Harm de Vries and
                  Leandro von Werra and
                  Arjun Guha and
                  Lingming Zhang},
  editor       = {Amir Globersons and
                  Lester Mackey and
                  Danielle Belgrave and
                  Angela Fan and
                  Ulrich Paquet and
                  Jakub M. Tomczak and
                  Cheng Zhang},
  title        = {SelfCodeAlign: Self-Alignment for Code Generation},
  booktitle    = {Advances in Neural Information Processing Systems 38: Annual Conference
                  on Neural Information Processing Systems 2024, NeurIPS 2024, Vancouver,
                  BC, Canada, December 10 - 15, 2024},
  year         = {2024},
  url          = {http://papers.nips.cc/paper\_files/paper/2024/hash/72da102da91a8042a0b2aa968429a9f9-Abstract-Conference.html},
  bibsource    = {dblp computer science bibliography, https://dblp.org}
}

@inproceedings{DBLP:conf/emnlp/Zhou0AN23,
  author       = {Shuyan Zhou and
                  Uri Alon and
                  Sumit Agarwal and
                  Graham Neubig},
  editor       = {Houda Bouamor and
                  Juan Pino and
                  Kalika Bali},
  title        = {CodeBERTScore: Evaluating Code Generation with Pretrained Models of
                  Code},
  booktitle    = {Proceedings of the 2023 Conference on Empirical Methods in Natural
                  Language Processing, {EMNLP} 2023, Singapore, December 6-10, 2023},
  pages        = {13921--13937},
  publisher    = {Association for Computational Linguistics},
  year         = {2023},
  url          = {https://doi.org/10.18653/v1/2023.emnlp-main.859},
  doi          = {10.18653/V1/2023.EMNLP-MAIN.859},
  bibsource    = {dblp computer science bibliography, https://dblp.org}
}

@article{Zhu2026CodeScalerSC,
  title={CodeScaler: Scaling Code LLM Training and Test-Time Inference via Execution-Free Reward Models},
  author={Xiao Zhu and Xinyu Zhou and Boyu Zhu and Hanxu Hu and Mingzhe Du and Haotian Zhang and Huiming Wang and Zhijiang Guo},
  journal={CoRR},
  volume={abs/2602.17684},
  year={2026},
  url={https://doi.org/10.48550/arXiv.2602.17684},
  doi={10.48550/arXiv.2602.17684},
  eprinttype={arXiv},
  eprint={2602.17684}
}

@article{DBLP:journals/corr/abs-2403-09032,
  author       = {Martin Weyssow and
                  Aton Kamanda and
                  Houari A. Sahraoui},
  title        = {CodeUltraFeedback: An LLM-as-a-Judge Dataset for Aligning Large Language
                  Models to Coding Preferences},
  journal      = {CoRR},
  volume       = {abs/2403.09032},
  year         = {2024},
  url          = {https://doi.org/10.48550/arXiv.2403.09032},
  doi          = {10.48550/ARXIV.2403.09032},
  eprinttype    = {arXiv},
  eprint       = {2403.09032},
  bibsource    = {dblp computer science bibliography, https://dblp.org}
}

@article{DBLP:journals/corr/abs-2505-10320,
  author       = {Chenxi Whitehouse and
                  Tianlu Wang and
                  Ping Yu and
                  Xian Li and
                  Jason Weston and
                  Ilia Kulikov and
                  Swarnadeep Saha},
  title        = {{J1:} Incentivizing Thinking in LLM-as-a-Judge via Reinforcement Learning},
  journal      = {CoRR},
  volume       = {abs/2505.10320},
  year         = {2025},
  url          = {https://doi.org/10.48550/arXiv.2505.10320},
  doi          = {10.48550/ARXIV.2505.10320},
  eprinttype    = {arXiv},
  eprint       = {2505.10320},
  bibsource    = {dblp computer science bibliography, https://dblp.org}
}

@article{DBLP:journals/corr/abs-2504-00050,
  author       = {Nuo Chen and
                  Zhiyuan Hu and
                  Qingyun Zou and
                  Jiaying Wu and
                  Qian Wang and
                  Bryan Hooi and
                  Bingsheng He},
  title        = {JudgeLRM: Large Reasoning Models as a Judge},
  journal      = {CoRR},
  volume       = {abs/2504.00050},
  year         = {2025},
  url          = {https://doi.org/10.48550/arXiv.2504.00050},
  doi          = {10.48550/ARXIV.2504.00050},
  eprinttype    = {arXiv},
  eprint       = {2504.00050},
  bibsource    = {dblp computer science bibliography, https://dblp.org}
}

@article{DBLP:journals/corr/abs-2506-03637,
  author       = {Zhuohao Yu and
                  Jiali Zeng and
                  Weizheng Gu and
                  Yidong Wang and
                  Jindong Wang and
                  Fandong Meng and
                  Jie Zhou and
                  Yue Zhang and
                  Shikun Zhang and
                  Wei Ye},
  title        = {RewardAnything: Generalizable Principle-Following Reward Models},
  journal      = {CoRR},
  volume       = {abs/2506.03637},
  year         = {2025},
  url          = {https://doi.org/10.48550/arXiv.2506.03637},
  doi          = {10.48550/ARXIV.2506.03637},
  eprinttype    = {arXiv},
  eprint       = {2506.03637},
  bibsource    = {dblp computer science bibliography, https://dblp.org}
}

@article{DBLP:journals/corr/abs-2505-11475,
  author       = {Zhilin Wang and
                  Jiaqi Zeng and
                  Olivier Delalleau and
                  Hoo{-}Chang Shin and
                  Felipe Soares and
                  Alexander Bukharin and
                  Ellie Evans and
                  Yi Dong and
                  Oleksii Kuchaiev},
  title        = {HelpSteer3-Preference: Open Human-Annotated Preference Data across
                  Diverse Tasks and Languages},
  journal      = {CoRR},
  volume       = {abs/2505.11475},
  year         = {2025},
  url          = {https://doi.org/10.48550/arXiv.2505.11475},
  doi          = {10.48550/ARXIV.2505.11475},
  eprinttype    = {arXiv},
  eprint       = {2505.11475},
  bibsource    = {dblp computer science bibliography, https://dblp.org}
}

@article{DBLP:journals/corr/abs-2403-16950,
  author       = {Yinhong Liu and
                  Han Zhou and
                  Zhijiang Guo and
                  Ehsan Shareghi and
                  Ivan Vulic and
                  Anna Korhonen and
                  Nigel Collier},
  title        = {Aligning with Human Judgement: The Role of Pairwise Preference in
                  Large Language Model Evaluators},
  journal      = {CoRR},
  volume       = {abs/2403.16950},
  year         = {2024},
  url          = {https://doi.org/10.48550/arXiv.2403.16950},
  doi          = {10.48550/ARXIV.2403.16950},
  eprinttype    = {arXiv},
  eprint       = {2403.16950},
  bibsource    = {dblp computer science bibliography, https://dblp.org}
}

@inproceedings{
zhu2024starlingb,
    title={Starling-7B: Improving Helpfulness and Harmlessness with {RLAIF}},
    author={Banghua Zhu and Evan Frick and Tianhao Wu and Hanlin Zhu and Karthik Ganesan and Wei-Lin Chiang and Jian Zhang and Jiantao Jiao},
    booktitle={First Conference on Language Modeling},
    year={2024},
    url={https://openreview.net/forum?id=GqDntYTTbk}
}

@inproceedings{DBLP:conf/naacl/QinJHZWYSLLMWB24,
  author       = {Zhen Qin and
                  Rolf Jagerman and
                  Kai Hui and
                  Honglei Zhuang and
                  Junru Wu and
                  Le Yan and
                  Jiaming Shen and
                  Tianqi Liu and
                  Jialu Liu and
                  Donald Metzler and
                  Xuanhui Wang and
                  Michael Bendersky},
  editor       = {Kevin Duh and
                  Helena G{\'{o}}mez{-}Adorno and
                  Steven Bethard},
  title        = {Large Language Models are Effective Text Rankers with Pairwise Ranking
                  Prompting},
  booktitle    = {Findings of the Association for Computational Linguistics: {NAACL}
                  2024, Mexico City, Mexico, June 16-21, 2024},
  series       = {Findings of {ACL}},
  volume       = {{NAACL} 2024},
  pages        = {1504--1518},
  publisher    = {Association for Computational Linguistics},
  year         = {2024},
  url          = {https://doi.org/10.18653/v1/2024.findings-naacl.97},
  doi          = {10.18653/V1/2024.FINDINGS-NAACL.97},
  bibsource    = {dblp computer science bibliography, https://dblp.org}
}

@inproceedings{DBLP:conf/nips/ZhaoLH23,
  author       = {Zirui Zhao and
                  Wee Sun Lee and
                  David Hsu},
  editor       = {Alice Oh and
                  Tristan Naumann and
                  Amir Globerson and
                  Kate Saenko and
                  Moritz Hardt and
                  Sergey Levine},
  title        = {Large Language Models as Commonsense Knowledge for Large-Scale Task
                  Planning},
  booktitle    = {Advances in Neural Information Processing Systems 36: Annual Conference
                  on Neural Information Processing Systems 2023, NeurIPS 2023, New Orleans,
                  LA, USA, December 10 - 16, 2023},
  year         = {2023},
  url          = {http://papers.nips.cc/paper\_files/paper/2023/hash/65a39213d7d0e1eb5d192aa77e77eeb7-Abstract-Conference.html},
  bibsource    = {dblp computer science bibliography, https://dblp.org}
}

@article{DBLP:journals/tosem/DongDJLLJ25,
  author       = {Yihong Dong and
                  Jiazheng Ding and
                  Xue Jiang and
                  Ge Li and
                  Zhuo Li and
                  Zhi Jin},
  title        = {CodeScore: Evaluating Code Generation by Learning Code Execution},
  journal      = {{ACM} Trans. Softw. Eng. Methodol.},
  volume       = {34},
  number       = {3},
  pages        = {77:1--77:22},
  year         = {2025},
  url          = {https://doi.org/10.1145/3695991},
  doi          = {10.1145/3695991},
  bibsource    = {dblp computer science bibliography, https://dblp.org}
}

@inproceedings{DBLP:conf/nips/ShinnCGNY23,
  author       = {Noah Shinn and
                  Federico Cassano and
                  Ashwin Gopinath and
                  Karthik Narasimhan and
                  Shunyu Yao},
  editor       = {Alice Oh and
                  Tristan Naumann and
                  Amir Globerson and
                  Kate Saenko and
                  Moritz Hardt and
                  Sergey Levine},
  title        = {Reflexion: language agents with verbal reinforcement learning},
  booktitle    = {Advances in Neural Information Processing Systems 36: Annual Conference
                  on Neural Information Processing Systems 2023, NeurIPS 2023, New Orleans,
                  LA, USA, December 10 - 16, 2023},
  year         = {2023},
  url          = {http://papers.nips.cc/paper\_files/paper/2023/hash/1b44b878bb782e6954cd888628510e90-Abstract-Conference.html},
  bibsource    = {dblp computer science bibliography, https://dblp.org}
}

@inproceedings{DBLP:conf/nips/MadaanTGHGW0DPY23,
  author       = {Aman Madaan and
                  Niket Tandon and
                  Prakhar Gupta and
                  Skyler Hallinan and
                  Luyu Gao and
                  Sarah Wiegreffe and
                  Uri Alon and
                  Nouha Dziri and
                  Shrimai Prabhumoye and
                  Yiming Yang and
                  Shashank Gupta and
                  Bodhisattwa Prasad Majumder and
                  Katherine Hermann and
                  Sean Welleck and
                  Amir Yazdanbakhsh and
                  Peter Clark},
  editor       = {Alice Oh and
                  Tristan Naumann and
                  Amir Globerson and
                  Kate Saenko and
                  Moritz Hardt and
                  Sergey Levine},
  title        = {Self-Refine: Iterative Refinement with Self-Feedback},
  booktitle    = {Advances in Neural Information Processing Systems 36: Annual Conference
                  on Neural Information Processing Systems 2023, NeurIPS 2023, New Orleans,
                  LA, USA, December 10 - 16, 2023},
  year         = {2023},
  url          = {http://papers.nips.cc/paper\_files/paper/2023/hash/91edff07232fb1b55a505a9e9f6c0ff3-Abstract-Conference.html},
  bibsource    = {dblp computer science bibliography, https://dblp.org}
}

@article{DBLP:journals/corr/abs-2502-09183,
  author       = {Changzhi Zhou and
                  Xinyu Zhang and
                  Dandan Song and
                  Xiancai Chen and
                  Wanli Gu and
                  Huipeng Ma and
                  Yuhang Tian and
                  Mengdi Zhang and
                  Linmei Hu},
  title        = {RefineCoder: Iterative Improving of Large Language Models via Adaptive
                  Critique Refinement for Code Generation},
  journal      = {CoRR},
  volume       = {abs/2502.09183},
  year         = {2025},
  url          = {https://doi.org/10.48550/arXiv.2502.09183},
  doi          = {10.48550/ARXIV.2502.09183},
  eprinttype    = {arXiv},
  eprint       = {2502.09183},
  bibsource    = {dblp computer science bibliography, https://dblp.org}
}

@inproceedings{DBLP:conf/acl/ZhangLLLJ23,
  author       = {Kechi Zhang and
                  Zhuo Li and
                  Jia Li and
                  Ge Li and
                  Zhi Jin},
  editor       = {Anna Rogers and
                  Jordan L. Boyd{-}Graber and
                  Naoaki Okazaki},
  title        = {Self-Edit: Fault-Aware Code Editor for Code Generation},
  booktitle    = {Proceedings of the 61st Annual Meeting of the Association for Computational
                  Linguistics (Volume 1: Long Papers), {ACL} 2023, Toronto, Canada,
                  July 9-14, 2023},
  pages        = {769--787},
  publisher    = {Association for Computational Linguistics},
  year         = {2023},
  url          = {https://doi.org/10.18653/v1/2023.acl-long.45},
  doi          = {10.18653/V1/2023.ACL-LONG.45},
  bibsource    = {dblp computer science bibliography, https://dblp.org}
}

@article{DBLP:journals/corr/abs-2303-16749,
  author       = {Angelica Chen and
                  J{\'{e}}r{\'{e}}my Scheurer and
                  Tomasz Korbak and
                  Jon Ander Campos and
                  Jun Shern Chan and
                  Samuel R. Bowman and
                  Kyunghyun Cho and
                  Ethan Perez},
  title        = {Improving Code Generation by Training with Natural Language Feedback},
  journal      = {CoRR},
  volume       = {abs/2303.16749},
  year         = {2023},
  url          = {https://doi.org/10.48550/arXiv.2303.16749},
  doi          = {10.48550/ARXIV.2303.16749},
  eprinttype    = {arXiv},
  eprint       = {2303.16749},
  bibsource    = {dblp computer science bibliography, https://dblp.org}
}

@article{DBLP:journals/corr/abs-2311-07215,
  author       = {Seungjun Moon and
                  Yongho Song and
                  Hyungjoo Chae and
                  Dongjin Kang and
                  Taeyoon Kwon and
                  Kai Tzu{-}iunn Ong and
                  Seung{-}won Hwang and
                  Jinyoung Yeo},
  title        = {Coffee: Boost Your Code LLMs by Fixing Bugs with Feedback},
  journal      = {CoRR},
  volume       = {abs/2311.07215},
  year         = {2023},
  url          = {https://doi.org/10.48550/arXiv.2311.07215},
  doi          = {10.48550/ARXIV.2311.07215},
  eprinttype    = {arXiv},
  eprint       = {2311.07215},
  bibsource    = {dblp computer science bibliography, https://dblp.org}
}

@inproceedings{DBLP:conf/nips/GuptaCCS20,
  author       = {Kavi Gupta and
                  Peter Ebert Christensen and
                  Xinyun Chen and
                  Dawn Song},
  editor       = {Hugo Larochelle and
                  Marc'Aurelio Ranzato and
                  Raia Hadsell and
                  Maria{-}Florina Balcan and
                  Hsuan{-}Tien Lin},
  title        = {Synthesize, Execute and Debug: Learning to Repair for Neural Program
                  Synthesis},
  booktitle    = {Advances in Neural Information Processing Systems 33: Annual Conference
                  on Neural Information Processing Systems 2020, NeurIPS 2020, December
                  6-12, 2020, virtual},
  year         = {2020},
  url          = {https://proceedings.neurips.cc/paper/2020/hash/cd0f74b5955dc87fd0605745c4b49ee8-Abstract.html},
  bibsource    = {dblp computer science bibliography, https://dblp.org}
}

@inproceedings{DBLP:conf/iclr/GouSGSYDC24,
  author       = {Zhibin Gou and
                  Zhihong Shao and
                  Yeyun Gong and
                  Yelong Shen and
                  Yujiu Yang and
                  Nan Duan and
                  Weizhu Chen},
  title        = {{CRITIC:} Large Language Models Can Self-Correct with Tool-Interactive
                  Critiquing},
  booktitle    = {The Twelfth International Conference on Learning Representations,
                  {ICLR} 2024, Vienna, Austria, May 7-11, 2024},
  publisher    = {OpenReview.net},
  year         = {2024},
  url          = {https://openreview.net/forum?id=Sx038qxjek},
  bibsource    = {dblp computer science bibliography, https://dblp.org}
}

@inproceedings{DBLP:conf/acl/WangWWMLZLWJL22,
  author       = {Xin Wang and
                  Yasheng Wang and
                  Yao Wan and
                  Fei Mi and
                  Yitong Li and
                  Pingyi Zhou and
                  Jin Liu and
                  Hao Wu and
                  Xin Jiang and
                  Qun Liu},
  editor       = {Smaranda Muresan and
                  Preslav Nakov and
                  Aline Villavicencio},
  title        = {Compilable Neural Code Generation with Compiler Feedback},
  booktitle    = {Findings of the Association for Computational Linguistics: {ACL} 2022,
                  Dublin, Ireland, May 22-27, 2022},
  series       = {Findings of {ACL}},
  volume       = {{ACL} 2022},
  pages        = {9--19},
  publisher    = {Association for Computational Linguistics},
  year         = {2022},
  url          = {https://doi.org/10.18653/v1/2022.findings-acl.2},
  doi          = {10.18653/V1/2022.FINDINGS-ACL.2},
  bibsource    = {dblp computer science bibliography, https://dblp.org}
}

@inproceedings{DBLP:conf/iclr/ChenLSZ24,
  author       = {Xinyun Chen and
                  Maxwell Lin and
                  Nathanael Sch{\"{a}}rli and
                  Denny Zhou},
  title        = {Teaching Large Language Models to Self-Debug},
  booktitle    = {The Twelfth International Conference on Learning Representations,
                  {ICLR} 2024, Vienna, Austria, May 7-11, 2024},
  publisher    = {OpenReview.net},
  year         = {2024},
  url          = {https://openreview.net/forum?id=KuPixIqPiq},
  bibsource    = {dblp computer science bibliography, https://dblp.org}
}

@inproceedings{DBLP:conf/icml/NiACDSSY24,
  author       = {Ansong Ni and
                  Miltiadis Allamanis and
                  Arman Cohan and
                  Yinlin Deng and
                  Kensen Shi and
                  Charles Sutton and
                  Pengcheng Yin},
  editor       = {Ruslan Salakhutdinov and
                  Zico Kolter and
                  Katherine A. Heller and
                  Adrian Weller and
                  Nuria Oliver and
                  Jonathan Scarlett and
                  Felix Berkenkamp},
  title        = {NExT: Teaching Large Language Models to Reason about Code Execution},
  booktitle    = {Forty-first International Conference on Machine Learning, {ICML} 2024,
                  Vienna, Austria, July 21-27, 2024},
  series       = {Proceedings of Machine Learning Research},
  volume       = {235},
  pages        = {37929--37956},
  publisher    = {{PMLR} / OpenReview.net},
  year         = {2024},
  url          = {https://proceedings.mlr.press/v235/ni24a.html},
  bibsource    = {dblp computer science bibliography, https://dblp.org}
}

@article{DBLP:journals/corr/abs-2410-03837,
  author       = {Jiawei Liu and
                  Thanh Nguyen and
                  Mingyue Shang and
                  Hantian Ding and
                  Xiaopeng Li and
                  Yu Yu and
                  Varun Kumar and
                  Zijian Wang},
  title        = {Learning Code Preference via Synthetic Evolution},
  journal      = {CoRR},
  volume       = {abs/2410.03837},
  year         = {2024},
  url          = {https://doi.org/10.48550/arXiv.2410.03837},
  doi          = {10.48550/ARXIV.2410.03837},
  eprinttype    = {arXiv},
  eprint       = {2410.03837},
  bibsource    = {dblp computer science bibliography, https://dblp.org}
}

@article{DBLP:journals/corr/abs-2502-01619,
  author       = {Archiki Prasad and
                  Elias Stengel{-}Eskin and
                  Justin Chih{-}Yao Chen and
                  Zaid Khan and
                  Mohit Bansal},
  title        = {Learning to Generate Unit Tests for Automated Debugging},
  journal      = {CoRR},
  volume       = {abs/2502.01619},
  year         = {2025},
  url          = {https://doi.org/10.48550/arXiv.2502.01619},
  doi          = {10.48550/ARXIV.2502.01619},
  eprinttype    = {arXiv},
  eprint       = {2502.01619},
  bibsource    = {dblp computer science bibliography, https://dblp.org}
}

@article{DBLP:journals/corr/abs-2411-02310,
  author       = {Shukai Liu and
                  Linzheng Chai and
                  Jian Yang and
                  Jiajun Shi and
                  He Zhu and
                  Liran Wang and
                  Ke Jin and
                  Wei Zhang and
                  Hualei Zhu and
                  Shuyue Guo and
                  Tao Sun and
                  Jiaheng Liu and
                  Yunlong Duan and
                  Yu Hao and
                  Liqun Yang and
                  Guanglin Niu and
                  Ge Zhang and
                  Zhoujun Li},
  title        = {MdEval: Massively Multilingual Code Debugging},
  journal      = {CoRR},
  volume       = {abs/2411.02310},
  year         = {2024},
  url          = {https://doi.org/10.48550/arXiv.2411.02310},
  doi          = {10.48550/ARXIV.2411.02310},
  eprinttype    = {arXiv},
  eprint       = {2411.02310},
  bibsource    = {dblp computer science bibliography, https://dblp.org}
}

@inproceedings{DBLP:conf/naacl/YangWLLYWGYLY25,
  author       = {Weiqing Yang and
                  Hanbin Wang and
                  Zhenghao Liu and
                  Xinze Li and
                  Yukun Yan and
                  Shuo Wang and
                  Yu Gu and
                  Minghe Yu and
                  Zhiyuan Liu and
                  Ge Yu},
  editor       = {Luis Chiruzzo and
                  Alan Ritter and
                  Lu Wang},
  title        = {{COAST:} Enhancing the Code Debugging Ability of LLMs through Communicative
                  Agent Based Data Synthesis},
  booktitle    = {Findings of the Association for Computational Linguistics: {NAACL}
                  2025, Albuquerque, New Mexico, USA, April 29 - May 4, 2025},
  series       = {Findings of {ACL}},
  volume       = {{NAACL} 2025},
  pages        = {2570--2585},
  publisher    = {Association for Computational Linguistics},
  year         = {2025},
  url          = {https://doi.org/10.18653/v1/2025.findings-naacl.139},
  doi          = {10.18653/V1/2025.FINDINGS-NAACL.139},
  bibsource    = {dblp computer science bibliography, https://dblp.org}
}

@article{DBLP:journals/corr/abs-2304-01102,
  author       = {Julian Aron Prenner and
                  Romain Robbes},
  title        = {RunBugRun - An Executable Dataset for Automated Program Repair},
  journal      = {CoRR},
  volume       = {abs/2304.01102},
  year         = {2023},
  url          = {https://doi.org/10.48550/arXiv.2304.01102},
  doi          = {10.48550/ARXIV.2304.01102},
  eprinttype    = {arXiv},
  eprint       = {2304.01102},
  bibsource    = {dblp computer science bibliography, https://dblp.org}
}

@inproceedings{DBLP:conf/emnlp/WaghjaleVWF24,
  author       = {Siddhant Waghjale and
                  Vishruth Veerendranath and
                  Zhiruo Wang and
                  Daniel Fried},
  editor       = {Yaser Al{-}Onaizan and
                  Mohit Bansal and
                  Yun{-}Nung Chen},
  title        = {{ECCO:} Can We Improve Model-Generated Code Efficiency Without Sacrificing
                  Functional Correctness?},
  booktitle    = {Proceedings of the 2024 Conference on Empirical Methods in Natural
                  Language Processing, {EMNLP} 2024, Miami, FL, USA, November 12-16,
                  2024},
  pages        = {15362--15376},
  publisher    = {Association for Computational Linguistics},
  year         = {2024},
  url          = {https://doi.org/10.18653/v1/2024.emnlp-main.859},
  doi          = {10.18653/V1/2024.EMNLP-MAIN.859},
  bibsource    = {dblp computer science bibliography, https://dblp.org}
}

@article{DBLP:journals/corr/abs-2408-06450,
  author       = {Jiawei Liu and
                  Songrun Xie and
                  Junhao Wang and
                  Yuxiang Wei and
                  Yifeng Ding and
                  Lingming Zhang},
  title        = {Evaluating Language Models for Efficient Code Generation},
  journal      = {CoRR},
  volume       = {abs/2408.06450},
  year         = {2024},
  url          = {https://doi.org/10.48550/arXiv.2408.06450},
  doi          = {10.48550/ARXIV.2408.06450},
  eprinttype    = {arXiv},
  eprint       = {2408.06450},
  bibsource    = {dblp computer science bibliography, https://dblp.org}
}

@article{DBLP:journals/corr/abs-2401-15963,
  author       = {Manav Singhal and
                  Tushar Aggarwal and
                  Abhijeet Awasthi and
                  Nagarajan Natarajan and
                  Aditya Kanade},
  title        = {NoFunEval: Funny How Code LMs Falter on Requirements Beyond Functional
                  Correctness},
  journal      = {CoRR},
  volume       = {abs/2401.15963},
  year         = {2024},
  url          = {https://doi.org/10.48550/arXiv.2401.15963},
  doi          = {10.48550/ARXIV.2401.15963},
  eprinttype    = {arXiv},
  eprint       = {2401.15963},
  bibsource    = {dblp computer science bibliography, https://dblp.org}
}

@inproceedings{DBLP:conf/msr/BuiSF22,
  author       = {Quang{-}Cuong Bui and
                  Riccardo Scandariato and
                  Nicol{\'{a}}s E. D{\'{\i}}az Ferreyra},
  title        = {Vul4J: {A} Dataset of Reproducible Java Vulnerabilities Geared Towards
                  the Study of Program Repair Techniques},
  booktitle    = {19th {IEEE/ACM} International Conference on Mining Software Repositories,
                  {MSR} 2022, Pittsburgh, PA, USA, May 23-24, 2022},
  pages        = {464--468},
  publisher    = {{ACM}},
  year         = {2022},
  url          = {https://doi.org/10.1145/3524842.3528482},
  doi          = {10.1145/3524842.3528482},
  bibsource    = {dblp computer science bibliography, https://dblp.org}
}

@article{DBLP:journals/corr/abs-2412-20787,
  author       = {Pengfei Jing and
                  Mengyun Tang and
                  Xiaorong Shi and
                  Xing Zheng and
                  Sen Nie and
                  Shi Wu and
                  Yong Yang and
                  Xiapu Luo},
  title        = {SecBench: {A} Comprehensive Multi-Dimensional Benchmarking Dataset
                  for LLMs in Cybersecurity},
  journal      = {CoRR},
  volume       = {abs/2412.20787},
  year         = {2024},
  url          = {https://doi.org/10.48550/arXiv.2412.20787},
  doi          = {10.48550/ARXIV.2412.20787},
  eprinttype    = {arXiv},
  eprint       = {2412.20787},
  bibsource    = {dblp computer science bibliography, https://dblp.org}
}

@inproceedings{DBLP:conf/acl/NguyenPSB25,
  author       = {Duy Nguyen and
                  Archiki Prasad and
                  Elias Stengel{-}Eskin and
                  Mohit Bansal},
  editor       = {Wanxiang Che and
                  Joyce Nabende and
                  Ekaterina Shutova and
                  Mohammad Taher Pilehvar},
  title        = {Multi-Attribute Steering of Language Models via Targeted Intervention},
  booktitle    = {Proceedings of the 63rd Annual Meeting of the Association for Computational
                  Linguistics (Volume 1: Long Papers), {ACL} 2025, Vienna, Austria,
                  July 27 - August 1, 2025},
  pages        = {20619--20634},
  publisher    = {Association for Computational Linguistics},
  year         = {2025},
  url          = {https://aclanthology.org/2025.acl-long.1007/},
  bibsource    = {dblp computer science bibliography, https://dblp.org}
}

@inproceedings{DBLP:conf/nips/YangLXHZA24,
  author       = {Kailai Yang and
                  Zhiwei Liu and
                  Qianqian Xie and
                  Jimin Huang and
                  Tianlin Zhang and
                  Sophia Ananiadou},
  editor       = {Amir Globersons and
                  Lester Mackey and
                  Danielle Belgrave and
                  Angela Fan and
                  Ulrich Paquet and
                  Jakub M. Tomczak and
                  Cheng Zhang},
  title        = {MetaAligner: Towards Generalizable Multi-Objective Alignment of Language
                  Models},
  booktitle    = {Advances in Neural Information Processing Systems 38: Annual Conference
                  on Neural Information Processing Systems 2024, NeurIPS 2024, Vancouver,
                  BC, Canada, December 10 - 15, 2024},
  year         = {2024},
  url          = {http://papers.nips.cc/paper\_files/paper/2024/hash/3d03800841fa1bb2f43ef1750aafcce4-Abstract-Conference.html},
  bibsource    = {dblp computer science bibliography, https://dblp.org}
}

@inproceedings{DBLP:conf/icml/0010PLQ00C24,
  author       = {Rui Yang and
                  Xiaoman Pan and
                  Feng Luo and
                  Shuang Qiu and
                  Han Zhong and
                  Dong Yu and
                  Jianshu Chen},
  editor       = {Ruslan Salakhutdinov and
                  Zico Kolter and
                  Katherine A. Heller and
                  Adrian Weller and
                  Nuria Oliver and
                  Jonathan Scarlett and
                  Felix Berkenkamp},
  title        = {Rewards-in-Context: Multi-objective Alignment of Foundation Models
                  with Dynamic Preference Adjustment},
  booktitle    = {Forty-first International Conference on Machine Learning, {ICML} 2024,
                  Vienna, Austria, July 21-27, 2024},
  series       = {Proceedings of Machine Learning Research},
  volume       = {235},
  pages        = {56276--56297},
  publisher    = {{PMLR} / OpenReview.net},
  year         = {2024},
  url          = {https://proceedings.mlr.press/v235/yang24q.html},
  bibsource    = {dblp computer science bibliography, https://dblp.org}
}

@inproceedings{DBLP:conf/acl/ZhouLS00O024,
  author       = {Zhanhui Zhou and
                  Jie Liu and
                  Jing Shao and
                  Xiangyu Yue and
                  Chao Yang and
                  Wanli Ouyang and
                  Yu Qiao},
  editor       = {Lun{-}Wei Ku and
                  Andre Martins and
                  Vivek Srikumar},
  title        = {Beyond One-Preference-Fits-All Alignment: Multi-Objective Direct Preference
                  Optimization},
  booktitle    = {Findings of the Association for Computational Linguistics, {ACL} 2024,
                  Bangkok, Thailand and virtual meeting, August 11-16, 2024},
  series       = {Findings of {ACL}},
  volume       = {{ACL} 2024},
  pages        = {10586--10613},
  publisher    = {Association for Computational Linguistics},
  year         = {2024},
  url          = {https://doi.org/10.18653/v1/2024.findings-acl.630},
  doi          = {10.18653/V1/2024.FINDINGS-ACL.630},
  bibsource    = {dblp computer science bibliography, https://dblp.org}
}

@article{DBLP:journals/tcyb/LiZW21,
  author       = {Kaiwen Li and
                  Tao Zhang and
                  Rui Wang},
  title        = {Deep Reinforcement Learning for Multiobjective Optimization},
  journal      = {{IEEE} Trans. Cybern.},
  volume       = {51},
  number       = {6},
  pages        = {3103--3114},
  year         = {2021},
  url          = {https://doi.org/10.1109/TCYB.2020.2977661},
  doi          = {10.1109/TCYB.2020.2977661},
  bibsource    = {dblp computer science bibliography, https://dblp.org}
}

@inproceedings{DBLP:conf/kbse/CoignionQR25,
  author       = {Tristan Coignion and
                  Cl{\'{e}}ment Quinton and
                  Romain Rouvoy},
  title        = {When Faster Isn't Greener: The Hidden Costs of LLM-Based Code
                  Optimization},
  booktitle    = {40th {IEEE/ACM} International Conference on Automated Software Engineering,
                  {ASE} 2025, Seoul, Korea, Republic of, November 16-20, 2025},
  pages        = {1655--1666},
  publisher    = {{IEEE}},
  year         = {2025},
  url          = {https://doi.org/10.1109/ASE63991.2025.00139},
  doi          = {10.1109/ASE63991.2025.00139},
  bibsource    = {dblp computer science bibliography, https://dblp.org}
}

@article{DBLP:journals/corr/abs-2506-08945,
  author       = {Simone Daniotti and
                  Johannes Wachs and
                  Xiangnan Feng and
                  Frank Neffke},
  title        = {Who is using {AI} to code? Global diffusion and impact of generative
                  {AI}},
  journal      = {CoRR},
  volume       = {abs/2506.08945},
  year         = {2025},
  url          = {https://doi.org/10.48550/arXiv.2506.08945},
  doi          = {10.48550/ARXIV.2506.08945},
  eprinttype    = {arXiv},
  eprint       = {2506.08945},
  bibsource    = {dblp computer science bibliography, https://dblp.org}
}

@article{DBLP:journals/corr/abs-2510-03029,
  author       = {Debalina Ghosh Paul and
                  Hong Zhu and
                  Ian Bayley},
  title        = {Investigating The Smells of {LLM} Generated Code},
  journal      = {CoRR},
  volume       = {abs/2510.03029},
  year         = {2025},
  url          = {https://doi.org/10.48550/arXiv.2510.03029},
  doi          = {10.48550/ARXIV.2510.03029},
  eprinttype    = {arXiv},
  eprint       = {2510.03029},
  bibsource    = {dblp computer science bibliography, https://dblp.org}
}

@article{DBLP:journals/cacm/PearceATDK25,
  author       = {Hammond Pearce and
                  Baleegh Ahmad and
                  Benjamin Tan and
                  Brendan Dolan{-}Gavitt and
                  Ramesh Karri},
  title        = {Asleep at the Keyboard? Assessing the Security of GitHub Copilot's
                  Code Contributions},
  journal      = {Commun. {ACM}},
  volume       = {68},
  number       = {2},
  pages        = {96--105},
  year         = {2025},
  url          = {https://doi.org/10.1145/3610721},
  doi          = {10.1145/3610721},
  bibsource    = {dblp computer science bibliography, https://dblp.org}
}

@article{DBLP:journals/corr/abs-2508-14419,
  author       = {Scott Blyth and
                  Sherlock A. Licorish and
                  Christoph Treude and
                  Markus Wagner},
  title        = {Static Analysis as a Feedback Loop: Enhancing LLM-Generated Code Beyond
                  Correctness},
  journal      = {CoRR},
  volume       = {abs/2508.14419},
  year         = {2025},
  url          = {https://doi.org/10.48550/arXiv.2508.14419},
  doi          = {10.48550/ARXIV.2508.14419},
  eprinttype    = {arXiv},
  eprint       = {2508.14419},
  bibsource    = {dblp computer science bibliography, https://dblp.org}
}

@article{Siddiq2022SecurityEvalDM,
  title={SecurityEval dataset: mining vulnerability examples to evaluate machine learning-based code generation techniques},
  author={Mohammed Latif Siddiq and Joanna C. S. Santos},
  journal={Proceedings of the 1st International Workshop on Mining Software Repositories Applications for Privacy and Security},
  year={2022},
  url={https://api.semanticscholar.org/CorpusID:252125684}
}

@article{DBLP:journals/corr/abs-2502-01853,
  author       = {Mohammed Kharma and
                  Soohyeon Choi and
                  Mohammed AlKhanafseh and
                  David Mohaisen},
  title        = {Security and Quality in LLM-Generated Code: {A} Multi-Language, Multi-Model
                  Analysis},
  journal      = {CoRR},
  volume       = {abs/2502.01853},
  year         = {2025},
  url          = {https://doi.org/10.48550/arXiv.2502.01853},
  doi          = {10.48550/ARXIV.2502.01853},
  eprinttype    = {arXiv},
  eprint       = {2502.01853},
  bibsource    = {dblp computer science bibliography, https://dblp.org}
}

@inproceedings{DBLP:conf/ccs/HeV23,
  author       = {Jingxuan He and
                  Martin T. Vechev},
  editor       = {Weizhi Meng and
                  Christian Damsgaard Jensen and
                  Cas Cremers and
                  Engin Kirda},
  title        = {Large Language Models for Code: Security Hardening and Adversarial
                  Testing},
  booktitle    = {Proceedings of the 2023 {ACM} {SIGSAC} Conference on Computer and
                  Communications Security, {CCS} 2023, Copenhagen, Denmark, November
                  26-30, 2023},
  pages        = {1865--1879},
  publisher    = {{ACM}},
  year         = {2023},
  url          = {https://doi.org/10.1145/3576915.3623175},
  doi          = {10.1145/3576915.3623175},
  bibsource    = {dblp computer science bibliography, https://dblp.org}
}

@article{DBLP:journals/corr/abs-2408-13855,
  author       = {Han Cui and
                  Menglei Xie and
                  Ting Su and
                  Chengyu Zhang and
                  Shin Hwei Tan},
  title        = {An Empirical Study of False Negatives and Positives of Static Code
                  Analyzers From the Perspective of Historical Issues},
  journal      = {CoRR},
  volume       = {abs/2408.13855},
  year         = {2024},
  url          = {https://doi.org/10.48550/arXiv.2408.13855},
  doi          = {10.48550/ARXIV.2408.13855},
  eprinttype    = {arXiv},
  eprint       = {2408.13855},
  bibsource    = {dblp computer science bibliography, https://dblp.org}
}

@article{DBLP:journals/corr/abs-2504-04699,
  author       = {Martin Weyssow and
                  Chengran Yang and
                  Junkai Chen and
                  Yikun Li and
                  Huihui Huang and
                  Ratnadira Widyasari and
                  Han Wei Ang and
                  Frank Liauw and
                  Eng Lieh Ouh and
                  Lwin Khin Shar and
                  David Lo},
  title        = {R2Vul: Learning to Reason about Software Vulnerabilities with Reinforcement
                  Learning and Structured Reasoning Distillation},
  journal      = {CoRR},
  volume       = {abs/2504.04699},
  year         = {2025},
  url          = {https://doi.org/10.48550/arXiv.2504.04699},
  doi          = {10.48550/ARXIV.2504.04699},
  eprinttype    = {arXiv},
  eprint       = {2504.04699},
  bibsource    = {dblp computer science bibliography, https://dblp.org}
}

@inproceedings{DBLP:conf/trustcom/QuLKLWYH25,
  author       = {Muzi Qu and
                  Jie Liu and
                  Liangyi Kang and
                  Shuyi Ling and
                  Shuai Wang and
                  Dan Ye and
                  Tao Huang},
  title        = {SCodeGen: {A} Real-Time Trustworthy Constrained Decoding Framework
                  for Secure Code Generation with LLMs},
  booktitle    = {24th {IEEE} International Conference on Trust, Security and Privacy
                  in Computing and Communications, TrustCom 2025, Guiyang, China, November
                  14-17, 2025},
  pages        = {492--503},
  publisher    = {{IEEE}},
  year         = {2025},
  url          = {https://doi.org/10.1109/Trustcom66490.2025.00061},
  doi          = {10.1109/TRUSTCOM66490.2025.00061},
  bibsource    = {dblp computer science bibliography, https://dblp.org}
}

@inproceedings{DBLP:conf/emnlp/WangLHWH25,
  author       = {Jiayou Wang and
                  Rundong Liu and
                  Yue Hu and
                  Huijia Wu and
                  Zhaofeng He},
  editor       = {Christos Christodoulopoulos and
                  Tanmoy Chakraborty and
                  Carolyn Rose and
                  Violet Peng},
  title        = {SecDecoding: Steerable Decoding for Safer {LLM} Generation},
  booktitle    = {Findings of the Association for Computational Linguistics: {EMNLP}
                  2025, Suzhou, China, November 4-9, 2025},
  pages        = {20504--20521},
  publisher    = {Association for Computational Linguistics},
  year         = {2025},
  url          = {https://aclanthology.org/2025.findings-emnlp.1118/},
  bibsource    = {dblp computer science bibliography, https://dblp.org}
}

@article{DBLP:journals/corr/abs-2405-00218,
  author       = {Yanjun Fu and
                  Ethan Baker and
                  Yizheng Chen},
  title        = {Constrained Decoding for Secure Code Generation},
  journal      = {CoRR},
  volume       = {abs/2405.00218},
  year         = {2024},
  url          = {https://doi.org/10.48550/arXiv.2405.00218},
  doi          = {10.48550/ARXIV.2405.00218},
  eprinttype    = {arXiv},
  eprint       = {2405.00218},
  bibsource    = {dblp computer science bibliography, https://dblp.org}
}

@article{Wu2026SecureCG,
  title={Secure Code Generation via Online Reinforcement Learning with Vulnerability Reward Model},
  author={Tianyi Wu and Mingzhe Du and Yue Liu and Cheng-Lin Yang and Terry Yue Zhuo and Jiaheng Zhang and See-kiong Ng},
  journal={CoRR},
  volume={abs/2602.07422},
  year={2026},
  url={https://doi.org/10.48550/arXiv.2602.07422},
  doi={10.48550/arXiv.2602.07422},
  eprinttype={arXiv},
  eprint={2602.07422}
}

@inproceedings{DBLP:conf/icml/XuSG0W025,
  author       = {Xiangzhe Xu and
                  Zian Su and
                  Jinyao Guo and
                  Kaiyuan Zhang and
                  Zhenting Wang and
                  Xiangyu Zhang},
  editor       = {Aarti Singh and
                  Maryam Fazel and
                  Daniel Hsu and
                  Simon Lacoste{-}Julien and
                  Felix Berkenkamp and
                  Tegan Maharaj and
                  Kiri Wagstaff and
                  Jerry Zhu},
  title        = {ProSec: Fortifying Code LLMs with Proactive Security Alignment},
  booktitle    = {Forty-second International Conference on Machine Learning, {ICML}
                  2025, Vancouver, BC, Canada, July 13-19, 2025},
  series       = {Proceedings of Machine Learning Research},
  volume       = {267},
  publisher    = {{PMLR} / OpenReview.net},
  year         = {2025},
  url          = {https://proceedings.mlr.press/v267/xu25aa.html},
  bibsource    = {dblp computer science bibliography, https://dblp.org}
}

@article{Quan2026LearningTG,
  title={Learning to Generate Secure Code via Token-Level Rewards},
  author={Jiazheng Quan and Xiaodong Li and Bin Wang and Guo An and Like Liu and Degen Huang and Lingfeng Liu and Chengbin Hou},
  journal={CoRR},
  volume={abs/2602.23407},
  year={2026},
  url={https://doi.org/10.48550/arXiv.2602.23407},
  doi={10.48550/arXiv.2602.23407},
  eprinttype={arXiv},
  eprint={2602.23407}
}

@inproceedings{DBLP:conf/issta/Li0ZLL0C024,
  author       = {Dong Li and
                  Meng Yan and
                  Yaosheng Zhang and
                  Zhongxin Liu and
                  Chao Liu and
                  Xiaohong Zhang and
                  Ting Chen and
                  David Lo},
  editor       = {Maria Christakis and
                  Michael Pradel},
  title        = {CoSec: On-the-Fly Security Hardening of Code LLMs via Supervised Co-decoding},
  booktitle    = {Proceedings of the 33rd {ACM} {SIGSOFT} International Symposium on
                  Software Testing and Analysis, {ISSTA} 2024, Vienna, Austria, September
                  16-20, 2024},
  pages        = {1428--1439},
  publisher    = {{ACM}},
  year         = {2024},
  url          = {https://doi.org/10.1145/3650212.3680371},
  doi          = {10.1145/3650212.3680371},
  bibsource    = {dblp computer science bibliography, https://dblp.org}
}

@inproceedings{DBLP:conf/icse/DingFISCAWR025,
  author       = {Yangruibo Ding and
                  Yanjun Fu and
                  Omniyyah Ibrahim and
                  Chawin Sitawarin and
                  Xinyun Chen and
                  Basel Alomair and
                  David A. Wagner and
                  Baishakhi Ray and
                  Yizheng Chen},
  title        = {Vulnerability Detection with Code Language Models: How Far are We?},
  booktitle    = {47th {IEEE/ACM} International Conference on Software Engineering,
                  {ICSE} 2025, Ottawa, ON, Canada, April 26 - May 6, 2025},
  pages        = {1729--1741},
  publisher    = {{IEEE}},
  year         = {2025},
  url          = {https://doi.org/10.1109/ICSE55347.2025.00038},
  doi          = {10.1109/ICSE55347.2025.00038},
  bibsource    = {dblp computer science bibliography, https://dblp.org}
}

@article{DBLP:journals/corr/abs-2507-16887,
  author       = {Youpeng Li and
                  Weiliang Qi and
                  Xuyu Wang and
                  Fuxun Yu and
                  Xinda Wang},
  title        = {Revisiting Pre-trained Language Models for Vulnerability Detection},
  journal      = {CoRR},
  volume       = {abs/2507.16887},
  year         = {2025},
  url          = {https://doi.org/10.48550/arXiv.2507.16887},
  doi          = {10.48550/ARXIV.2507.16887},
  eprinttype    = {arXiv},
  eprint       = {2507.16887},
  bibsource    = {dblp computer science bibliography, https://dblp.org}
}

@article{DBLP:journals/corr/abs-2507-19060,
  author       = {Jiawei Liu and
                  Nirav Diwan and
                  Zhe Wang and
                  Haoyu Zhai and
                  Xiaona Zhou and
                  Kiet A. Nguyen and
                  Tianjiao Yu and
                  Muntasir Wahed and
                  Yinlin Deng and
                  Hadjer Benkraouda and
                  Yuxiang Wei and
                  Lingming Zhang and
                  Ismini Lourentzou and
                  Gang Wang},
  title        = {PurpCode: Reasoning for Safer Code Generation},
  journal      = {CoRR},
  volume       = {abs/2507.19060},
  year         = {2025},
  url          = {https://doi.org/10.48550/arXiv.2507.19060},
  doi          = {10.48550/ARXIV.2507.19060},
  eprinttype    = {arXiv},
  eprint       = {2507.19060},
  bibsource    = {dblp computer science bibliography, https://dblp.org}
}

@inproceedings{DBLP:conf/emnlp/ZhangDT0CCWY24,
  author       = {Boyu Zhang and
                  Tianyu Du and
                  Junkai Tong and
                  Xuhong Zhang and
                  Kingsum Chow and
                  Sheng Cheng and
                  Xun Wang and
                  Jianwei Yin},
  editor       = {Yaser Al{-}Onaizan and
                  Mohit Bansal and
                  Yun{-}Nung Chen},
  title        = {SecCoder: Towards Generalizable and Robust Secure Code Generation},
  booktitle    = {Proceedings of the 2024 Conference on Empirical Methods in Natural
                  Language Processing, {EMNLP} 2024, Miami, FL, USA, November 12-16,
                  2024},
  pages        = {14557--14571},
  publisher    = {Association for Computational Linguistics},
  year         = {2024},
  url          = {https://doi.org/10.18653/v1/2024.emnlp-main.806},
  doi          = {10.18653/V1/2024.EMNLP-MAIN.806},
  bibsource    = {dblp computer science bibliography, https://dblp.org}
}

@inproceedings{DBLP:conf/kbse/Zhang0FFL24,
  author       = {Beiqi Zhang and
                  Peng Liang and
                  Qiong Feng and
                  Yujia Fu and
                  Zengyang Li},
  editor       = {Vladimir Filkov and
                  Baishakhi Ray and
                  Minghui Zhou},
  title        = {Copilot-in-the-Loop: Fixing Code Smells in Copilot-Generated Python
                  Code using Copilot},
  booktitle    = {Proceedings of the 39th {IEEE/ACM} International Conference on Automated
                  Software Engineering, {ASE} 2024, Sacramento, CA, USA, October 27
                  - November 1, 2024},
  pages        = {2230--2234},
  publisher    = {{ACM}},
  year         = {2024},
  url          = {https://doi.org/10.1145/3691620.3695290},
  doi          = {10.1145/3691620.3695290},
  bibsource    = {dblp computer science bibliography, https://dblp.org}
}

@inproceedings{DBLP:conf/nips/LeSZXS24,
  author       = {Hung Le and
                  Doyen Sahoo and
                  Yingbo Zhou and
                  Caiming Xiong and
                  Silvio Savarese},
  editor       = {Amir Globersons and
                  Lester Mackey and
                  Danielle Belgrave and
                  Angela Fan and
                  Ulrich Paquet and
                  Jakub M. Tomczak and
                  Cheng Zhang},
  title        = {{INDICT:} Code Generation with Internal Dialogues of Critiques for
                  Both Security and Helpfulness},
  booktitle    = {Advances in Neural Information Processing Systems 38: Annual Conference
                  on Neural Information Processing Systems 2024, NeurIPS 2024, Vancouver,
                  BC, Canada, December 10 - 15, 2024},
  year         = {2024},
  url          = {http://papers.nips.cc/paper\_files/paper/2024/hash/9b812ee4b831c21e14156ced8659197c-Abstract-Conference.html},
  bibsource    = {dblp computer science bibliography, https://dblp.org}
}

@article{DBLP:journals/corr/abs-2601-01184,
  author       = {Suryansh Singh Sijwali and
                  Suman Saha},
  title        = {SecureCodeRL: Security-Aware Reinforcement Learning for Code Generation
                  with Partial-Credit Rewards},
  journal      = {CoRR},
  volume       = {abs/2601.01184},
  year         = {2026},
  url          = {https://doi.org/10.48550/arXiv.2601.01184},
  doi          = {10.48550/ARXIV.2601.01184},
  eprinttype    = {arXiv},
  eprint       = {2601.01184},
  bibsource    = {dblp computer science bibliography, https://dblp.org}
}

@article{DBLP:journals/corr/abs-2401-07031,
  author       = {Nafis Tanveer Islam and
                  Mohammad Bahrami Karkevandi and
                  Peyman Najafirad},
  title        = {Code Security Vulnerability Repair Using Reinforcement Learning with
                  Large Language Models},
  journal      = {CoRR},
  volume       = {abs/2401.07031},
  year         = {2024},
  url          = {https://doi.org/10.48550/arXiv.2401.07031},
  doi          = {10.48550/ARXIV.2401.07031},
  eprinttype    = {arXiv},
  eprint       = {2401.07031},
  bibsource    = {dblp computer science bibliography, https://dblp.org}
}

@inproceedings{DBLP:conf/icml/HeVKV24,
  author       = {Jingxuan He and
                  Mark Vero and
                  Gabriela Krasnopolska and
                  Martin T. Vechev},
  editor       = {Ruslan Salakhutdinov and
                  Zico Kolter and
                  Katherine A. Heller and
                  Adrian Weller and
                  Nuria Oliver and
                  Jonathan Scarlett and
                  Felix Berkenkamp},
  title        = {Instruction Tuning for Secure Code Generation},
  booktitle    = {Forty-first International Conference on Machine Learning, {ICML} 2024,
                  Vienna, Austria, July 21-27, 2024},
  series       = {Proceedings of Machine Learning Research},
  volume       = {235},
  pages        = {18043--18062},
  publisher    = {{PMLR} / OpenReview.net},
  year         = {2024},
  url          = {https://proceedings.mlr.press/v235/he24k.html},
  bibsource    = {dblp computer science bibliography, https://dblp.org}
}

@article{DBLP:journals/corr/abs-2508-00910,
  author       = {Terry Yue Zhuo and
                  Dingmin Wang and
                  Hantian Ding and
                  Varun Kumar and
                  Zijian Wang},
  title        = {Cyber-Zero: Training Cybersecurity Agents without Runtime},
  journal      = {CoRR},
  volume       = {abs/2508.00910},
  year         = {2025},
  url          = {https://doi.org/10.48550/arXiv.2508.00910},
  doi          = {10.48550/ARXIV.2508.00910},
  eprinttype    = {arXiv},
  eprint       = {2508.00910},
  bibsource    = {dblp computer science bibliography, https://dblp.org}
}

@article{DBLP:journals/corr/abs-2507-12415,
  author       = {Xinyi He and
                  Qian Liu and
                  Mingzhe Du and
                  Lin Yan and
                  Zhijie Fan and
                  Yiming Huang and
                  Zejian Yuan and
                  Zejun Ma},
  title        = {SWE-Perf: Can Language Models Optimize Code Performance on Real-World
                  Repositories?},
  journal      = {CoRR},
  volume       = {abs/2507.12415},
  year         = {2025},
  url          = {https://doi.org/10.48550/arXiv.2507.12415},
  doi          = {10.48550/ARXIV.2507.12415},
  eprinttype    = {arXiv},
  eprint       = {2507.12415},
  bibsource    = {dblp computer science bibliography, https://dblp.org}
}

@article{DBLP:journals/corr/abs-2509-09853,
  author       = {Zhiyu Fan and
                  Kirill Vasilevski and
                  Dayi Lin and
                  Boyuan Chen and
                  Yihao Chen and
                  Zhiqing Zhong and
                  Jie M. Zhang and
                  Pinjia He and
                  Ahmed E. Hassan},
  title        = {SWE-Effi: Re-Evaluating Software {AI} Agent System Effectiveness Under
                  Resource Constraints},
  journal      = {CoRR},
  volume       = {abs/2509.09853},
  year         = {2025},
  url          = {https://doi.org/10.48550/arXiv.2509.09853},
  doi          = {10.48550/ARXIV.2509.09853},
  eprinttype    = {arXiv},
  eprint       = {2509.09853},
  bibsource    = {dblp computer science bibliography, https://dblp.org}
}

@article{DBLP:journals/corr/abs-2511-06090,
  author       = {Jeffrey Jian Ma and
                  Milad Hashemi and
                  Amir Yazdanbakhsh and
                  Kevin Swersky and
                  Ofir Press and
                  Enhui Li and
                  Vijay Janapa Reddi and
                  Parthasarathy Ranganathan},
  title        = {SWE-fficiency: Can Language Models Optimize Real-World Repositories
                  on Real Workloads?},
  journal      = {CoRR},
  volume       = {abs/2511.06090},
  year         = {2025},
  url          = {https://doi.org/10.48550/arXiv.2511.06090},
  doi          = {10.48550/ARXIV.2511.06090},
  eprinttype    = {arXiv},
  eprint       = {2511.06090},
  bibsource    = {dblp computer science bibliography, https://dblp.org}
}

@inproceedings{DBLP:conf/ics/LamouriAKB25,
  author       = {Djamel Rassem Lamouri and
                  Iheb Nassim Aouadj and
                  Smail Kourta and
                  Riyadh Baghdadi},
  title        = {Pearl: Automatic Code Optimization Using Deep Reinforcement Learning},
  booktitle    = {Proceedings of the 39th {ACM} International Conference on Supercomputing,
                  {ICS} 2025, Salt Lake City, UT, USA, June 8-11, 2025},
  pages        = {959--974},
  publisher    = {{ACM}},
  year         = {2025},
  url          = {https://doi.org/10.1145/3721145.3725766},
  doi          = {10.1145/3721145.3725766},
  bibsource    = {dblp computer science bibliography, https://dblp.org}
}

@article{DBLP:journals/corr/abs-2508-20124,
  author       = {Yunlong Feng and
                  Yang Xu and
                  Xiao Xu and
                  Binyuan Hui and
                  Junyang Lin},
  title        = {Towards Better Correctness and Efficiency in Code Generation},
  journal      = {CoRR},
  volume       = {abs/2508.20124},
  year         = {2025},
  url          = {https://doi.org/10.48550/arXiv.2508.20124},
  doi          = {10.48550/ARXIV.2508.20124},
  eprinttype    = {arXiv},
  eprint       = {2508.20124},
  bibsource    = {dblp computer science bibliography, https://dblp.org}
}

@article{DBLP:journals/corr/abs-2512-22827,
  author       = {Yue Wu and
                  Minghao Han and
                  Ruiyin Li and
                  Peng Liang and
                  Amjed Tahir and
                  Zengyang Li and
                  Qiong Feng and
                  Mojtaba Shahin},
  title        = {FasterPy: An LLM-based Code Execution Efficiency Optimization Framework},
  journal      = {CoRR},
  volume       = {abs/2512.22827},
  year         = {2025},
  url          = {https://doi.org/10.48550/arXiv.2512.22827},
  doi          = {10.48550/ARXIV.2512.22827},
  eprinttype    = {arXiv},
  eprint       = {2512.22827},
  bibsource    = {dblp computer science bibliography, https://dblp.org}
}

@article{DBLP:journals/corr/abs-2502-18489,
  author       = {Tong Ye and
                  Weigang Huang and
                  Xuhong Zhang and
                  Tengfei Ma and
                  Peiyu Liu and
                  Jianwei Yin and
                  Wenhai Wang},
  title        = {{LLM4EFFI:} Leveraging Large Language Models to Enhance Code Efficiency
                  and Correctness},
  journal      = {CoRR},
  volume       = {abs/2502.18489},
  year         = {2025},
  url          = {https://doi.org/10.48550/arXiv.2502.18489},
  doi          = {10.48550/ARXIV.2502.18489},
  eprinttype    = {arXiv},
  eprint       = {2502.18489},
  bibsource    = {dblp computer science bibliography, https://dblp.org}
}

@inproceedings{DBLP:conf/icml/0005ZDLWQCG025,
  author       = {Dong Huang and
                  Guangtao Zeng and
                  Jianbo Dai and
                  Meng Luo and
                  Han Weng and
                  Yuhao Qing and
                  Heming Cui and
                  Zhijiang Guo and
                  Jie Zhang},
  editor       = {Aarti Singh and
                  Maryam Fazel and
                  Daniel Hsu and
                  Simon Lacoste{-}Julien and
                  Felix Berkenkamp and
                  Tegan Maharaj and
                  Kiri Wagstaff and
                  Jerry Zhu},
  title        = {EffiCoder: Enhancing Code Generation in Large Language Models through
                  Efficiency-Aware Fine-tuning},
  booktitle    = {Forty-second International Conference on Machine Learning, {ICML}
                  2025, Vancouver, BC, Canada, July 13-19, 2025},
  series       = {Proceedings of Machine Learning Research},
  volume       = {267},
  publisher    = {{PMLR} / OpenReview.net},
  year         = {2025},
  url          = {https://proceedings.mlr.press/v267/huang25as.html},
  bibsource    = {dblp computer science bibliography, https://dblp.org}
}

@article{DBLP:journals/corr/abs-2512-14018,
  author       = {Jiuding Yang and
                  Shengyao Lu and
                  Hongxuan Liu and
                  Shayan Shirahmad Gale Bagi and
                  Zahra Fazel and
                  Tomasz Czajkowski and
                  Di Niu},
  title        = {PerfCoder: Large Language Models for Interpretable Code Performance
                  Optimization},
  journal      = {CoRR},
  volume       = {abs/2512.14018},
  year         = {2025},
  url          = {https://doi.org/10.48550/arXiv.2512.14018},
  doi          = {10.48550/ARXIV.2512.14018},
  eprinttype    = {arXiv},
  eprint       = {2512.14018},
  bibsource    = {dblp computer science bibliography, https://dblp.org}
}

@inproceedings{DBLP:conf/forge/PengGLXSS25,
  author       = {Yun Peng and
                  Akhilesh Deepak Gotmare and
                  Michael R. Lyu and
                  Caiming Xiong and
                  Silvio Savarese and
                  Doyen Sahoo},
  title        = {PerfCodeGen: Improving Performance of {LLM} Generated Code with Execution
                  Feedback},
  booktitle    = {{IEEE/ACM} Second International Conference on {AI} Foundation Models
                  and Software Engineering, Forge@ICSE 2025, Ottawa, ON, Canada, April
                  27-28, 2025},
  pages        = {1--13},
  publisher    = {{IEEE}},
  year         = {2025},
  url          = {https://doi.org/10.1109/Forge66646.2025.00008},
  doi          = {10.1109/FORGE66646.2025.00008},
  bibsource    = {dblp computer science bibliography, https://dblp.org}
}

@article{DBLP:journals/corr/abs-2505-23387,
  author       = {Mingzhe Du and
                  Luu Tuan Tuan and
                  Yue Liu and
                  Yuhao Qing and
                  Dong Huang and
                  Xinyi He and
                  Qian Liu and
                  Zejun Ma and
                  See{-}Kiong Ng},
  title        = {Afterburner: Reinforcement Learning Facilitates Self-Improving Code
                  Efficiency Optimization},
  journal      = {CoRR},
  volume       = {abs/2505.23387},
  year         = {2025},
  url          = {https://doi.org/10.48550/arXiv.2505.23387},
  doi          = {10.48550/ARXIV.2505.23387},
  eprinttype    = {arXiv},
  eprint       = {2505.23387},
  bibsource    = {dblp computer science bibliography, https://dblp.org}
}

@article{DBLP:journals/corr/abs-2410-10209,
  author       = {Dong Huang and
                  Guangtao Zeng and
                  Jianbo Dai and
                  Meng Luo and
                  Han Weng and
                  Yuhao Qing and
                  Heming Cui and
                  Zhijiang Guo and
                  Jie M. Zhang},
  title        = {Effi-Code: Unleashing Code Efficiency in Language Models},
  journal      = {CoRR},
  volume       = {abs/2410.10209},
  year         = {2024},
  url          = {https://doi.org/10.48550/arXiv.2410.10209},
  doi          = {10.48550/ARXIV.2410.10209},
  eprinttype    = {arXiv},
  eprint       = {2410.10209},
  bibsource    = {dblp computer science bibliography, https://dblp.org}
}

@article{DBLP:journals/corr/abs-2601-01215,
  author       = {Prateek Rajput and
                  Yewei Song and
                  Abdoul Aziz Bonkoungou and
                  Iyiola E. Olatunji and
                  Abdoul Kader Kabor{\'{e}} and
                  Jacques Klein and
                  Tegawend{\'{e}} F. Bissyand{\'{e}}},
  title        = {Correctness isnt Efficiency: Runtime Memory Divergence in LLM-Generated
                  Code},
  journal      = {CoRR},
  volume       = {abs/2601.01215},
  year         = {2026},
  url          = {https://doi.org/10.48550/arXiv.2601.01215},
  doi          = {10.48550/ARXIV.2601.01215},
  eprinttype    = {arXiv},
  eprint       = {2601.01215},
  bibsource    = {dblp computer science bibliography, https://dblp.org}
}

@article{DBLP:journals/corr/abs-2505-19442,
  author       = {Dutao Zhang and
                  Sergey V. Kovalchuk and
                  YuLong He},
  title        = {Style2Code: {A} Style-Controllable Code Generation Framework with
                  Dual-Modal Contrastive Representation Learning},
  journal      = {CoRR},
  volume       = {abs/2505.19442},
  year         = {2025},
  url          = {https://doi.org/10.48550/arXiv.2505.19442},
  doi          = {10.48550/ARXIV.2505.19442},
  eprinttype    = {arXiv},
  eprint       = {2505.19442},
  bibsource    = {dblp computer science bibliography, https://dblp.org}
}

@article{DBLP:journals/corr/abs-2409-12866,
  author       = {Lezhi Ma and
                  Shangqing Liu and
                  Lei Bu and
                  Shangru Li and
                  Yida Wang and
                  Yang Liu},
  title        = {SpecEval: Evaluating Code Comprehension in Large Language Models via
                  Program Specifications},
  journal      = {CoRR},
  volume       = {abs/2409.12866},
  year         = {2024},
  url          = {https://doi.org/10.48550/arXiv.2409.12866},
  doi          = {10.48550/ARXIV.2409.12866},
  eprinttype    = {arXiv},
  eprint       = {2409.12866},
  bibsource    = {dblp computer science bibliography, https://dblp.org}
}

@inproceedings{DBLP:conf/saner/NunesFSNFS25,
  author       = {Henrique Gomes Nunes and
                  Eduardo Figueiredo and
                  Larissa Rocha Soares and
                  Sarah Nadi and
                  Fischer Ferreira and
                  Geanderson E. dos Santos},
  title        = {Evaluating the Effectiveness of LLMs in Fixing Maintainability Issues
                  in Real-World Projects},
  booktitle    = {{IEEE} International Conference on Software Analysis, Evolution and
                  Reengineering, {SANER} 2025, Montreal, QC, Canada, March 4-7, 2025},
  pages        = {669--680},
  publisher    = {{IEEE}},
  year         = {2025},
  url          = {https://doi.org/10.1109/SANER64311.2025.00069},
  doi          = {10.1109/SANER64311.2025.00069},
  bibsource    = {dblp computer science bibliography, https://dblp.org}
}

@inproceedings{DBLP:conf/iclr/DaiPSJXL0024,
  author       = {Josef Dai and
                  Xuehai Pan and
                  Ruiyang Sun and
                  Jiaming Ji and
                  Xinbo Xu and
                  Mickel Liu and
                  Yizhou Wang and
                  Yaodong Yang},
  title        = {Safe {RLHF:} Safe Reinforcement Learning from Human Feedback},
  booktitle    = {The Twelfth International Conference on Learning Representations,
                  {ICLR} 2024, Vienna, Austria, May 7-11, 2024},
  publisher    = {OpenReview.net},
  year         = {2024},
  url          = {https://openreview.net/forum?id=TyFrPOKYXw},
  bibsource    = {dblp computer science bibliography, https://dblp.org}
}

@inproceedings{DBLP:conf/iclr/JacobSFA24,
  author       = {Athul Paul Jacob and
                  Yikang Shen and
                  Gabriele Farina and
                  Jacob Andreas},
  title        = {The Consensus Game: Language Model Generation via Equilibrium Search},
  booktitle    = {The Twelfth International Conference on Learning Representations,
                  {ICLR} 2024, Vienna, Austria, May 7-11, 2024},
  publisher    = {OpenReview.net},
  year         = {2024},
  url          = {https://openreview.net/forum?id=n9xeGcI4Yg},
  bibsource    = {dblp computer science bibliography, https://dblp.org}
}

@inproceedings{DBLP:conf/nips/0001L023,
  author       = {Yuanhao Wang and
                  Qinghua Liu and
                  Chi Jin},
  editor       = {Alice Oh and
                  Tristan Naumann and
                  Amir Globerson and
                  Kate Saenko and
                  Moritz Hardt and
                  Sergey Levine},
  title        = {Is {RLHF} More Difficult than Standard RL? {A} Theoretical Perspective},
  booktitle    = {Advances in Neural Information Processing Systems 36: Annual Conference
                  on Neural Information Processing Systems 2023, NeurIPS 2023, New Orleans,
                  LA, USA, December 10 - 16, 2023},
  year         = {2023},
  url          = {http://papers.nips.cc/paper\_files/paper/2023/hash/efb9629755e598c4f261c44aeb6fde5e-Abstract-Conference.html},
  bibsource    = {dblp computer science bibliography, https://dblp.org}
}

@inproceedings{DBLP:conf/icml/MunosVCARGTGMFM24,
  author       = {R{\'{e}}mi Munos and
                  Michal Valko and
                  Daniele Calandriello and
                  Mohammad Gheshlaghi Azar and
                  Mark Rowland and
                  Daniel Guo and
                  Yunhao Tang and
                  Matthieu Geist and
                  Thomas Mesnard and
                  C{\^{o}}me Fiegel and
                  Andrea Michi and
                  Marco Selvi and
                  Sertan Girgin and
                  Nikola Momchev and
                  Olivier Bachem and
                  Daniel J. Mankowitz and
                  Doina Precup and
                  Bilal Piot},
  editor       = {Ruslan Salakhutdinov and
                  Zico Kolter and
                  Katherine A. Heller and
                  Adrian Weller and
                  Nuria Oliver and
                  Jonathan Scarlett and
                  Felix Berkenkamp},
  title        = {Nash Learning from Human Feedback},
  booktitle    = {Forty-first International Conference on Machine Learning, {ICML} 2024,
                  Vienna, Austria, July 21-27, 2024},
  series       = {Proceedings of Machine Learning Research},
  volume       = {235},
  pages        = {36743--36768},
  publisher    = {{PMLR} / OpenReview.net},
  year         = {2024},
  url          = {https://proceedings.mlr.press/v235/munos24a.html},
  bibsource    = {dblp computer science bibliography, https://dblp.org}
}

@inproceedings{DBLP:conf/iclr/YangKCPT24,
  author       = {Kevin Yang and
                  Dan Klein and
                  Asli Celikyilmaz and
                  Nanyun Peng and
                  Yuandong Tian},
  title        = {{RLCD:} Reinforcement Learning from Contrastive Distillation for {LM}
                  Alignment},
  booktitle    = {The Twelfth International Conference on Learning Representations,
                  {ICLR} 2024, Vienna, Austria, May 7-11, 2024},
  publisher    = {OpenReview.net},
  year         = {2024},
  url          = {https://openreview.net/forum?id=v3XXtxWKi6},
  bibsource    = {dblp computer science bibliography, https://dblp.org}
}

@article{DBLP:journals/corr/abs-2506-09183,
  author       = {Mingkang Wu and
                  Devin White and
                  Evelyn Rose and
                  Vernon Lawhern and
                  Nicholas R. Waytowich and
                  Yongcan Cao},
  title        = {Multi-Task Reward Learning from Human Ratings},
  journal      = {CoRR},
  volume       = {abs/2506.09183},
  year         = {2025},
  url          = {https://doi.org/10.48550/arXiv.2506.09183},
  doi          = {10.48550/ARXIV.2506.09183},
  eprinttype    = {arXiv},
  eprint       = {2506.09183},
  bibsource    = {dblp computer science bibliography, https://dblp.org}
}

@article{DBLP:journals/corr/abs-2502-06876,
  author       = {Jinluan Yang and
                  Dingnan Jin and
                  Anke Tang and
                  Li Shen and
                  Didi Zhu and
                  Zhengyu Chen and
                  Daixin Wang and
                  Qing Cui and
                  Zhiqiang Zhang and
                  Jun Zhou and
                  Fei Wu and
                  Kun Kuang},
  title        = {Mix Data or Merge Models? Balancing the Helpfulness, Honesty, and
                  Harmlessness of Large Language Model via Model Merging},
  journal      = {CoRR},
  volume       = {abs/2502.06876},
  year         = {2025},
  url          = {https://doi.org/10.48550/arXiv.2502.06876},
  doi          = {10.48550/ARXIV.2502.06876},
  eprinttype    = {arXiv},
  eprint       = {2502.06876},
  bibsource    = {dblp computer science bibliography, https://dblp.org}
}

@article{DBLP:journals/corr/abs-2503-06358,
  author       = {Idan Shenfeld and
                  Felix Faltings and
                  Pulkit Agrawal and
                  Aldo Pacchiano},
  title        = {Language Model Personalization via Reward Factorization},
  journal      = {CoRR},
  volume       = {abs/2503.06358},
  year         = {2025},
  url          = {https://doi.org/10.48550/arXiv.2503.06358},
  doi          = {10.48550/ARXIV.2503.06358},
  eprinttype    = {arXiv},
  eprint       = {2503.06358},
  bibsource    = {dblp computer science bibliography, https://dblp.org}
}

@inproceedings{DBLP:conf/icml/LinJXCC25,
  author       = {Baijiong Lin and
                  Weisen Jiang and
                  Yuancheng Xu and
                  Hao Chen and
                  Ying{-}Cong Chen},
  editor       = {Aarti Singh and
                  Maryam Fazel and
                  Daniel Hsu and
                  Simon Lacoste{-}Julien and
                  Felix Berkenkamp and
                  Tegan Maharaj and
                  Kiri Wagstaff and
                  Jerry Zhu},
  title        = {{PARM:} Multi-Objective Test-Time Alignment via Preference-Aware Autoregressive
                  Reward Model},
  booktitle    = {Forty-second International Conference on Machine Learning, {ICML}
                  2025, Vancouver, BC, Canada, July 13-19, 2025},
  series       = {Proceedings of Machine Learning Research},
  volume       = {267},
  publisher    = {{PMLR} / OpenReview.net},
  year         = {2025},
  url          = {https://proceedings.mlr.press/v267/lin25h.html},
  bibsource    = {dblp computer science bibliography, https://dblp.org}
}

@inproceedings{DBLP:conf/iclr/WinataASKW25,
  author       = {Genta Indra Winata and
                  David Anugraha and
                  Lucky Susanto and
                  Garry Kuwanto and
                  Derry Tanti Wijaya},
  title        = {MetaMetrics: Calibrating Metrics for Generation Tasks Using Human
                  Preferences},
  booktitle    = {The Thirteenth International Conference on Learning Representations,
                  {ICLR} 2025, Singapore, April 24-28, 2025},
  publisher    = {OpenReview.net},
  year         = {2025},
  url          = {https://openreview.net/forum?id=slO3xTt4CG},
  bibsource    = {dblp computer science bibliography, https://dblp.org}
}

@article{DBLP:journals/corr/abs-2310-11564,
  author       = {Joel Jang and
                  Seungone Kim and
                  Bill Yuchen Lin and
                  Yizhong Wang and
                  Jack Hessel and
                  Luke Zettlemoyer and
                  Hannaneh Hajishirzi and
                  Yejin Choi and
                  Prithviraj Ammanabrolu},
  title        = {Personalized Soups: Personalized Large Language Model Alignment via
                  Post-hoc Parameter Merging},
  journal      = {CoRR},
  volume       = {abs/2310.11564},
  year         = {2023},
  url          = {https://doi.org/10.48550/arXiv.2310.11564},
  doi          = {10.48550/ARXIV.2310.11564},
  eprinttype    = {arXiv},
  eprint       = {2310.11564},
  bibsource    = {dblp computer science bibliography, https://dblp.org}
}

@inproceedings{DBLP:conf/icml/ChakrabortyQYKM24,
  author       = {Souradip Chakraborty and
                  Jiahao Qiu and
                  Hui Yuan and
                  Alec Koppel and
                  Dinesh Manocha and
                  Furong Huang and
                  Amrit S. Bedi and
                  Mengdi Wang},
  editor       = {Ruslan Salakhutdinov and
                  Zico Kolter and
                  Katherine A. Heller and
                  Adrian Weller and
                  Nuria Oliver and
                  Jonathan Scarlett and
                  Felix Berkenkamp},
  title        = {MaxMin-RLHF: Alignment with Diverse Human Preferences},
  booktitle    = {Forty-first International Conference on Machine Learning, {ICML} 2024,
                  Vienna, Austria, July 21-27, 2024},
  series       = {Proceedings of Machine Learning Research},
  volume       = {235},
  pages        = {6116--6135},
  publisher    = {{PMLR} / OpenReview.net},
  year         = {2024},
  url          = {https://proceedings.mlr.press/v235/chakraborty24b.html},
  bibsource    = {dblp computer science bibliography, https://dblp.org}
}

@inproceedings{DBLP:conf/emnlp/VuKATFS24,
  author       = {Tu Vu and
                  Kalpesh Krishna and
                  Salaheddin Alzubi and
                  Chris Tar and
                  Manaal Faruqui and
                  Yun{-}Hsuan Sung},
  editor       = {Yaser Al{-}Onaizan and
                  Mohit Bansal and
                  Yun{-}Nung Chen},
  title        = {Foundational Autoraters: Taming Large Language Models for Better Automatic
                  Evaluation},
  booktitle    = {Proceedings of the 2024 Conference on Empirical Methods in Natural
                  Language Processing, {EMNLP} 2024, Miami, FL, USA, November 12-16,
                  2024},
  pages        = {17086--17105},
  publisher    = {Association for Computational Linguistics},
  year         = {2024},
  url          = {https://doi.org/10.18653/v1/2024.emnlp-main.949},
  doi          = {10.18653/V1/2024.EMNLP-MAIN.949},
  bibsource    = {dblp computer science bibliography, https://dblp.org}
}

@inproceedings{DBLP:conf/fat/ChristianKTSD25,
  author       = {Brian Christian and
                  Hannah Rose Kirk and
                  Jessica A. F. Thompson and
                  Christopher Summerfield and
                  Tsvetomira Dumbalska},
  title        = {Reward Model Interpretability via Optimal and Pessimal Tokens},
  booktitle    = {Proceedings of the 2025 {ACM} Conference on Fairness, Accountability,
                  and Transparency, FAccT 2025, Athens, Greece, June 23-26, 2025},
  pages        = {1048--1059},
  publisher    = {{ACM}},
  year         = {2025},
  url          = {https://doi.org/10.1145/3715275.3732068},
  doi          = {10.1145/3715275.3732068},
  bibsource    = {dblp computer science bibliography, https://dblp.org}
}

@article{DBLP:journals/corr/abs-2601-15968,
  author       = {Xin Xie and
                  Jiaxian Guo and
                  Dong Gong},
  title        = {HyperAlign: Hypernetwork for Efficient Test-Time Alignment of Diffusion
                  Models},
  journal      = {CoRR},
  volume       = {abs/2601.15968},
  year         = {2026},
  url          = {https://doi.org/10.48550/arXiv.2601.15968},
  doi          = {10.48550/ARXIV.2601.15968},
  eprinttype    = {arXiv},
  eprint       = {2601.15968},
  bibsource    = {dblp computer science bibliography, https://dblp.org}
}

@inproceedings{DBLP:conf/emnlp/00030X0024,
  author       = {Haoxiang Wang and
                  Wei Xiong and
                  Tengyang Xie and
                  Han Zhao and
                  Tong Zhang},
  editor       = {Yaser Al{-}Onaizan and
                  Mohit Bansal and
                  Yun{-}Nung Chen},
  title        = {Interpretable Preferences via Multi-Objective Reward Modeling and
                  Mixture-of-Experts},
  booktitle    = {Findings of the Association for Computational Linguistics: {EMNLP}
                  2024, Miami, Florida, USA, November 12-16, 2024},
  series       = {Findings of {ACL}},
  volume       = {{EMNLP} 2024},
  pages        = {10582--10592},
  publisher    = {Association for Computational Linguistics},
  year         = {2024},
  url          = {https://doi.org/10.18653/v1/2024.findings-emnlp.620},
  doi          = {10.18653/V1/2024.FINDINGS-EMNLP.620},
  bibsource    = {dblp computer science bibliography, https://dblp.org}
}

@inproceedings{DBLP:conf/emnlp/CalderonER25,
  author       = {Nitay Calderon and
                  Liat Ein{-}Dor and
                  Roi Reichart},
  editor       = {Christos Christodoulopoulos and
                  Tanmoy Chakraborty and
                  Carolyn Rose and
                  Violet Peng},
  title        = {Multi-Domain Explainability of Preferences},
  booktitle    = {Proceedings of the 2025 Conference on Empirical Methods in Natural
                  Language Processing, {EMNLP} 2025, Suzhou, China, November 4-9, 2025},
  pages        = {14542--14575},
  publisher    = {Association for Computational Linguistics},
  year         = {2025},
  url          = {https://doi.org/10.18653/v1/2025.emnlp-main.736},
  doi          = {10.18653/V1/2025.EMNLP-MAIN.736},
  bibsource    = {dblp computer science bibliography, https://dblp.org}
}

@inproceedings{DBLP:conf/nips/LeePKS24,
  author       = {Seongyun Lee and
                  Sue Hyun Park and
                  Seungone Kim and
                  Minjoon Seo},
  editor       = {Amir Globersons and
                  Lester Mackey and
                  Danielle Belgrave and
                  Angela Fan and
                  Ulrich Paquet and
                  Jakub M. Tomczak and
                  Cheng Zhang},
  title        = {Aligning to Thousands of Preferences via System Message Generalization},
  booktitle    = {Advances in Neural Information Processing Systems 38: Annual Conference
                  on Neural Information Processing Systems 2024, NeurIPS 2024, Vancouver,
                  BC, Canada, December 10 - 15, 2024},
  year         = {2024},
  url          = {http://papers.nips.cc/paper\_files/paper/2024/hash/86c9df30129f7663ad4d429b6f80d461-Abstract-Conference.html},
  bibsource    = {dblp computer science bibliography, https://dblp.org}
}

@inproceedings{DBLP:conf/iclr/SunSZZCCYG24,
  author       = {Zhiqing Sun and
                  Yikang Shen and
                  Hongxin Zhang and
                  Qinhong Zhou and
                  Zhenfang Chen and
                  David Daniel Cox and
                  Yiming Yang and
                  Chuang Gan},
  title        = {{SALMON:} Self-Alignment with Instructable Reward Models},
  booktitle    = {The Twelfth International Conference on Learning Representations,
                  {ICLR} 2024, Vienna, Austria, May 7-11, 2024},
  publisher    = {OpenReview.net},
  year         = {2024},
  url          = {https://openreview.net/forum?id=xJbsmB8UMx},
  bibsource    = {dblp computer science bibliography, https://dblp.org}
}

@article{DBLP:journals/corr/abs-2212-08073,
  author       = {Yuntao Bai and
                  Saurav Kadavath and
                  Sandipan Kundu and
                  Amanda Askell and
                  Jackson Kernion and
                  Andy Jones and
                  Anna Chen and
                  Anna Goldie and
                  Azalia Mirhoseini and
                  Cameron McKinnon and
                  Carol Chen and
                  Catherine Olsson and
                  Christopher Olah and
                  Danny Hernandez and
                  Dawn Drain and
                  Deep Ganguli and
                  Dustin Li and
                  Eli Tran{-}Johnson and
                  Ethan Perez and
                  Jamie Kerr and
                  Jared Mueller and
                  Jeffrey Ladish and
                  Joshua Landau and
                  Kamal Ndousse and
                  Kamile Lukosiute and
                  Liane Lovitt and
                  Michael Sellitto and
                  Nelson Elhage and
                  Nicholas Schiefer and
                  Noem{\'{\i}} Mercado and
                  Nova DasSarma and
                  Robert Lasenby and
                  Robin Larson and
                  Sam Ringer and
                  Scott Johnston and
                  Shauna Kravec and
                  Sheer El Showk and
                  Stanislav Fort and
                  Tamera Lanham and
                  Timothy Telleen{-}Lawton and
                  Tom Conerly and
                  Tom Henighan and
                  Tristan Hume and
                  Samuel R. Bowman and
                  Zac Hatfield{-}Dodds and
                  Ben Mann and
                  Dario Amodei and
                  Nicholas Joseph and
                  Sam McCandlish and
                  Tom Brown and
                  Jared Kaplan},
  title        = {Constitutional {AI:} Harmlessness from {AI} Feedback},
  journal      = {CoRR},
  volume       = {abs/2212.08073},
  year         = {2022},
  url          = {https://doi.org/10.48550/arXiv.2212.08073},
  doi          = {10.48550/ARXIV.2212.08073},
  eprinttype    = {arXiv},
  eprint       = {2212.08073},
  bibsource    = {dblp computer science bibliography, https://dblp.org}
}

@inproceedings{DBLP:conf/acl/WangLXYDQZZ24,
  author       = {Haoxiang Wang and
                  Yong Lin and
                  Wei Xiong and
                  Rui Yang and
                  Shizhe Diao and
                  Shuang Qiu and
                  Han Zhao and
                  Tong Zhang},
  editor       = {Lun{-}Wei Ku and
                  Andre Martins and
                  Vivek Srikumar},
  title        = {Arithmetic Control of LLMs for Diverse User Preferences: Directional
                  Preference Alignment with Multi-Objective Rewards},
  booktitle    = {Proceedings of the 62nd Annual Meeting of the Association for Computational
                  Linguistics (Volume 1: Long Papers), {ACL} 2024, Bangkok, Thailand,
                  August 11-16, 2024},
  pages        = {8642--8655},
  publisher    = {Association for Computational Linguistics},
  year         = {2024},
  url          = {https://doi.org/10.18653/v1/2024.acl-long.468},
  doi          = {10.18653/V1/2024.ACL-LONG.468},
  bibsource    = {dblp computer science bibliography, https://dblp.org}
}

@inproceedings{DBLP:conf/emnlp/DongWSWK23,
  author       = {Yi Dong and
                  Zhilin Wang and
                  Makesh Narsimhan Sreedhar and
                  Xianchao Wu and
                  Oleksii Kuchaiev},
  editor       = {Houda Bouamor and
                  Juan Pino and
                  Kalika Bali},
  title        = {SteerLM: Attribute Conditioned {SFT} as an (User-Steerable) Alternative
                  to {RLHF}},
  booktitle    = {Findings of the Association for Computational Linguistics: {EMNLP}
                  2023, Singapore, December 6-10, 2023},
  series       = {Findings of {ACL}},
  volume       = {{EMNLP} 2023},
  pages        = {11275--11288},
  publisher    = {Association for Computational Linguistics},
  year         = {2023},
  url          = {https://doi.org/10.18653/v1/2023.findings-emnlp.754},
  doi          = {10.18653/V1/2023.FINDINGS-EMNLP.754},
  bibsource    = {dblp computer science bibliography, https://dblp.org}
}

@article{DBLP:journals/corr/abs-2507-07375,
  author       = {Zhiwei Zhang and
                  Hui Liu and
                  Xiaomin Li and
                  Zhenwei Dai and
                  Jingying Zeng and
                  Fali Wang and
                  Minhua Lin and
                  Ramraj Chandradevan and
                  Zheng Li and
                  Chen Luo and
                  Xianfeng Tang and
                  Qi He and
                  Suhang Wang},
  title        = {Bradley-Terry and Multi-Objective Reward Modeling Are Complementary},
  journal      = {CoRR},
  volume       = {abs/2507.07375},
  year         = {2025},
  url          = {https://doi.org/10.48550/arXiv.2507.07375},
  doi          = {10.48550/ARXIV.2507.07375},
  eprinttype    = {arXiv},
  eprint       = {2507.07375},
  bibsource    = {dblp computer science bibliography, https://dblp.org}
}

@inproceedings{DBLP:conf/emnlp/WangXZXJ25,
  author       = {Peifeng Wang and
                  Austin Xu and
                  Yilun Zhou and
                  Caiming Xiong and
                  Shafiq Joty},
  editor       = {Christos Christodoulopoulos and
                  Tanmoy Chakraborty and
                  Carolyn Rose and
                  Violet Peng},
  title        = {Direct Judgement Preference Optimization},
  booktitle    = {Proceedings of the 2025 Conference on Empirical Methods in Natural
                  Language Processing, {EMNLP} 2025, Suzhou, China, November 4-9, 2025},
  pages        = {1979--2009},
  publisher    = {Association for Computational Linguistics},
  year         = {2025},
  url          = {https://doi.org/10.18653/v1/2025.emnlp-main.103},
  doi          = {10.18653/V1/2025.EMNLP-MAIN.103},
  bibsource    = {dblp computer science bibliography, https://dblp.org}
}

@inproceedings{DBLP:conf/iclr/KwonXBS23,
  author       = {Minae Kwon and
                  Sang Michael Xie and
                  Kalesha Bullard and
                  Dorsa Sadigh},
  title        = {Reward Design with Language Models},
  booktitle    = {The Eleventh International Conference on Learning Representations,
                  {ICLR} 2023, Kigali, Rwanda, May 1-5, 2023},
  publisher    = {OpenReview.net},
  year         = {2023},
  url          = {https://openreview.net/forum?id=10uNUgI5Kl},
  bibsource    = {dblp computer science bibliography, https://dblp.org}
}

@article{DBLP:journals/corr/abs-2505-14674,
  author       = {Jiaxin Guo and
                  Zewen Chi and
                  Li Dong and
                  Qingxiu Dong and
                  Xun Wu and
                  Shaohan Huang and
                  Furu Wei},
  title        = {Reward Reasoning Model},
  journal      = {CoRR},
  volume       = {abs/2505.14674},
  year         = {2025},
  url          = {https://doi.org/10.48550/arXiv.2505.14674},
  doi          = {10.48550/ARXIV.2505.14674},
  eprinttype    = {arXiv},
  eprint       = {2505.14674},
  bibsource    = {dblp computer science bibliography, https://dblp.org}
}

@article{DBLP:journals/corr/abs-2509-06822,
  author       = {Chenyang Zhu and
                  Spencer Hong and
                  Jingyu Wu and
                  Kushal Chawla and
                  Charlotte Tang and
                  Youbing Yin and
                  Nathan Wolfe and
                  Erin Babinsky and
                  Daben Liu},
  title        = {{RAFFLES:} Reasoning-based Attribution of Faults for {LLM} Systems},
  journal      = {CoRR},
  volume       = {abs/2509.06822},
  year         = {2025},
  url          = {https://doi.org/10.48550/arXiv.2509.06822},
  doi          = {10.48550/ARXIV.2509.06822},
  eprinttype    = {arXiv},
  eprint       = {2509.06822},
  bibsource    = {dblp computer science bibliography, https://dblp.org}
}

@article{DBLP:journals/corr/abs-2509-21500,
  author       = {Junkai Zhang and
                  Zihao Wang and
                  Lin Gui and
                  Swarnashree Mysore Sathyendra and
                  Jaehwan Jeong and
                  Victor Veitch and
                  Wei Wang and
                  Yunzhong He and
                  Bing Liu and
                  Lifeng Jin},
  title        = {Chasing the Tail: Effective Rubric-based Reward Modeling for Large
                  Language Model Post-Training},
  journal      = {CoRR},
  volume       = {abs/2509.21500},
  year         = {2025},
  url          = {https://doi.org/10.48550/arXiv.2509.21500},
  doi          = {10.48550/ARXIV.2509.21500},
  eprinttype    = {arXiv},
  eprint       = {2509.21500},
  bibsource    = {dblp computer science bibliography, https://dblp.org}
}

@inproceedings{DBLP:conf/emnlp/Zhong0YMJLZJH22,
  author       = {Ming Zhong and
                  Yang Liu and
                  Da Yin and
                  Yuning Mao and
                  Yizhu Jiao and
                  Pengfei Liu and
                  Chenguang Zhu and
                  Heng Ji and
                  Jiawei Han},
  editor       = {Yoav Goldberg and
                  Zornitsa Kozareva and
                  Yue Zhang},
  title        = {Towards a Unified Multi-Dimensional Evaluator for Text Generation},
  booktitle    = {Proceedings of the 2022 Conference on Empirical Methods in Natural
                  Language Processing, {EMNLP} 2022, Abu Dhabi, United Arab Emirates,
                  December 7-11, 2022},
  pages        = {2023--2038},
  publisher    = {Association for Computational Linguistics},
  year         = {2022},
  url          = {https://doi.org/10.18653/v1/2022.emnlp-main.131},
  doi          = {10.18653/V1/2022.EMNLP-MAIN.131},
  bibsource    = {dblp computer science bibliography, https://dblp.org}
}

@inproceedings{DBLP:conf/naacl/LiuSX0CKGH24,
  author       = {Minqian Liu and
                  Ying Shen and
                  Zhiyang Xu and
                  Yixin Cao and
                  Eunah Cho and
                  Vaibhav Kumar and
                  Reza Ghanadan and
                  Lifu Huang},
  editor       = {Kevin Duh and
                  Helena G{\'{o}}mez{-}Adorno and
                  Steven Bethard},
  title        = {X-Eval: Generalizable Multi-aspect Text Evaluation via Augmented Instruction
                  Tuning with Auxiliary Evaluation Aspects},
  booktitle    = {Proceedings of the 2024 Conference of the North American Chapter of
                  the Association for Computational Linguistics: Human Language Technologies
                  (Volume 1: Long Papers), {NAACL} 2024, Mexico City, Mexico, June 16-21,
                  2024},
  pages        = {8560--8579},
  publisher    = {Association for Computational Linguistics},
  year         = {2024},
  url          = {https://doi.org/10.18653/v1/2024.naacl-long.473},
  doi          = {10.18653/V1/2024.NAACL-LONG.473},
  bibsource    = {dblp computer science bibliography, https://dblp.org}
}

@article{DBLP:journals/corr/abs-2504-02495,
  author       = {Zijun Liu and
                  Peiyi Wang and
                  Runxin Xu and
                  Shirong Ma and
                  Chong Ruan and
                  Peng Li and
                  Yang Liu and
                  Yu Wu},
  title        = {Inference-Time Scaling for Generalist Reward Modeling},
  journal      = {CoRR},
  volume       = {abs/2504.02495},
  year         = {2025},
  url          = {https://doi.org/10.48550/arXiv.2504.02495},
  doi          = {10.48550/ARXIV.2504.02495},
  eprinttype    = {arXiv},
  eprint       = {2504.02495},
  bibsource    = {dblp computer science bibliography, https://dblp.org}
}

@article{DBLP:journals/corr/abs-2505-13388,
  author       = {David Anugraha and
                  Zilu Tang and
                  Lester James V. Miranda and
                  Hanyang Zhao and
                  Mohammad Rifqi Farhansyah and
                  Garry Kuwanto and
                  Derry Wijaya and
                  Genta Indra Winata},
  title        = {{R3:} Robust Rubric-Agnostic Reward Models},
  journal      = {CoRR},
  volume       = {abs/2505.13388},
  year         = {2025},
  url          = {https://doi.org/10.48550/arXiv.2505.13388},
  doi          = {10.48550/ARXIV.2505.13388},
  eprinttype    = {arXiv},
  eprint       = {2505.13388},
  bibsource    = {dblp computer science bibliography, https://dblp.org}
}

@article{DBLP:journals/corr/abs-2511-10507,
  author       = {Yun He and
                  Wenzhe Li and
                  Hejia Zhang and
                  Songlin Li and
                  Karishma Mandyam and
                  Sopan Khosla and
                  Yuanhao Xiong and
                  Nanshu Wang and
                  Xiaoliang Peng and
                  Beibin Li and
                  Shengjie Bi and
                  Shishir G. Patil and
                  Qi Qi and
                  Shengyu Feng and
                  Julian Katz{-}Samuels and
                  Richard Yuanzhe Pang and
                  Sujan Gonugondla and
                  Hunter Lang and
                  Yue Yu and
                  Yundi Qian and
                  Maryam Fazel{-}Zarandi and
                  Licheng Yu and
                  Amine Benhalloum and
                  Hany Awadalla and
                  Manaal Faruqui},
  title        = {AdvancedIF: Rubric-Based Benchmarking and Reinforcement Learning for
                  Advancing {LLM} Instruction Following},
  journal      = {CoRR},
  volume       = {abs/2511.10507},
  year         = {2025},
  url          = {https://doi.org/10.48550/arXiv.2511.10507},
  doi          = {10.48550/ARXIV.2511.10507},
  eprinttype    = {arXiv},
  eprint       = {2511.10507},
  bibsource    = {dblp computer science bibliography, https://dblp.org}
}

@article{DBLP:journals/corr/abs-2510-07743,
  author       = {Tianci Liu and
                  Ran Xu and
                  Tony Yu and
                  Ilgee Hong and
                  Carl Yang and
                  Tuo Zhao and
                  Haoyu Wang},
  title        = {OpenRubrics: Towards Scalable Synthetic Rubric Generation for Reward
                  Modeling and {LLM} Alignment},
  journal      = {CoRR},
  volume       = {abs/2510.07743},
  year         = {2025},
  url          = {https://doi.org/10.48550/arXiv.2510.07743},
  doi          = {10.48550/ARXIV.2510.07743},
  eprinttype    = {arXiv},
  eprint       = {2510.07743},
  bibsource    = {dblp computer science bibliography, https://dblp.org}
}

@article{DBLP:journals/tmlr/JiangLZHLC24,
  author       = {Dongfu Jiang and
                  Yishan Li and
                  Ge Zhang and
                  Wenhao Huang and
                  Bill Yuchen Lin and
                  Wenhu Chen},
  title        = {TIGERScore: Towards Building Explainable Metric for All Text Generation
                  Tasks},
  journal      = {Trans. Mach. Learn. Res.},
  volume       = {2024},
  year         = {2024},
  url          = {https://openreview.net/forum?id=EE1CBKC0SZ},
  bibsource    = {dblp computer science bibliography, https://dblp.org}
}

@article{DBLP:journals/corr/abs-2508-12790,
  author       = {Zenan Huang and
                  Yihong Zhuang and
                  Guoshan Lu and
                  Zeyu Qin and
                  Haokai Xu and
                  Tianyu Zhao and
                  Ru Peng and
                  Jiaqi Hu and
                  Zhanming Shen and
                  Xiaomeng Hu and
                  Xijun Gu and
                  Peiyi Tu and
                  Jiaxin Liu and
                  Wenyu Chen and
                  Yuzhuo Fu and
                  Zhiting Fan and
                  Yanmei Gu and
                  Yuanyuan Wang and
                  Zhengkai Yang and
                  Jianguo Li and
                  Junbo Zhao},
  title        = {Reinforcement Learning with Rubric Anchors},
  journal      = {CoRR},
  volume       = {abs/2508.12790},
  year         = {2025},
  url          = {https://doi.org/10.48550/arXiv.2508.12790},
  doi          = {10.48550/ARXIV.2508.12790},
  eprinttype    = {arXiv},
  eprint       = {2508.12790},
  bibsource    = {dblp computer science bibliography, https://dblp.org}
}

@article{DBLP:journals/corr/abs-2505-02387,
  author       = {Xiusi Chen and
                  Gaotang Li and
                  Ziqi Wang and
                  Bowen Jin and
                  Cheng Qian and
                  Yu Wang and
                  Hongru Wang and
                  Yu Zhang and
                  Denghui Zhang and
                  Tong Zhang and
                  Hanghang Tong and
                  Heng Ji},
  title        = {{RM-R1:} Reward Modeling as Reasoning},
  journal      = {CoRR},
  volume       = {abs/2505.02387},
  year         = {2025},
  url          = {https://doi.org/10.48550/arXiv.2505.02387},
  doi          = {10.48550/ARXIV.2505.02387},
  eprinttype    = {arXiv},
  eprint       = {2505.02387},
  bibsource    = {dblp computer science bibliography, https://dblp.org}
}

@inproceedings{DBLP:conf/icml/Saha0GWW25,
  author       = {Swarnadeep Saha and
                  Xian Li and
                  Marjan Ghazvininejad and
                  Jason E. Weston and
                  Tianlu Wang},
  editor       = {Aarti Singh and
                  Maryam Fazel and
                  Daniel Hsu and
                  Simon Lacoste{-}Julien and
                  Felix Berkenkamp and
                  Tegan Maharaj and
                  Kiri Wagstaff and
                  Jerry Zhu},
  title        = {Learning to Plan {\&} Reason for Evaluation with Thinking-LLM-as-a-Judge},
  booktitle    = {Forty-second International Conference on Machine Learning, {ICML}
                  2025, Vancouver, BC, Canada, July 13-19, 2025},
  series       = {Proceedings of Machine Learning Research},
  volume       = {267},
  publisher    = {{PMLR} / OpenReview.net},
  year         = {2025},
  url          = {https://proceedings.mlr.press/v267/saha25b.html},
  bibsource    = {dblp computer science bibliography, https://dblp.org}
}

@inproceedings{Zhou2026AutoChecklistCP,
  title={AutoChecklist: Composable Pipelines for Checklist Generation and Scoring with LLM-as-a-Judge},
  author={Karen Zhou and Chenhao Tan},
  booktitle={Proceedings of the 64th Annual Meeting of the Association for Computational Linguistics (Volume 3: System Demonstrations), {ACL} 2026},
  pages={515--525},
  publisher={Association for Computational Linguistics},
  year={2026},
  url={https://doi.org/10.18653/v1/2026.acl-demo.51}
}

@inproceedings{DBLP:conf/emnlp/SaadFalconVBNFVSKM25,
  author       = {Jon Saad{-}Falcon and
                  Rajan Vivek and
                  William Berrios and
                  Nandita Shankar Naik and
                  Matija Franklin and
                  Bertie Vidgen and
                  Amanpreet Singh and
                  Douwe Kiela and
                  Shikib Mehri},
  editor       = {Christos Christodoulopoulos and
                  Tanmoy Chakraborty and
                  Carolyn Rose and
                  Violet Peng},
  title        = {{LMUNIT:} Fine-grained Evaluation with Natural Language Unit Tests},
  booktitle    = {Findings of the Association for Computational Linguistics: {EMNLP}
                  2025, Suzhou, China, November 4-9, 2025},
  pages        = {3303--3324},
  publisher    = {Association for Computational Linguistics},
  year         = {2025},
  url          = {https://aclanthology.org/2025.findings-emnlp.176/},
  bibsource    = {dblp computer science bibliography, https://dblp.org}
}

@article{DBLP:journals/corr/abs-2408-11791,
  author       = {Zachary Ankner and
                  Mansheej Paul and
                  Brandon Cui and
                  Jonathan D. Chang and
                  Prithviraj Ammanabrolu},
  title        = {Critique-out-Loud Reward Models},
  journal      = {CoRR},
  volume       = {abs/2408.11791},
  year         = {2024},
  url          = {https://doi.org/10.48550/arXiv.2408.11791},
  doi          = {10.48550/ARXIV.2408.11791},
  eprinttype    = {arXiv},
  eprint       = {2408.11791},
  bibsource    = {dblp computer science bibliography, https://dblp.org}
}

@inproceedings{DBLP:conf/emnlp/KimSLLSWNL0S24,
  author       = {Seungone Kim and
                  Juyoung Suk and
                  Shayne Longpre and
                  Bill Yuchen Lin and
                  Jamin Shin and
                  Sean Welleck and
                  Graham Neubig and
                  Moontae Lee and
                  Kyungjae Lee and
                  Minjoon Seo},
  editor       = {Yaser Al{-}Onaizan and
                  Mohit Bansal and
                  Yun{-}Nung Chen},
  title        = {Prometheus 2: An Open Source Language Model Specialized in Evaluating
                  Other Language Models},
  booktitle    = {Proceedings of the 2024 Conference on Empirical Methods in Natural
                  Language Processing, {EMNLP} 2024, Miami, FL, USA, November 12-16,
                  2024},
  pages        = {4334--4353},
  publisher    = {Association for Computational Linguistics},
  year         = {2024},
  url          = {https://doi.org/10.18653/v1/2024.emnlp-main.248},
  doi          = {10.18653/V1/2024.EMNLP-MAIN.248},
  bibsource    = {dblp computer science bibliography, https://dblp.org}
}

@inproceedings{DBLP:conf/emnlp/LiFZZLXWJH25,
  author       = {Zihao Li and
                  Feihao Fang and
                  Xitong Zhang and
                  Jiaru Zou and
                  Zhining Liu and
                  Wei Xiong and
                  Ziwei Wu and
                  Baoyu Jing and
                  Jingrui He},
  editor       = {Christos Christodoulopoulos and
                  Tanmoy Chakraborty and
                  Carolyn Rose and
                  Violet Peng},
  title        = {Not All Voices Are Rewarded Equally: Probing and Repairing Reward
                  Models across Human Diversity},
  booktitle    = {Findings of the Association for Computational Linguistics: {EMNLP}
                  2025, Suzhou, China, November 4-9, 2025},
  pages        = {3426--3455},
  publisher    = {Association for Computational Linguistics},
  year         = {2025},
  url          = {https://aclanthology.org/2025.findings-emnlp.183/},
  bibsource    = {dblp computer science bibliography, https://dblp.org}
}

@article{DBLP:journals/corr/abs-2401-06080,
  author       = {Binghai Wang and
                  Rui Zheng and
                  Lu Chen and
                  Yan Liu and
                  Shihan Dou and
                  Caishuang Huang and
                  Wei Shen and
                  Senjie Jin and
                  Enyu Zhou and
                  Chenyu Shi and
                  Songyang Gao and
                  Nuo Xu and
                  Yuhao Zhou and
                  Xiaoran Fan and
                  Zhiheng Xi and
                  Jun Zhao and
                  Xiao Wang and
                  Tao Ji and
                  Hang Yan and
                  Lixing Shen and
                  Zhan Chen and
                  Tao Gui and
                  Qi Zhang and
                  Xipeng Qiu and
                  Xuanjing Huang and
                  Zuxuan Wu and
                  Yu{-}Gang Jiang},
  title        = {Secrets of {RLHF} in Large Language Models Part {II:} Reward Modeling},
  journal      = {CoRR},
  volume       = {abs/2401.06080},
  year         = {2024},
  url          = {https://doi.org/10.48550/arXiv.2401.06080},
  doi          = {10.48550/ARXIV.2401.06080},
  eprinttype    = {arXiv},
  eprint       = {2401.06080},
  bibsource    = {dblp computer science bibliography, https://dblp.org}
}

@inproceedings{DBLP:conf/nips/KirkWRBMGCBW0VH24,
  author       = {Hannah Rose Kirk and
                  Alexander Whitefield and
                  Paul R{\"{o}}ttger and
                  Andrew M. Bean and
                  Katerina Margatina and
                  Rafael Mosquera G{\'{o}}mez and
                  Juan Ciro and
                  Max Bartolo and
                  Adina Williams and
                  He He and
                  Bertie Vidgen and
                  Scott Hale},
  editor       = {Amir Globersons and
                  Lester Mackey and
                  Danielle Belgrave and
                  Angela Fan and
                  Ulrich Paquet and
                  Jakub M. Tomczak and
                  Cheng Zhang},
  title        = {The {PRISM} Alignment Dataset: What Participatory, Representative
                  and Individualised Human Feedback Reveals About the Subjective and
                  Multicultural Alignment of Large Language Models},
  booktitle    = {Advances in Neural Information Processing Systems 38: Annual Conference
                  on Neural Information Processing Systems 2024, NeurIPS 2024, Vancouver,
                  BC, Canada, December 10 - 15, 2024},
  year         = {2024},
  url          = {http://papers.nips.cc/paper\_files/paper/2024/hash/be2e1b68b44f2419e19f6c35a1b8cf35-Abstract-Datasets\_and\_Benchmarks\_Track.html},
  bibsource    = {dblp computer science bibliography, https://dblp.org}
}

@inproceedings{DBLP:conf/iclr/ZhouZWXDBSXFMZG25,
  author       = {Enyu Zhou and
                  Guodong Zheng and
                  Binghai Wang and
                  Zhiheng Xi and
                  Shihan Dou and
                  Rong Bao and
                  Wei Shen and
                  Limao Xiong and
                  Jessica Fan and
                  Yurong Mou and
                  Rui Zheng and
                  Tao Gui and
                  Qi Zhang and
                  Xuanjing Huang},
  title        = {{RMB:} Comprehensively benchmarking reward models in {LLM} alignment},
  booktitle    = {The Thirteenth International Conference on Learning Representations,
                  {ICLR} 2025, Singapore, April 24-28, 2025},
  publisher    = {OpenReview.net},
  year         = {2025},
  url          = {https://openreview.net/forum?id=kmgrlG9TR0},
  bibsource    = {dblp computer science bibliography, https://dblp.org}
}

@article{DBLP:journals/corr/abs-2510-18596,
  author       = {Haojia Lin and
                  Xiaoyu Tan and
                  Yulei Qin and
                  Zihan Xu and
                  Yuchen Shi and
                  Zongyi Li and
                  Gang Li and
                  Shaofei Cai and
                  Siqi Cai and
                  Chaoyou Fu and
                  Ke Li and
                  Xing Sun},
  title        = {CUARewardBench: {A} Benchmark for Evaluating Reward Models on Computer-using
                  Agent},
  journal      = {CoRR},
  volume       = {abs/2510.18596},
  year         = {2025},
  url          = {https://doi.org/10.48550/arXiv.2510.18596},
  doi          = {10.48550/ARXIV.2510.18596},
  eprinttype    = {arXiv},
  eprint       = {2510.18596},
  bibsource    = {dblp computer science bibliography, https://dblp.org}
}

@inproceedings{Wen2026IFRewardBenchBJ,
  title={IF-RewardBench: Benchmarking Judge Models for Instruction-Following Evaluation},
  author={Bosi Wen and Yilin Niu and Cunxiang Wang and Xiaoying Ling and Ying Zhang and Pei Ke and Hongning Wang and Minlie Huang},
  booktitle={Proceedings of the 64th Annual Meeting of the Association for Computational Linguistics (Volume 1: Long Papers), {ACL} 2026},
  pages={23816--23843},
  publisher={Association for Computational Linguistics},
  year={2026},
  url={https://doi.org/10.18653/v1/2026.acl-long.1092}
}

@inproceedings{DBLP:conf/cvpr/LiW0YSWALLLKL25,
  author       = {Lei Li and
                  Yuancheng Wei and
                  Zhihui Xie and
                  Xuqing Yang and
                  Yifan Song and
                  Peiyi Wang and
                  Chenxin An and
                  Tianyu Liu and
                  Sujian Li and
                  Bill Yuchen Lin and
                  Lingpeng Kong and
                  Qi Liu},
  title        = {VL-RewardBench: {A} Challenging Benchmark for Vision-Language Generative
                  Reward Models},
  booktitle    = {{IEEE/CVF} Conference on Computer Vision and Pattern Recognition,
                  {CVPR} 2025, Nashville, TN, USA, June 11-15, 2025},
  pages        = {24657--24668},
  publisher    = {Computer Vision Foundation / {IEEE}},
  year         = {2025},
  url          = {https://openaccess.thecvf.com/content/CVPR2025/html/Li\_VL-RewardBench\_A\_Challenging\_Benchmark\_for\_Vision-Language\_Generative\_Reward\_Models\_CVPR\_2025\_paper.html},
  doi          = {10.1109/CVPR52734.2025.02296},
  bibsource    = {dblp computer science bibliography, https://dblp.org}
}

@inproceedings{DBLP:conf/emnlp/MattonSATAHMVGG24,
  author       = {Alexandre Matton and
                  Tom Sherborne and
                  Dennis Aumiller and
                  Elena Tommasone and
                  Milad Alizadeh and
                  Jingyi He and
                  Raymond Ma and
                  Maxime Voisin and
                  Ellen Gilsenan{-}McMahon and
                  Matthias Gall{\'{e}}},
  editor       = {Yaser Al{-}Onaizan and
                  Mohit Bansal and
                  Yun{-}Nung Chen},
  title        = {On Leakage of Code Generation Evaluation Datasets},
  booktitle    = {Findings of the Association for Computational Linguistics: {EMNLP}
                  2024, Miami, Florida, USA, November 12-16, 2024},
  series       = {Findings of {ACL}},
  volume       = {{EMNLP} 2024},
  pages        = {13215--13223},
  publisher    = {Association for Computational Linguistics},
  year         = {2024},
  url          = {https://doi.org/10.18653/v1/2024.findings-emnlp.772},
  doi          = {10.18653/V1/2024.FINDINGS-EMNLP.772},
  bibsource    = {dblp computer science bibliography, https://dblp.org}
}

@article{DBLP:journals/tacl/NiYZRFSYLYXJZRCC24,
  author       = {Ansong Ni and
                  Pengcheng Yin and
                  Yilun Zhao and
                  Martin Riddell and
                  Troy Feng and
                  Rui Shen and
                  Stephen Yin and
                  Ye Liu and
                  Semih Yavuz and
                  Caiming Xiong and
                  Shafiq Joty and
                  Yingbo Zhou and
                  Dragomir Radev and
                  Arman Cohan},
  title        = {\texttt{L2CEval}: Evaluating Language-to-Code Generation Capabilities
                  of Large Language Models},
  journal      = {Trans. Assoc. Comput. Linguistics},
  volume       = {12},
  pages        = {1311--1329},
  year         = {2024},
  url          = {https://doi.org/10.1162/tacl\_a\_00705},
  doi          = {10.1162/TACL\_A\_00705},
  bibsource    = {dblp computer science bibliography, https://dblp.org}
}

@article{DBLP:journals/corr/abs-2601-12186,
  author       = {Vatsal Venkatkrishna and
                  Indraneil Paul and
                  Iryna Gurevych},
  title        = {Aletheia: What Makes {RLVR} For Code Verifiers Tick?},
  journal      = {CoRR},
  volume       = {abs/2601.12186},
  year         = {2026},
  url          = {https://doi.org/10.48550/arXiv.2601.12186},
  doi          = {10.48550/ARXIV.2601.12186},
  eprinttype    = {arXiv},
  eprint       = {2601.12186},
  bibsource    = {dblp computer science bibliography, https://dblp.org}
}

@article{DBLP:journals/corr/abs-2506-01937,
  author       = {Saumya Malik and
                  Valentina Pyatkin and
                  Sander Land and
                  Jacob Morrison and
                  Noah A. Smith and
                  Hannaneh Hajishirzi and
                  Nathan Lambert},
  title        = {RewardBench 2: Advancing Reward Model Evaluation},
  journal      = {CoRR},
  volume       = {abs/2506.01937},
  year         = {2025},
  url          = {https://doi.org/10.48550/arXiv.2506.01937},
  doi          = {10.48550/ARXIV.2506.01937},
  eprinttype    = {arXiv},
  eprint       = {2506.01937},
  bibsource    = {dblp computer science bibliography, https://dblp.org}
}

@inproceedings{DBLP:conf/iclr/WenL0LXLHHZ025,
  author       = {Xueru Wen and
                  Jie Lou and
                  Yaojie Lu and
                  Hongyu Lin and
                  XingYu and
                  Xinyu Lu and
                  Ben He and
                  Xianpei Han and
                  Debing Zhang and
                  Le Sun},
  title        = {Rethinking Reward Model Evaluation: Are We Barking up the Wrong Tree?},
  booktitle    = {The Thirteenth International Conference on Learning Representations,
                  {ICLR} 2025, Singapore, April 24-28, 2025},
  publisher    = {OpenReview.net},
  year         = {2025},
  url          = {https://openreview.net/forum?id=Cnwz9jONi5},
  bibsource    = {dblp computer science bibliography, https://dblp.org}
}

@inproceedings{DBLP:conf/acl/KimKKC0Y25,
  author       = {Sunghwan Kim and
                  Dongjin Kang and
                  Taeyoon Kwon and
                  Hyungjoo Chae and
                  Dongha Lee and
                  Jinyoung Yeo},
  editor       = {Wanxiang Che and
                  Joyce Nabende and
                  Ekaterina Shutova and
                  Mohammad Taher Pilehvar},
  title        = {Rethinking Reward Model Evaluation Through the Lens of Reward Overoptimization},
  booktitle    = {Proceedings of the 63rd Annual Meeting of the Association for Computational
                  Linguistics (Volume 1: Long Papers), {ACL} 2025, Vienna, Austria,
                  July 27 - August 1, 2025},
  pages        = {13252--13280},
  publisher    = {Association for Computational Linguistics},
  year         = {2025},
  url          = {https://aclanthology.org/2025.acl-long.649/},
  bibsource    = {dblp computer science bibliography, https://dblp.org}
}

@article{DBLP:journals/corr/abs-2505-16222,
  author       = {Jiwon Moon and
                  Yerin Hwang and
                  Dongryeol Lee and
                  Taegwan Kang and
                  Yongil Kim and
                  Kyomin Jung},
  title        = {Don't Judge Code by Its Cover: Exploring Biases in {LLM} Judges
                  for Code Evaluation},
  journal      = {CoRR},
  volume       = {abs/2505.16222},
  year         = {2025},
  url          = {https://doi.org/10.48550/arXiv.2505.16222},
  doi          = {10.48550/ARXIV.2505.16222},
  eprinttype    = {arXiv},
  eprint       = {2505.16222},
  bibsource    = {dblp computer science bibliography, https://dblp.org}
}

@inproceedings{DBLP:conf/acl/Jiang0L23,
  author       = {Dongfu Jiang and
                  Xiang Ren and
                  Bill Yuchen Lin},
  editor       = {Anna Rogers and
                  Jordan L. Boyd{-}Graber and
                  Naoaki Okazaki},
  title        = {LLM-Blender: Ensembling Large Language Models with Pairwise Ranking
                  and Generative Fusion},
  booktitle    = {Proceedings of the 61st Annual Meeting of the Association for Computational
                  Linguistics (Volume 1: Long Papers), {ACL} 2023, Toronto, Canada,
                  July 9-14, 2023},
  pages        = {14165--14178},
  publisher    = {Association for Computational Linguistics},
  year         = {2023},
  url          = {https://doi.org/10.18653/v1/2023.acl-long.792},
  doi          = {10.18653/V1/2023.ACL-LONG.792},
  bibsource    = {dblp computer science bibliography, https://dblp.org}
}

@inproceedings{DBLP:conf/iclr/MuennighoffLZZH24,
  author       = {Niklas Muennighoff and
                  Qian Liu and
                  Armel Randy Zebaze and
                  Qinkai Zheng and
                  Binyuan Hui and
                  Terry Yue Zhuo and
                  Swayam Singh and
                  Xiangru Tang and
                  Leandro von Werra and
                  Shayne Longpre},
  title        = {OctoPack: Instruction Tuning Code Large Language Models},
  booktitle    = {The Twelfth International Conference on Learning Representations,
                  {ICLR} 2024, Vienna, Austria, May 7-11, 2024},
  publisher    = {OpenReview.net},
  year         = {2024},
  url          = {https://openreview.net/forum?id=mw1PWNSWZP},
  bibsource    = {dblp computer science bibliography, https://dblp.org}
}

@article{DBLP:journals/corr/abs-2308-03281,
  author       = {Zehan Li and
                  Xin Zhang and
                  Yanzhao Zhang and
                  Dingkun Long and
                  Pengjun Xie and
                  Meishan Zhang},
  title        = {Towards General Text Embeddings with Multi-stage Contrastive Learning},
  journal      = {CoRR},
  volume       = {abs/2308.03281},
  year         = {2023},
  url          = {https://doi.org/10.48550/arXiv.2308.03281},
  doi          = {10.48550/ARXIV.2308.03281},
  eprinttype   = {arXiv},
  eprint       = {2308.03281},
  bibsource    = {dblp computer science bibliography, https://dblp.org}
}

@article{DBLP:journals/corr/abs-2505-12697,
  author       = {Chaofan Li and
                  Jianlyu Chen and
                  Yingxia Shao and
                  Defu Lian and
                  Zheng Liu},
  title        = {Towards {A} Generalist Code Embedding Model Based On Massive Data
                  Synthesis},
  journal      = {CoRR},
  volume       = {abs/2505.12697},
  year         = {2025},
  url          = {https://doi.org/10.48550/arXiv.2505.12697},
  doi          = {10.48550/ARXIV.2505.12697},
  eprinttype   = {arXiv},
  eprint       = {2505.12697},
  bibsource    = {dblp computer science bibliography, https://dblp.org}
}

@article{DBLP:journals/corr/abs-2410-18451,
  author       = {Chris Yuhao Liu and
                  Liang Zeng and
                  Jiacai Liu and
                  Rui Yan and
                  Jujie He and
                  Chaojie Wang and
                  Shuicheng Yan and
                  Yang Liu and
                  Yahui Zhou},
  title        = {Skywork-Reward: Bag of Tricks for Reward Modeling in LLMs},
  journal      = {CoRR},
  volume       = {abs/2410.18451},
  year         = {2024},
  url          = {https://doi.org/10.48550/arXiv.2410.18451},
  doi          = {10.48550/ARXIV.2410.18451},
  eprinttype   = {arXiv},
  eprint       = {2410.18451},
  bibsource    = {dblp computer science bibliography, https://dblp.org}
}

@inproceedings{DBLP:conf/nips/HanREJL00D24,
  author       = {Seungju Han and
                  Kavel Rao and
                  Allyson Ettinger and
                  Liwei Jiang and
                  Bill Yuchen Lin and
                  Nathan Lambert and
                  Yejin Choi and
                  Nouha Dziri},
  editor       = {Amir Globersons and
                  Lester Mackey and
                  Danielle Belgrave and
                  Angela Fan and
                  Ulrich Paquet and
                  Jakub M. Tomczak and
                  Cheng Zhang},
  title        = {WildGuard: Open One-stop Moderation Tools for Safety Risks, Jailbreaks,
                  and Refusals of LLMs},
  booktitle    = {Advances in Neural Information Processing Systems 38: Annual Conference
                  on Neural Information Processing Systems 2024, NeurIPS 2024, Vancouver,
                  BC, Canada, December 10 - 15, 2024},
  year         = {2024},
  url          = {http://papers.nips.cc/paper\_files/paper/2024/hash/0f69b4b96a46f284b726fbd70f74fb3b-Abstract-Datasets\_and\_Benchmarks\_Track.html},
  bibsource    = {dblp computer science bibliography, https://dblp.org}
}

@inproceedings{DBLP:conf/emnlp/ParkJRKC24,
  author       = {Junsoo Park and
                  Seungyeon Jwa and
                  Meiying Ren and
                  Daeyoung Kim and
                  Sanghyuk Choi},
  editor       = {Yaser Al{-}Onaizan and
                  Mohit Bansal and
                  Yun{-}Nung Chen},
  title        = {OffsetBias: Leveraging Debiased Data for Tuning Evaluators},
  booktitle    = {Findings of the Association for Computational Linguistics: {EMNLP}
                  2024, Miami, Florida, USA, November 12-16, 2024},
  series       = {Findings of {ACL}},
  pages        = {1043--1067},
  publisher    = {Association for Computational Linguistics},
  year         = {2024},
  url          = {https://doi.org/10.18653/v1/2024.findings-emnlp.57},
  doi          = {10.18653/V1/2024.FINDINGS-EMNLP.57},
  bibsource    = {dblp computer science bibliography, https://dblp.org}
}

@inproceedings{DBLP:conf/iclr/XuJNDP0L25,
  author       = {Zhangchen Xu and
                  Fengqing Jiang and
                  Luyao Niu and
                  Yuntian Deng and
                  Radha Poovendran and
                  Yejin Choi and
                  Bill Yuchen Lin},
  title        = {Magpie: Alignment Data Synthesis from Scratch by Prompting Aligned
                  LLMs with Nothing},
  booktitle    = {The Thirteenth International Conference on Learning Representations,
                  {ICLR} 2025, Singapore, April 24-28, 2025},
  publisher    = {OpenReview.net},
  year         = {2025},
  url          = {https://openreview.net/forum?id=Pnk7vMbznK},
  bibsource    = {dblp computer science bibliography, https://dblp.org}
}

@article{DBLP:journals/corr/abs-2406-08673,
  author       = {Zhilin Wang and
                  Yi Dong and
                  Olivier Delalleau and
                  Jiaqi Zeng and
                  Gerald Shen and
                  Daniel Egert and
                  Jimmy J. Zhang and
                  Makesh Narsimhan Sreedhar and
                  Oleksii Kuchaiev},
  title        = {HelpSteer2: Open-source dataset for training top-performing reward
                  models},
  journal      = {CoRR},
  volume       = {abs/2406.08673},
  year         = {2024},
  url          = {https://doi.org/10.48550/arXiv.2406.08673},
  doi          = {10.48550/ARXIV.2406.08673},
  eprinttype   = {arXiv},
  eprint       = {2406.08673},
  bibsource    = {dblp computer science bibliography, https://dblp.org}
}

@misc{tdcommonsUsingCode,
	author = {Emmanouil Koukoumidis},
	title = {{U}sing {C}ode {R}eview {R}epositories and {C}hangelists to {T}rain {L}arge {L}anguage {M}odels for {C}ode {G}eneration --- tdcommons.org},
	howpublished = {\url{https://www.tdcommons.org/dpubs\_series/6027/}},
	year = {2023},
}
\bibliographystyle{tmlr}

\clearpage

\appendix

\section{Architecture And Training Details}
\label{appdx:1_Arch_Train}

\begin{table}[htbp!]
    \centering
    \caption{Architectural heritage and training attributes of the \texttt{Themis-RM} suite of models across the preference model pre-training and preference modeling stages. The training objective is described in \cref{sec:Themis_RM} and the training data composition for both stages is detailed in \Cref{appdx:2_Data_Collection}.}
    \label{tab:Arch_And_Training}
    \renewcommand{\arraystretch}{1.12}
    \scalebox{0.6}{
    \begin{tabular}{lrrrrrr}
        \toprule
        
        \textbf{\texttt{Attribute}} & \textbf{\texttt{Themis-RM-0.6B}} & 
        \textbf{\texttt{Themis-RM-1.7B}} &
        \textbf{\texttt{Themis-RM-4B}} &
        \textbf{\texttt{Themis-RM-8B}} &
        \textbf{\texttt{Themis-RM-14B}} &
        \textbf{\texttt{Themis-RM-32B}}\\

        \midrule
    
        & \multicolumn{6}{c}{\textbf{\texttt{Architecture Attributes}}} \\
        
        \midrule

        \texttt{Backbone Architecture} & \hfmodel{Qwen/Qwen3-0.6B} & \hfmodel{Qwen/Qwen3-1.7B} & \hfmodel{Qwen/Qwen3-4B} & \hfmodel{Qwen/Qwen3-8B} & \hfmodel{Qwen/Qwen3-14B} &  \hfmodel{Qwen/Qwen3-32B}\\

        \midrule
    
        & \multicolumn{6}{c}{\textbf{\texttt{Training Attributes: Preference Model Pre-Training (PT)}}} \\
        
        \midrule
        \texttt{Training Dataset} & \multicolumn{6}{c}{\texttt{Themis-GeneralPreference}} \\
        \texttt{Criteria Following} & \multicolumn{6}{c}{\texttt{True}} \\
        \texttt{Reward Centering Coefficient ($\mu$)} & \multicolumn{6}{c}{\texttt{0.01}} \\
        \texttt{Behavior Cloning Coefficient ($\lambda$)} & \multicolumn{6}{c}{\texttt{0.4}} \\
        \texttt{Scheduler Type} & \multicolumn{6}{c}{\texttt{Cosine}} \\
        \texttt{Scheduler Warmup Proportion} & \multicolumn{6}{c}{\texttt{0.05}} \\
        \texttt{Optimizer Type} & \multicolumn{6}{c}{\texttt{AdamW-Fused}} \\
        \texttt{Peak Learning Rate} & \multicolumn{6}{c}{\texttt{2e-5}} \\
        \texttt{Terminal Learning Rate} & \multicolumn{6}{c}{\texttt{1e-5}} \\
        \texttt{Beta} & \multicolumn{6}{c}{\texttt{\{0.9, 0.95\}}} \\
        \texttt{Epsilon} & \multicolumn{6}{c}{\texttt{1e-8}} \\
        \texttt{Gradient Clipping} & \multicolumn{6}{c}{\texttt{2.0}} \\
        \texttt{Gradient Checkpointing} & \multicolumn{6}{c}{\texttt{True}} \\
        \texttt{Weight Decay} & \multicolumn{6}{c}{\texttt{0.1}} \\
        \texttt{FSDP Variant} & \multicolumn{6}{c}{\texttt{Version 1}} \\
        \texttt{FSDP Param Offload} & \multicolumn{6}{c}{\texttt{True}} \\
        \texttt{Flash Attention Variant} & \multicolumn{6}{c}{\texttt{Version 2}} \\
        \texttt{Flash Attention RMS Norm} & \multicolumn{6}{c}{\texttt{True}} \\
        \texttt{Flash Attention Fuse QKV} & \multicolumn{6}{c}{\texttt{True}} \\
        \texttt{Flash Attention Fuse MLP} & \multicolumn{6}{c}{\texttt{True}} \\
        \texttt{Liger Kernels} & \multicolumn{6}{c}{\texttt{True}} \\
        \texttt{Liger RoPE} & \multicolumn{6}{c}{\texttt{True}} \\
        \texttt{Liger RMS Norm} & \multicolumn{6}{c}{\texttt{True}} \\
        \texttt{Liger GLU Activation} & \multicolumn{6}{c}{\texttt{True}} \\
        \texttt{Model Datatype} & \multicolumn{6}{c}{\texttt{bfloat16}} \\
        \texttt{Softmax Datatype} & \multicolumn{6}{c}{\texttt{float32}} \\
        \texttt{AllReduce Datatype} & \multicolumn{6}{c}{\texttt{float32}} \\
        \texttt{Training Sequence Length} & \multicolumn{6}{c}{\texttt{2560}} \\
        \texttt{Global Batch Size} & \multicolumn{6}{c}{\texttt{1024}} \\
        \texttt{Training Epochs} & \multicolumn{6}{c}{\texttt{2.0}} \\
        \texttt{Training GPU Count} & \texttt{8.0} & \texttt{8.0} & \texttt{16.0} & \texttt{32.0} & \texttt{32.0} & \texttt{64.0} \\

        \midrule
    
        & \multicolumn{6}{c}{\textbf{\texttt{Training Attributes: Preference Modeling (PM)}}} \\
        
        \midrule
        \texttt{Training Dataset} & \multicolumn{6}{c}{\texttt{Themis-CodePreference}} \\
        \texttt{Criteria Following} & \multicolumn{6}{c}{\texttt{True}} \\
        \texttt{Reward Centering Coefficient ($\mu$)} & \multicolumn{6}{c}{\texttt{0.001}} \\
        \texttt{Behavior Cloning Coefficient ($\lambda$)} & \multicolumn{6}{c}{\texttt{0.25}} \\
        \texttt{Scheduler Type} & \multicolumn{6}{c}{\texttt{Cosine}} \\
        \texttt{Scheduler Warmup Proportion} & \multicolumn{6}{c}{\texttt{0.05}} \\
        \texttt{Optimizer Type} & \multicolumn{6}{c}{\texttt{AdamW-Fused}} \\
        \texttt{Peak Learning Rate} & \multicolumn{6}{c}{\texttt{1e-5}} \\
        \texttt{Terminal Learning Rate} & \multicolumn{6}{c}{\texttt{5e-7}} \\
        \texttt{Beta} & \multicolumn{6}{c}{\texttt{\{0.9, 0.95\}}} \\
        \texttt{Epsilon} & \multicolumn{6}{c}{\texttt{1e-8}} \\
        \texttt{Gradient Clipping} & \multicolumn{6}{c}{\texttt{1.5}} \\
        \texttt{Gradient Checkpointing} & \multicolumn{6}{c}{\texttt{True}} \\
        \texttt{Weight Decay} & \multicolumn{6}{c}{\texttt{0.1}} \\
        \texttt{FSDP Variant} & \multicolumn{6}{c}{\texttt{Version 1}} \\
        \texttt{FSDP Param Offload} & \multicolumn{6}{c}{\texttt{True}} \\
        \texttt{Flash Attention Variant} & \multicolumn{6}{c}{\texttt{Version 2}} \\
        \texttt{Flash Attention RMS Norm} & \multicolumn{6}{c}{\texttt{True}} \\
        \texttt{Flash Attention Fuse QKV} & \multicolumn{6}{c}{\texttt{True}} \\
        \texttt{Flash Attention Fuse MLP} & \multicolumn{6}{c}{\texttt{True}} \\
        \texttt{Liger Kernels} & \multicolumn{6}{c}{\texttt{True}} \\
        \texttt{Liger RoPE} & \multicolumn{6}{c}{\texttt{True}} \\
        \texttt{Liger RMS Norm} & \multicolumn{6}{c}{\texttt{True}} \\
        \texttt{Liger GLU Activation} & \multicolumn{6}{c}{\texttt{True}} \\
        \texttt{Model Datatype} & \multicolumn{6}{c}{\texttt{bfloat16}} \\
        \texttt{Softmax Datatype} & \multicolumn{6}{c}{\texttt{float32}} \\
        \texttt{AllReduce Datatype} & \multicolumn{6}{c}{\texttt{float32}} \\
        \texttt{Training Sequence Length} & \multicolumn{6}{c}{\texttt{4096}} \\
        \texttt{Global Batch Size} & \multicolumn{6}{c}{\texttt{512}} \\
        \texttt{Training Epochs} & \multicolumn{6}{c}{\texttt{1.0}} \\
        \texttt{Training GPU Count} & \texttt{8.0} & \texttt{8.0} & \texttt{16.0} & \texttt{32.0} & \texttt{64.0} & \texttt{128.0} \\

        \bottomrule
    \end{tabular}
    }
\end{table}

\section{Data Collection}
\label{appdx:2_Data_Collection}

\subsection{BigQuery SQL Query For Commit Mining}
\label{subsec:BigQuery_SQL}

\begin{codelisting}[H]
\begin{codeframe}[title={\bqrepo{github/github-repos} BigQuery Single-File Commit Mining}]
\begin{lstlisting}[style=sqlstyle]
SELECT
  c.commit,
  c.subject,
  c.message,
  STRING_AGG(DISTINCT unnested_repo_name) AS repos,
  l.license,
  d.old_path AS old_file,
  d.new_path AS new_file,
  c.committer.time_sec AS unix_time
FROM
  `bigquery-public-data.github_repos.languages` AS lang_table,
  UNNEST(language) AS lang
JOIN `bigquery-public-data.github_repos.licenses` AS l
  ON l.repo_name = lang_table.repo_name
JOIN (
  SELECT *, unnested_repo_name
  FROM `bigquery-public-data.github_repos.commits`, 
  UNNEST(repo_name) AS unnested_repo_name
) c
  ON c.unnested_repo_name = lang_table.repo_name,
  UNNEST(c.difference) AS d
WHERE
  l.license IN (
    'mit', 'artistic-2.0', 'isc', 'cc0-1.0', 'epl-1.0', 'mpl-2.0',
    'unlicense', 'apache-2.0', 'bsd-3-clause', 'agpl-3.0', 'lgpl-2.1', 'bsd-2-clause'
  )
  AND lang.name IN (
    'Python', 'Java', 'JavaScript', 'C', 'C#', 'C++', 'TypeScript', 'Go', 'Ruby',
  )
  AND LENGTH(c.message) > 10 AND LENGTH(c.message) < 15000
  AND LOWER(c.message) NOT IN (
    'update readme.md', 'initial commit', 'update', 'mirroring from micro.blog.',
    'update data.json', 'update data.js', 'add files via upload', 'update readme',
    "can't you see i'm updating the time?", 'dummy', 'update index.html', 'first commit', 
    'create readme.md', 'heartbeat update', 'updated readme', 'update log', 'test', 
    'no message', 'readme', 'wip', 'updates', 'commit', 'update _config.yaml', 'testing', 
    'tweak', 'tweaks', 'modified', 'edited', 'yolo commit', 'yolo', 'made it work', 
    'work in progress', 'fixing', 'for review', 'my changes', 'revised', 'addressed comments', 
    'placeholder', 'test commit', 'trying something', 'experimental changes', 'hack',
    'do not merge', 'various updates', 'stuff'
  )
  AND LOWER(c.message) NOT LIKE '%pi push%' AND LOWER(c.message) NOT LIKE '%push pi%'
  AND LOWER(c.message) NOT LIKE 'merge%' AND d.old_path = d.new_path
  AND d.old_path IS NOT NULL AND d.new_path IS NOT NULL
GROUP BY
  c.commit, c.subject, c.message, l.license, d.old_path, d.new_path, c.committer.time_sec
HAVING COUNT(DISTINCT d.old_path) = 1
\end{lstlisting}
\end{codeframe}
\caption{Streamlined GoogleSQL query for mining single-file-changing commits in openly licensed repositories, modified from the BigQuery GitHub dataset query pipeline presented in \citet{DBLP:conf/iclr/MuennighoffLZZH24}.}
\end{codelisting}

\subsection{Search Terms For Aspect-Specific Commit Classifier Training}
\label{subsec:Commit_Mining}

\newlength{\totalwidth}
\newlength{\colone}
\newlength{\coltwo}

\setlength{\totalwidth}{\textwidth * \real{1.6667}}
\setlength{\colone}{0.1\totalwidth}
\setlength{\coltwo}{\totalwidth - \colone - 4\tabcolsep}

\def\arraystretch{1.12}
\begin{table}[ht]
  \centering
  \caption{Per-criteria search terms used to recall positives for training the commit classifier. Refer to \cref{sec:Code_Reward_Bench} for a description of the complete commit preference mining pipeline.}
  \label{tab:Mining_Terms}
  \scalebox{0.6}{%
    \begin{tabular}{p{\colone} p{\coltwo}}
      \toprule
      \textbf{\texttt{Criteria}} & \textbf{\texttt{Terms}} \\
      \midrule
      \textbf{\texttt{Functional Correctness}} &
        \texttt{adjust logic, align, boundary case, bug fix, class misus, corner case, correct algorithm, correct infinite, correct layout, correct logic, edge case, edgecase, enhancement, error handling, error recovery, faulty index, faulty logic, faulty loop, feature, fix broken, fix bug, fix cache, fix case, fix concurre, fix defect, fix do, fix duplicate, fix else, fix error, fix exception, fix fault, fix for, fix if, fix infinite, fix issue, fix iteration, fix logic, fix loop, fix null, fix problem, fix recursion, fix regres, fix switch, fix ui, fix ux, fix while, function misus, handle error, handle exception, handle failure, handle fault, handle null, handling error, handling exception, handling failure, illegal command, illegal comment, illegal condition, illegal declaration, illegal definition, illegal directive, illegal expression, illegal field, illegal indentation, illegal index, illegal key, illegal offset, illegal variable, input handling, integer overflow, library misus, logic error, missing command, missing comment, missing condition, missing declaration, missing definition, repair broken, repair bug, repair error, repair issue, resolve logic, self assign, self-assign, unclosed brace, unclosed bracket, unclosed comment, unclosed parenthesis, unclosed string, undefined class, undefined constant, undefined field, undefined function, undefined index, undefined key, undefined member, undefined method, undefined object, undefined offset, undefined property, undefined value, undefined variable, uninitialized class, uninitialized constant, uninitialized field, uninitialized function, uninitialized index, uninitialized key, uninitialized member, uninitialized method, uninitialized object, uninitialized offset, uninitialized property, uninitialized value, uninitialized variable, unmatched brace, unmatched bracket, wrong logic, wrong loop, wrong variable, wrongly assign} \\
    \midrule
      \textbf{\texttt{Execution Efficiency}} &
        \texttt{add async, asynchronous, avoid unnecessary computation, batch api, batch operation, batch proc, boost efficiency, boost performance, cache frequently, cache function, cache results, cache the, cache values, cache variables, cache with memoization, caching, code optimization, cold path, concurrent, constant time, constant-time, continuous batch, decrease runtime, dynamic batch, early termination, efficiency improvement, efficient access, efficient algorithm, efficient code, efficient execution, efficient implementation, efficient iteration, efficient processing, efficient structures, enhance bit manipulation, enhance bit operations, enhance bit twiddling, enhance bitwise, enhance efficiency, enhance latency, enhance processing speed, enhance sort, faster, fix latency, fix stalling, fix startup time, fix time complexity, hash, hot path, implement async operations, implement connection pooling, implement efficient data serialization, implement lazy evaluation, implement lazy loading, implement multithreading, implement parallel processing, implement producer-consumer pattern, branch prediction, improve cache coherency, cache hit, cache locality, cache performance, improve cache utilization, improve speed, inplace, latency sensitive, load data in chunks, lru, make code more efficient, make code more performant, make code more responsive, make code run quicker, memoize, memory map, o(1) lookup, o(nlogn) sorting algorithm, optimize data, optimize file, optimize i/o, optimize import, optimize initialization, optimize latency, optimize list, optimize lookahead, optimize lookup, optimize loop, optimize memoization, optimize merge step, optimize network requests, optimize numpy operations, optimize path cost calculation, optimize pathfinding, optimize performance, overhead reduc, oversynchroni, parallelize, perf improvement, performance improvement, precompute expensive operations, precompute frequently used results, precompute frequently used values, precompute lookup tables, precompute results, precompute values, prune unnecessary, quickness improvement, reduce algorithm complexity, reduce computational overhead, reduce function call, reduce i/o operations, reduce lookup time, reduce overhead, reduce recursion, reduce recursive call, reduce runtime, time complexity, remove overhead, remove slow, replace slow, reduce latency, speed fix, speed up, speedup, tail call optimization, unroll loops, use bisect for binary search, use built-in, use builtin, use in-place, use list comprehension for speed, use map, use numpy vectorization, use optimized, use set, utilize multiprocessing, vectorize, work steal} \\
    \midrule
      \textbf{\texttt{Memory Efficiency}} &
        \texttt{fix memory management issue, fix memory, garbage collection, high memory, improve memory allocation, improve memory bound, improve memory usage, improve memory management, lazy eval, lazy load, memory bloat, memory consumption, memory efficien, memory footprint, memory leak fix, memory leak, memory map, memory optimization, memory pool, memory wastage, memory-efficien, resource leak, unnecessary malloc, unnecessary memory alloc, use less memory, zero copy} \\
    \midrule
      \textbf{\texttt{Readability And Maintainability}} &
        \texttt{abstract, anti-pattern, antipattern, appropriate nam, big class, big method, break down, break up, builder pattern, chain of respons, circular dependency, clarify, clarity, clean up, cleanup, code formatting, code layout, code path, code quality, code smell, code style, coding standards, coding style, cohesion, cohesive, command pattern, complex condition, consistent nam, control coupling, control dependency, control flow, convention, cyclomatic, data clump, data coupling, data dependency, dead assignment, dead class, dead code, dead function, dead method, dead parameter, dead variable, decouple, deep nest, deeply nest, dependency inversion, descriptive nam, dry, duplication, halstead, inconsistent doc, inconsistent format, interface segregation, intimacy violation, large class, large method, large parameter list, liskov substitution, long class, long method, long parameter list, maintainability, maintainable, message chain, modernize, modular, normalize, omitted default, omitted else, omitted error, omitted exception, omitted if, omitted switch, open/closed, parallel inheritance, pep 8, pep8, portability, primitive obsession, readability, readable, reduce complexity, reduce coupling, reduce dependency, reduce depth, reduce nest, refactor, refused bequest, remove goto, remove hardcod, separate, separation, shotgun surgery, simplify, single responsibility, solid, split function, style guide, tech debt, technical debt, temporary field, test coverage, testability, testable, unconditional branch, unconditional control, unconditional jump, unconditional statement, unused assignment, unused class, unused code, unused function, unused method, unused parameter, unused variable, visitor pattern}\\
    \midrule
      \textbf{\texttt{Security Hardness}} &
        \texttt{access control, android injection, auth bypass, auth forgery, authorization bypass, avoid directory traversal, avoid dos attack, avoid file inclusion, avoid format string vulnerability, avoid overflow, avoid vuln, avoid weak encryption, avoid weak hashing, clear text pass, clear text secret, clear text sess, clear-text cookie, clear-text cred, clear-text pass, clear-text secret, clear-text sess, cmd injection, code injection, collision attack, collision resist, command injection, critical patch, critical vuln, cross path injection, cross site scripting, cross-path injection, cross-site scripting, csrf, cve, cwe, data forgery, data validation, denial of service, deprecated hash, deprecate cipher, deprecate encryption, deprecate vuln, deprecated crypto, dns prefetch, double dealloc, double free, double-free, dynamic rege, dynamic require, enhance safety, eval injection, exploit, expression injection, file injection, fix deseriali, fix entropy, fix hash, fix key entropy, fix key length, fix key randomness, fix key size, fix key strength, fix overflow, fix randomness, fix safety, fix sensitive, fix secret, fix seriali, fix snyk, fix vuln, fix weak cipher, fix weak encryption, fix weak hash, fixed entropy, fixed nonce, fixed prng, fixed random, fixed seed, fixed-random, fortify, fragment injection, hard-coded secret, hard-coded key, hard-coded password, hard-coded token, hardcoded key, hardcoded password, hardcoded secret, hardcoded token, harden, header injection, heap corrupt, heap overflow, http response split, imap injection, improper authentication, improper authorization, improper validation, improve safety, input validation, insufficient entropy, ldap injection, log injection, meet-in-the-middle attack, mem align, memory abuse, memory alignment, memory corrup, memory misuse, module injection, non-constant rege, non-constant require, nvd, open redirect, origin validat, periodic random, permission bypass, permission escalation, permission exploit, permission injection, permission override, permission poisoning, plaintext cookie, plaintext cred, plaintext pass, plaintext secret, plaintext sess, pointer abuse, pointer misma, pointer misuse, pre-image attack, preimage attack, preimage resist, privilege elevation, privilege escalation, privilege exploit, random byte, random-byterce, remove deprecated encryption, remove deprecated hash, remove overflow, remove secret, security, sess fix, sess hijack, sess poison, sess repla, session fixation, session hijacking, session poisoning, session replay, shell injection, signal error, signal exploit, signal injection, smtp injection, sql injection, sqli, timing attack, type forgery, type injection, type validation, uaf, unauthori, unsafe, unsalted hash, untrusted content, untrusted data, untrusted deseriali, untrusted host, untrusted input, untrusted origin, untrusted source, untrusted user, unverified data, unverified host, unverified input, unverified origin, unverified source, unverified user, update hash, upgrade hash, use after dealloc, use after free, use-after-free, validate auth, variable require, vuln fix, vuln patch, vulnerability fix, vulnerability patch, xml external entit, xml injection, xpath injection, xquery injection, xss, xxe, zero day, zero-day}\\
      \bottomrule
    \end{tabular}%
  }
\end{table}

\subsection{\texttt{Themis-RM} Training Data Mixture And Filtering}
\label{subsec:Data_Comp}

\paragraph{The \texttt{Themis-GeneralPreference} collection:}

We curate \texttt{Themis-GeneralPreference} from a mixture of pre-existing preference datasets that pertain to LM helpfulness and harmlessness in natural language. We also repurpose a small number of relevance preferences from code retrieval datasets. Designed to teach RMs the nuances of general human preferences, this \texttt{110k+} sample dataset is composed as follows:

\begin{enumerate}[leftmargin=*]

\dataset
  {CodeR-Pile}
  {\deanon{https://huggingface.co/datasets/nebula2025/CodeR-Pile}}
  {nebula2025/CodeR-Pile}
  {41924}
  {We create preference pairs from the CodeR-Pile~\citep{DBLP:journals/corr/abs-2505-12697} code retrieval dataset. Specifically, we leverage the code augmentation, exemplar, integration, refinement, simplification, pseudo-code, tutorial, and web query subsets of the dataset. We curate the chosen response from code parsed in the positive document. Similarly, we parse the rejected response from a document chosen via Zipfian sampling over the top-10 ranked list of mined hard negatives.}
  {\criterialine{0}{0}{0}{0}{0}{1}{0}}
  {\langline{1}{1}{1}{1}{1}{1}{1}{1}{0}}

\dataset
  {Skywork-Preference}
  {\deanon{https://huggingface.co/datasets/Skywork/Skywork-Reward-Preference-80K-v0.2}}
  {Skywork/Skywork-Reward-Preference-80K-v0.2}
  {30146}
  {We source preference pairs from the Skywork Preference~\citep{DBLP:journals/corr/abs-2410-18451} dataset. Specifically, we select and further filter the WildGuard~\citep{DBLP:conf/nips/HanREJL00D24}, OffsetBias~\citep{DBLP:conf/emnlp/ParkJRKC24}, Magpie~\citep{DBLP:conf/iclr/XuJNDP0L25}, and HelpSteer2~\citep{DBLP:journals/corr/abs-2406-08673} subsets.}
  {\criterialine{0}{0}{0}{0}{0}{1}{1}}
  {\langline{0}{0}{0}{0}{0}{0}{0}{0}{1}}

\dataset
  {Tulu-IF}
  {\deanon{https://huggingface.co/datasets/allenai/tulu-3-pref-personas-instruction-following}}
  {allenai/tulu-3-pref-personas-instruction-following}
  {13705}
  {We source response pairs that follow and violate a pre-specified set of instructions present in the Tulu v3~\citep{DBLP:journals/corr/abs-2411-15124} instruction-tuning dataset.}
  {\criterialine{0}{0}{0}{0}{0}{1}{0}}
  {\langline{0}{0}{0}{0}{0}{0}{0}{0}{1}}

\dataset
  {H4-Stackexchange}
  {\deanon{https://huggingface.co/datasets/HuggingFaceH4/stack-exchange-preferences}}
  {HuggingFaceH4/stack-exchange-preferences}
  {12139}
  {We follow~\citet{DBLP:journals/corr/abs-2112-00861} and filter StackExchange technical forums, selecting highly voted accepted answers as the chosen response and well-formed but low-scoring answers as the rejected ones.}
  {\criterialine{0}{0}{0}{0}{0}{1}{0}}
  {\langline{0}{0}{0}{0}{0}{0}{0}{0}{1}}

\dataset
  {Arena-HumanPreference}
  {\deanon{https://huggingface.co/datasets/lmarena-ai/arena-human-preference-140k}}
  {lmarena-ai/arena-human-preference-140k}
  {7068}
  {Human voting validated preferences from filtered data dumps of model head-to-head contests on LMArena~\citep{DBLP:conf/icml/ChiangZ0ALLZ0JG24}.}
  {\criterialine{0}{0}{0}{0}{0}{1}{0}}
  {\langline{0}{0}{0}{0}{0}{0}{0}{0}{1}}

\dataset
  {Prometheus-Preference}
  {\deanon{https://huggingface.co/datasets/prometheus-eval/Preference-Collection}}
  {prometheus-eval/Preference-Collection}
  {2864}
  {We mine response pairs from instruction clusters corresponding to helpfulness and harmlessness preferences from the Prometheus Preference~\citep{DBLP:conf/iclr/KimS0JLLYSKTS24} data.}
  {\criterialine{0}{0}{0}{0}{0}{1}{1}}
  {\langline{0}{0}{0}{0}{0}{0}{0}{0}{1}}

\dataset
  {HelpSteer3}
  {\deanon{https://huggingface.co/datasets/nvidia/HelpSteer3}}
  {nvidia/HelpSteer3}
  {1452}
  {High-margin preference pairs from the HelpSteer3 Preference~\citep{DBLP:journals/corr/abs-2505-11475} data. Specifically, we incorporate the general and stem subsets.}
  {\criterialine{0}{0}{0}{0}{0}{1}{0}}
  {\langline{0}{0}{0}{0}{0}{0}{0}{0}{1}}

\dataset
  {Argilla-DPO}
  {\deanon{https://huggingface.co/datasets/argilla/distilabel-math-preference-dpo}}
  {argilla/distilabel-math-preference-dpo}
  {1042}
  {High-margin math answer preference pairs from the Argilla Preference data. We synthetically filter for pairs in which both the chosen and rejected responses converge on the same answer, thereby selecting for stylistic preferences that capture aspects beyond correctness.}
  {\criterialine{0}{0}{0}{0}{0}{1}{0}}
  {\langline{0}{0}{0}{0}{0}{0}{0}{0}{1}}

\dataset
  {Truthy-DPO}
  {\deanon{https://huggingface.co/datasets/jondurbin/truthy-dpo-v0.1}}
  {jondurbin/truthy-dpo-v0.1}
  {377}
  {Synthetically labeled truthfulness preferences mined from existing instruction-tuning data.}
  {\criterialine{0}{0}{0}{0}{0}{1}{0}}
  {\langline{0}{0}{0}{0}{0}{0}{0}{0}{1}}

\end{enumerate}

\paragraph{The \texttt{Themis-CodePreference} collection:}

We primarily curate \texttt{Themis-CodePreference} from code preference datasets we construct. Our collection centers on sourcing diverse preference scenarios from GitHub commits and from synthetically bugged instruction-tuning data. We further augment our data mixture with training sets from pre-existing code-preference and retrieval datasets. Primarily designed to teach RMs the nuances of scoring code along functional and non-functional axes, this \texttt{350k+} sample dataset is composed as follows:

\begin{enumerate}[leftmargin=*]

\dataset
  {Commit-Preference}
  {}
  {}
  {126586}
  {We source code preferences from non-reverted single-file GitHub commits verified to be part of subsequent successfully merged pull requests. Preference strength is validated via consensus between multiple frontier LMs, and pairs are selected for the presence of a single intent (e.g., improving runtime efficiency). A detailed description of the commit preference collection is outlined in \cref{sec:Code_Reward_Bench} and \cref{appdx:2_Data_Collection,appdx:3_Data_Prompts}.}
  {\criterialine{1}{1}{1}{1}{1}{0}{0}}
  {\langline{1}{1}{1}{1}{1}{1}{1}{1}{0}}

\dataset
  {CodeR-Pile}
  {\deanon{https://huggingface.co/datasets/nebula2025/CodeR-Pile}}
  {nebula2025/CodeR-Pile}
  {77153}
  {We create preference pairs from the CodeR-Pile~\citep{DBLP:journals/corr/abs-2505-12697} code retrieval dataset. Specifically, we leverage the code augmentation, exemplar, integration, refinement, simplification, pseudo-code, tutorial, and web query subsets of the dataset. We curate the chosen response from code parsed in the positive document. Similarly, we parse the rejected response from a document chosen via Zipfian sampling over the top-10 ranked list of mined hard negatives.}
  {\criterialine{1}{0}{0}{0}{0}{0}{0}}
  {\langline{1}{1}{1}{1}{1}{1}{1}{1}{0}}

\dataset
  {Bugged-Instruct}
  {}
  {}
  {54969}
  {We repurpose instruction-tuning datasets by creating rejected responses via model-based introduction of algorithmic or syntactic bugs. We incorporate data from a mix of diverse extant datasets, including OSS-Instruct~\citep{DBLP:conf/icml/0003W0D024}, Inverse-Instruct~\citep{DBLP:conf/aaai/WuHSWPGLNYZZDGY25}, McEval-Instruct~\citep{DBLP:conf/iclr/ChaiL0YJLS0RGWW25}, and Package-Instruct~\citep{DBLP:conf/acl/HuangCLXHSXYLZC25}. The prompt used is outlined in \cref{subsec:Prompt_Bugging}.}
  {\criterialine{1}{0}{0}{0}{0}{0}{0}}
  {\langline{1}{1}{1}{1}{1}{1}{1}{1}{0}}

\dataset
  {ProSec}
  {\deanon{https://huggingface.co/datasets/prosecalign/prosec-mixed-clm7b-inst}}
  {prosecalign/prosec-mixed-clm7b-inst}
  {37690}
  {We procure synthetically generated pairs of vulnerability fixes from ProSec~\citep{DBLP:conf/icml/XuSG0W025}.}
  {\criterialine{0}{0}{0}{0}{1}{0}{0}}
  {\langline{1}{0}{1}{0}{1}{1}{1}{0}{0}}

\dataset
  {Venus}
  {\deanon{https://huggingface.co/datasets/Elfsong/Venus}}
  {Elfsong/Venus}
  {21784}
  {We source runtime and memory usage code preferences from Venus~\citep{DBLP:journals/corr/abs-2505-23387}, selecting samples where the chosen completion is at least \texttt{5x} as efficient as the rejected one.}
  {\criterialine{0}{1}{1}{0}{0}{0}{0}}
  {\langline{0}{0}{0}{0}{0}{0}{1}{0}{0}}

\dataset
  {CodeNet}
  {\deanon{https://huggingface.co/datasets/iNeil77/CodeNet}}
  {iNeil77/CodeNet}
  {16819}
  {We source runtime and memory usage code preferences per submitter from CodeNet~\citep{DBLP:conf/nips/Puri0JZDZD0CDTB21}, selecting samples where the chosen completion is at least \texttt{5x} as efficient as the rejected one.}
  {\criterialine{0}{1}{1}{0}{0}{0}{0}}
  {\langline{1}{1}{1}{1}{1}{1}{1}{1}{0}}

\dataset
  {RunBugRun}
  {\deanon{https://huggingface.co/datasets/ASSERT-KTH/RunBugRun-Final}}
  {ASSERT-KTH/RunBugRun-Final}
  {6931}
  {We procure code contest program-repair pairs from the train set of RunBugRun~\citep{DBLP:journals/corr/abs-2304-01102}.}
  {\criterialine{1}{0}{0}{0}{0}{0}{0}}
  {\langline{0}{0}{0}{0}{0}{0}{1}{0}{0}}

\dataset
  {ECCO}
  {\deanon{https://huggingface.co/datasets/CodeEff/ECCO}}
  {CodeEff/ECCO}
  {6188}
  {We source code contest runtime preferences from the train set of ECCO~\citep{DBLP:conf/emnlp/WaghjaleVWF24}.}
  {\criterialine{0}{1}{0}{0}{0}{0}{0}}
  {\langline{0}{0}{0}{0}{0}{0}{1}{0}{0}}

\dataset
  {CodeScaleR}
  {\deanon{https://huggingface.co/datasets/LARK-Lab/CodeScalerPair-51K}}
  {LARK-Lab/CodeScalerPair-51K}
  {2896}
  {Model-generated correctness preference pairs procured from CodeScaleR~\citep{Zhu2026CodeScalerSC}. We obtain solutions to coding problems with a focus on difficult preference pairs where the rejected sample passes most but not all test-cases.}
  {\criterialine{1}{0}{0}{0}{0}{0}{0}}
  {\langline{0}{0}{0}{0}{0}{0}{1}{0}{0}}

\dataset
  {Pie4Perf}
  {\deanon{https://huggingface.co/datasets/iNeil77/pie4perf-Train}}
  {iNeil77/pie4perf-Train}
  {1640}
  {We source execution runtime preferences from the Pie4Perf~\citep{DBLP:conf/iclr/ShypulaMZ0GYHNR24} dataset. Specifically, we use the HQ-Train subset, which contains preference pairs derived from successful submissions by non-spam authors with highly divergent runtimes.}
  {\criterialine{0}{1}{0}{0}{0}{0}{0}}
  {\langline{0}{0}{0}{0}{0}{0}{0}{0}{0}}

\dataset
  {Cybernative-DPO}
  {\deanon{https://huggingface.co/datasets/CyberNative/Code_Vulnerability_Security_DPO}}
  {CyberNative/Code\_Vulnerability\_Security\_DPO}
  {1354}
  {We source security-fix preferences created via synthetic fixes to vulnerable code.}
  {\criterialine{0}{0}{0}{0}{1}{0}{0}}
  {\langline{0}{1}{1}{1}{1}{1}{1}{1}{0}}

\end{enumerate}

\paragraph{The \texttt{Themis-GeneralPreference} and \texttt{Themis-CodePreference} filtering procedure:}

We thoroughly clean and decontaminate our training data for the preference model pre-training (PT) and the preference modeling (PM) phases via the following steps:

\begin{enumerate}[leftmargin=*]

\item We first ensure that all samples in \texttt{Themis-GeneralPreference} and \texttt{Themis-CodePreference} are no longer than \texttt{2560} and \texttt{4096} tokens, respectively.\footnote{As measured by the \texttt{Themis-RM} tokenizer.} Subsequently, we filter out samples with trivial code responses whose syntax tree is shallower than \texttt{3} levels. Additionally, we ensure that all GitHub commit preference data we train on is sourced no later than \texttt{March 2019}.

\item We next leverage the GlotLID~\citep{DBLP:conf/emnlp/KargaranIYS23} language classifier to discard samples with non-English prompts, followed by filtering out samples with prompt perplexities greater than \texttt{1200}, as measured by a KenLM~\citep{DBLP:conf/wmt/Heafield11} model trained on the OSCAR EN corpus~\citep{DBLP:conf/lrec/AbadjiSRS22}.

\item Next, we run a dataset-level (i.e., \texttt{Themis-GeneralPreference} and \texttt{Themis-CodePreference} separately) near-deduplication step using a MinHash~\citep{DBLP:conf/sequences/Broder97} filter with a shingle size of \texttt{20} and a similarity threshold of \texttt{0.75}. Finally, following prior work~\citep{DBLP:conf/nips/BrownMRSKDNSSAA20, DBLP:conf/iclr/ElazarBMRSSWGS024}, we decontaminate our training data by removing any sample whose prompt registers a \texttt{13-gram} overlap with a prompt in \texttt{Themis-CodeRewardBench}, RewardBench V1~\citep{DBLP:conf/naacl/LambertPMMLCDKZCSH25}, RewardBench V2~\citep{DBLP:journals/corr/abs-2506-01937}, JudgeBench~\citep{DBLP:conf/iclr/TanZMTC0PS25} or RM-Bench~\citep{DBLP:conf/iclr/Liu0M00L25}.

\end{enumerate}

\section{Data Acquisition And Filtering Prompts}
\label{appdx:3_Data_Prompts}

\subsection{Commit Saliency Rubric}
\label{subsec:Commit_Saliency}

\begin{codelisting}[H]
\begin{codeframetop}[title={System Prompt}]
\begin{lstlisting}[style=termstyle]
You are an experienced Principal Software Engineer asked to critically review the code changes by a colleague tasked with improving the production readiness of a critical service's codebase.
\end{lstlisting}
\end{codeframetop}%
\nointerlineskip
\begin{codeframebottom}[title={User Prompt}]
\begin{lstlisting}[style=termstyle,escapeinside={(*}{*)}]
Code Change Review: (*\color{orange}{\{\{criteria\}\}}*)

This specific change edits a single source file written in (*\color{orange}{\{\{programming\_language\}\}}*). The faithfulness of the change description to the actual code changes is not guaranteed and should be verified by you via careful examination of the code.

[OLD_CODE](*\color{orange}{\{\{old\_file\_contents\}\}}*)[/OLD_CODE]

[NEW_CODE](*\color{orange}{\{\{new\_file\_contents\}\}}*)[/NEW_CODE]

[CHANGES](*\color{orange}{\{\{commit\_message\}\}}*)[/CHANGES]

Provide a factual summary of the specific functional and technical changes made by your colleague. You are free to use the provided file contents to support your summary, but ensure that you do not copy-paste your colleague's change description verbatim. This summary must be enclosed within [SUMMARY] and [/SUMMARY]. Subsequently, score the specific functional and technical changes made by your colleague and score the quality of the code changes on a scale of 1 to 5 (inclusive) based on whether they improve the code along the axis under consideration and whether the changes are specific to (*\color{orange}{\{\{criteria\}\}}*) within [RATING] and [/RATING] tags. The scoring rubric follows:

1. The change doesn't improve the code's (*\color{orange}{\{\{criteria\}\}}*) or degrades it overall. The change may or may not also contain unrelated edits that are not specific to (*\color{orange}{\{\{criteria\}\}}*). The change may also introduce other issues or bugs unrelated to (*\color{orange}{\{\{criteria\}\}}*).

2. The code change is unnecessary and does not have any discernible effect on the code's (*\color{orange}{\{\{criteria\}\}}*), but does not degrade its (*\color{orange}{\{\{criteria\}\}}*) either. The change may or may not also contain unrelated edits that are not specific to (*\color{orange}{\{\{criteria\}\}}*).

3. The code change makes the code slightly better with respect to (*\color{orange}{\{\{criteria\}\}}*) but largely leaves it the same. The change might also contain unnecessary edits unrelated to (*\color{orange}{\{\{criteria\}\}}*), but the majority of the changes are specific to (*\color{orange}{\{\{criteria\}\}}*).

4. The code change makes the code significantly better with respect to (*\color{orange}{\{\{criteria\}\}}*). Sporadic edits unrelated to (*\color{orange}{\{\{criteria\}\}}*) may exist, but the majority of the changes are specific to (*\color{orange}{\{\{criteria\}\}}*). The change does not introduce any new issues or bugs unrelated to (*\color{orange}{\{\{criteria\}\}}*).

5. The code change greatly improves the code's (*\color{orange}{\{\{criteria\}\}}*), making it a must-have feature or addition. The change is also well implemented and specific, i.e., not a generic suggestion that could apply to any codebase. The incidence of unnecessary edits that are unrelated to (*\color{orange}{\{\{criteria\}\}}*) is minimal or non-existent in the change. The change does not introduce any new issues or bugs unrelated to (*\color{orange}{\{\{criteria\}\}}*).
\end{lstlisting}
\end{codeframebottom}
\caption{The scoring rubric used to evaluate the criteria-level preference strength present in single-file GitHub commits. Refer to \cref{sec:Code_Reward_Bench,sec:Themis_RM} for more details on commit preference mining.}
\end{codelisting}

\subsection{Commit Inverse Instruction Generation}
\label{subsec:Commit_Inverse}

\begin{codelisting}[H]
\begin{codeframetop}[title={System Prompt}]
\begin{lstlisting}[style=termstyle]
You are a veteran problem-setter for a popular programming contest and coding interview preparation platform. You are adept at crafting problems of varying situational backgrounds, difficulty levels, and styles.
\end{lstlisting}
\end{codeframetop}%
\nointerlineskip
\begin{codeframebottom}[title={User Prompt}]
\begin{lstlisting}[style=termstyle,escapeinside={(*}{*)}]
(*\color{white!60!black}{\#\# The \{\{programming\_language\}\} and \{\{content\_style\}\} can be one of the following tuples
\begin{itemize}[leftmargin=*]
\item Graduate Course Assignment: Problems that resemble assignments given in a graduate-level computer science course in their level of detail, complexity, and structure.

\item Quora Question: Problems that are reflective of the kind of questions asked on the popular question-and-answer platform Quora in their style.

\item Stackoverflow Question: Problems that are reflective of the kind of questions asked on the popular question-and-answer platform Stackoverflow in their level of detail, complexity, structure, and style.

\item Google Search Query: Problems that are reflective of the kind of queries entered into the popular search engine Google in their level of detail, complexity, structure, and style.

\item Programming Contest Problem: Problems that are reflective of the kind of problems set in a programming contest in their level of detail, structure, and style.
\end{itemize}
}*)

You are given two snippets in (*\color{orange}{\{\{programming\_language\}\}}*) that answer a (*\color{orange}{\{\{problem\_style\}\}}*) (). These are specified between the [EXAMPLE1] and [\EXAMPLE1] and the [EXAMPLE2] and [\EXAMPLE2] tags below:

[EXAMPLE1](*\color{orange}{\{\{old\_file\_contents\}\}}*)[/EXAMPLE1]

[EXAMPLE2](*\color{orange}{\{\{new\_file\_contents\}\}}*)[/EXAMPLE2]

While we have the resulting solution code snippets, your first order of business is to inspect the reference solutions and detail what they accomplish in full. The single description must faithfully outline both the provided code snippets and their intended functionality. This description must be enclosed within [DESCRIPTION] and [/DESCRIPTION]

Secondly, you must craft a clear and concise problem statement that fulfils the following criteria: 

1. Is consistent in meaning with the provided reference solutions.

2. Would sufficiently identifiably lead a mid-tier to experienced developer to plausibly converge on either of the reference solutions (EXAMPLE1 or EXAMPLE2) with equal likelihood.

3. Resembles a (*\color{orange}{\{\{content\_style\}\}}*) in its level of detail, complexity, structure, and style. 

4. Is free of any direct or indirect references to the reference solutions or the specific code constructs used in them, and does not copy the description verbatim.

Provide the instruction you craft between [INSTRUCTION] and [\INSTRUCTION] tags.

\end{lstlisting}
\end{codeframebottom}
\caption{The inverse instruction creation prompts for crafting realistic queries for code change pairs mined from GitHub. Refer to \cref{sec:Code_Reward_Bench,sec:Themis_RM} for more details on commit preference mining.}
\end{codelisting}

\subsection{\texttt{Themis-RM Training And Evaluation System Prompt}}
\label{subsec:Themis_Principles}

\begin{codelisting}[H]
\begin{codeframe}[title={Themis-RM Evaluation Principles}]
\begin{lstlisting}[style=termstyle,escapeinside={(*}{*)}]
You are an experienced AI judge and evaluator, specifically designed and trained to assess the overall quality of responses to human prompts. The prompts can be of various types, including but not limited to open-ended questions, coding tasks, creative writing, and requests for information. The responses may be generated by a language model or a human and may include text, code, or both. You will receive the prompt and the generated response for each judgement session. You must score the response based on a pre-specified set of criteria. 

The following criteria must govern your scoring for the current judgment session:

(*\color{white!60!black}{\#\# General Principles}*)
1. **Helpfulness**: How well does the response address the prompt? Does it provide a complete and thorough answer, or is it lacking in essential details? This tenet applies to evaluating all types of responses. Helpful responses must score higher than unhelpful ones.

2. **Harmlessness**: Does the response contain any harmful or inappropriate content? This includes hate speech, violence, self-harm, or other harmful content. This tenet applies to evaluating all types of responses. You must score harmless responses higher than harmful ones.

(*\color{white!60!black}{\#\# Criteria-Specific Code-Only Principles }*)
3. **Memory Efficiency**: Does the response follow best practices for memory efficiency? Examples include using efficient data structures, minimizing memory usage, avoiding memory leaks, and effectively pooling/managing resources, among others. This tenet applies to evaluating code responses. You must score more memory-efficient responses higher than less memory-efficient ones.

4. **Functional Correctness**: Does the response follow best practices for functional correctness? Examples include algorithmic correctness, specifications adherence, and edge-case handling, among others. This tenet applies to evaluating code responses. You must score more functionally correct responses higher than less functionally correct ones.

5. **Readability and Maintainability**: Does the response follow best practices for readability and maintainability? Examples include using clear, descriptive names, following consistent formatting and style guidelines, modularizing code, and providing comments and documentation where necessary. This tenet applies to evaluating code responses. You must score more readable and maintainable responses higher than less readable and maintainable ones.

6. **Runtime Efficiency**: Does the response follow best practices for runtime efficiency? Examples include using efficient algorithms and data structures, minimizing time complexity, avoiding unnecessary computations, caching results, and leveraging parallel processing or asynchronous programming techniques where appropriate. This tenet applies to evaluating code responses. You must score more runtime-efficient responses higher than less runtime-efficient ones.

7. **Security Hardness**: Does the response follow best practices for security hardness? Examples include input validation, output encoding, proper error handling, and secure coding practices. This tenet applies to evaluating code responses. You must score more secure, less vulnerable responses higher than less secure, more vulnerable ones.
\end{lstlisting}
\end{codeframe}
\caption{The list of evaluation principles used for training and evaluating \texttt{Themis-RM}. Refer to \cref{subsec:Experiments_RQ2} for analysis on how specifying scoring criteria improves RMs.}
\end{codelisting}

\subsection{Synthetic Bug-Laden Solution Generation}
\label{subsec:Prompt_Bugging}

\begin{codelisting}[H]
\begin{codeframetop}[title={System Prompt}]
\begin{lstlisting}[style=termstyle,escapeinside={(*}{*)}]
You are an experienced (*\color{orange}{\{\{programming\_language\}\}}*) developer who also works part-time at a contest- and interview-preparation platform. You incorporate your knowledge of common coding patterns and best practices to suggest challenging coding problems that require a deep understanding of algorithms and data structures.
\end{lstlisting}
\end{codeframetop}%
\nointerlineskip
\begin{codeframebottom}[title={User Prompt}]
\begin{lstlisting}[style=termstyle,escapeinside={(*}{*)}]
You are tasked with creating a coding problem for a contest in (*\color{orange}{\{\{programming\_language\}\}}*). In the interest of raising the quality of the problems, you decide to make the problems involve an open-ended question (PROBLEM) with a buggy code snippet (BUGGY_CODE). The contestants will have to fix the bug and match the reference solution to answer the question successfully. To scale this problem-setting process, you decide to base your problems on pre-existing validated (PROBLEM, REFERENCE_SOLUTION) pairs and generate the buggy code (BUGGY_CODE) yourself. The buggy code must follow the following constraints:

1. It must be a modification of the reference solution that introduces only functional, logical, and algorithmic bugs. 

2. The introduction of small syntax and grammatical errors is also allowed. However, you must try to maintain the code's surface-level structure as much as possible.

3. The buggy code must not allude to the original problem or the reference solution in any way. The problem statement and the reference solution are provided to you as part of the input, but you must not use them in your output.

4. The buggy code must not allude to the introduced bugs in any way. Variables, functions, classes, and other identifiers should not be named in a way that suggests the presence of bugs. Similarly, the comments and documentation should not hint at the bugs.

5. The addition of new features or the removal of existing ones is out of scope for this task.

6. The introduction of security vulnerabilities, memory leaks, or other non-functional bugs is out of scope for this task.

Below is a validated (PROBLEM, REFERENCE_SOLUTION) pair that you can use to generate the buggy code snippet. The problem is enclosed between the tags [PROBLEM] and [\PROBLEM]. The reference solution is enclosed between the tags [REFERENCE_SOLUTION] and [\REFERENCE_SOLUTION].

[PROBLEM](*\color{orange}{\{\{problem\_description\}\}}*)[/PROBLEM]

[REFERENCE_SOLUTION](*\color{orange}{\{\{solution\_code\}\}}*)[/REFERENCE_SOLUTION]

Firstly, develop a buggy code snippet (BUGGY_CODE) that modifies the reference solution. The buggy code must follow the constraints mentioned above and be enclosed between the [BUGGY_CODE] and [\BUGGY_CODE] tags. Secondly, outline and explain the bugs that you have introduced in the buggy code snippet. The explanation must be enclosed between the tags [BUG_EXPLANATION] and [\BUG_EXPLANATION].
\end{lstlisting}
\end{codeframebottom}
\caption{The prompt for generating buggy solutions conditioned on pre-existing valid solutions. Such samples enable the creation of diverse pairs of functional correctness preferences, which we use to enhance the robustness of \texttt{Themis-RM} in \cref{sec:Themis_RM}.}
\end{codelisting}

\section{Detailed Results}
\label{appdx:4_Detailed_Results}

\subsection{Functional Correctness (FC) Dataset-Level Accuracy}
\label{subsec:Detailed_FC}

\def\arraystretch{1.12}
\begin{table}[ht]
\centering
\caption{Detailed dataset-level preference accuracy of extant RMs and the \texttt{Themis-RM} suite on the \texttt{Functional Correctness (FC)} split of \texttt{Themis-CodeRewardBench}. Observe the marked drop in performance of extant RMs at judging partially correct solutions (\texttt{MBPP+Fix (Hard)}) as well as on open-domain GitHub commit preferences (\texttt{CommitPref}). For a detailed discussion of comparative results and the split-level average, refer to \cref{subsec:Experiments_RQ1}. Refer to \cref{subsec:Experiments_RQ2} for experiments on how RMs trained on functional preferences fare when tested on non-functional preferences.}
\label{tab:FC_Results}
\scalebox{0.6}{%
\begin{tabular}{@{}c l l r c cccccc@{}}
\toprule
& \multirow{2}{*}{\texttt{\textbf{Model}}} &  & \multirow{2}{*}{\texttt{\textbf{Size}}} & \multicolumn{6}{c}{\textbf{\texttt{Functional Correctness (FC)}}} \\
\cmidrule(lr){5-10} 
&  & &  & \textbf{\texttt{HEPack}} & \textbf{\texttt{MBPP+Fix (Hard)}} & \textbf{\texttt{MDEval}} & \textbf{\texttt{DebugEval}} & \textbf{\texttt{RunBugRun}} & \textbf{\texttt{CommitPref}}  \\
\midrule
\multirow{13}{*}{\rotatebox[origin=c]{90}{\textbf{\texttt{XL}}}}
& \hfmodel{Qwen/Qwen2.5-Math-RM-72B}         & \iconMath & \texttt{72B}  & \mrbhl{98.41} & \mrbhl{81.08} & \mrbhl{83.58} & \mrbhl{83.70} & \mrbhl{79.30} & \mrbhl{63.52} \\
& \hfmodel{nvidia/AceMath-72B-RM}     & \iconMath & \texttt{72B} & \mrbhl{99.20} & \mrbhl{70.27} & \mrbhl{94.78} & \mrbhl{96.55} & \mrbhl{86.26} & \mrbhl{58.06} \\
& \hfmodel{Qwen/WorldPM-72B-RLHFLow}     &              & \texttt{72B}  & \mrbhl{98.73} & \mrbhl{54.05} & \mrbhl{93.28} & \mrbhl{96.27} & \mrbhl{85.39} & \mrbhl{69.45} \\
& \hfmodel{ContextualAI/LMUnit-qwen2.5-72b}         & \iconAuxiliary\;\iconGenerative\;\iconPrinciple\;\iconReasoning & \texttt{72B} & \mrbhl{84.24} & \mrbhl{40.54} & \mrbhl{63.43} & \mrbhl{76.38} & \mrbhl{24.28} & \mrbhl{11.64} \\
& \hfmodel{nvidia/Llama-3.3-Nemotron-70B-Reward}          &  & \texttt{70B}  & \mrbhl{98.57} & \mrbhl{56.76} & \mrbhl{94.03} & \mrbhl{97.93} & \mrbhl{88.11} & \mrbhl{75.76} \\
& \hfmodel{infly/INF-ORM-Llama3.1-70B}         & \iconGenerative   & \texttt{70B}  & \mrbhl{99.20} & \mrbhl{51.35} & \mrbhl{97.01} & \mrbhl{95.30} & \mrbhl{80.58} & \mrbhl{65.45} \\
& \hfmodel{allenai/Llama-3.1-70B-Instruct-RM-RB2}     &             & \texttt{70B}  & \mrbhl{99.20} & \mrbhl{43.34} & \mrbhl{93.28} & \mrbhl{96.55} & \mrbhl{84.81} & \mrbhl{72.24} \\
& \hfmodel{allenai/Llama-3.1-Tulu-3-70B-SFT-RM-RB2}     &  & \texttt{70B}  &\mrbhl{99.04} & \mrbhl{45.95} & \mrbhl{95.52} & \mrbhl{96.96} & \mrbhl{84.12} & \mrbhl{73.58} \\
& \hfmodel{Nexusflow/Athene-RM-70B}     &              & \texttt{70B}  & \mrbhl{98.89} & \mrbhl{56.76} & \mrbhl{95.52} & \mrbhl{95.99} & \mrbhl{85.02} & \mrbhl{79.88} \\
& \hfmodel{Nexusflow/Starling-RM-34B}         &  & \texttt{34B} & \mrbhl{96.02} & \mrbhl{21.62} & \mrbhl{91.04} & \mrbhl{93.78} & \mrbhl{73.21} & \mrbhl{63.52} \\
& \textbf{\texttt{Themis-RM-32B}}          & \iconAuxiliary\;\iconCode\;\iconPrinciple & \texttt{32B}  & \mrbhl{99.84} & \mrbhl{89.19} & \mrbhl{95.52} & \mrbhl{98.34} & \mrbhl{91.98} & \mrbhl{93.21} \\
& \hfmodel{TIGER-Lab/AceCodeRM-32B}         & \iconCode   & \texttt{32B}  & \mrbhl{97.93} & \mrbhl{64.86} & \mrbhl{90.30} & \mrbhl{95.03} & \mrbhl{70.62} & \mrbhl{43.15} \\
& \hfmodel{nvidia/Qwen3-Nemotron-32B-GenRM-Principle}     &   \iconGenerative\;\iconPrinciple           & \texttt{32B}  & \mrbhl{96.34} & \mrbhl{78.38} & \mrbhl{81.34} & \mrbhl{94.75} & \mrbhl{77.98} & \mrbhl{56.61} \\
\midrule
\multirow{7}{*}{\rotatebox[origin=c]{90}{\textbf{\texttt{L}}}}
& \hfmodel{nicolinho/QRM-Gemma-2-27B}         & \iconAuxiliary & \texttt{27B} & \mrbhl{64.01} & \mrbhl{56.76} & \mrbhl{60.25} & \mrbhl{61.19} & \mrbhl{50.53} & \mrbhl{48.48} \\
& \hfmodel{ShikaiChen/LDL-Reward-Gemma-2-27B-v0.1}     & \iconAuxiliary & \texttt{27B}  & \mrbhl{97.45} & \mrbhl{45.95} & \mrbhl{92.54} & \mrbhl{96.96} & \mrbhl{84.16} & \mrbhl{63.88} \\
& \hfmodel{Skywork/Skywork-Reward-Gemma-2-27B-v0.2}     &              & \texttt{27B}  & \mrbhl{68.31} & \mrbhl{78.38} & \mrbhl{65.67} & \mrbhl{65.47} & \mrbhl{52.59} & \mrbhl{50.06} \\
& \hfmodel{internlm/internlm2-20b-reward}         & \iconAuxiliary\;\iconPrinciple & \texttt{20B} & \mrbhl{97.93} & \mrbhl{35.14} & \mrbhl{90.30} & \mrbhl{94.47} & \mrbhl{78.40} & \mrbhl{66.75} \\
& \textbf{\texttt{Themis-RM-14B}}          & \iconAuxiliary\;\iconCode\;\iconPrinciple & \texttt{14B}  & \mrbhl{98.57} & \mrbhl{72.97} & \mrbhl{94.03} & \mrbhl{97.24} & \mrbhl{90.70} & \mrbhl{91.03} \\
& \hfmodel{rubricreward/R3-Qwen3-14B-14k}         & \iconGenerative\;\iconPrinciple\;\iconReasoning   & \texttt{14B}  & \mrbhl{87.10} & \mrbhl{56.76} & \mrbhl{66.42} & \mrbhl{73.48} & \mrbhl{53.37} & \mrbhl{27.84} \\
& \hfmodel{openbmb/UltraRM-13b}     &             & \texttt{13B}  & \mrbhl{88.85} & \mrbhl{16.22} & \mrbhl{91.79} & \mrbhl{83.29} & \mrbhl{67.94} & \mrbhl{73.09} \\
\midrule
\multirow{19}{*}{\rotatebox[origin=c]{90}{\textbf{\texttt{M}}}}
& \textbf{\texttt{Themis-RM-8B}}          & \iconAuxiliary\;\iconCode\;\iconPrinciple & \texttt{8B}  & \mrbhl{98.57} & \mrbhl{62.16} & \mrbhl{94.03} & \mrbhl{97.24} & \mrbhl{89.51} & \mrbhl{89.33} \\
& \hfmodel{LARK-Lab/CodeScaler-8B}     & \iconCode & \texttt{8B}  & \mrbhl{99.04} & \mrbhl{43.24} & \mrbhl{94.78} & \mrbhl{97.51} & \mrbhl{85.60} & \mrbhl{75.88} \\
& \hfmodel{rubricreward/R3-Qwen3-8B-14k}         & \iconGenerative\;\iconPrinciple\;\iconReasoning & \texttt{8B} & \mrbhl{82.80} & \mrbhl{43.24} & \mrbhl{56.72} & \mrbhl{72.79} & \mrbhl{48.15} & \mrbhl{26.55} \\
& \hfmodel{Skywork/Skywork-Reward-V2-Qwen3-8B}     &              & \texttt{8B}  & \mrbhl{98.89} & \mrbhl{37.84} & \mrbhl{94.78} & \mrbhl{97.51} & \mrbhl{84.86} & \mrbhl{77.45} \\
& \hfmodel{RLHFlow/ArmoRM-Llama3-8B-v0.1}         & \iconAuxiliary & \texttt{8B} & \mrbhl{97.61} & \mrbhl{45.95} & \mrbhl{85.07} & \mrbhl{93.92} & \mrbhl{76.83} & \mrbhl{62.06} \\
& \hfmodel{nicolinho/QRM-Llama3.1-8B-v2}     & \iconAuxiliary & \texttt{8B}  & \mrbhl{95.86} & \mrbhl{45.95} & \mrbhl{94.78} & \mrbhl{94.34} & \mrbhl{76.50} & \mrbhl{65.70} \\
& \hfmodel{allenai/Llama-3.1-8B-Base-RM-RB2}     &             & \texttt{8B}  & \mrbhl{97.77} & \mrbhl{32.43} & \mrbhl{97.01} & \mrbhl{94.89} & \mrbhl{78.68} & \mrbhl{71.52} \\
& \hfmodel{Ray2333/GRM-llama3-8B-sftreg}          & \iconAuxiliary & \texttt{8B} & \mrbhl{96.97} & \mrbhl{24.32} & \mrbhl{83.58} & \mrbhl{93.78} & \mrbhl{75.23} & \mrbhl{65.45} \\
& \hfmodel{LxzGordon/URM-LLaMa-3.1-8B}     & \iconAuxiliary & \texttt{8B}  & \mrbhl{95.86} & \mrbhl{48.65} & \mrbhl{94.78} & \mrbhl{95.30} & \mrbhl{75.84} & \mrbhl{65.21} \\
& \hfmodel{NCSOFT/Llama-3-OffsetBias-RM-8B}     &              & \texttt{8B}  & \mrbhl{96.82} & \mrbhl{40.54} & \mrbhl{91.79} & \mrbhl{92.68} & \mrbhl{76.54} & \mrbhl{64.97} \\
& \hfmodel{sfairXC/FsfairX-LLaMA3-RM-v0.1}         &  & \texttt{8B} & \mrbhl{97.45} & \mrbhl{32.43} & \mrbhl{91.04} & \mrbhl{94.34} & \mrbhl{77.12} & \mrbhl{68.85} \\
& \hfmodel{Nexusflow/Athene-RM-8B}     &  & \texttt{8B}  & \mrbhl{94.90} & \mrbhl{37.84} & \mrbhl{95.52} & \mrbhl{91.71} & \mrbhl{78.52} & \mrbhl{74.18} \\
& \hfmodel{TIGER-Lab/AceCodeRM-7B}     & \iconCode  & \texttt{7B}  & \mrbhl{96.18} & \mrbhl{59.46} & \mrbhl{91.79} & \mrbhl{95.86} & \mrbhl{80.70} & \mrbhl{65.09} \\
& \hfmodel{eth-dl-rewards/internlm2-7b-reward-code-100k}          & \iconCode & \texttt{7B} & \mrbhl{96.66} & \mrbhl{21.62} & \mrbhl{85.82} & \mrbhl{93.92} & \mrbhl{75.06} & \mrbhl{65.09} \\
& \hfmodel{eth-dl-rewards/internlm2-7b-reward-math-100k}     & \iconMath & \texttt{7B}  & \mrbhl{95.22} & \mrbhl{29.73} & \mrbhl{85.07} & \mrbhl{89.50} & \mrbhl{74.77} & \mrbhl{63.88} \\
& \hfmodel{reciprocate/mistral-7b-gsm8k-code-rm}        & \iconMath & \texttt{7B} & \mrbhl{88.22} & \mrbhl{29.73} & \mrbhl{85.82} & \mrbhl{88.26} & \mrbhl{66.91} & \mrbhl{58.18} \\
& \hfmodel{nvidia/AceMath-7B-RM}     & \iconMath             & \texttt{7B}  & \mrbhl{95.70} & \mrbhl{64.86} & \mrbhl{82.09} & \mrbhl{93.37} & \mrbhl{75.14} & \mrbhl{55.88} \\
& \hfmodel{openbmb/Eurus-RM-7b}     &  & \texttt{7B}  & \mrbhl{89.97} & \mrbhl{24.32} & \mrbhl{74.63} & \mrbhl{91.44} & \mrbhl{69.14} & \mrbhl{63.52} \\
& \hfmodel{internlm/internlm2-7b-reward}     &    \iconAuxiliary\;\iconPrinciple         & \texttt{7B}  & \mrbhl{94.59} & \mrbhl{32.43} & \mrbhl{87.21} & \mrbhl{92.13} & \mrbhl{76.09} & \mrbhl{66.67} \\
\midrule
\multirow{6}{*}{\rotatebox[origin=c]{90}{\textbf{\texttt{S}}}}
& \textbf{\texttt{Themis-RM-4B}}          & \iconAuxiliary\;\iconCode\;\iconPrinciple & \texttt{4B} &  \mrbhl{98.29} & \mrbhl{67.57} & \mrbhl{91.04} & \mrbhl{94.20} & \mrbhl{86.54} & \mrbhl{89.45} \\
& \hfmodel{LARK-Lab/CodeScaler-4B}     & \iconCode & \texttt{4B}  & \mrbhl{98.09} & \mrbhl{48.65} & \mrbhl{95.52} & \mrbhl{97.38} & \mrbhl{83.05} & \mrbhl{73.21} \\
& \hfmodel{PKU-ONELab/CE-RM-4B}     & \iconAuxiliary\;\iconGenerative\;\iconPrinciple\;\iconReasoning             & \texttt{4B}  & \mrbhl{91.08} & \mrbhl{51.35} & \mrbhl{75.37} & \mrbhl{86.74} & \mrbhl{61.69} & \mrbhl{44.24} \\
& \hfmodel{rubricreward/R3-Qwen3-4B-14k}         & \iconGenerative\;\iconPrinciple\;\iconReasoning & \texttt{4B} & \mrbhl{81.21} & \mrbhl{54.05} & \mrbhl{59.70} & \mrbhl{68.78} & \mrbhl{45.76} & \mrbhl{23.52} \\
& \hfmodel{Skywork/Skywork-Reward-V2-Qwen3-4B}     &  & \texttt{4B}  & \mrbhl{98.73} & \mrbhl{45.95} & \mrbhl{96.27} & \mrbhl{97.79} & \mrbhl{83.54} & \mrbhl{75.64} \\
& \hfmodel{Ray2333/GRM-llama3.2-3B-sftreg}     &  \iconAuxiliary           & \texttt{3B}  & \mrbhl{94.90} & \mrbhl{35.14} & \mrbhl{93.28} & \mrbhl{93.09} & \mrbhl{74.86} & \mrbhl{62.30} \\
\midrule
\multirow{4}{*}{\rotatebox[origin=c]{90}{\textbf{\texttt{XS}}}}
& \hfmodel{internlm/internlm2-1\_8b-reward}  & \iconAuxiliary\;\iconPrinciple & \texttt{1.8B}  & \mrbhl{81.69} & \mrbhl{21.62} & \mrbhl{79.10} & \mrbhl{72.38} & \mrbhl{62.80} & \mrbhl{63.88} \\
& \textbf{\texttt{Themis-RM-1.7B}}  & \iconAuxiliary\;\iconCode\;\iconPrinciple & \texttt{1.7B}   & \mrbhl{93.47} & \mrbhl{43.24} & \mrbhl{90.30} & \mrbhl{88.81} & \mrbhl{78.85} & \mrbhl{81.70} \\
& \hfmodel{LARK-Lab/CodeScaler-1.7B}  & \iconCode & \texttt{1.7B}  & \mrbhl{92.68} & \mrbhl{29.73} & \mrbhl{90.30} & \mrbhl{93.92} & \mrbhl{77.37} & \mrbhl{71.39} \\
& \hfmodel{Skywork/Skywork-Reward-V2-Qwen3-1.7B}  &  & \texttt{1.7B}   & \mrbhl{95.70} & \mrbhl{32.43} & \mrbhl{94.03} & \mrbhl{93.70} & \mrbhl{78.68} & \mrbhl{73.33} \\
\midrule
\multirow{2}{*}{\rotatebox[origin=c]{90}{\textbf{\texttt{XXS}}}}
& \textbf{\texttt{Themis-RM-0.6B}}   & \iconAuxiliary\;\iconCode\;\iconPrinciple            & \texttt{0.6B}  & \mrbhl{90.61} & \mrbhl{45.95} & \mrbhl{83.58} & \mrbhl{76.80} & \mrbhl{73.05} & \mrbhl{78.79} \\
&  \hfmodel{Skywork/Skywork-Reward-V2-Qwen3-0.6B}   &  & \texttt{0.6B}  & \mrbhl{93.95} & \mrbhl{35.14} & \mrbhl{89.55} & \mrbhl{92.13} & \mrbhl{73.25} & \mrbhl{70.42} \\
\bottomrule
\end{tabular}%
}
\scalebox{0.75}{\footnotesize
\iconAuxiliary\;\textbf{\texttt{Auxiliary Training Objectives}} \quad \iconCode\;\textbf{\texttt{Code RM}} \quad \iconPrinciple\;\textbf{\texttt{Criteria-Following RM}} \quad \iconMath\;\textbf{\texttt{Math RM}} \quad \iconGenerative\;\textbf{\texttt{Generative RM}} \quad \iconReasoning\;\textbf{\texttt{Reasoning RM}}}
\end{table}
\vfill
\pagebreak

\subsection{Execution Efficiency (EE) And Memory Efficiency (ME) Dataset-Level Accuracy}
\label{subsec:Detailed_EM}

\def\arraystretch{1.12}
\begin{table}[ht]
\centering
\caption{Detailed dataset-level preference accuracy of extant RMs and the \texttt{Themis-RM} suite on the \texttt{Execution Efficiency (EE)} and \texttt{Memory Efficiency (ME)} splits of \texttt{Themis-CodeRewardBench}. Observe how most extant scalar RMs degenerate to near-random performance on non-functional criteria and how the low-resolution scoring of generative RMs can render them unusable in such settings. For a detailed discussion of comparative results and the split-level averages, refer to \cref{subsec:Experiments_RQ1}. Refer to \cref{subsec:Experiments_RQ2} for experiments on how RMs trained on functional preferences fare when tested on non-functional preferences.}
\label{tab:EM_Results}
\scalebox{0.6}{%
\begin{tabular}{@{}c l l r c cccccc@{}}
\toprule
& \multirow{2}{*}{\texttt{\textbf{Model}}} &  & \multirow{2}{*}{\texttt{\textbf{Size}}} & \multicolumn{4}{c}{\textbf{\texttt{Execution Efficiency (EE)}}} & \multicolumn{2}{c}{\textbf{\texttt{Memory Efficiency (ME)}}} \\
\cmidrule(lr){5-8} \cmidrule(lr){9-10}
&  & &  & \textbf{\texttt{Pie4Perf}} & \textbf{\texttt{EvalPerf}} & \textbf{\texttt{ECCO}} & \textbf{\texttt{CommitPref}} & \textbf{\texttt{NoFunEval}} & \textbf{\texttt{CommitPref}}  \\
\midrule
\multirow{13}{*}{\rotatebox[origin=c]{90}{\textbf{\texttt{XL}}}}
& \hfmodel{Qwen/Qwen2.5-Math-RM-72B}         & \iconMath & \texttt{72B}  & \mrbhl{51.74} & \mrbhl{50.47} & \mrbhl{62.41} & \mrbhl{57.98} & \mrbhl{51.35} & \mrbhl{57.94} \\
& \hfmodel{nvidia/AceMath-72B-RM}     & \iconMath & \texttt{72B} & \mrbhl{60.43} & \mrbhl{64.62} & \mrbhl{62.16} & \mrbhl{60.92} & \mrbhl{51.35} & \mrbhl{53.57} \\
& \hfmodel{Qwen/WorldPM-72B-RLHFLow}     &              & \texttt{72B}  & \mrbhl{63.26} & \mrbhl{65.57} & \mrbhl{51.13} & \mrbhl{67.65} & \mrbhl{56.76} & \mrbhl{64.68} \\
& \hfmodel{ContextualAI/LMUnit-qwen2.5-72b}         & \iconAuxiliary\;\iconGenerative\;\iconPrinciple\;\iconReasoning & \texttt{72B} & \mrbhl{11.74} & \mrbhl{66.04} & \mrbhl{15.29} & \mrbhl{3.78} & \mrbhl{8.11} & \mrbhl{6.75} \\
& \hfmodel{nvidia/Llama-3.3-Nemotron-70B-Reward}          &  & \texttt{70B}  & \mrbhl{59.13} & \mrbhl{66.98} & \mrbhl{56.64} & \mrbhl{69.33} & \mrbhl{64.86} & \mrbhl{66.67} \\
& \hfmodel{infly/INF-ORM-Llama3.1-70B}         & \iconGenerative   & \texttt{70B}  & \mrbhl{60.00} & \mrbhl{70.75} & \mrbhl{59.40} & \mrbhl{62.61} & \mrbhl{51.35} & \mrbhl{64.29} \\
& \hfmodel{allenai/Llama-3.1-70B-Instruct-RM-RB2}     &             & \texttt{70B}  & \mrbhl{62.61} & \mrbhl{69.34} & \mrbhl{61.90} & \mrbhl{65.55} & \mrbhl{75.68} & \mrbhl{63.10} \\
& \hfmodel{allenai/Llama-3.1-Tulu-3-70B-SFT-RM-RB2}     &  & \texttt{70B}  & \mrbhl{61.74} & \mrbhl{67.92} & \mrbhl{62.91} & \mrbhl{73.11} & \mrbhl{56.76} & \mrbhl{68.65} \\
& \hfmodel{Nexusflow/Athene-RM-70B}     &              & \texttt{70B}  & \mrbhl{63.04} & \mrbhl{74.53} & \mrbhl{63.91} & \mrbhl{73.53} & \mrbhl{70.27} & \mrbhl{78.17} \\
& \hfmodel{Nexusflow/Starling-RM-34B}         &  & \texttt{34B} & \mrbhl{47.17} & \mrbhl{65.09} & \mrbhl{51.13} & \mrbhl{65.13} & \mrbhl{45.95} & \mrbhl{66.27} \\
& \textbf{\texttt{Themis-RM-32B}}          & \iconAuxiliary\;\iconCode\;\iconPrinciple & \texttt{32B}  & \mrbhl{82.83} & \mrbhl{86.32} & \mrbhl{83.46} & \mrbhl{98.34} & \mrbhl{89.19} & \mrbhl{96.03} \\
& \hfmodel{TIGER-Lab/AceCodeRM-32B}         & \iconCode   & \texttt{32B}  & \mrbhl{61.74} & \mrbhl{60.85} & \mrbhl{60.15} & \mrbhl{44.12} & \mrbhl{43.24} & \mrbhl{40.08} \\
& \hfmodel{nvidia/Qwen3-Nemotron-32B-GenRM-Principle}     &   \iconGenerative\;\iconPrinciple           & \texttt{32B}  & \mrbhl{64.57} & \mrbhl{74.06} & \mrbhl{68.17} & \mrbhl{62.18} & \mrbhl{59.46} & \mrbhl{69.44} \\
\midrule
\multirow{7}{*}{\rotatebox[origin=c]{90}{\textbf{\texttt{L}}}}
& \hfmodel{nicolinho/QRM-Gemma-2-27B}         & \iconAuxiliary & \texttt{27B} & \mrbhl{52.61} & \mrbhl{60.85} & \mrbhl{51.63} & \mrbhl{47.48} & \mrbhl{59.46} & \mrbhl{45.24} \\
& \hfmodel{ShikaiChen/LDL-Reward-Gemma-2-27B-v0.1}     & \iconAuxiliary & \texttt{27B}  & \mrbhl{62.61} & \mrbhl{73.58} & \mrbhl{65.16} & \mrbhl{64.29} & \mrbhl{64.86} & \mrbhl{62.30} \\
& \hfmodel{Skywork/Skywork-Reward-Gemma-2-27B-v0.2}     &              & \texttt{27B}  & \mrbhl{52.39} & \mrbhl{58.96} & \mrbhl{49.12} & \mrbhl{49.16} & \mrbhl{45.95} & \mrbhl{49.60} \\
& \hfmodel{internlm/internlm2-20b-reward}         & \iconAuxiliary\;\iconPrinciple & \texttt{20B} & \mrbhl{55.65} & \mrbhl{64.15} & \mrbhl{57.64} & \mrbhl{69.75} & \mrbhl{64.86} & \mrbhl{65.87} \\
& \textbf{\texttt{Themis-RM-14B}}          & \iconAuxiliary\;\iconCode\;\iconPrinciple & \texttt{14B}  & \mrbhl{85.00} & \mrbhl{85.85} & \mrbhl{85.46} & \mrbhl{89.50} & \mrbhl{91.89} & \mrbhl{94.84} \\
& \hfmodel{rubricreward/R3-Qwen3-14B-14k}         & \iconGenerative\;\iconPrinciple\;\iconReasoning   & \texttt{14B}  & \mrbhl{40.65} & \mrbhl{53.77} & \mrbhl{41.35} & \mrbhl{34.03} & \mrbhl{16.22} & \mrbhl{34.13} \\
& \hfmodel{openbmb/UltraRM-13b}     &             & \texttt{13B}  & \mrbhl{57.61} & \mrbhl{53.77} & \mrbhl{46.87} & \mrbhl{65.97} & \mrbhl{78.38} & \mrbhl{64.68} \\
\midrule
\multirow{19}{*}{\rotatebox[origin=c]{90}{\textbf{\texttt{M}}}}
& \textbf{\texttt{Themis-RM-8B}}          & \iconAuxiliary\;\iconCode\;\iconPrinciple & \texttt{8B}  & \mrbhl{82.17} & \mrbhl{82.55} & \mrbhl{84.21} & \mrbhl{86.55} & \mrbhl{81.08} & \mrbhl{93.65} \\
& \hfmodel{LARK-Lab/CodeScaler-8B}     & \iconCode & \texttt{8B}  & \mrbhl{53.48} & \mrbhl{72.17} & \mrbhl{59.90} & \mrbhl{68.91} & \mrbhl{62.16} & \mrbhl{73.81} \\
& \hfmodel{rubricreward/R3-Qwen3-8B-14k}         & \iconGenerative\;\iconPrinciple\;\iconReasoning & \texttt{8B} & \mrbhl{35.65} & \mrbhl{52.83} & \mrbhl{45.11} & \mrbhl{27.73} & \mrbhl{29.73} & \mrbhl{28.97} \\
& \hfmodel{Skywork/Skywork-Reward-V2-Qwen3-8B}     &              & \texttt{8B}  & \mrbhl{58.91} & \mrbhl{70.75} & \mrbhl{64.91} & \mrbhl{69.75} & \mrbhl{67.57} & \mrbhl{72.22} \\
& \hfmodel{RLHFlow/ArmoRM-Llama3-8B-v0.1}         & \iconAuxiliary & \texttt{8B} & \mrbhl{64.78} & \mrbhl{66.04} & \mrbhl{54.89} & \mrbhl{64.71} & \mrbhl{43.24} & \mrbhl{60.32} \\
& \hfmodel{nicolinho/QRM-Llama3.1-8B-v2}     & \iconAuxiliary & \texttt{8B}  & \mrbhl{53.04} & \mrbhl{65.57} & \mrbhl{59.15} & \mrbhl{68.49} & \mrbhl{64.86} & \mrbhl{67.06} \\
& \hfmodel{allenai/Llama-3.1-8B-Base-RM-RB2}     &             & \texttt{8B}  & \mrbhl{55.87} & \mrbhl{70.75} & \mrbhl{58.40} & \mrbhl{68.91} & \mrbhl{62.16} & \mrbhl{71.03} \\
& \hfmodel{Ray2333/GRM-llama3-8B-sftreg}          & \iconAuxiliary & \texttt{8B} & \mrbhl{63.26} & \mrbhl{53.30} & \mrbhl{52.13} & \mrbhl{65.97} & \mrbhl{56.76} & \mrbhl{57.54} \\
& \hfmodel{LxzGordon/URM-LLaMa-3.1-8B}     & \iconAuxiliary & \texttt{8B}  & \mrbhl{60.22} & \mrbhl{65.57} & \mrbhl{54.64} & \mrbhl{67.65} & \mrbhl{56.76} & \mrbhl{66.27} \\
& \hfmodel{NCSOFT/Llama-3-OffsetBias-RM-8B}     &              & \texttt{8B}  & \mrbhl{61.52} & \mrbhl{62.26} & \mrbhl{51.38} & \mrbhl{64.29} & \mrbhl{56.75} & \mrbhl{59.52} \\
& \hfmodel{sfairXC/FsfairX-LLaMA3-RM-v0.1}         &  & \texttt{8B} & \mrbhl{61.30} & \mrbhl{62.74} & \mrbhl{50.08} & \mrbhl{68.91} & \mrbhl{54.05} & \mrbhl{69.84} \\
& \hfmodel{Nexusflow/Athene-RM-8B}     &  & \texttt{8B}  & \mrbhl{56.30} & \mrbhl{73.11} & \mrbhl{59.65} & \mrbhl{73.53} & \mrbhl{59.46} & \mrbhl{71.83} \\
& \hfmodel{TIGER-Lab/AceCodeRM-7B}     & \iconCode  & \texttt{7B}  & \mrbhl{52.17} & \mrbhl{60.38} & \mrbhl{57.64} & \mrbhl{54.20} & \mrbhl{70.27} & \mrbhl{59.13} \\
& \hfmodel{eth-dl-rewards/internlm2-7b-reward-code-100k}          & \iconCode & \texttt{7B} & \mrbhl{57.83} & \mrbhl{58.02} & \mrbhl{50.88} & \mrbhl{59.66} & \mrbhl{64.86} & \mrbhl{58.33} \\
& \hfmodel{eth-dl-rewards/internlm2-7b-reward-math-100k}     & \iconMath & \texttt{7B}  & \mrbhl{55.65} & \mrbhl{59.43} & \mrbhl{48.87} & \mrbhl{66.39} & \mrbhl{51.35} & \mrbhl{58.73} \\
& \hfmodel{reciprocate/mistral-7b-gsm8k-code-rm}        & \iconMath & \texttt{7B} & \mrbhl{47.61} & \mrbhl{48.11} & \mrbhl{47.62} & \mrbhl{52.10} & \mrbhl{48.65} & \mrbhl{53.97} \\
& \hfmodel{nvidia/AceMath-7B-RM}     & \iconMath             & \texttt{7B}  & \mrbhl{65.65} & \mrbhl{60.38} & \mrbhl{56.14} & \mrbhl{51.26} & \mrbhl{45.95} & \mrbhl{54.76} \\
& \hfmodel{openbmb/Eurus-RM-7b}     &  & \texttt{7B}  & \mrbhl{54.57} & \mrbhl{50.94} & \mrbhl{51.63} & \mrbhl{64.71} & \mrbhl{62.16} & \mrbhl{57.54} \\
& \hfmodel{internlm/internlm2-7b-reward}     &      \iconAuxiliary\;\iconPrinciple       & \texttt{7B}  & \mrbhl{58.48} & \mrbhl{59.43} & \mrbhl{49.62} & \mrbhl{66.39} & \mrbhl{75.68} & \mrbhl{60.32} \\
\midrule
\multirow{6}{*}{\rotatebox[origin=c]{90}{\textbf{\texttt{S}}}}
& \textbf{\texttt{Themis-RM-4B}}          & \iconAuxiliary\;\iconCode\;\iconPrinciple & \texttt{4B} & \mrbhl{82.17} & \mrbhl{81.13} & \mrbhl{83.96} & \mrbhl{88.24} & \mrbhl{89.19} & \mrbhl{92.86} \\
& \hfmodel{LARK-Lab/CodeScaler-4B}     & \iconCode & \texttt{4B}  & \mrbhl{60.87} & \mrbhl{68.40} & \mrbhl{61.40} & \mrbhl{70.17} & \mrbhl{51.35} & \mrbhl{74.21} \\
& \hfmodel{PKU-ONELab/CE-RM-4B}     & \iconAuxiliary\;\iconGenerative\;\iconPrinciple\;\iconReasoning             & \texttt{4B}  & \mrbhl{58.91} & \mrbhl{50.47} & \mrbhl{54.14} & \mrbhl{40.76} & \mrbhl{35.14} & \mrbhl{39.68} \\
& \hfmodel{rubricreward/R3-Qwen3-4B-14k}         & \iconGenerative\;\iconPrinciple\;\iconReasoning & \texttt{4B} & \mrbhl{43.91} & \mrbhl{53.77} & \mrbhl{43.11} & \mrbhl{21.01} & \mrbhl{16.22} & \mrbhl{22.62} \\
& \hfmodel{Skywork/Skywork-Reward-V2-Qwen3-4B}     &  & \texttt{4B}  & \mrbhl{60.87} & \mrbhl{68.87} & \mrbhl{62.91} & \mrbhl{71.85} & \mrbhl{67.57} & \mrbhl{73.41} \\
& \hfmodel{Ray2333/GRM-llama3.2-3B-sftreg}     &  \iconAuxiliary           & \texttt{3B}  & \mrbhl{60.22} & \mrbhl{54.72} & \mrbhl{54.64} & \mrbhl{57.14} & \mrbhl{62.16} & \mrbhl{63.10} \\
\midrule
\multirow{4}{*}{\rotatebox[origin=c]{90}{\textbf{\texttt{XS}}}}
& \hfmodel{internlm/internlm2-1\_8b-reward}  & \iconAuxiliary\;\iconPrinciple & \texttt{1.8B}  & \mrbhl{53.04} & \mrbhl{57.08} & \mrbhl{45.11} & \mrbhl{60.08} & \mrbhl{51.35} & \mrbhl{63.89} \\
& \textbf{\texttt{Themis-RM-1.7B}}  & \iconAuxiliary\;\iconCode\;\iconPrinciple & \texttt{1.7B}   & \mrbhl{84.13} & \mrbhl{79.72} & \mrbhl{79.70} & \mrbhl{83.19} & \mrbhl{75.68} & \mrbhl{88.49} \\
& \hfmodel{LARK-Lab/CodeScaler-1.7B}  & \iconCode & \texttt{1.7B}  & \mrbhl{58.48} & \mrbhl{65.09} & \mrbhl{59.65} & \mrbhl{68.07} & \mrbhl{59.46} & \mrbhl{71.43} \\
& \hfmodel{Skywork/Skywork-Reward-V2-Qwen3-1.7B}  &  & \texttt{1.7B}   & \mrbhl{60.43} & \mrbhl{65.57} & \mrbhl{64.66} & \mrbhl{69.33} & \mrbhl{54.05} & \mrbhl{71.43} \\
\midrule
\multirow{2}{*}{\rotatebox[origin=c]{90}{\textbf{\texttt{XXS}}}}
& \textbf{\texttt{Themis-RM-0.6B}}   & \iconAuxiliary\;\iconCode\;\iconPrinciple            & \texttt{0.6B}  & \mrbhl{82.39} & \mrbhl{69.81} & \mrbhl{82.46} & \mrbhl{80.25} & \mrbhl{70.27} & \mrbhl{89.29} \\
&  \hfmodel{Skywork/Skywork-Reward-V2-Qwen3-0.6B}   &  & \texttt{0.6B}  & \mrbhl{60.65} & \mrbhl{64.62} & \mrbhl{61.65} & \mrbhl{67.23} & \mrbhl{40.54} & \mrbhl{66.27} \\
\bottomrule
\end{tabular}%
}
\scalebox{0.75}{\footnotesize
\iconAuxiliary\;\textbf{\texttt{Auxiliary Training Objectives}} \quad \iconCode\;\textbf{\texttt{Code RM}} \quad \iconPrinciple\;\textbf{\texttt{Criteria-Following RM}} \quad \iconMath\;\textbf{\texttt{Math RM}} \quad \iconGenerative\;\textbf{\texttt{Generative RM}} \quad \iconReasoning\;\textbf{\texttt{Reasoning RM}}}
\end{table}
\vfill
\pagebreak

\subsection{Readability And Maintainability (R\&M) And Security Hardness (SH) Dataset-Level Accuracy}
\label{subsec:Detailed_RS}

\def\arraystretch{1.12}
\begin{table}[ht]
\centering
\caption{Detailed dataset-level preference accuracy of extant RMs and the \texttt{Themis-RM} suite on the \texttt{Readability And Maintainability (R\&M)} and \texttt{Security Hardness (SH)} splits of \texttt{Themis-CodeRewardBench}. Observe how most extant scalar RMs degenerate to near-random performance on non-functional criteria and how the low-resolution scoring of generative RMs can render them unusable in such settings. For a detailed discussion of comparative results and the split-level averages, refer to \cref{subsec:Experiments_RQ1}. Refer to \cref{subsec:Experiments_RQ2} for experiments on how RMs trained on functional preferences fare when tested on non-functional preferences.}
\label{tab:RS_Results}
\scalebox{0.6}{%
\begin{tabular}{@{}c l l r c ccccccc@{}}
\toprule
& \multirow{2}{*}{\texttt{\textbf{Model}}} &  & \multirow{2}{*}{\texttt{\textbf{Size}}} & \multicolumn{2}{c}{\textbf{\texttt{Maintainability (R\&M)}}} & \multicolumn{5}{c}{\textbf{\texttt{Security Hardness (SH)}}} \\
\cmidrule(lr){5-6} \cmidrule(lr){7-11}
&  & &  & \textbf{\texttt{NoFunEval}} & \textbf{\texttt{CommitPref}} & \textbf{\texttt{CodePrefBench}} & \textbf{\texttt{Vul4J}} & \textbf{\texttt{SecBench}} & \textbf{\texttt{NoFunEval}}  & \textbf{\texttt{CommitPref}} \\
\midrule
\multirow{13}{*}{\rotatebox[origin=c]{90}{\textbf{\texttt{XL}}}}
& \hfmodel{Qwen/Qwen2.5-Math-RM-72B}         & \iconMath & \texttt{72B}  & \mrbhl{46.09} & \mrbhl{58.50} & \mrbhl{49.71} & \mrbhl{62.50} & \mrbhl{64.29} & \mrbhl{77.78} & \mrbhl{55.53} \\
& \hfmodel{nvidia/AceMath-72B-RM}     & \iconMath & \texttt{72B} & \mrbhl{50.00} & \mrbhl{60.10} & \mrbhl{52.02} & \mrbhl{75.00} & \mrbhl{57.14} & \mrbhl{66.67} & \mrbhl{56.57} \\
& \hfmodel{Qwen/WorldPM-72B-RLHFLow}     &              & \texttt{72B}  & \mrbhl{59.38} & \mrbhl{71.41} & \mrbhl{66.47} & \mrbhl{37.50} & \mrbhl{92.86} & \mrbhl{88.89} & \mrbhl{68.14} \\
& \hfmodel{ContextualAI/LMUnit-qwen2.5-72b}         & \iconAuxiliary\;\iconGenerative\;\iconPrinciple\;\iconReasoning & \texttt{72B} & \mrbhl{6.25} & \mrbhl{12.33} & \mrbhl{23.12} & \mrbhl{12.50} & \mrbhl{28.57} & \mrbhl{25.93} & \mrbhl{11.96} \\
& \hfmodel{nvidia/Llama-3.3-Nemotron-70B-Reward}          &  & \texttt{70B}  & \mrbhl{50.78} & \mrbhl{67.18} & \mrbhl{60.12} & \mrbhl{87.50} & \mrbhl{71.43} & \mrbhl{92.59} & \mrbhl{75.42} \\
& \hfmodel{infly/INF-ORM-Llama3.1-70B}         & \iconGenerative   & \texttt{70B}  & \mrbhl{45.31} & \mrbhl{70.31} & \mrbhl{63.58} & \mrbhl{37.50} & \mrbhl{78.57} & \mrbhl{88.89} & \mrbhl{66.58} \\
& \hfmodel{allenai/Llama-3.1-70B-Instruct-RM-RB2}     &             & \texttt{70B}  & \mrbhl{56.25} & \mrbhl{72.65} & \mrbhl{71.68} & \mrbhl{62.50} & \mrbhl{78.57} & \mrbhl{92.59} & \mrbhl{72.56} \\
& \hfmodel{allenai/Llama-3.1-Tulu-3-70B-SFT-RM-RB2}     &  & \texttt{70B}  & \mrbhl{52.34} & \mrbhl{75.57} & \mrbhl{73.41} & \mrbhl{62.50} & \mrbhl{64.29} & \mrbhl{88.89} & \mrbhl{73.60} \\
& \hfmodel{Nexusflow/Athene-RM-70B}     &              & \texttt{70B}  & \mrbhl{53.91} & \mrbhl{77.46} & \mrbhl{80.92} & \mrbhl{50.00} & \mrbhl{85.71} & \mrbhl{92.59} & \mrbhl{77.37} \\
& \hfmodel{Nexusflow/Starling-RM-34B}         &  & \texttt{34B} & \mrbhl{48.44} & \mrbhl{66.08} & \mrbhl{69.36} & \mrbhl{75.00} & \mrbhl{28.57} & \mrbhl{81.48} & \mrbhl{68.79} \\
& \textbf{\texttt{Themis-RM-32B}}          & \iconAuxiliary\;\iconCode\;\iconPrinciple & \texttt{32B}  & \mrbhl{59.38} & \mrbhl{90.23} & \mrbhl{95.95} & \mrbhl{100.00} & \mrbhl{92.86} & \mrbhl{100.00} & \mrbhl{94.02} \\
& \hfmodel{TIGER-Lab/AceCodeRM-32B}         & \iconCode   & \texttt{32B}  & \mrbhl{29.69} & \mrbhl{47.92} & \mrbhl{52.02} & \mrbhl{37.50} & \mrbhl{50.00} & \mrbhl{88.89} & \mrbhl{47.72} \\
& \hfmodel{nvidia/Qwen3-Nemotron-32B-GenRM-Principle}     &   \iconGenerative\;\iconPrinciple           & \texttt{32B}  & \mrbhl{49.22} & \mrbhl{57.91} & \mrbhl{77.46} & \mrbhl{62.50} & \mrbhl{64.29} & \mrbhl{81.48} & \mrbhl{60.47} \\
\midrule
\multirow{7}{*}{\rotatebox[origin=c]{90}{\textbf{\texttt{L}}}}
& \hfmodel{nicolinho/QRM-Gemma-2-27B}         & \iconAuxiliary & \texttt{27B} & \mrbhl{49.22} & \mrbhl{50.69} & \mrbhl{54.34} & \mrbhl{62.50} & \mrbhl{42.86} & \mrbhl{44.44} & \mrbhl{49.02} \\
& \hfmodel{ShikaiChen/LDL-Reward-Gemma-2-27B-v0.1}     & \iconAuxiliary & \texttt{27B}  & \mrbhl{44.53} & \mrbhl{69.44} & \mrbhl{46.82} & \mrbhl{50.00} & \mrbhl{71.43} & \mrbhl{81.48} & \mrbhl{66.19} \\
& \hfmodel{Skywork/Skywork-Reward-Gemma-2-27B-v0.2}     &              & \texttt{27B}  & \mrbhl{55.47} & \mrbhl{50.33} & \mrbhl{59.54} & \mrbhl{12.50} & \mrbhl{50.00} & \mrbhl{44.44} & \mrbhl{50.20} \\
& \hfmodel{internlm/internlm2-20b-reward}         & \iconAuxiliary\;\iconPrinciple & \texttt{20B} & \mrbhl{42.97} & \mrbhl{66.08} & \mrbhl{67.63} & \mrbhl{62.50} & \mrbhl{78.57} & \mrbhl{66.67} & \mrbhl{65.76} \\
& \textbf{\texttt{Themis-RM-14B}}          & \iconAuxiliary\;\iconCode\;\iconPrinciple & \texttt{14B}  & \mrbhl{60.94} & \mrbhl{89.79} & \mrbhl{96.53} & \mrbhl{87.50} & \mrbhl{92.86} & \mrbhl{96.30} & \mrbhl{95.19} \\
& \hfmodel{rubricreward/R3-Qwen3-14B-14k}         & \iconGenerative\;\iconPrinciple\;\iconReasoning   & \texttt{14B}  & \mrbhl{22.66} & \mrbhl{25.16} & \mrbhl{25.43} & \mrbhl{50.00} & \mrbhl{28.57} & \mrbhl{44.44} & \mrbhl{30.56} \\
& \hfmodel{openbmb/UltraRM-13b}     &             & \texttt{13B}  & \mrbhl{49.22} & \mrbhl{73.23} & \mrbhl{72.25} & \mrbhl{50.00} & \mrbhl{64.29} & \mrbhl{81.48} & \mrbhl{72.56} \\
\midrule
\multirow{19}{*}{\rotatebox[origin=c]{90}{\textbf{\texttt{M}}}}
& \textbf{\texttt{Themis-RM-8B}}          & \iconAuxiliary\;\iconCode\;\iconPrinciple & \texttt{8B}  & \mrbhl{60.94} & \mrbhl{88.40} & \mrbhl{94.80} & \mrbhl{100.00} & \mrbhl{92.86} & \mrbhl{100.00} & \mrbhl{92.72} \\
& \hfmodel{LARK-Lab/CodeScaler-8B}     & \iconCode & \texttt{8B}  & \mrbhl{57.03} & \mrbhl{74.98} & \mrbhl{63.01} & \mrbhl{100.00} & \mrbhl{85.71} & \mrbhl{92.59} & \mrbhl{74.38} \\
& \hfmodel{rubricreward/R3-Qwen3-8B-14k}         & \iconGenerative\;\iconPrinciple\;\iconReasoning & \texttt{8B} & \mrbhl{27.34} & \mrbhl{27.21} & \mrbhl{27.17} & \mrbhl{12.50} & \mrbhl{35.71} & \mrbhl{48.15} & \mrbhl{29.26} \\
& \hfmodel{Skywork/Skywork-Reward-V2-Qwen3-8B}     &              & \texttt{8B}  & \mrbhl{54.69} & \mrbhl{76.95} & \mrbhl{65.90} & \mrbhl{87.50} & \mrbhl{78.57} & \mrbhl{92.59} & \mrbhl{76.20} \\
& \hfmodel{RLHFlow/ArmoRM-Llama3-8B-v0.1}         & \iconAuxiliary & \texttt{8B} & \mrbhl{55.47} & \mrbhl{65.57} & \mrbhl{66.47} & \mrbhl{37.50} & \mrbhl{85.71} & \mrbhl{85.19} & \mrbhl{59.43} \\
& \hfmodel{nicolinho/QRM-Llama3.1-8B-v2}     & \iconAuxiliary & \texttt{8B}  & \mrbhl{49.22} & \mrbhl{64.26} & \mrbhl{62.43} & \mrbhl{0.00} & \mrbhl{57.14} & \mrbhl{85.19} & \mrbhl{62.29} \\
& \hfmodel{allenai/Llama-3.1-8B-Base-RM-RB2}     &             & \texttt{8B}  & \mrbhl{52.34} & \mrbhl{72.50} & \mrbhl{71.68} & \mrbhl{62.50} & \mrbhl{71.43} & \mrbhl{81.48} & \mrbhl{70.87} \\
& \hfmodel{Ray2333/GRM-llama3-8B-sftreg}          & \iconAuxiliary & \texttt{8B} & \mrbhl{51.56} & \mrbhl{67.47} & \mrbhl{72.83} & \mrbhl{25.00} & \mrbhl{57.14} & \mrbhl{85.19} & \mrbhl{65.67} \\
& \hfmodel{LxzGordon/URM-LLaMa-3.1-8B}     & \iconAuxiliary & \texttt{8B}  & \mrbhl{45.31} & \mrbhl{63.89} & \mrbhl{56.07} & \mrbhl{37.50} & \mrbhl{57.14} & \mrbhl{85.19} & \mrbhl{62.94} \\
& \hfmodel{NCSOFT/Llama-3-OffsetBias-RM-8B}     &              & \texttt{8B}  & \mrbhl{46.09} & \mrbhl{65.57} & \mrbhl{64.16} & \mrbhl{50.00} & \mrbhl{78.57} & \mrbhl{92.59} & \mrbhl{64.50} \\
& \hfmodel{sfairXC/FsfairX-LLaMA3-RM-v0.1}         &  & \texttt{8B} & \mrbhl{50.00} & \mrbhl{68.93} & \mrbhl{65.90} & \mrbhl{50.00} & \mrbhl{85.71} & \mrbhl{81.48} & \mrbhl{66.06} \\
& \hfmodel{Nexusflow/Athene-RM-8B}     &  & \texttt{8B}  & \mrbhl{57.03} & \mrbhl{74.69} & \mrbhl{73.41} & \mrbhl{50.00} & \mrbhl{71.43} & \mrbhl{92.59} & \mrbhl{74.77} \\
& \hfmodel{TIGER-Lab/AceCodeRM-7B}     & \iconCode  & \texttt{7B}  & \mrbhl{52.34} & \mrbhl{60.10} & \mrbhl{46.24} & \mrbhl{50.00} & \mrbhl{64.29} & \mrbhl{81.48} & \mrbhl{59.30} \\
& \hfmodel{eth-dl-rewards/internlm2-7b-reward-code-100k}          & \iconCode & \texttt{7B} & \mrbhl{45.31} & \mrbhl{66.74} & \mrbhl{56.07} & \mrbhl{50.00} & \mrbhl{87.57} & \mrbhl{62.96} & \mrbhl{61.25} \\
& \hfmodel{eth-dl-rewards/internlm2-7b-reward-math-100k}     & \iconMath & \texttt{7B}  & \mrbhl{42.97} & \mrbhl{68.34} & \mrbhl{60.69} & \mrbhl{87.50} & \mrbhl{76.57} & \mrbhl{55.56} & \mrbhl{63.20} \\
& \hfmodel{reciprocate/mistral-7b-gsm8k-code-rm}        & \iconMath & \texttt{7B} & \mrbhl{53.91} & \mrbhl{59.37} & \mrbhl{50.87} & \mrbhl{87.50} & \mrbhl{71.43} & \mrbhl{66.67} & \mrbhl{54.49} \\
& \hfmodel{nvidia/AceMath-7B-RM}     & \iconMath             & \texttt{7B}  & \mrbhl{43.75} & \mrbhl{54.12} & \mrbhl{60.12} & \mrbhl{50.00} & \mrbhl{85.71} & \mrbhl{62.96} & \mrbhl{53.71} \\
& \hfmodel{openbmb/Eurus-RM-7b}     &  & \texttt{7B}  & \mrbhl{54.69} & \mrbhl{62.07} & \mrbhl{56.07} & \mrbhl{75.00} & \mrbhl{64.29} & \mrbhl{77.78} & \mrbhl{63.98} \\
& \hfmodel{internlm/internlm2-7b-reward}     &     \iconAuxiliary\;\iconPrinciple        & \texttt{7B}  & \mrbhl{44.53} & \mrbhl{69.51} & \mrbhl{63.58} & \mrbhl{62.50} & \mrbhl{85.71} & \mrbhl{77.78} & \mrbhl{62.68} \\
\midrule
\multirow{6}{*}{\rotatebox[origin=c]{90}{\textbf{\texttt{S}}}}
& \textbf{\texttt{Themis-RM-4B}}          & \iconAuxiliary\;\iconCode\;\iconPrinciple & \texttt{4B} &  \mrbhl{61.72} & \mrbhl{86.65} & \mrbhl{95.38} & \mrbhl{100.00} & \mrbhl{85.71} & \mrbhl{92.59} & \mrbhl{92.46} \\
& \hfmodel{LARK-Lab/CodeScaler-4B}     & \iconCode & \texttt{4B}  & \mrbhl{50.00} & \mrbhl{74.11} & \mrbhl{68.21} & \mrbhl{50.00} & \mrbhl{92.86} & \mrbhl{88.89} & \mrbhl{70.48} \\
& \hfmodel{PKU-ONELab/CE-RM-4B}     & \iconAuxiliary\;\iconGenerative\;\iconPrinciple\;\iconReasoning             & \texttt{4B}  & \mrbhl{38.28} & \mrbhl{44.71} & \mrbhl{38.15} & \mrbhl{37.50} & \mrbhl{42.86} & \mrbhl{70.37} & \mrbhl{42.13} \\
& \hfmodel{rubricreward/R3-Qwen3-4B-14k}         & \iconGenerative\;\iconPrinciple\;\iconReasoning & \texttt{4B} & \mrbhl{17.19} & \mrbhl{24.58} & \mrbhl{26.59} & \mrbhl{25.00} & \mrbhl{7.14} & \mrbhl{37.04} & \mrbhl{27.44} \\
& \hfmodel{Skywork/Skywork-Reward-V2-Qwen3-4B}     &  & \texttt{4B}  & \mrbhl{51.56} & \mrbhl{76.51} & \mrbhl{71.68} & \mrbhl{37.50} & \mrbhl{78.57} & \mrbhl{88.89} & \mrbhl{73.47} \\
& \hfmodel{Ray2333/GRM-llama3.2-3B-sftreg}     &  \iconAuxiliary           & \texttt{3B}  & \mrbhl{49.22} & \mrbhl{67.54} & \mrbhl{66.47} & \mrbhl{37.50} & \mrbhl{78.57} & \mrbhl{81.48} & \mrbhl{66.32} \\
\midrule
\multirow{4}{*}{\rotatebox[origin=c]{90}{\textbf{\texttt{XS}}}}
& \hfmodel{internlm/internlm2-1\_8b-reward}  & \iconAuxiliary\;\iconPrinciple & \texttt{1.8B}  & \mrbhl{67.19} & \mrbhl{64.48} & \mrbhl{55.49} & \mrbhl{62.50} & \mrbhl{50.00} & \mrbhl{74.07} & \mrbhl{63.72} \\
& \textbf{\texttt{Themis-RM-1.7B}}  & \iconAuxiliary\;\iconCode\;\iconPrinciple & \texttt{1.7B}   & \mrbhl{55.47} & \mrbhl{82.35} & \mrbhl{94.80} & \mrbhl{62.50} & \mrbhl{78.57} & \mrbhl{92.59} & \mrbhl{88.04} \\
& \hfmodel{LARK-Lab/CodeScaler-1.7B}  & \iconCode & \texttt{1.7B}  & \mrbhl{45.31} & \mrbhl{69.44} & \mrbhl{86.21} & \mrbhl{37.50} & \mrbhl{71.43} & \mrbhl{77.78} & \mrbhl{65.67} \\
& \hfmodel{Skywork/Skywork-Reward-V2-Qwen3-1.7B}  &  & \texttt{1.7B}   & \mrbhl{47.66} & \mrbhl{70.75} & \mrbhl{73.41} & \mrbhl{50.00} & \mrbhl{71.43} & \mrbhl{81.48} & \mrbhl{69.44} \\
\midrule
\multirow{2}{*}{\rotatebox[origin=c]{90}{\textbf{\texttt{XXS}}}}
& \textbf{\texttt{Themis-RM-0.6B}}   & \iconAuxiliary\;\iconCode\;\iconPrinciple            & \texttt{0.6B}  & \mrbhl{65.62} & \mrbhl{80.01} & \mrbhl{86.71} & \mrbhl{100.00} & \mrbhl{85.71} & \mrbhl{92.59} & \mrbhl{87.65} \\
&  \hfmodel{Skywork/Skywork-Reward-V2-Qwen3-0.6B}   &  & \texttt{0.6B}  & \mrbhl{48.44} & \mrbhl{71.41} & \mrbhl{64.16} & \mrbhl{62.50} & \mrbhl{64.29} & \mrbhl{66.32} & \mrbhl{66.32} \\
\bottomrule
\end{tabular}%
}
\scalebox{0.75}{\footnotesize
\iconAuxiliary\;\textbf{\texttt{Auxiliary Training Objectives}} \quad \iconCode\;\textbf{\texttt{Code RM}} \quad \iconPrinciple\;\textbf{\texttt{Criteria-Following RM}} \quad \iconMath\;\textbf{\texttt{Math RM}} \quad \iconGenerative\;\textbf{\texttt{Generative RM}} \quad \iconReasoning\;\textbf{\texttt{Reasoning RM}}}
\end{table}

\clearpage
\subsection{General-Domain Reward Modeling Performance}
\label{subsec:External_Evals}

\def\arraystretch{1.12}
\begin{table}[ht]
\centering
\caption{External general-domain reward modeling evaluation of the \texttt{Themis-RM} suite on RewardBench V1~\citep{DBLP:conf/naacl/LambertPMMLCDKZCSH25}, RewardBench V2~\citep{DBLP:journals/corr/abs-2506-01937}, and JudgeBench~\citep{DBLP:conf/iclr/TanZMTC0PS25}. We benchmark against a selection of the highest-scoring extant RMs on \texttt{Themis-CodeRewardBench}.}
\label{tab:External_Evals}
\scalebox{0.6}{%
\begin{tabular}{@{}c l l r ccc@{}}
\toprule
& \multirow{2}{*}{\texttt{\textbf{Model}}} & & \multirow{2}{*}{\texttt{\textbf{Size}}} & \multicolumn{3}{c}{\textbf{\texttt{External Benchmark Accuracy}}} \\
\cmidrule(lr){5-7}
&  &  &  & \textbf{\texttt{RewardBench V1}} & \textbf{\texttt{RewardBench V2}} & \textbf{\texttt{JudgeBench}} \\
\midrule
\multirow{5}{*}{\rotatebox[origin=c]{90}{\textbf{\texttt{XL}}}}
& \hfmodel{Qwen/WorldPM-72B-RLHFLow}     &              & \texttt{72B}  & \mrbhl{90.88} & \mrbhl{67.92} & \mrbhl{55.21} \\
& \hfmodel{nvidia/Llama-3.3-Nemotron-70B-Reward}          &  & \texttt{70B}  & \mrbhl{93.88} & \mrbhl{70.49} & \mrbhl{73.47} \\
& \hfmodel{Nexusflow/Athene-RM-70B}     &              & \texttt{70B}  & \mrbhl{91.22} & \mrbhl{68.76} & \mrbhl{63.45} \\
& \textbf{\texttt{Themis-RM-32B}}          & \iconAuxiliary\;\iconCode\;\iconPrinciple & \texttt{32B} & \mrbhl{94.89} & \mrbhl{72.34} & \mrbhl{71.65} \\
& \hfmodel{TIGER-Lab/AceCodeRM-32B}         & \iconCode   & \texttt{32B}  & \mrbhl{23.58} & \mrbhl{67.98} & \mrbhl{66.77} \\
\midrule
\multirow{1}{*}{\rotatebox[origin=c]{90}{\textbf{\texttt{L}}}}
& \textbf{\texttt{Themis-RM-14B}}          & \iconAuxiliary\;\iconCode\;\iconPrinciple & \texttt{14B} & \mrbhl{94.11} & \mrbhl{71.44} & \mrbhl{70.85} \\
\midrule
\multirow{5}{*}{\rotatebox[origin=c]{90}{\textbf{\texttt{M}}}}
& \textbf{\texttt{Themis-RM-8B}}          & \iconAuxiliary\;\iconCode\;\iconPrinciple & \texttt{8B} & \mrbhl{93.69} & \mrbhl{65.87} & \mrbhl{69.97} \\
& \hfmodel{LARK-Lab/CodeScaler-8B}     & \iconCode & \texttt{8B}  & \mrbhl{94.66} & \mrbhl{76.51} & \mrbhl{70.05} \\
& \hfmodel{Skywork/Skywork-Reward-V2-Qwen3-8B}     &              & \texttt{8B}  & \mrbhl{94.76} & \mrbhl{76.93} & \mrbhl{67.90} \\
& \hfmodel{Nexusflow/Athene-RM-8B}     &  & \texttt{8B}  & \mrbhl{87.48} & \mrbhl{62.96} & \mrbhl{61.12} \\
& \hfmodel{TIGER-Lab/AceCodeRM-7B}     & \iconCode  & \texttt{7B}  & \mrbhl{22.74} & \mrbhl{63.16} & \mrbhl{61.09} \\
\midrule
\multirow{3}{*}{\rotatebox[origin=c]{90}{\textbf{\texttt{S}}}}
& \textbf{\texttt{Themis-RM-4B}}          & \iconAuxiliary\;\iconCode\;\iconPrinciple & \texttt{4B} & \mrbhl{92.46} & \mrbhl{63.81} & \mrbhl{68.02} \\
& \hfmodel{LARK-Lab/CodeScaler-4B}     & \iconCode & \texttt{4B}  & \mrbhl{94.32} & \mrbhl{75.13} & \mrbhl{68.44} \\
& \hfmodel{Skywork/Skywork-Reward-V2-Qwen3-4B}     &  & \texttt{4B}  & \mrbhl{94.06} & \mrbhl{74.26} & \mrbhl{65.43} \\
\midrule
\multirow{3}{*}{\rotatebox[origin=c]{90}{\textbf{\texttt{XS}}}}
& \textbf{\texttt{Themis-RM-1.7B}}  & \iconAuxiliary\;\iconCode\;\iconPrinciple & \texttt{1.7B} & \mrbhl{89.17} & \mrbhl{56.22} & \mrbhl{63.29} \\
& \hfmodel{LARK-Lab/CodeScaler-1.7B}  & \iconCode & \texttt{1.7B} & \mrbhl{91.13} & \mrbhl{68.44} & \mrbhl{66.17} \\
& \hfmodel{Skywork/Skywork-Reward-V2-Qwen3-1.7B}  &  & \texttt{1.7B}   & \mrbhl{91.64} & \mrbhl{67.71} & \mrbhl{66.48} \\
\midrule
\multirow{2}{*}{\rotatebox[origin=c]{90}{\textbf{\texttt{XXS}}}}
& \textbf{\texttt{Themis-RM-0.6B}}   & \iconAuxiliary\;\iconCode\;\iconPrinciple            & \texttt{0.6B} & \mrbhl{83.41} & \mrbhl{49.61} & \mrbhl{63.84} \\
&  \hfmodel{Skywork/Skywork-Reward-V2-Qwen3-0.6B}   &  & \texttt{0.6B} & \mrbhl{86.32} & \mrbhl{60.83} & \mrbhl{63.65} \\
\bottomrule
\end{tabular}%
}
\scalebox{0.75}{\footnotesize
\iconAuxiliary\;\textbf{\texttt{Auxiliary Training Objectives}} \quad \iconCode\;\textbf{\texttt{Code RM}} \quad \iconPrinciple\;\textbf{\texttt{Criteria-Following RM}} \quad \iconMath\;\textbf{\texttt{Math RM}} \quad \iconGenerative\;\textbf{\texttt{Generative RM}} \quad \iconReasoning\;\textbf{\texttt{Reasoning RM}}}
\end{table}

\deanon{%
\section{\texttt{Project Themis} Artifacts}
\label{appdx:5_Released_Artifacts}

\def\arraystretch{1.12}
\begin{table}[ht]
\centering
\caption{Complete list of publicly released artifacts accompanying \texttt{Project Themis}.}
\label{tab:Artifacts}
\scalebox{0.6}{%
\begin{tabular}{@{}c l l p{8cm}@{}}
\toprule
& \texttt{\textbf{Artifact}} & \texttt{\textbf{Type}} & \texttt{\textbf{Description}} \\
\midrule
\multirow{1}{*}{\rotatebox[origin=c]{90}{\textbf{\texttt{Eval}}}}
& \hfmodel{project-themis/Themis-CodeRewardBench}         & \texttt{Benchmark}  & 8,866 preference pairs spanning 5 quality dimensions, 8 languages, and 19 source subsets. \\
\midrule
& \hfmodel{project-themis/Themis-CodePreference}          & \texttt{Dataset}    & 354k code preference pairs across 5 quality criteria and 8 programming languages. Training data for the PM stage. \\
\multirow{-2}{*}{\rotatebox[origin=c]{90}{\textbf{\texttt{Datasets}}}}
& \hfmodel{project-themis/Themis-GeneralPreference}       & \texttt{Dataset}    & 110k+ general-domain and code retrieval preferences. Training data for the PT stage. \\
& \hfmodel{project-themis/git-commits-merged}             & \texttt{Dataset}    & ${\sim}$3M single-file commits from merged PRs in permissively licensed repos, cross-referenced with GHTorrent. 24 languages. \\
& \hfmodel{project-themis/git-commits}                    & \texttt{Dataset}    & ${\sim}$28M raw single-file commits mined from permissively licensed repos via BigQuery. Full unfiltered pool. \\
\midrule
& \hfmodel{project-themis/Themis-RM-32B}                  & \texttt{Model}      & \multirow{6}{8cm}{Reward models ranging from 0.6B to 32B parameters. Preference modeling (PM) stage checkpoints.} \\
& \hfmodel{project-themis/Themis-RM-14B}                  & \texttt{Model}      & \\
& \hfmodel{project-themis/Themis-RM-8B}                   & \texttt{Model}      & \\
\multirow{-4}{*}{\rotatebox[origin=c]{90}{\textbf{\texttt{PM Models}}}}
& \hfmodel{project-themis/Themis-RM-4B}                   & \texttt{Model}      & \\
& \hfmodel{project-themis/Themis-RM-1.7B}                 & \texttt{Model}      & \\
& \hfmodel{project-themis/Themis-RM-0.6B}                 & \texttt{Model}      & \\
\midrule
& \hfmodel{project-themis/Themis-RM-32B-PMP}              & \texttt{Model}      & \multirow{6}{8cm}{Reward models ranging from 0.6B to 32B parameters. Preference model pre-training (PT) stage checkpoints.} \\
& \hfmodel{project-themis/Themis-RM-14B-PMP}              & \texttt{Model}      & \\
& \hfmodel{project-themis/Themis-RM-8B-PMP}               & \texttt{Model}      & \\
\multirow{-4}{*}{\rotatebox[origin=c]{90}{\textbf{\texttt{PMP Models}}}}
& \hfmodel{project-themis/Themis-RM-4B-PMP}               & \texttt{Model}      & \\
& \hfmodel{project-themis/Themis-RM-1.7B-PMP}             & \texttt{Model}      & \\
& \hfmodel{project-themis/Themis-RM-0.6B-PMP}             & \texttt{Model}      & \\
\midrule
\multirow{-1}{*}{\rotatebox[origin=c]{90}{\textbf{\texttt{Infra}}}}
& \gitrepo{iNeil77/Themis}                                & \texttt{Code}       & Source code: dataset construction pipeline, distributed training scripts, and evaluation suite. \\
& \dockerrepo{ineil77/themis}                              & \texttt{Container}  & Training container image: PyTorch, CUDA, NCCL, high-speed networking, and Liger kernels. \\
\bottomrule
\end{tabular}%
}
\end{table}

}

\end{document}